\documentclass[prd,amsmath,amssymb,nofootinbib]{revtex4_modifiedtoc}%revtex4-1}

\pdfoutput=1

\usepackage[pdftex
,bookmarks=true
,bookmarksopen=true
,bookmarksnumbered=true
,nesting=false
,colorlinks=false
]{hyperref}

\usepackage[sort&compress]{natbib}
\usepackage{enumerate}
\usepackage{ifthen}
\usepackage[normalem]{ulem}

\usepackage{float}
\floatstyle{plaintop}
\restylefloat{table}
\usepackage[a4paper,centering,vcentering,marginratio=1:1,hscale=0.75]{geometry}
\usepackage[section]{placeins}
\usepackage{graphicx}
\usepackage{amssymb}
\usepackage{amsfonts, color, float,  amsmath}
\usepackage{subfigure} 
\usepackage{url}
\usepackage{array}
\usepackage{upgreek}
\usepackage{slashed}
\usepackage[svgnames]{xcolor} %colours with names
\usepackage{verbatim}
\usepackage[small,bf,sf]{caption}  % better captions, style of the captions. Compatible with sidecap.
\usepackage{microtype}
\usepackage{textcomp} 
\usepackage{latexsym} 
\usepackage[english]{babel} 
\usepackage{booktabs}         %Fancy tables: use toprule, midrule, and bottomrule.
\newcommand{\beq}{\begin{equation}}
\newcommand{\eeq}{\end{equation}}
\newcommand{\bea}{\begin{eqnarray}}
\newcommand{\eea}{\end{eqnarray}}

\newcommand{\gsim}{\lower.7ex\hbox{$
\;\stackrel{\textstyle>}{\sim}\;$}}
\newcommand{\lsim}{\lower.7ex\hbox{$
\;\stackrel{\textstyle<}{\sim}\;$}}
\renewcommand{\Im}{{\rm Im}\,}
\newcommand{\ket}[1]{\left|#1\right\rangle}

\newcommand{\sand}[3]{\langle#1|#2|#3 \rangle}

\renewcommand{\Im}{{\rm Im}}

\def\beq {\begin{equation}}
\def\eeq {\end{equation}}
\def\bea {\begin{eqnarray}}
\def\eea {\end{eqnarray}}
\def\G{\Gamma}
\def\ni {\noindent}
\def\nn {\nonumber}

\def\A {\mathcal{A}}

\def\DKpp{D^+ \to K^- \pi^+ \pi^+}

\def \babar{B{\sc a}B{\sc ar}}
\def\beq{\begin{equation}}
\def\eeq{\end{equation}}
\def\bea{\begin{eqnarray}}
\def\eea{\end{eqnarray}}
\def\nn{\nonumber}
\def\roughly#1{\mathrel{\raise.3ex\hbox
{$#1$\kern-.75em\lower1ex\hbox{$\sim$}}}}
\def\lsim{\roughly<}
\def\gsim{\roughly>}

\def\kbar{{\bar K}^0}
\def\bd{B^0}

\def\btod{{\bar b} \to {\bar d}}
\def\btos{{\bar b} \to {\bar s}}

\def\btokpipi{B \to K \pi \pi}
\def\btokkk{B \to KK{\bar K}}
\def\bs{B_s^0}

\newcommand{\tmat}[2]{\mathcal{T}_{#1#2}^{(L_{#1},L_{#2})}}
\newcommand{\rse}{\mathcal{R}}
\newcommand{\bes}[1]{j^{#1}_{L_{#1}}}
\newcommand{\han}[1]{h^{(1)#1}_{L_{#1}}}
\newcommand{\One}{1\hspace{-1.6mm}1}

 \newcommand{\ba}{\begin{eqnarray}} 
 \newcommand{\ea}{\end{eqnarray}}

\def\sp {\!+\!}
\def\sm {\!-\!}
\def\cK {{\cal{K}}}
\def\Ob {\bar{\Omega}}
\def\S{\Sigma}
\def\lb {\left[ }
\def\rb {\right] }

\def\pb {\bar{p}}
\def\r{\rho}
\def\Kb {\bar{K}}
\def\T {\Theta}
\def\bea {\begin{eqnarray}}
\def\eea {\end{eqnarray}}

\def\m {\mu}
\def\nn {\nonumber}
\def\lp {\left( }
\def\rp {\right) }
\def\lc {\left\{ }
\def\rc {\right\} }

\def\a {\alpha }

\def\d {\delta}
\def\D {\Delta}
\def\e {\epsilon}

\def\g {\gamma}
\def\G {\Gamma}
\def\l {\lambda }

\def\m {\mu}
\def\n {\nu}

\def\p {\pi}

\def\r {\rho}
\def\s {\sigma}
\def\S {\Sigma}

\def\cM {{\cal{M}}}
\newcommand{\rpp}{\rho^0 \to \pi^+\pi^-}

\newcommand{\bppp}{B^+ \to  \pi^-\pi^+\pi^+}
\def\p {\pi}
\def\ni {\noindent}
\def\rar {\rightarrow}

\newcommand{\jpsi}{\ensuremath{J/\psi}}
\newcommand{\Dbar}{\ensuremath{\overline{D}}}
\newcommand{\mev}{\ensuremath{\mathrm{MeV}}}
\begin{document}

\newcommand{\final}{0} %% <<< change this to 1 to remove comments and integrate changes fully into the text

\ifthenelse{\final=1}{
\newcommand{\changed}[1]{#1}
\newcommand{\change}[2]{#1}
\newcommand{\remark}[1]{}
}{
\newcommand{\changed}[1]{{\color{red}#1}}
\newcommand{\change}[2]{{\color{red}\sout{#1}}{\color{blue}#2}}
\newcommand{\remark}[1]{{\color{green}\{Comment: #1\}}}
}

\pagestyle{plain} % restore page numbers for the main text
\setcounter{page}{1}
\pagenumbering{arabic}

\newcommand{\intoc}{1}
\newcommand{\quantoc}{1}

\newcommand{\aSection}[2]{%
\section{#1 {\protect \ifx \protect \intoc \protect \quantoc {by #2} \protect \fi}}%
%\section{{#2: #1}}%
%\section{{\ifthenelse{\equal{\intoc}{1}}{#2}{}: #1}}%
}

\newcounter{abc}
\setcounter{abc}{0}

\def\afli{a}
\newcommand{\defineAffiliation}[2]{%
\stepcounter{abc}
\newcounter{#1count}
\setcounter{#1count}{\value{abc}}
\expandafter\def\csname f#1\endcsname{\alph{#1count}}
\expandafter\def\csname #1sym\endcsname{\ensuremath{^{\csname f#1\endcsname}}}
\expandafter\def\csname #1txt\endcsname{#2}%
}

\defineAffiliation{ITA}{Instituto Tecnol\'ogico de Aeron\'autica (Brazil)}
\defineAffiliation{UFRJ}{Univ. Federal do Rio de Janeiro (Brazil)}
\defineAffiliation{TUM}{Technische Universit\"at M\"unchen (Germany)}
\defineAffiliation{PUC}{Pontificia Universidade Catolica (Brazil)}
\defineAffiliation{CBPF}{Centro Brasileiro de Pesquisas F\'isicas (Brazil)}
\defineAffiliation{CNRS}{Centre National de la Recherche Scientifique (France)}
\defineAffiliation{Coimbra}{Universidade de Coimbra (Portugal)}
\defineAffiliation{Mont}{Universit\'e de Montreal (Canada)}
\defineAffiliation{ND}{University of Notre Dame (USA)}
\defineAffiliation{SaoP}{Universidade de S\~ao Paulo (Brazil)}
\defineAffiliation{Man}{University of Manchester (UK)}
\defineAffiliation{Warw}{University of Warwick (UK)}
\defineAffiliation{BRS}{University of Bristol (UK)}
\defineAffiliation{LPNHE}{LPNHE (UPMC, Paris, France)}
\defineAffiliation{UFSC}{Universidade Federal de S\~ao Carlos (Brazil)}
\defineAffiliation{Cin}{University of Cincinnati (USA)}
\defineAffiliation{Col}{Universidad Nacional de Colombia (Colombia)}
\defineAffiliation{Val}{Universitat de Val\`encia (Spain)}
\defineAffiliation{UERJ}{Universidade do Estado do Rio de Janeiro (Brazil)}
\defineAffiliation{Madrid}{Universidad Complutense de Madrid (Spain)}
\defineAffiliation{LAL}{Laboratoire de l'Accelerateur Lineaire (France)}
\defineAffiliation{ISTUL}{Universidade de Lisboa (Portugal)}
\defineAffiliation{Zuer}{Universit\"at Z\"urich (Switzerland)}
\defineAffiliation{Syr}{Syracuse University (USA)}
\defineAffiliation{IU}{Indiana University, Bloomington (USA)}

\newcommand{\authorlist}{%
Jorge 	H. Alvarenga Nogueira\ITAsym,
Sandra Amato\UFRJsym,
Alexander Austregesilo\TUMsym,
Clarissa Baesso\PUCsym,
Ignacio Bediaga Hickman\CBPFsym,
Eli Ben Haim\CNRSsym,
Eef van~Beveren\Coimbrasym,
Bhubanjyoti Bhattacharya\Montsym,
Ikaros Bigi\NDsym,
Diogo Boito\SaoPsym,
Jolanta Brodzicka\Mansym,
Marcelo Campos\PUCsym,
Ana B\`arbara R. Cavalcante\CBPFsym,
Alberto Correa dos~Reis\CBPFsym,
Daniel Charles Craik\Warwsym,
Melissa Maria Cruz Torres\PUCsym,
Jeremy Dalseno\BRSsym,
Daniel Evangelho Vieira\UFRJsym,
Fernando Luiz Ferreira Rodrigues\CBPFsym,
Tobias Frederico\ITAsym,
Timothy Gershon\Warwsym,
Carla G\"obel\PUCsym,
Daniel Greenwald\TUMsym,
Samuel Thomas Harnew\BRSsym,
Louis Henry\CNRSsym,
Adlene 	Hicheur\UFRJsym,
Thomas Edward Latham\Warwsym,
Benoit Loiseau\LPNHEsym,
David London\Montsym,
Helder Lopes\UFRJsym,
Odilon Louren\c{c}o\UFSCsym,
Patricia Magalh\~{a}es\CBPFsym,
Jussara Marques de~Miranda\CBPFsym,
Danielle Martins Tostes\UFRJsym,
Andre Massafferri Rodrigues\CBPFsym,
Abhijit Mathad\Warwsym,
Brian Meadows\Cinsym,
Diego Milanes Carreno\Colsym,
Josue Danilo Molina Rodriguez\PUCsym,
Danielle Moraes\ITAsym,
Irina Nasteva\UFRJsym,
Fernando Navarra\SaoPsym,
Marina 	Nielsen\SaoPsym,
Daniel Patrick 	O'Hanlon\Warwsym,
Eulogio Oset\Valsym,
Bruno 	Osorio Rodrigues\UERJsym,
Juan Martin Otalora Goicochea\UFRJsym,
Stephan Paul\TUMsym,
Jos\'e R. Pelaez\Madridsym,
Erica 	Polycarpo\UFRJsym,
Claire 	Prouv\'e\BRSsym,
Renato 	Quagliani\BRSsym{}\LALsym,
Jonas 	Rademacker\BRSsym,
Manoel 	Robilotta\SaoPsym,
Jairo Alexis Rodriguez\Colsym,
George Rupp\ISTULsym,
Rafael Silva Coutinho\Zuersym,
Tomasz Skwarnicki\Syrsym,
Liang Sun\Cinsym,
Adam Szczepaniak\IUsym,
Rafael Tourinho  Aoude\PUCsym,
Charlotte Wallace\Warwsym,
Mark Whitehead\Warwsym
\vspace{1ex}\\
\textbf{Editors:} Jonas Rademacker\BRSsym\ and  Alberto C. dos Reis\CBPFsym
\vspace{1ex}\\
}

\newcommand{\afl}{%
\count1=0
\loop
ump  % \the\toks\count0\par
\ifnum\count1<5 
  \advance \count1 by 1
  \repeat
}

%\message{DDDDDDDDDDDD \meaning\affiliationlist}

\newcommand{\affiliationlist}{%
\ITAsym\ITAtxt,
\UFRJsym\UFRJtxt,
\TUMsym\TUMtxt,
\PUCsym\PUCtxt,
\CBPFsym\CBPFtxt,
\CNRSsym\CNRStxt,
\Coimbrasym\Coimbratxt,
\Montsym\Monttxt,
\NDsym\NDtxt,
\SaoPsym\SaoPtxt,
\Mansym\Mantxt,
\Warwsym\Warwtxt,
\BRSsym\BRStxt,
\LPNHEsym\LPNHEtxt,
\UFSCsym\UFSCtxt,
\Cinsym\Cintxt,
\Colsym\Coltxt,
\Valsym\Valtxt,
\UERJsym\UERJtxt,
\Madridsym\Madridtxt,
\LALsym\LALtxt,
\ISTULsym\ISTULtxt,
\Zuersym\Zuertxt,
\Syrsym\Syrtxt,
\IUsym\IUtxt
}

\title{Summary of the 2015 LHCb workshop on multi-body decays of $D$ and $B$ mesons} 

\author{%Jonas Rademacker$^a$ and Alberto C. dos Reis$^b$ (editors)
\authorlist
}
\affiliation{\affiliationlist}
%$^a$H. H. Wills Physics Laboratory, University of Bristol, \\
%Bristol, United Kingdom\\
%$^b$Centro Brasileiro de Pesquisas F\'isicas  --  CBPF\\
%Rio de Janeiro, Brazil

\begin{abstract}
%\afl
This document contains a summary of the LHCb workshop on multi-body decays of 
$D$ and $B$ mesons, held at CBPF, Rio de Janeiro, in July 2015. The workshop was
focused on issues related to amplitude analysis of three- and four-body hadronic
decays. In addition to selected LHCb results, contributions from guest theorists are
included.
\end{abstract}
\maketitle

\newpage
\renewcommand{\intoc}{1}
\tableofcontents
\renewcommand{\intoc}{0}
\newpage
\aSection{Introduction}{J. Rademacker, A. C. do Reis}
\begin{center}
Jonas Rademacker$^a$ and Alberto C. dos Reis$^b$ \\
\vskip .2cm
{\small $^a$H. H. Wills Physics Laboratory, University of Bristol, \\
Bristol, United Kingdom}
\vskip .2cm
{\small $^b$Centro Brasileiro de Pesquisas F\'isicas  --  CBPF\\
Rio de Janeiro, Brazil}
\end{center}

The study of multi-body hadronic decays of $D$ and $B$ mesons is very important
in many respects. The very large samples available after LHCb allow for precision
measurements, opening interesting possibilities in fields such as {\em CP} violation, 
mixing, heavy and light meson spectroscopy. 

The analysis of these data and the interpretation of the results, however, involve
rather intricate aspects, since multi-body hadronic decays involve a complex
interplay between weak and low-energy strong interactions. Traditional approaches
techniques and tools are no longer satisfactory. Improved models are required for 
extracting the most information from the new  data. With this motivation, a LHCb 
workshop on amplitude analysis of multi-body took place in July, 2015, with the 
participation of  invited theorists.

To trigger the discussions, a number of questions were proposed:

\begin{itemize}
\item Rescattering is an important ingredient which is not included in the usual decay models.
How do one describes it in $D$ and $B$ decays? What role could combined analyses in multiple 
decay channels play?

\item  Quarks and hadrons --- To what extent can one think of penguin diagrams when dealing with
exclusive channels? When is it not useful anymore to think 
in terms of quark-level diagrams? What should one do instead?
Ideally, one should explore the connection between the quark and hadron degrees of freedom 
to built sound parameterisations of nonresonant amplitudes. This is obviously a rather intricate 
problem. Is there any alternative approach?

\item  Symmetries --- To what extend is it useful to use approximate symmetries like 
isospin, U-spin, V-spin, SU(3)-flavour? How can we quantify the effects of the breaking 
of those symmetries? Fundamental principles, such as unitarity, analyticity and CPT conservation, 
are widely used in $2\to 2$ processes to constrain and to relate the scattering amplitudes. 
How could one implement such constraints in the context of a hadronic weak decay of a heavy  meson?

\item  Translatable results --- 
What are the limitations and opportunities for using $2\to 2$ scattering results for describing $1\to 3$ decays? What about $1\to 4$ decays? What are the limitations of invoking Watson theorem in this context?
Can the (model-independent) results for $\pi\pi$ and other line shapes obtained by COMPASS be directly applied to modeling the (weak) multi-body decays of $D$ and $B$ mesons?
To what degree are mass/width results (such as those from Breit Wigner-based analyses, but also others) translatable between different experiments?

\item  The statistical precision in charm Dalitz analyses is unprecedented. What to do with these extremely clean, huge data samples (some with tens of millions of events), if one does not 
have models that are anywhere near precise enough to describe them?

\item  Threshold data provide a way of measuring phases across Dalitz plots in a model-independent
fashion. As a matter of fact, this was used in model-independent analyses. Should one uses 
this unique information instead to build better models?
Why is the CP-odd content of $D^0\to \pi\pi\pi^0$ ~100\%? What does the CP-odd content of ~75\% in other investigated $D^0$ decay modes, such as $D^0\to 4\pi$, $D^0\to KK\pi^0$, tell us? 
Can this be used to test models?

\item  New resonances --- How can one determine whether an observed structure corresponds to
a new state or to some kind of threshold effect? Does an Argand plot with the phase motion 
(and its direction) tell us anything about that?

\end{itemize}

This document  contains a summary of the discussions held at the LHCb workshop,
with individual contributions from the guest theorists addressing the above and related 
questions. A non-exhaustive set of LHCb results on amplitude analysis is also presented.
\newpage

\aSection{LHCb - summary of $B \to Dhh'$ analyses}{M. Whitehead for LHCb}
\begin{center}
Mark Whitehead, for the LHCb collaboration \\
\vskip .2cm
{\small Department of Physics, University of Warwick, \\
Coventry, United Kingdom}
\end{center}
\subsection{Introduction}
Results from the following analyses were presented at the workshop $B^0_s \to \overline{D}{}^{0} K^- \pi^+$~\cite{lhcb3}, 
$B^+ \to D^- K^+ \pi^-$~\cite{lhcb4}, $B^0 \to \overline{D}{}^{0} \pi^+ \pi^-$~\cite{lhcb5}, 
$B^0 \to \overline{D}{}^{0} K^+ \pi^-$~\cite{lhcb6}, $B^0 \to D K^+ \pi^-$ and $\Lambda^0_b \to D^0 p \pi^-$.
Of these, the first four are published results whilst the last two analyses are currently active.
These decays modes can be used to study spectroscopy in the $DK$, $D\pi$, $Dp$, $K\pi$, $\pi\pi$ and $p\pi$ systems, 
and so have a wide and important role in the understanding of such resonant states.
Additionally, amplitude analyses of some $B \to Dhh'$ channels can be used to measure $CP$ violation. The decay 
$B^0 \to D K^+ \pi^-$, with $D \to K^+ K^-\, , \pi^+ \pi^-$, can be used to measure the CKM angle $\gamma$~\cite{tim1,tim2}  
and the CKM angle $\beta$ can be measured using $B^0 \to \overline{D}{}^{0} \pi^+ \pi^-$ decays~\cite{tom,beta}.

\subsection{Summary of results}
\begin{figure}[!htb]
\centering
\includegraphics[scale=0.35]{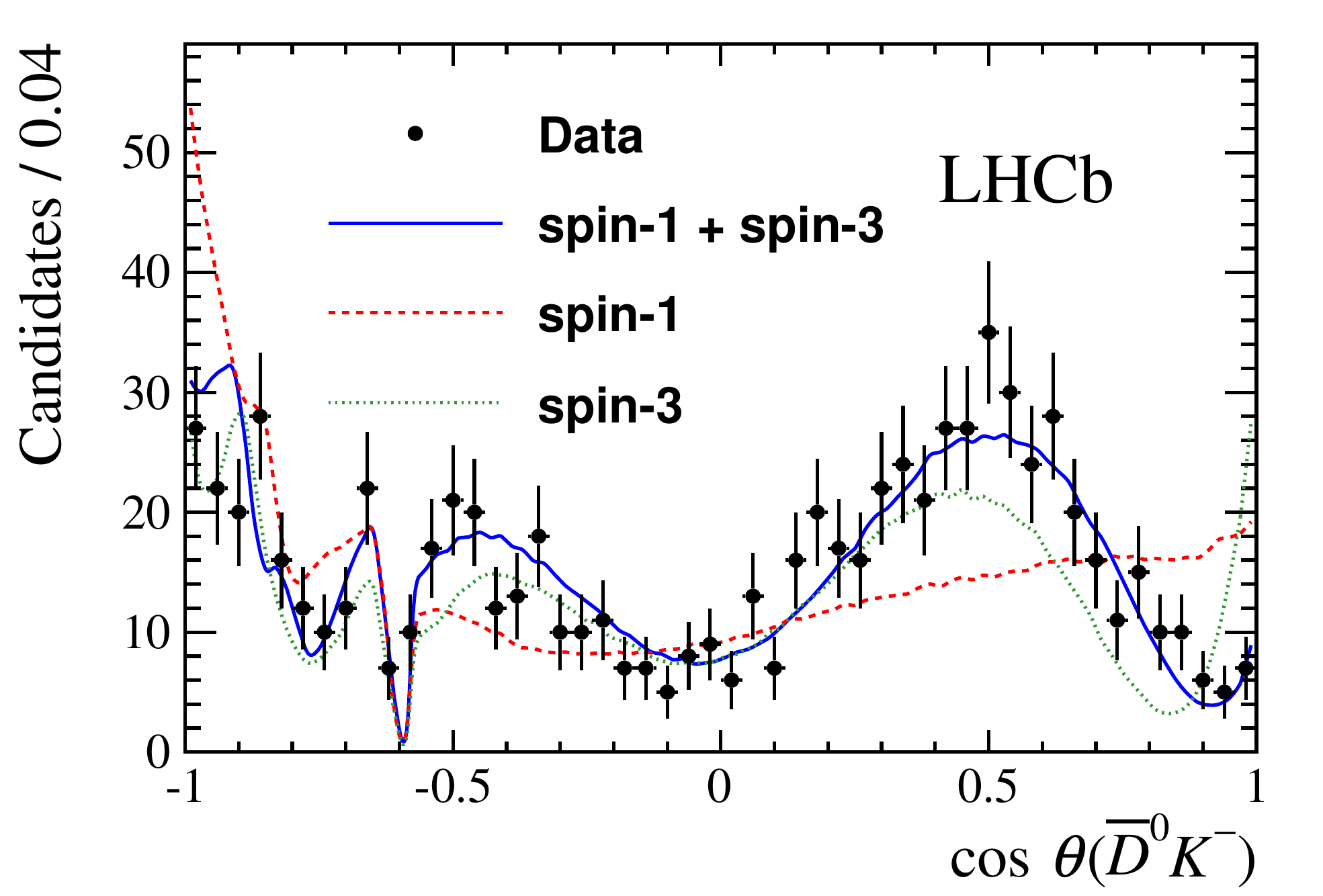}
\caption{\small Projections of data and fit models on to the cosine of the helicity angle of the $DK$ system for 
$B^0_s \to \overline{D}{}^{0} K^- \pi^+$ decays in the mass window $2770 < m(DK) < 2910$ MeV/$c^2$. The data 
are black points and the fit models are described in the legend.}
\label{fig:dsj} 
\end{figure}

The Dalitz plot analysis of $B^0_s \to \overline{D}{}^{0} K^- \pi^+$ decays found that a resonance previously seen 
in inclusive $DK$ events, the so called $D^*_{sJ}(2860)^+$, is actually comprised of two overlapping states of 
different spin. 
Figure~\ref{fig:dsj} shows the cosine of the helicity angle of the $DK$ system in the region $2770 < m(DK) < 2910$ MeV/$c^2$, 
with fits including spin 1, spin 3 and both spin 1 and spin 3 contributions overlaid.
The reported masses and widths of these states are 
\begin{eqnarray*}
m(D^*_{s1}(2860)^+) &=& 2859 \pm 12 \pm 6 \pm 23 \, {\rm MeV/c^2}, \\
\Gamma(D^*_{s1}(2860)^+) &=& 159 \pm 23 \pm 27 \pm 72 \, {\rm MeV/c^2}, \\
m(D^*_{s3}(2860)^+) &=& 2860.5 \pm 2.6 \pm 2.5 \pm 6.0 \, {\rm MeV/c^2}, \\
\Gamma(D^*_{s3}(2860)^+) &=& 53 \pm 7 \pm 4 \pm 6 \, {\rm MeV/c^2},
\end{eqnarray*}
where the uncertainties are statistical, experimental systematic and model systematic, respectively.

\begin{figure}[!htb]
\centering
\includegraphics[scale=0.35]{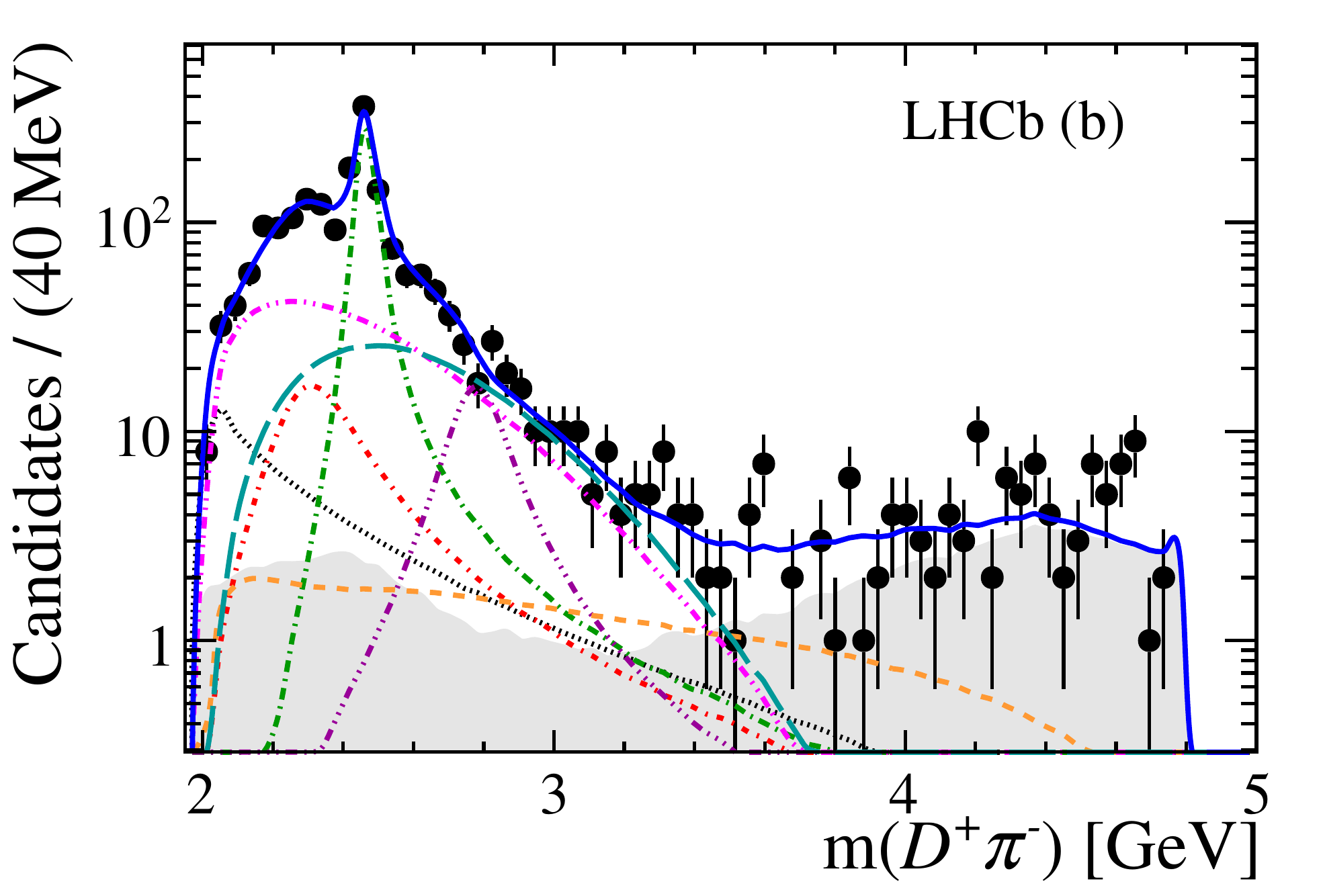}
\includegraphics[scale=0.35]{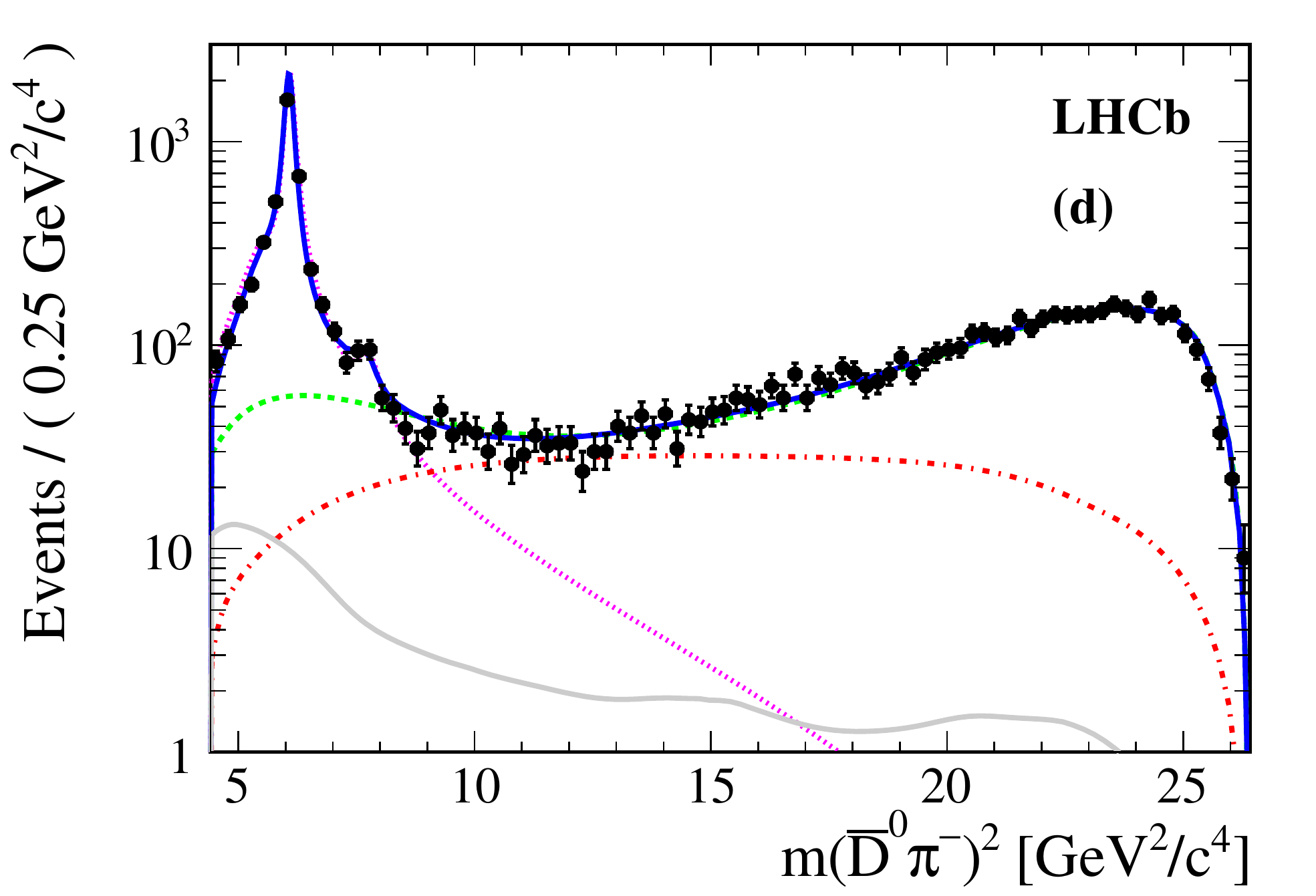}
\includegraphics[scale=0.35]{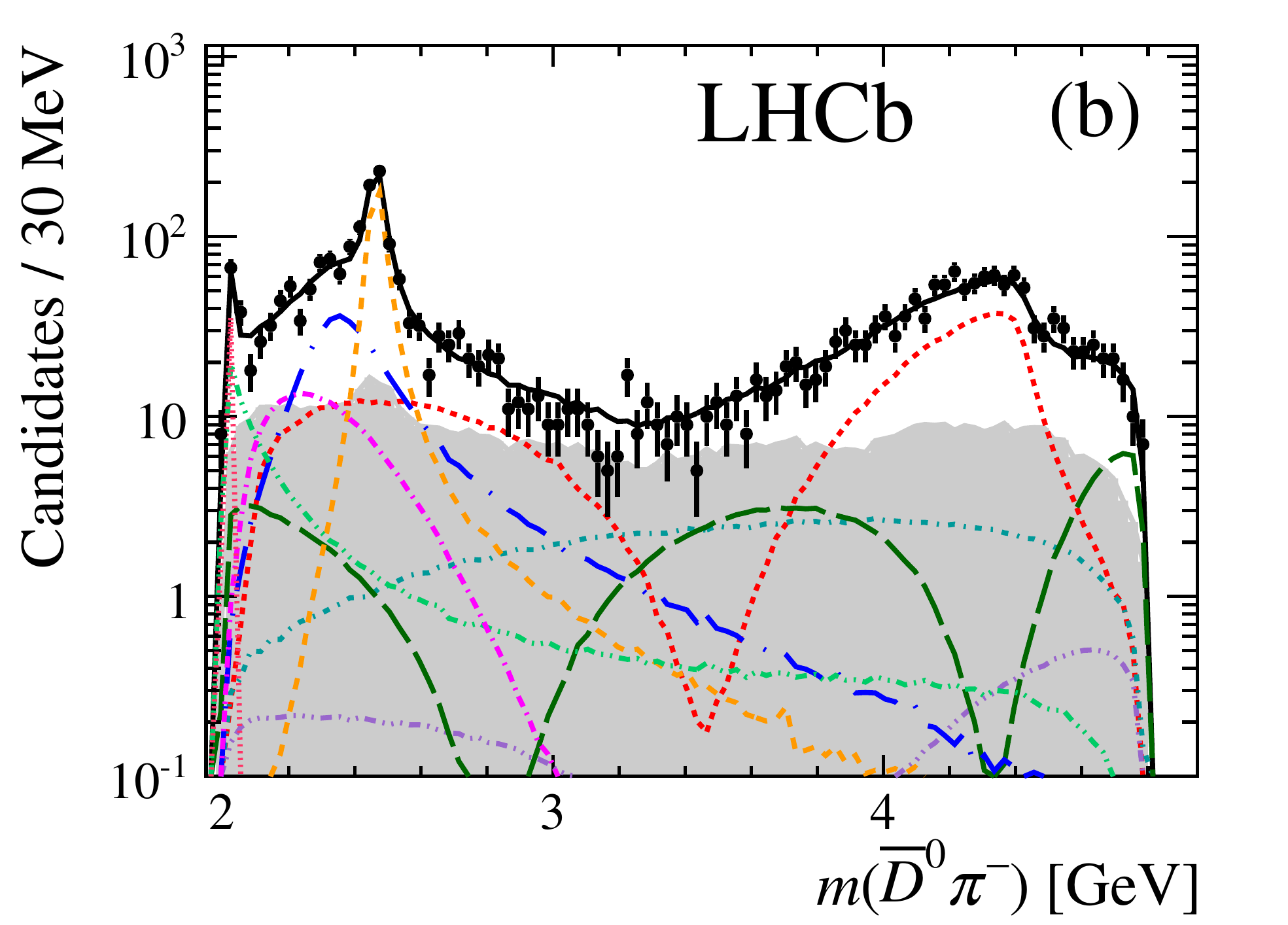}
\caption{\small Projections of the data and fit models on $m(D\pi)$ for the (left) $B^+ \to D^- K^+ \pi^-$, 
(right) $B^0 \to \overline{D}{}^{0} \pi^+ \pi^-$ and (bottom) $B^0 \to \overline{D}{}^{0} K^+ \pi^-$ samples.
Note that it is actually $m^2(D\pi)$ shown for $B^0 \to \overline{D}{}^{0} \pi^+ \pi^-$ decays.}
\label{fig:dpi}
\end{figure}

Analyses of $B^+ \to D^- K^+ \pi^-$, $B^0 \to \overline{D}{}^{0} \pi^+ \pi^-$ and $B^0 \to \overline{D}{}^{0} K^+ \pi^-$
decays showed the equivalent results for $D\pi$ resonances. The previously seen $D^*_J(2760)$ was found to be spin 1 
using $B^+ \to D^- K^+ \pi^-$ decays while a spin 3 state was observed in the $B^0 \to \overline{D}{}^{0} \pi^+ \pi^-$ 
final state. Note that the $B^+$ analysis sees neutral charm mesons while the $B^0$ final state is sensitive to charged 
charm states. For $B^0 \to \overline{D}{}^{0} K^+ \pi^-$ decays no evidence for a $D^*_J(2760)$ state 
was found. 
Figure~\ref{fig:dpi} shows the distribution of $m(D\pi)$ for $B^+ \to D^- K^+ \pi^-$, $B^0 \to \overline{D}{}^{0} \pi^+ \pi^-$
 and $B^0 \to \overline{D}{}^{0} K^+ \pi^-$ final states together with the amplitude fit projections.
The masses and widths of the $D^*_J(2760)$ states are
\begin{eqnarray*}
m(D^*_{1}(2760)^0) &=& 2781 \pm 18 \pm 11 \pm 6 \, {\rm MeV/c^2}, \\
\Gamma(D^*_{1}(2760)^0) &=& 177 \pm 32 \pm 20 \pm 7 \, {\rm MeV/c^2}, \\
m(D^*_{3}(2760)^+) &=& 2798 \pm 7 \pm 1 \pm 7 \, {\rm MeV/c^2}, \\
\Gamma(D^*_{3}(2760)^+) &=& 105 \pm 18 \pm 6 \pm 23 \, {\rm MeV/c^2},
\end{eqnarray*}
where the uncertainties are as before. More data will be required to search for the $D^*_{1}(2760)^+$ and 
$D^*_{3}(2760)^0$ states, neither of which were ruled out in the analyses described above.

\subsection{Questions and challenges}
The amplitude analysis $\Lambda^0_b \to D^0 p \pi^-$ decays is currently in progress. There was some discussion about the 
use of model independent partial waves to model the large non-resonant contributions to $m(Dp)$. It was felt that while this 
approach would work, and could potentially describe the amplitude well, the results would not be portable to other analyses.
This portability would be desirable when performing an amplitude analysis of the suppressed $\Lambda^0_b \to D^0 p K^-$ decay, which is sensitive to the CKM angle $\gamma$.

This is also an approach being considered for the $D\pi$ S-wave in high statistics modes such as $B^+ \to D^- \pi^+ \pi^+$. 
The $D\pi$ S-wave has been described by exponential form factors in $B^+ \to D^- K^+ \pi^-$ and 
$B^0 \to \overline{D}{}^{0} \pi^+ \pi^-$ analyses, while the $B^0 \to \overline{D}{}^{0} K^+ \pi^-$ analysis used the 
dabba model~\cite{dabba}. In all cases an additional component was included for the $D^*_0(2400)$ resonance, described by a relativistic 
Breit-Wigner function. 
The combination of these functions violates unitarity, this problem could be solved with either model independent partial waves, 
a LASS-like approach, a K-matrix approach or the use of Veneziano-like models~\cite{ven} as suggested by Adam Szczepaniak.

The $\pi\pi$ S-wave in $B^0 \to \overline{D}{}^{0} \pi^+ \pi^-$ was modelled using two separate methods, the isobar model 
and the K-matrix approach. While both of these gave good quality fits, and the results agree well between the two, it was 
reiterated by the theory colleagues that there are well-known problems using the isobar model the $\pi\pi$ S-wave, in 
particular for the $\sigma$ meson. It was also mentioned that it would be interesting to try a K-matrix approach to model 
the $K\pi$ S-wave.

\newpage

\aSection{LHCb - {\em CP} violation in  charmless  $B^{\pm}$ three-body decays}{A. C. dos Reis, for LHCb}
\begin{center}
Alberto C. dos Reis, for the LHCb collaboration \\
\vskip .2cm
{\small Centro Brasileiro de Pesquisas F\'isicas  --  CBPF\\
Rio de Janeiro - Brazil}
\end{center}

\subsection{Introduction}

The study of {\em CP} violation ({\em CP}V) is one the main topics in contemporary 
particle physics. In the Standard Model (SM) {\em CP}V arises due to an irreducible 
phase in the Cabibbo-Kobayashi-Maskawa (CKM) matrix. All existing measurements 
of {\em CP} violation in decays of  flavoured mesons are consistent with the SM 
predictions. However,  the CKM mechanism for  {\em CP}V fails to account for the
baryon number density of the Universe by many orders of magnitude. Other sources 
of {\em CP}V must 
exist. The study of {\em CP}V, therefore, offers unique opportunities to search for new Physics,
being one of the main purposes of the LHCb experiment.

The study of {\em CP}V in charmless $B$ decays is an important part of the LHCb physics 
programme. In this section, decays of $B^{\pm}$ mesons into three light pseudo-scalars 
are discussed. Four decay channels are analyzed~\cite{expnew}:  $B^+\to \pi^+\pi^+\pi^-$, 
$B^+\to K^+\pi^+\pi^-$, $B^+\to K^+\pi^+K^-$ and $B^+\to K^+K^+K^-$.
These  processes are a primary source for  studies of {\em CP}V in decay, since mixing is 
not possible for charged $B$ mesons.

{\em CP}V in decay occurs when a given final state may be reached by two interfering 
amplitudes with different weak and strong phases:

\[
A = a_1 + a_2\ e^{i(\delta+\gamma)},\hskip .4cm \overline{A} = a_1 + a_2\ e^{i(\delta-\gamma)}\ ,
\]
where $\gamma$ and $\delta$ are the weak ({\em CP} odd) and strong phases ({\em CP} even) and
$a_1$ and $a_2$ are real numbers.
Given that $\Gamma(\bar B\to \bar f)\propto|A|^2$ and $\Gamma(\bar B\to \bar f)
\propto |\overline{A}|^2$, the $\mathcal{A}_{CP}^{\mathrm{dir}}$ observable is

\[
\mathcal{A}_{CP}^{\mathrm{dir}} \equiv \frac{\Gamma(B\to f)-
\Gamma(\bar B\to \bar f)}{\Gamma(B\to f)+\Gamma(\bar B\to \bar f)}
= \frac{2a_1a_2\sin\!\gamma \sin\!\delta}
{a_1^2+a_2^2+2a_1a_2\cos\!\gamma \cos\!\delta}.
\]
\vskip .2cm

The weak phase difference arises from the amplitudes for
the $b\to u\bar u s$ and $b\to s \bar q q$ transitions, in the case of final states 
with net strangeness; and $b\to u\bar u d$ and $b\to d \bar q q$, in the case of final 
states with zero strangeness. The advantage of charmless three-body decays results
from the fact that most final states have a rich resonant structure. 
The interference between resonances plus the final state interactions (FSI) at the
hadron level provide the required strong phase differences. These phase differences 
vary across the phase space and can be very large, giving rise to large {\em CP} 
asymmetries in specific regions of the Dalitz plot~\cite{babar,miranda1,miranda2}.

\subsection{The Dalitz plots}

\begin{figure}[h]
\begin{center}
\includegraphics[width=6.5cm]{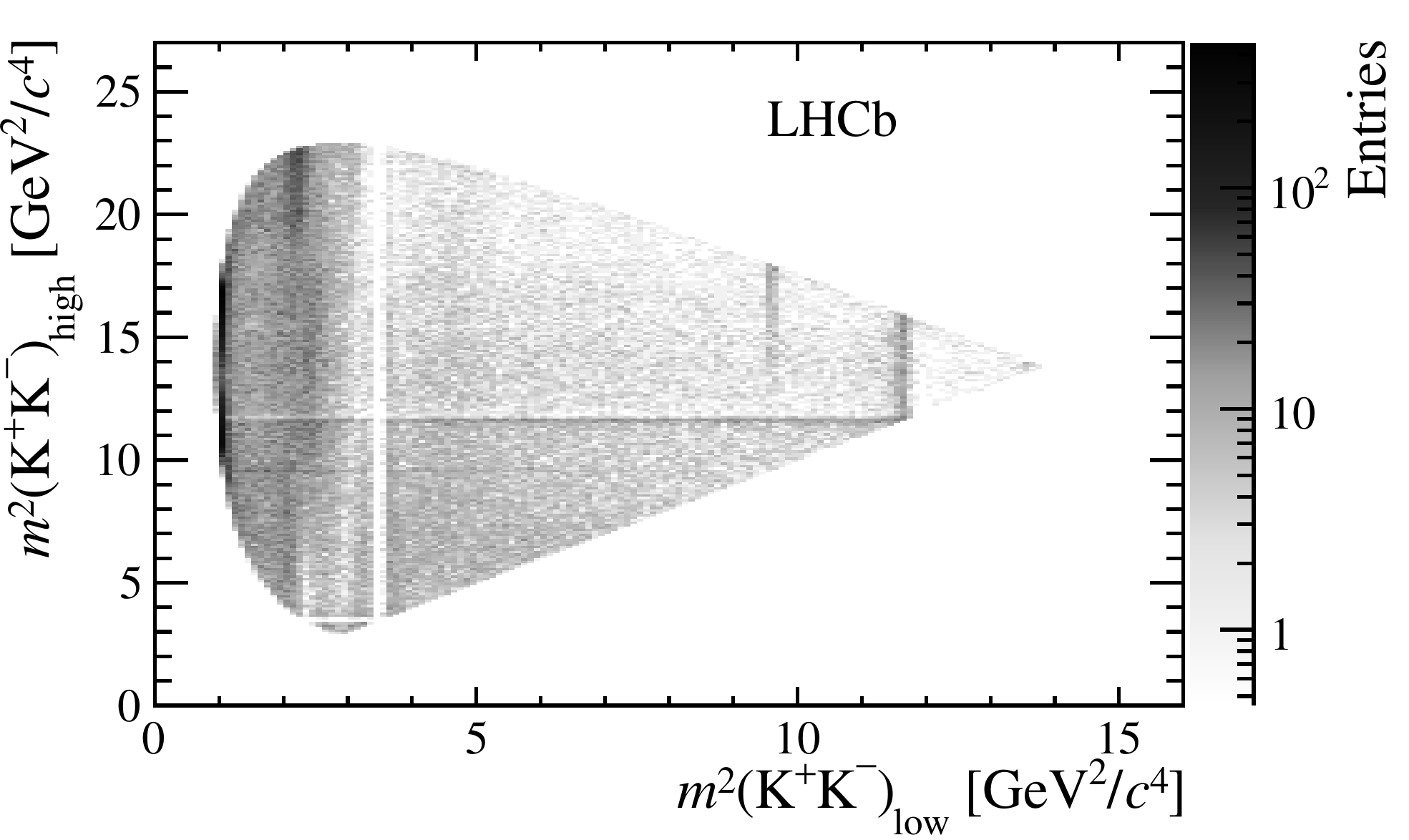}\includegraphics[width=6.5cm]{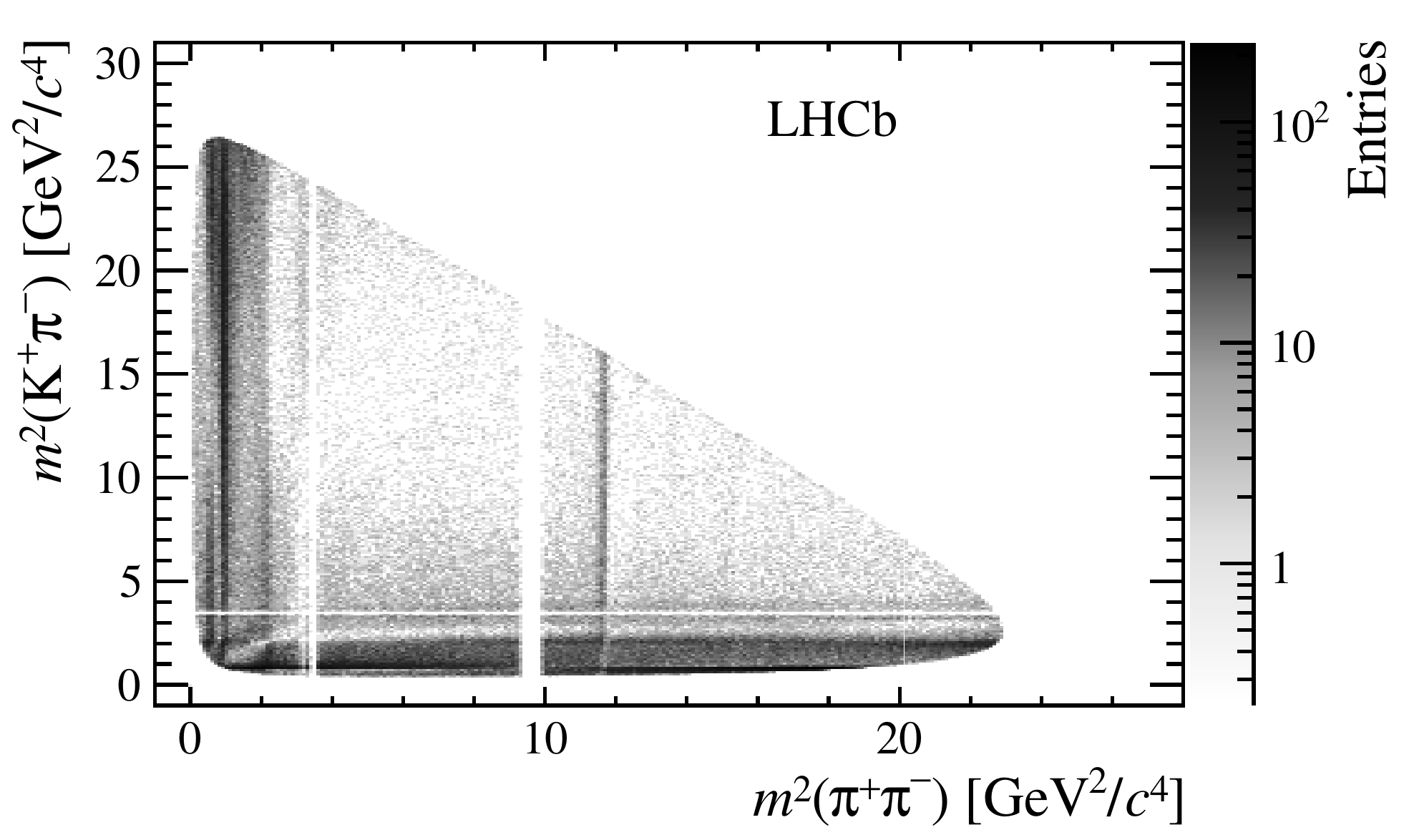}

\includegraphics[width=6.5cm]{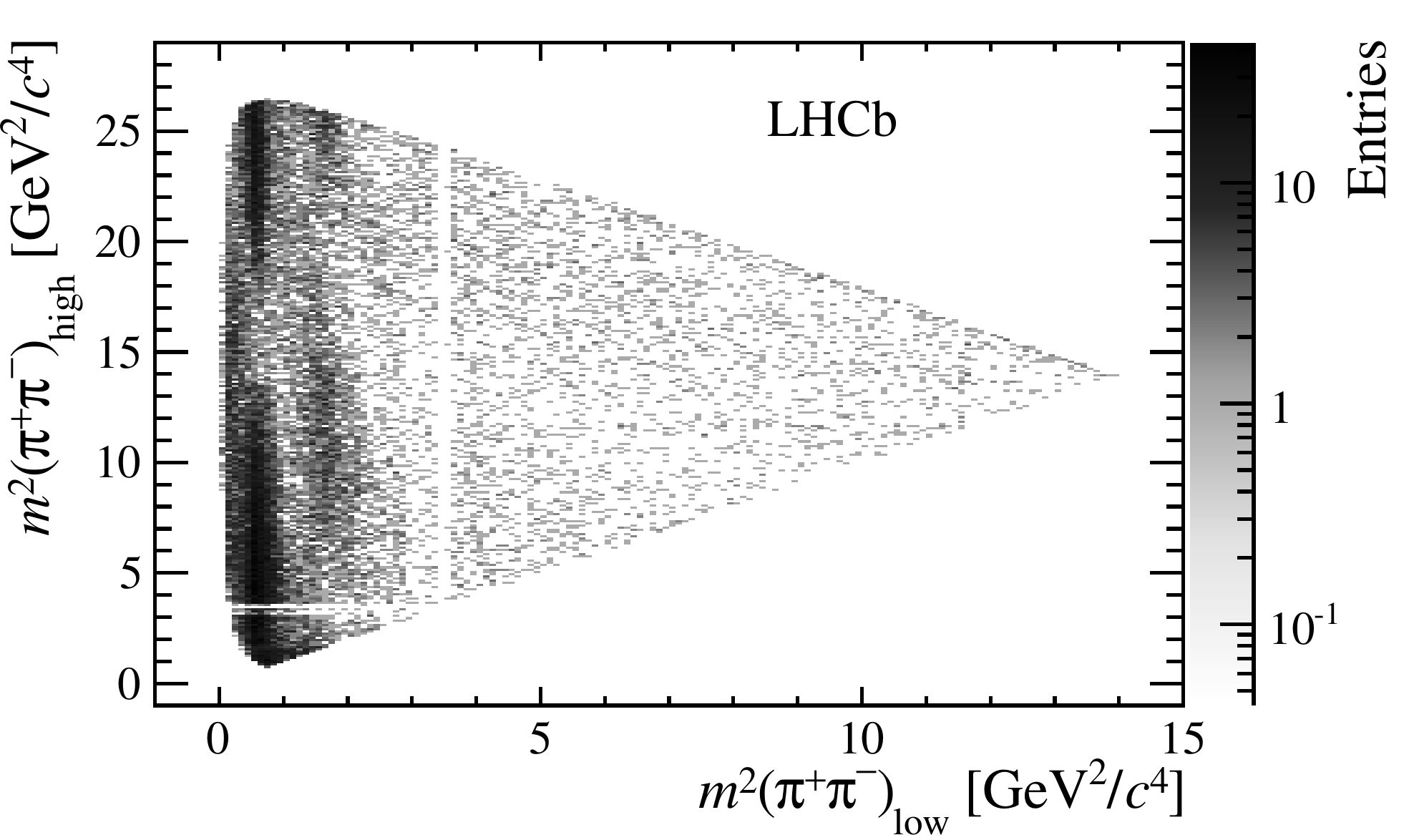}\includegraphics[width=6.5cm]{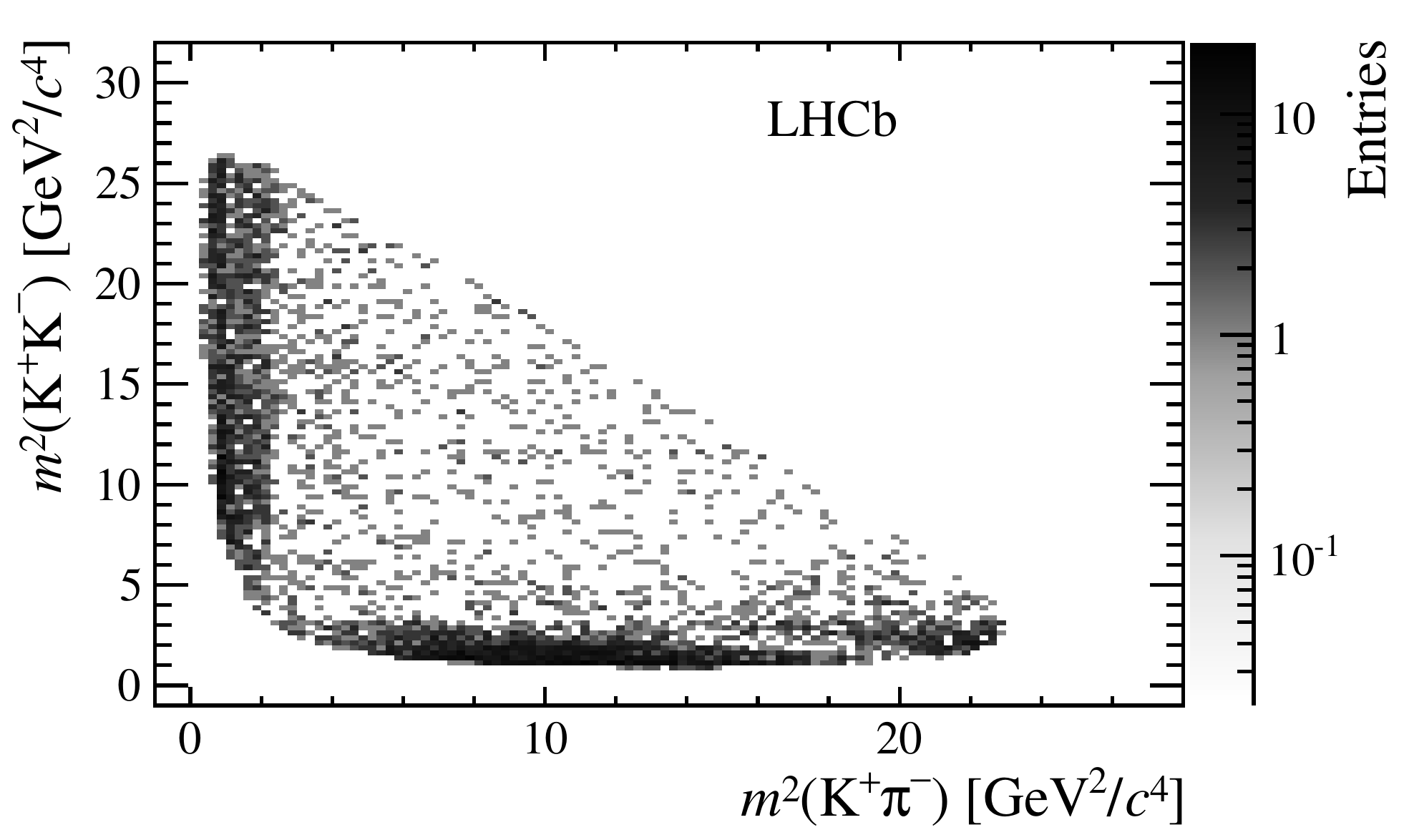}
\caption{Dalitz plots of (top left)  $K^+K^+K^-$, (top right) $K^+\pi^+\pi^-$, (bottom left)
 $\pi^+\pi^+\pi^-$ and (bottom right) $K^+\pi^+K^-$ candidates.}
 \label{dpB2hhh}
\end{center}
\end{figure}

The Dalitz plots for the four decay channels are shown in Fig. \ref{dpB2hhh}. In all cases
a dense, rich resonance structure is observed at low $\pi^+\pi^-$, $K^+\pi^-$ and 
$K^+K^-$ invariant mass squared. Since the phase space is so large, in all
cases there is also a significant non-resonant component.

\subsection{{\em CP} asymmetry across the Dalitz plot}

The {\em CP} asymmetry is computed from observed signal yields:

\[
A_{\mathrm{obs}} = \frac{N_{B^-}-N_{B^+}}{N_{B^-}+N_{B^+}}.
\]

The {\em CP} asymmetry is obtained correcting  $A_{\mathrm{obs}}$  for the $B^{\pm}$
production asymmetry and asymmetry in the detection of unpaired hadron  
($B^{\pm}\to K^{\pm}h^+h^-$, $B^{\pm}\to \pi^{\pm}h^+h^-$), 
\[
\mathcal{A}_{CP} = A_{\mathrm{obs}} - A_{\mathrm{prod}}^B - A _{\mathrm{det}}^h\ .
\]

The correction factors $A_{\mathrm{prod}}^B$ and  $A _{\mathrm{det}}^h$ 
are determined using data-driven methods. Typical values are
of the order of 1\%, and are small compared to  $A_{\mathrm{obs}}$.
In all plots that follow we assume $A_{\mathrm{obs}}\simeq\!\mathcal{A}_{CP}$.

\begin{figure}[ht]
\begin{center}
\includegraphics[width=6.5cm]{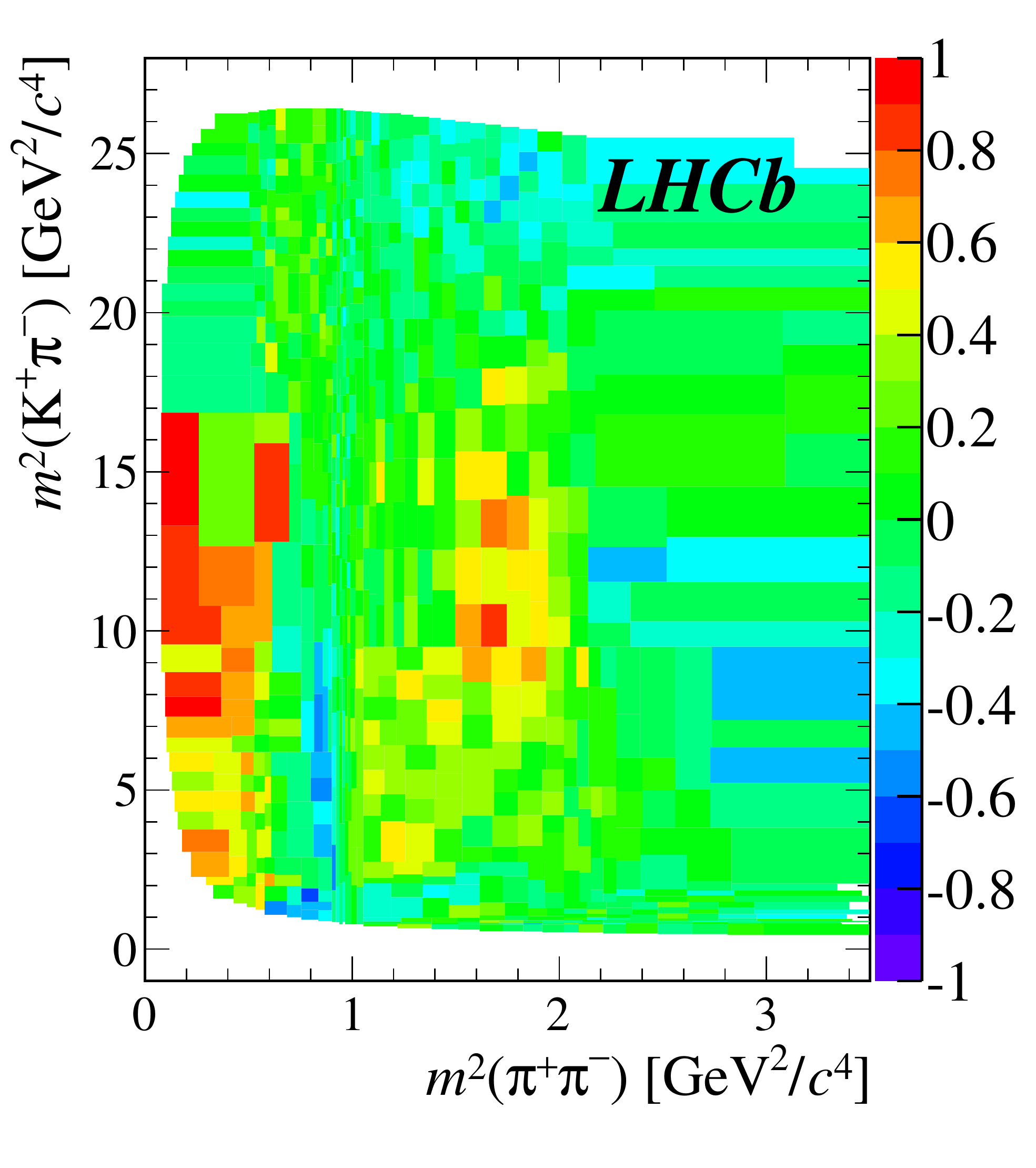}\includegraphics[width=6.5cm]{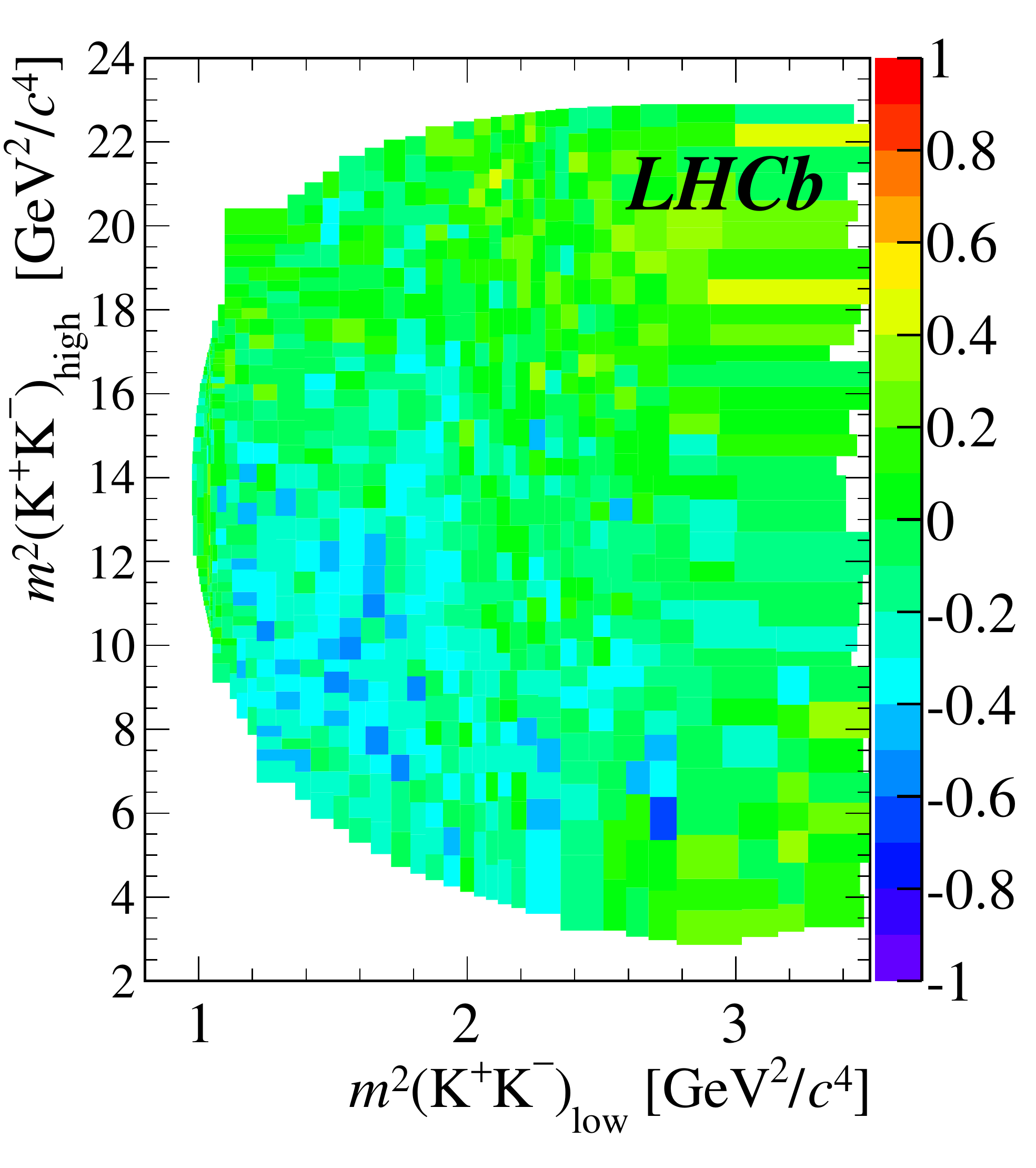}

\includegraphics[width=6.5cm]{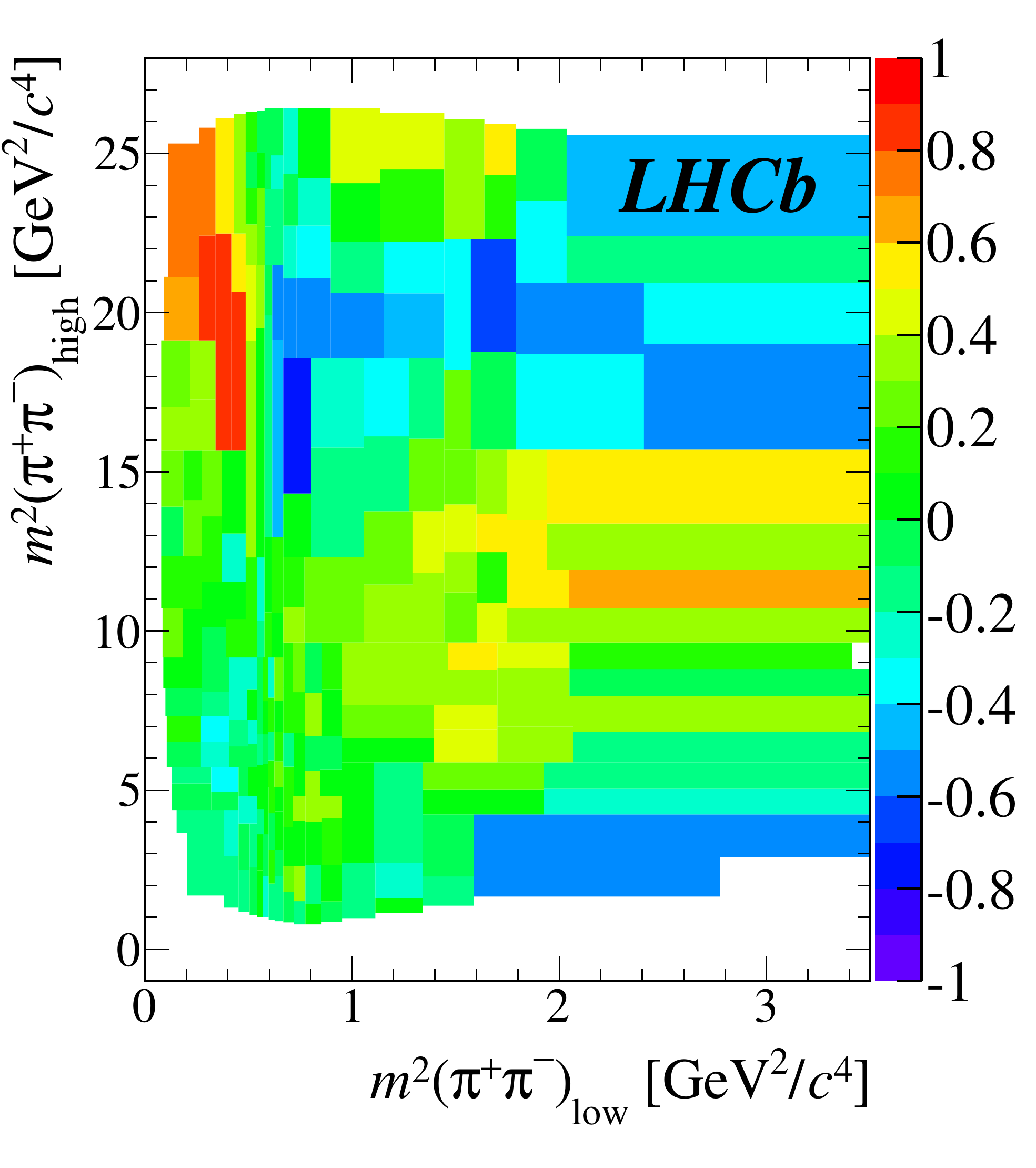}\includegraphics[width=6.5cm]{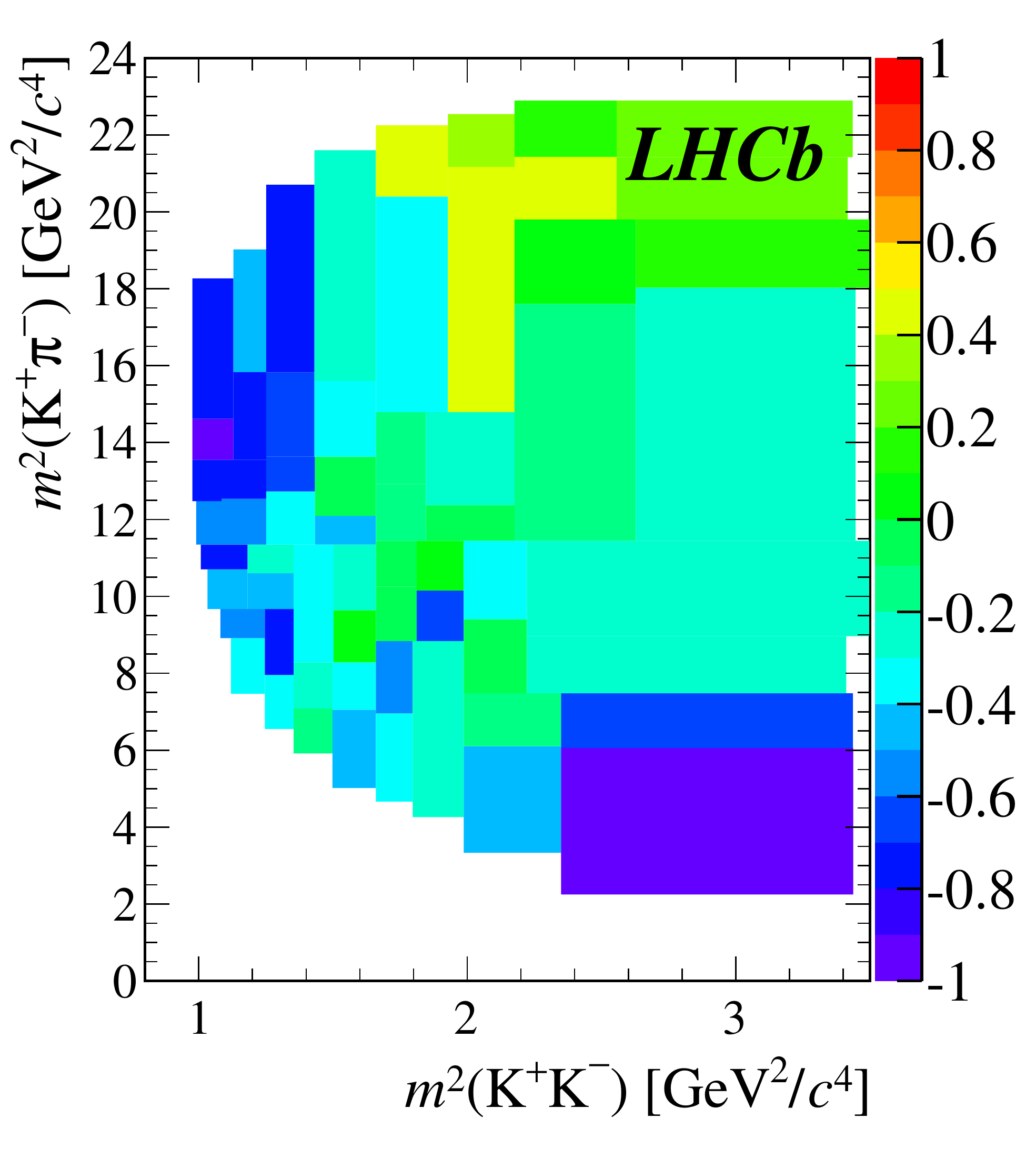}
\caption{A zoom of the Dalitz plots of the $K^+\pi^+\pi^-$ (top left), $K^+K^+K^-$ (top right),
$\pi^+\pi^+\pi^-$ (bottom left), $K^+\pi^+K^-$ (bottom right) decays,
showing the distribution of $\mathcal{A}_{CP}$.}
\label{acpB2hhh}
\end{center}
\end{figure}

The Dalitz plots of the four channels are divided into bins with the same population. In each bin the
value of $\mathcal{A}_{CP}$ is computed. The distribution of $\mathcal{A}_{CP}$ values in the
regions $m^2(h^+h^-)\!<\!3.5$ GeV$^2/c^4$ is shown in Fig. \ref{acpB2hhh}.

The distribution of $\mathcal{A}_{CP}$ shown in Fig. \ref{acpB2hhh} exhibits some
interesting features. In some regions bellow 1 GeV$^2/c^4$ one observes values of the {\em CP} 
asymmetry as high as 80\%, which are rather unusual. Moving from threshold to 1 GeV$^2/c^4$
in the $K^+\pi^+\pi^-$ final state, the {\em CP} asymmetry changes sign as one crosses the
$\rho(770)$ mass. A similar effect is visible in the $\pi^+\pi^+\pi^-$ Dalitz plot.

This effect may be caused by the interference between the various resonances. This hypothesis
is tested with a fast simulation exercise, where the decay amplitude is represented by a sum of an S-
and a P-wave component. This simple model yields a distribution of $\mathcal{A}_{CP}$ values
in qualitative agreement with the experimental observations.    

In the lower part of the $K^+\pi^+\pi^-$ Dalitz plot, the region between 1 and 2.2 
GeV$^2/c^4$ has mostly positive $\mathcal{A}_{CP}$ values. In the same region in the 
$K^+K^+K^-$ Dalitz plot one sees mostly negative $\mathcal{A}_{CP}$ values. A similar
effect is observed in the $\pi^+\pi^+\pi^-$ and $K^+\pi^+K^-$ channels.

A possible interpretation of the pattern observed in the region between 1 and 2.2 GeV$^2/c^4$ 
invokes {\em CPT} invariance. {\em CPT} symmetry implies equal partial widths for particle 
and antiparticle.  As a consequence, in a family of final states sharing the same quantum 
numbers one must have 
$\sum \Gamma_i(B\!\to\!f_i) = \sum \Gamma_i (\overline{B}\!\to\!\bar f_i)$. The $K^+\pi^+\pi^-$ 
and $K^+K^+K^-$ channels are connected by $\pi\pi\to KK$ re-scattering, and so are the
$\pi^+\pi^+\pi^-$ and $K^+\pi^+K^-$ final states. Recall that
the $\pi\pi$ interaction becomes inelastic with the opening of the $K\overline{K}$ channel, 
and up to a center-of-mass energy of $\sim 1.5$ GeV all the inelasticity of the $\pi\pi$ 
interaction goes into the $K\overline{K}$ channel.

In the absence of {\em CP} violation there would be a balance between the outgoing and the 
ingoing flux in each channel. But if {\em CP} is violated this balance would be broken, 
giving rise to a positive asymmetry in one channel which must be compensated by a negative 
asymmetry in other channels. Of course in a comprehensive analysis of such effect one should  
also consider the neutral modes ($\pi^0\pi^0$ and $K^0\overline{K}^0$). But if this is the 
underlying mechanism of the observed {\em CP} asymmetries, then the $\pi\pi \leftrightarrow KK$ 
re-scattering would provide the source of the strong phase difference, a unique feature of
multi-body decays. A recent studies of the impact of the $\pi\pi \leftrightarrow KK$ re-scattering
can be found in Refs. \cite{BedPRD14,NogPRD15}.

\subsection{Challenges for the amplitude analysis}

The study of {\em CP} violation in $B^{\pm}$ decays reveals an unusual pattern with
large asymmetries in specific regions of the Dalitz plot. A full amplitude analysis 
is the necessary step towards identifying the underlying mechanisms 
responsible for the observed $\mathcal{A}_{CP}$
distribution. The data suggest  that the hadronic degrees of freedom are the source 
of strong phase difference. The inclusion of the these degrees of freedom in a consistent
manner is one of the main challenge of the Dalitz plot analysis of the charmless 
three-body $B^{\pm}$ decays.

The vast phase space of charmless three-body $B$ decays is populated by a non-resonant
component. The Dalitz plot analyses of $B$ decays adopt empirical parameterizations
of the non-resonant amplitude. In all cases the S-waves are a significant
part of the decay amplitude. In general the S-wave includes broad structures which 
interfere with the smooth non-resonant component, often resulting in very large
interference terms or, in other words, a sum of fit fractions that largely exceeds 100\%.

A theoretical description the three-body FSI is not possible from first principles.
Calculations of these FSI in the $D^+\!\to\!K^-\pi^+\pi^+$ have been performed~\cite{dkpipi1,dkpipi2,
dkpipi3},
showing that this is an important effect and that a slowly varying phase is introduced. 
To our knowledge, no such calculation have been performed for $B$ decays.

The inclusion of re-scattering effects in the decay amplitude is a far from trivial task.
This effect is entangled with the three-body re-scattering. This is another instance where
input from theory is much needed.

\newpage

\aSection{LHCb - Three-body decays of charged  $D$ mesons}{A. C. dos Reis, for LHCb}
\begin{center}
Alberto C. dos Reis, for the LHCb collaboration \\
\vskip .2cm
{\small Centro Brasileiro de Pesquisas F\'isicas  --  CBPF\\
Rio de Janeiro - Brazil}
\end{center}

\subsection{Introduction}

Light meson spectroscopy from hadronic charm meson decays is one of the
main lines of the LHCb programe on charm physics. In this section,  
ongoing Dalitz plot analyses of charged $D$ decays are discussed.

The very large samples of $D^+$ and $D^+_s$ decays into three charged
light mesons offer an unique opportunity for low-energy hadron physics 
studies. The main motivation is the understanding of the strong dynamics 
of the final state, with particular emphasis on the S-wave amplitudes ---
a key input for {\em CP} violation studies performed with decays of $B$ 
mesons. 

S-wave amplitudes are the dominant component in three-body $D$ decays having
a pair of identical particles in the final state, such as
the $D^+ \to K^-\pi^+\pi^+$ and $D^+_{(s)} \to \pi^-\pi^+\pi^+$ 
decays~\cite{kppe791,d3pi-e791,ds3pi-e791,d3pi-cleo}. In addition to the
abundant contribution of scalar resonances, there is an unique feature of 
$D$ decays: the $\pi\pi$, $K\pi$ and $K\bar K$ scattering amplitudes can be 
measured throughout the whole elastic part of the spectrum, starting from 
threshold. 

The analysis of very large data sets ($\mathcal{O}(10^7)$ reconstructed decays 
in Cabibbo-suppressed modes) is rather challenging. Most Dalitz plot analyses are 
performed in the framework of the isobar model~\cite{asner}. This approach has 
known limitations, especially when scalar particles are involved. Going beyond 
the simple isobar model is indeed the major challenge. Some alternative formulations
are based on the assumption that the dynamics of the final states is entirely 
driven by two-body interactions, disregarding any possible effect related to the 
third particle. It is now clear that corrections to these two-body re-scattering must 
be incorporated in a consistent way, using a maximum of theoretical constraints from 
unitarity, analycity and crossing symmetries.

\subsection{The $D^+_{(s)} \to \pi^-\pi^+\pi^+$ decays}

An interesting aspect of the $\pi^-\pi^+\pi^+$  final state is that  the 
$\pi^-\pi^+$ S-wave amplitude  in the  $D^+$ decay is very different from that of the $D^+_s$.
The $D^+ \to \pi^-\pi^+\pi^+$ decay is Cabibbo-suppressed, has a large contribution
of the $\sigma$ meson and some  $f_0(980)$. The $D^+_s \to \pi^-\pi^+\pi^+$ decay is 
Cabibbo allowed, with a large contribution of $f_0(980)$ and no $\sigma$. In both cases, 
however,  there is a significant contribution of a scalar particle, referred to as ``$f_0(X)$",
with mass between 1.4 and 1.5 GeV and width between 100 and 200 MeV (natural units 
are adopted). Previous determinations of the $f_0(X)$ parameters~\cite{ds3pi-e791,focus-3pi} 
indicate that this state is not consistent with neither the $f_0(1500)$ nor the $f_0(1370)$.

The combined study of the  $D^+$ and $D^+_s$ decays provide, therefore, inputs to the 
physics of the light scalars, from  the $\sigma$  to the $f_0(1710)$.  The basic challenge 
is to extract this information. There is a consensus that one should not represent the 
broad, overlapping  scalars as a sum of Breit-Wigner functions. Alternative approaches
assume a {\em 2+1} picture, where the dynamics of the final state is entirely determined 
by the two-body interaction, the third particle playing no role. One example of this
approach is the recent analysis of the $B^0 \to \overline{D}{}^0 \pi^+\pi^-$ decay by 
LHCb~\cite{lhcb5}. In this work the  K-matrix formalism is used for the S-wave amplitude,  
where the poles are obtained from a fit to the  $\pi\pi$ scattering data~\cite{sarantsev}
and fixed in the Dalitz plot fit. A good description of the data is obtained, but
one would like to go the other way around, that is, to extract information about  $\pi\pi$ 
interaction from the $D$ and $B$  decays.

Another alternative is to perform a minimal model-dependent analysis, extracting the 
S-wave amplitude from the data making no assumption about its composition.
This technique was developed by the E791 collaboration~\cite{brian}, and herein is
referred to as ''MIPWA". In the  ''MIPWA", the S-wave amplitude is parameterized 
as a generic complex function, $\mathcal{A}_S=a(s)e^{i \phi(s)}$ ($s\equiv m^2(\pi^-\pi^+)$). 
The real functions $a(s)$ and $\phi(s)$ are obtained directly from the data, 
dividing the $\pi^-\pi^+$ spectrum($2m_{\pi}\!<\!m_{\pi\pi}\!<\!M_D-m_{\pi}$) into bins .
At each bin edge, two fit parameters, $a_k$ and $\phi_k$, determine the S-wave amplitude.
A spline interpolation yields the S-wave amplitude at any $m_{\pi\pi}$ value in the $\pi^-\pi^+$
spectrum. This approach was used  by the BaBar collaboration in the analysis of the 
$D^+_s \to\pi^-\pi^+\pi^+$ decay~\cite{antimo}.

\begin{figure}[h]
\begin{center}
\includegraphics[width=6.5cm]{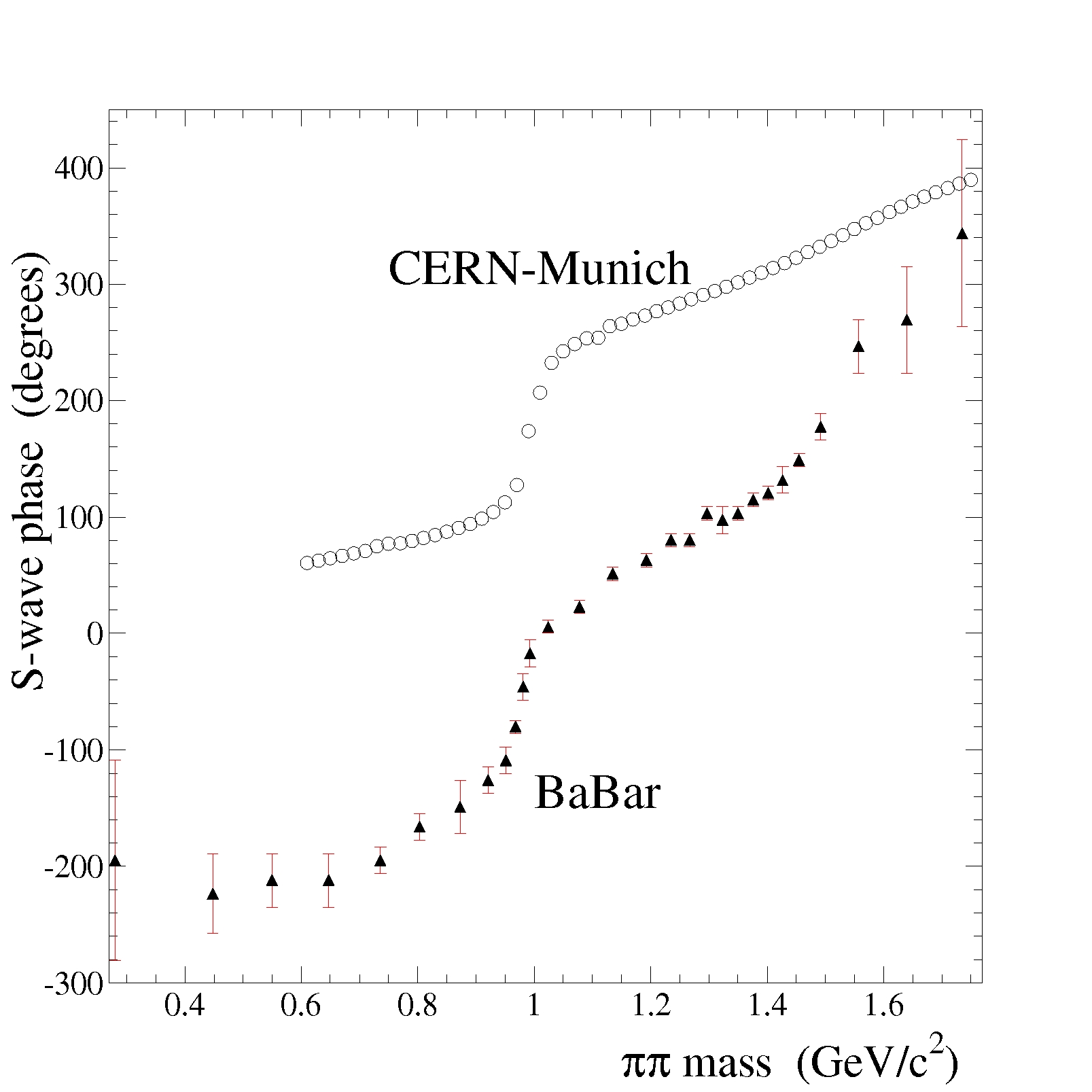}\includegraphics[width=6.5cm]{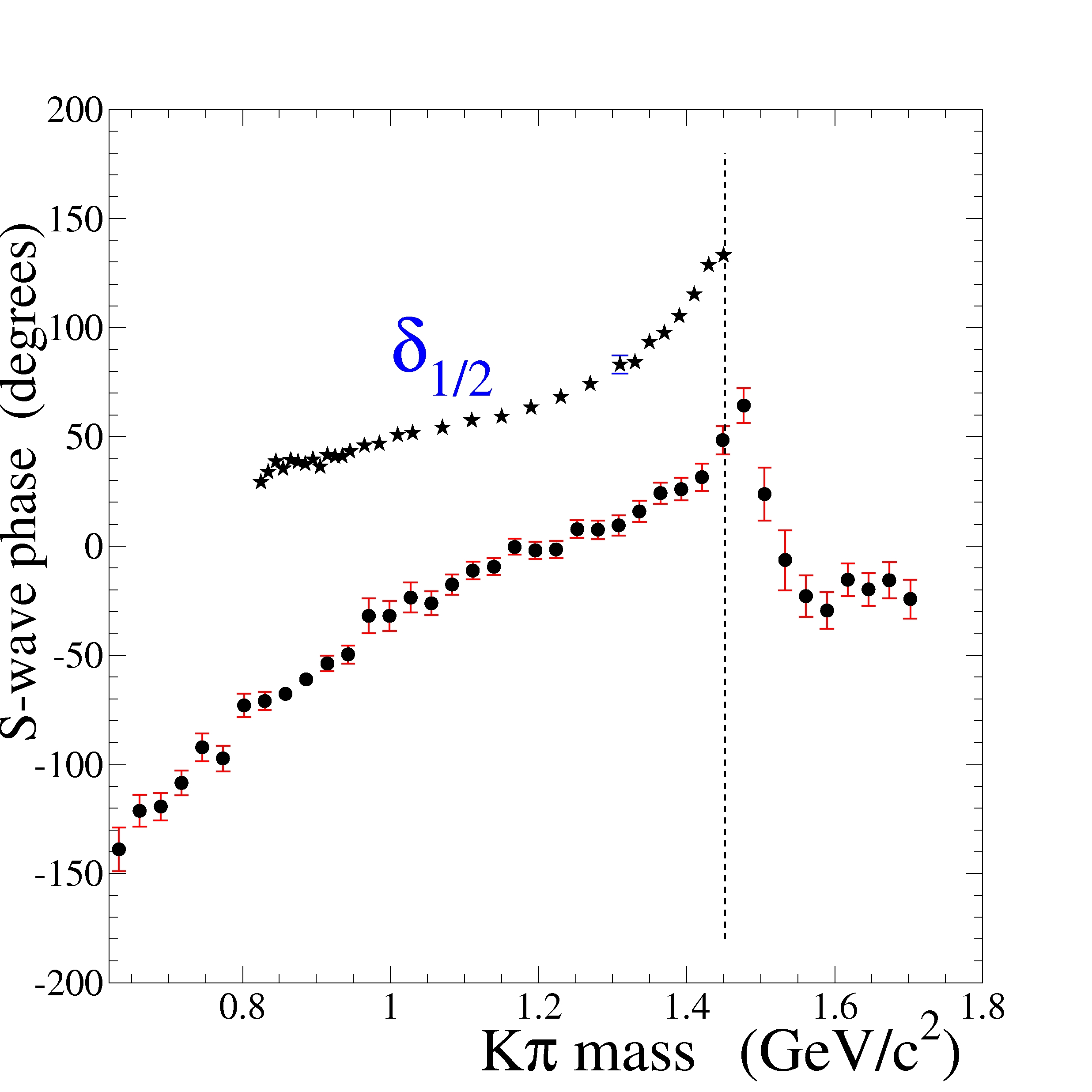}
\caption{Left: S-wave phase from $D^+_s \to\pi^-\pi^+\pi^+$~\cite{antimo} and from
the CERN-Munich collaboration~\cite{cern-munich1, *cern-munich2}. Right: S-wave phase from the FOCUS
Dalitz plot analysis~\cite{focuspwa} of the
$D^+ \to K^-\pi^+\pi^+$ decay and from the LASS ($I\!=\!1/2$) experiment~\cite{lass}. 
The $K^-\pi^+ \to K^-\pi^+$ cross section is elastic up to the $K\eta'$ threshold, 
indicated by vertical line.}
\label{swaves}
\end{center}
\end{figure}

The MIPWA is probably the approach with less model dependence, relying on the
assumption is that the P- and D-waves are well represented by a sum of Breit-Wigner 
functions. A comparison between the phases obtained by the MIPWA and those
from scattering data  is shown in Fig. \ref{swaves}.
In the left panel,  the S-wave phases obtained by the BaBar measurement and by 
the CERN-Munich collaboration~\cite{cern-munich1, *cern-munich2} ($\pi^+\pi^- \to \pi^+\pi^-$) are 
displayed as a function of the $\pi^+\pi^-$ mass.  On the right, a similar
measurement of the $K^-\pi^+$  S-wave phase performed by the FOCUS 
collaboration~\cite{focuspwa} in the $D^+ \to K^-\pi^+\pi^+$ decay is superimposed to
the {\em I}=1/2 \ S-wave phase from $K^-\pi^+ \to K^-\pi^+$ scattering obtained by the
LASS collaboration~\cite{lass}.

The differences between the S-wave phases obtained from $D$ decays and from scattering
data are evident, which seems to indicate deviations from Watson’s final-state theorem\cite{watson}.
In addition to an overall shift, in both $\pi^+\pi^-$ and $K^-\pi^+$
systems there is a clear mismatch between the phase motion from decay and scattering. 
This discrepancy indicates that other effects must be taken into account.

One can picture the decay of a $D$ meson as a time-ordered sequence of the
short-distance $c$ quark transition, resulting in the ''final state quarks", from which the
long-distance hadronization occurs. The different configuration of the final state quarks
would explain the differences in the S-wave composition in the $D^+$ and $D_s^+$ decays 
described above. Strong phases, possibly with some smooth energy dependence, may be 
generated in the "weak vertex", and would depend on the nature of the initial $D$ meson. 
The hadrons emerging from the weak vertex re-scatter in all possible ways before reaching 
the detector: $D\to K K\bar \pi \to \pi\pi\pi$, $D\to K^0 \pi^0 \pi \to K^- \pi^+\pi$, 
three-body interactions. Studies performed with the $D^+ \to K^-\pi^+\pi^+$ 
decay~\cite{dkpipi1,dkpipi2,dkpipi3,pat} show that FSI involving all three particles in hadronic
loops are an important effect, explaining part of the differences in the S-wave phases
observed in Fig. \ref{swaves}.

The S-wave phases measured with the MIPWA technique  must be understood as
a convolution of the pure, universal $\pi^+\pi^-$/$K^-\pi^+$ scattering phases with 
those inherited from the weak vertices  and with phases resulting from three-body 
FSI. Moreover, in $D$ decays one cannot disentangle the contribution from the different 
isospin amplitudes. In addition, given the large number of free parameters, any defect
on the representation of the other waves will leak into the MIPWA phase.

A comparative study of the  MIPWA $\pi^+\pi^-$ S-wave amplitude in 
$D^+$ and $D^+_s \to\pi^-\pi^+\pi^+$ decays is very important. The mass difference 
between the $D^+$ and the $D^+_s$ is only 100 MeV. This may be  seen 
as a bonus: three-body FSI should be similar in the $D^+$ and the $D^+_s$ decays. 
A comparison between measurements of the raw S-wave phases would provide 
information about the $D^+$ and $D^+_s$ weak vertices. This should be taken with
a grain of salt, though, since it is not really possible to draw a clear line separating classes 
of FSI. 

In a model-dependent measurement, different line shapes of the $\sigma$ meson can be 
tested. A simultaneous study of the S-wave amplitude from the $D^+$ and $D^+_s$ decays 
will give information about the scalar meson at $\sim$1.4 GeV.
Isospin-breaking processes can also be addressed in model-dependent measurements.
A precise determination of the $f_0(980)$ line shape, possible in the $D^+_s$ decay,
is related to the physics of the $f_0-a_0$ mixing. This effect is enhanced in the 
region between the $K^+K^-$ and $K^0\bar K^0$ thresholds~\cite{achasov}. These 
thresholds are 8 MeV apart, while the LHCb mass resolution is of the order of 2 MeV. 
The $\rho-\omega$ mixing can be studied in the $D^+$ decay, where a large 
$\rho(770)^0\pi^+$ contribution is observed.
  
Model-dependent and MIPWA measurements are under way in LHCb.
For the $D^+$ decay the MIPWA measurements of the $\pi^+\pi^-$ S-wave amplitude 
will be made for the first time. For the $D^+_s$ decay a significant improvement in
precision is expected, since the LHCb sample is one order of magnitude 
larger than that of BaBar. 

\subsection{The $D^+_{(s)} \to K^-K^+\pi^+$ decays}

Another ongoing effort in the LHCb charm physics programme is the study of the
$D^+_{(s)} \to K^-K^+\pi^+$ decays. The main goal is the MIPWA
measurement of the $K^+K^-$ S-wave
amplitude, an important input to {\em CP} violation studies in the $B$ meson system.
Statistics is not an issue: the selected $D^+_{(s)} \to K^-K^+\pi^+$ sample is 
approximately ten times larger than that of $D^+_{(s)} \to \pi^-\pi^+\pi^+$, 
with much higher purity due to the presence of two kaons. Compared to BaBar, the
LHCb $D^+_{(s)} \to K^-K^+\pi^+$ signals are two orders of magnitude larger.
The major challenge for the MIPWA measurement of the $K^+K^-$ S-wave is the 
presence of an S-wave component also in the $K^-\pi^+$ system. Both amplitudes 
populate the whole phase space. The best strategy for this measurement needs still to be
defined. As in the case of the $\pi^-\pi^+\pi^+$ final state, a model-dependent analysis 
is also underway.

The $D^+ \to K^-K^+\pi^+$ is a Cabibbo-suppressed decay. There are two 
tree-level diagrams leading to $(K\pi)^0K^+$ and to $(K\bar K)^0\pi^+$ intermediate 
states. Contributions from resonances coupling to both $K^-\pi^+$ and $K^-K^+$ are
therefore expected.

The  $D^+_{s} \to K^-K^+\pi^+$  decay is Cabibbo-allowed. As for the $D^+$, there are two 
tree-level diagrams. The colour suppressed can lead to both $K^-\pi^+$ and $K^-K^+$
resonances, whereas the colour allowed (external W-radiation) leads mainly to $K^+K^-$ 
resonances. The latter is the same as for the $D^+_s \to \pi^-\pi^+\pi^+$, so a strong 
contribution of the $f_0(980)\pi^+$ is expected, as well as a large $\phi\pi^+$ component.
In the model-dependent analysis, the coupling of the $f_0(980)$ to $K\bar K$ can
be measured independently in the $K^-K^+\pi^+$ and $\pi^-\pi^+\pi^+$ final states.

In both $D^+$ and $D^+_{s}$ decays there is a tree-level diagram with an $s \bar s$
pair which could form resonances of the $f_0$ family with mass above 1 GeV, including
the $f_0(X)$ state. The situation
concerning these resonances and their contribution to $D$ and $B$ decays is still rather 
confuse. The measurement of the $K^-K^+$ S-wave amplitude will help to understand the
role --- and the nature --- of these states in heavy flavour decays.

\newpage

\aSection{LHCb - Exotic hadron spectroscopy}{T. Skwarnicki, for LHCb}
\begin{center}
Tomasz Skwarnicki, for the LHCb collaboration \\
\vskip .2cm
{\small Syracuse University\\
Syracuse, New York, USA}
\end{center}
%\section{Exotic Hadron Spectroscopy}
\newlength{\figsize}
\setlength{\figsize}{\hsize}

The LHCb experiment offers unique opportunities in the field of exotic hadron spectroscopy. 
Production rates of $B$ mesons are 3 orders of magnitude larger 
than previously available at the $e^+e^-$ $B$ factories (Belle and BaBar). 
Even after correcting for the smaller reconstruction efficiencies, LHCb has typically collected 
10 times bigger data samples of $B$ meson decays to $J/\psi$ and light-quark hadrons
in all charged particle final states during Run I at LHC (2011-12, 3 fb$^{-1}$). 
The long visible lifetime of the lightest $b$-hadrons provides for excellent suppression of combinatorial backgrounds
and two RICH detectors provide further background suppression for final states with at least one charged
kaon. Thus, signal-to-background ratios have been slightly better than those at the $e^+e^-$ $B$ factories.
The RICH detectors, and the large trigger bandwidth to tape (up to 5 kHz at Run I) devoted almost entirely to
the flavor physics offer a competitive edge over the general purpose detectors, ATLAS and CMS. The larger $b\bar b$
cross-section at the LHC as compared to the Tevatron, offers additional advantage over the CDF and D0 data samples.
A unique advantage over the $e^+e^-$ $B$ factories comes with simultaneous productions of $B$, $B_s$, $B_c$ mesons
and $b-$quark baryons ($B_s$ samples require dedicated beam time at the $B$ factories, while $B_c$ and $b-$baryons are 
not accessible). 
The efficient trigger on final states with the $\jpsi\to\mu^+\mu^-$ decays led to several high impact results in the
field of exotic hadrons related to the charmonium family. 
The same final states offer also an opportunity to study conventional and exotic 
light-hadron spectroscopy in a relatively clean environment as illustrated in Ref.~\cite{Aaij:2014siy}.%%LHCb-PAPER-2014-012}.
While this potential of the LHCb experiment has gone largely unexplored, we concentrate here on exotic
spectroscopy with heavy charmed quarks inside. The dominant weak decays turn a $b$ quark to a $c$ quark, while 
a companion $\bar c$ comes from the associated $W\to \bar{c}s$ vertex. 

Three LHCb amplitude analyses are worth invoking here.
The use of full angular phase-space (5D) in the analysis of 
the $X(3872)$ decays to $\rho^0\jpsi$, with $\rho^0\to\pi^+\pi^-$ and $\jpsi\to\mu^+\mu^-$, 
with the fully polarized $X(3872)$ states 
produced in $B^+\to X(3872) K^+$ decays, led to the first decisive determination of $X(3872)$ spin 
and parity ($J^{P}=1^+$)  \cite{LHCb-PAPER-2013-001}, 
in a sample not much larger than previously available in Belle, where an opportunity to achieve such a goal earlier 
was missed due to the limitations of three 1D angular fits which were employed.
This example underscores the importance of studying full decay dynamics especially when number of signal events is limited.  
An update to this analysis published recently, confirmed these results without any
assumptions about the orbital angular momentum in the $X(3872)\to\rho^0\jpsi$ decays 
and set a tight upper limit on
the $D$-wave fraction \cite{LHCb-PAPER-2015-015}. 
This assignment of quantum numbers ruled out the closest-in-mass $c\bar c$ hypothesis ($\eta_c(2^3D_{2^{-+}})$).
The remaining explanations invoke either a loosely bound $D^0\Dbar^{*0}$ molecule, 
which would explain the mass coincidence with the $D^0\Dbar^{*0}$ threshold, 
the $\chi_c(2^3P_{1^{++}})$ conventional meson or a tightly bound $(cu)(\bar{c}\bar{d})$ tetraquark,
attracted to the threshold via quantum effects \cite{Bugg:2008wu} (for a recent review see Ref.~\cite{Esposito:2014rxa}).
The $X(3872)$ is too narrow to be simply a $D^0\Dbar^{*0}$ cusp, with no bound state contribution \cite{Bugg:2008wu}.

\begin{figure}[tbhp]
  \begin{center}
%  \ifthenelse{\boolean{pdflatex}}{ 
        \includegraphics*[width=\figsize]{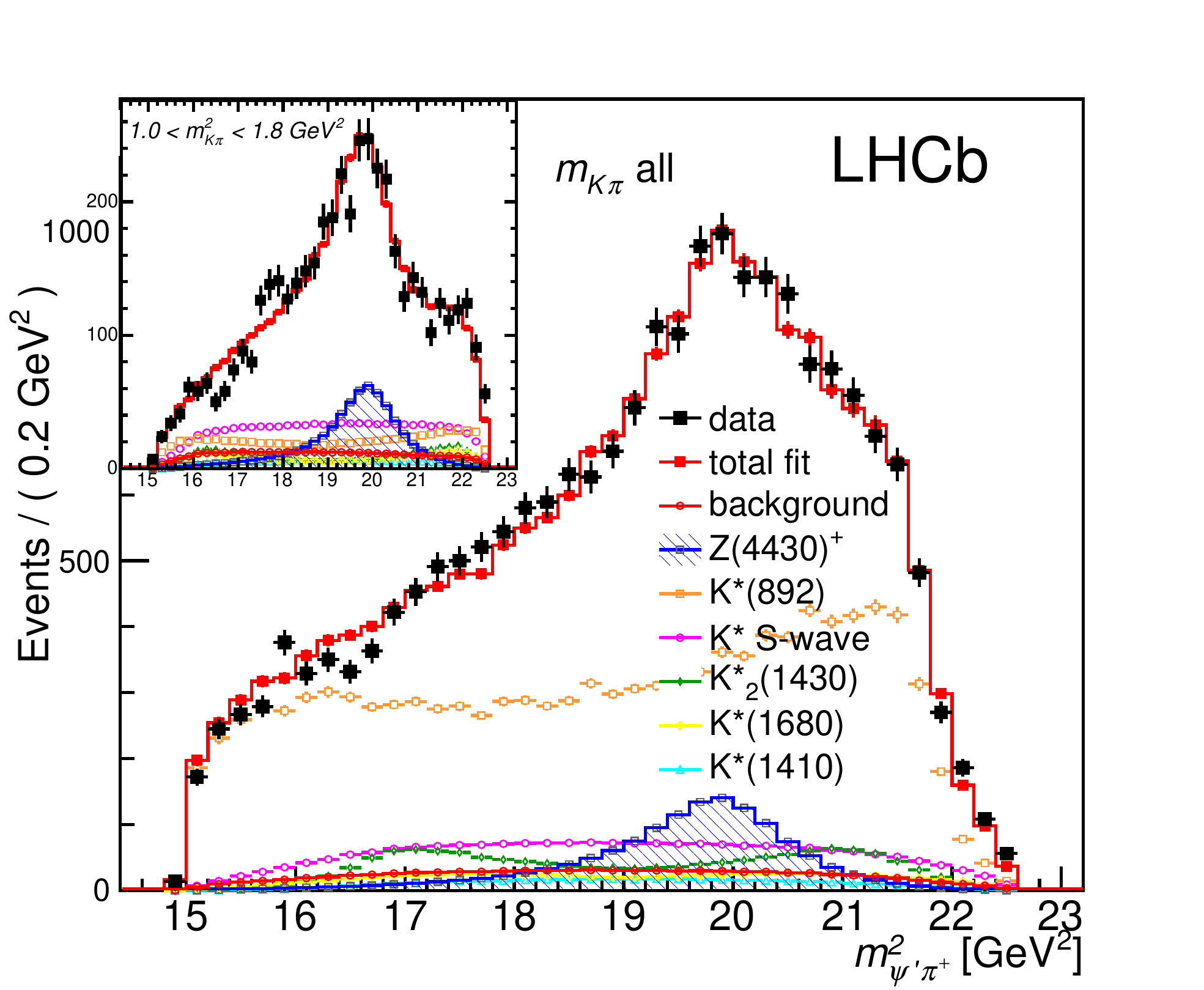}
%   }{
%        \includegraphics*[width=\figsize]{figs/c3_Z_1070_inset.pdf}
%   } 
  \end{center}
  \vskip-0.1cm\caption{\small 
    Distribution of invariant mass of $\psi'\pi^+$ squared for
    $B^0\to\psi'\pi^+K^-$, $\psi'\to\mu^+\mu^-$ candidates 
    together with the projections of the 4D fit of $K^{*0}\to K^-\pi^+$ and
    $Z(4430)^+\to\psi'\pi^+$ amplitudes.   
    The distribution for the $K^-\pi^+$ mass slice in which the $K^{*0}$ contributions 
    are smaller is shown in the inset.     
  \label{fig:z4430}
  }
\end{figure}

\begin{figure}[bthp]
  \begin{center}
%  \ifthenelse{\boolean{pdflatex}}{ 
        \includegraphics*[width=0.5\figsize]{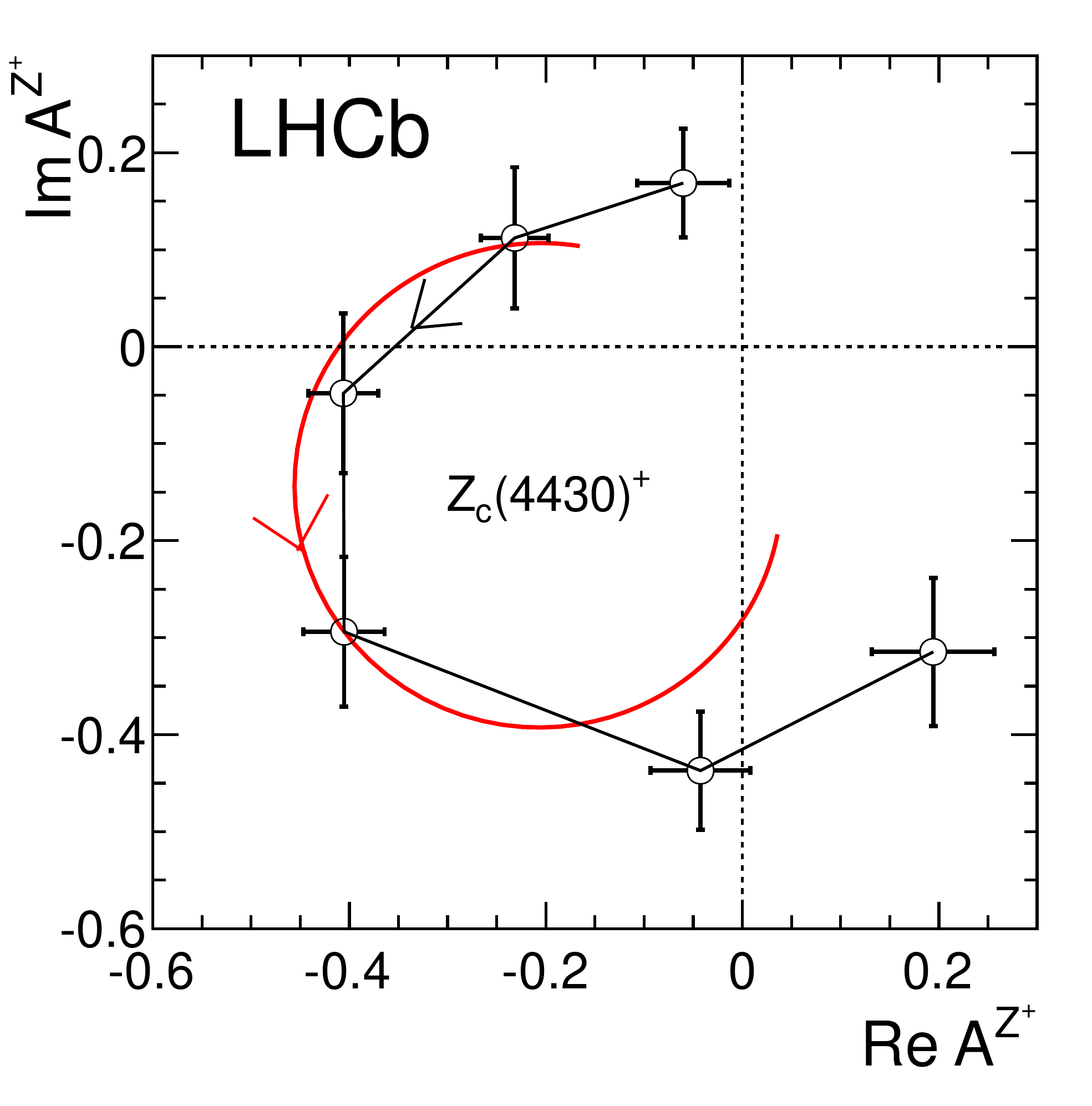}
%   }{
%        \includegraphics*[width=0.5\figsize]{figs/Argand-Z.pdf}
%   } 
  \end{center}
  \vskip-0.1cm\caption{\small 
Fitted values of the real and imaginary parts of the amplitude for the $Z(4430)^+$ state
divided into six $m_{\psi'\pi}^2$ bins of equal width between $18.0$ and $21.5$ GeV$^2$ range, 
shown in the Argand diagram
as connected points with the error bars ($m_{\psi'\pi}^2$ increases counterclockwise). 
The solid (red) curve is the prediction
from the Breit-Wigner formula with a resonance mass (width) of 4475 (172) \mev
and magnitude scaled to intersect the bin with the largest magnitude centered at (4477 MeV)$^2$.
The phase convention assumes the helicity-zero $K^*(892)$ amplitude to be real.
Systematic uncertainties are not included.
  \label{fig:ArgandZ}
  }
\end{figure}

The 4D amplitude analysis of $B^0\to\psi' \pi^+ K^-$ decays, had less angular dimensions to work with, 
but many conventional $K^{*0}$ states contributing to the observed $\pi^+K^-$ mass distribution. 
A successful amplitude model was found, which required a very significant ($>13\sigma$) 
contribution of the exotic $Z(4430)^+$ state decaying
to $\psi'\pi^+$ (see Fig.~\ref{fig:z4430}) \cite{Aaij:2014jqa}.%%LHCb-PAPER-2014-014}. 
This state was first observed by Belle \cite{Choi:2007wga},
but was in experimental limbo for many years after the BaBar experiment was not able to confirm it  \cite{Aubert:2008aa}.
The LHCb results agree very well with the results of the earlier Belle 4D amplitude analysis 
of the same channel \cite{Chilikin:2013tch} on
the $Z(4430)^+$ mass, width and fit fraction, while improving their determination.
The $3.4\sigma$ evidence for $J^P=1^+$ assignment to $Z(4430)^+$ from Belle, was also confirmed beyond any doubt.
Last but not least, thanks to a sample 10 times larger than that available at Belle,
LHCb was able to probe $Z(4430)^+$ amplitude dependence on $\psi'\pi^+$ mass without the Breit-Wigner assumption.
The resulting Argand diagram (Fig.~\ref{fig:ArgandZ}) shows a quick counter-clockwise 
change of the amplitude phase at the peak of its
magnitude, in agreement with the resonant hypothesis. 
This diagram rules out the rescattering model by Pakhlov-Uglov \cite{Pakhlov:2014qva}, 
offered as an explanation for the $Z(4430)^+$ mass peak,
since it predicts the phase mass running in the opposite direction.
  
\begin{figure}[tbp]
  \begin{center}
%  \ifthenelse{\boolean{pdflatex}}{ 
        \includegraphics*[width=\figsize]{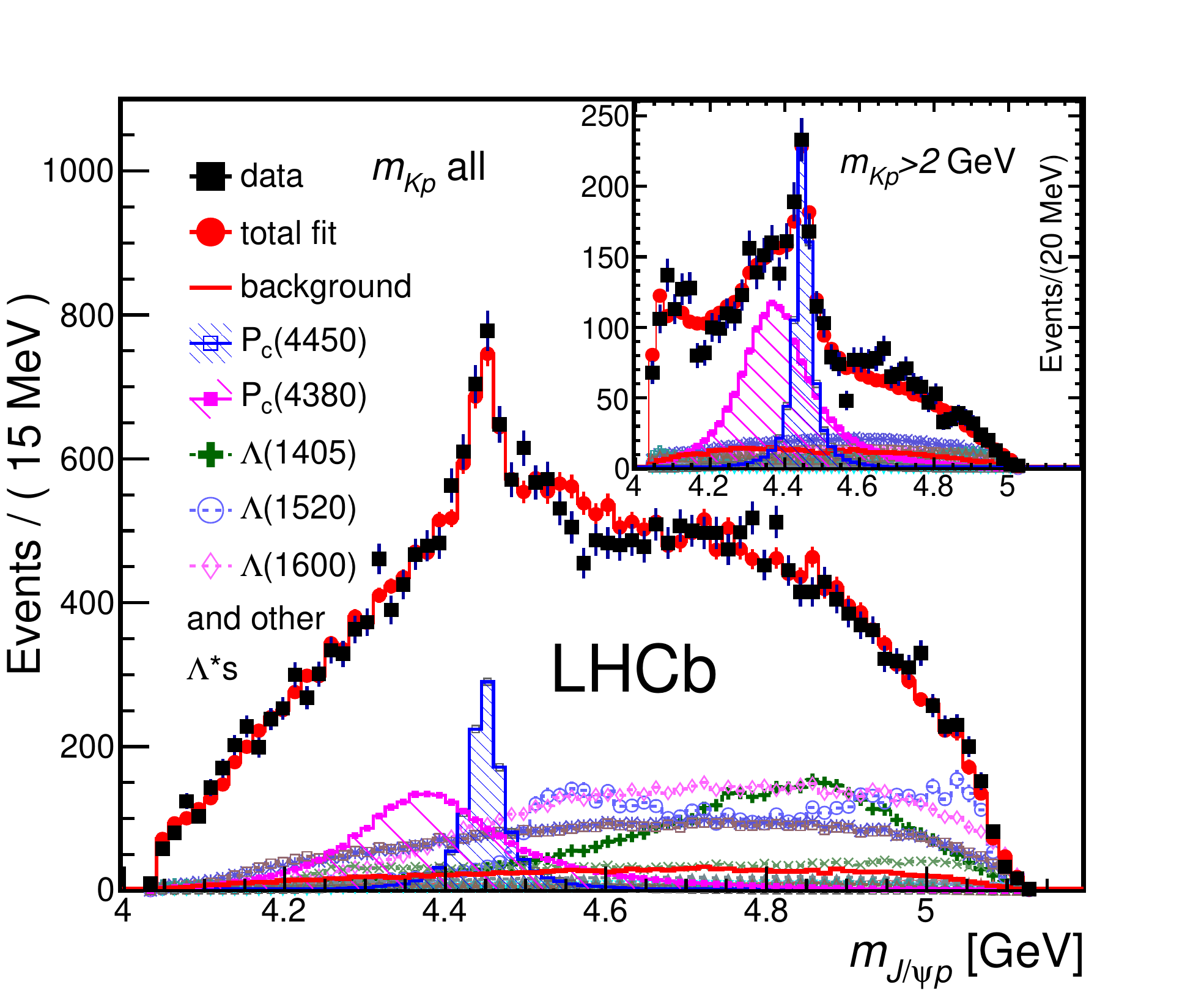}
%   }{
%        \includegraphics*[width=\figsize]{figs/mjpsip-default-inset-all.pdf}
%   } 
  \end{center}
  \vskip-0.1cm\caption{\small 
    Distribution of invariant mass of $\jpsi p$ for
    $B^0\to\jpsi p K^-$, $\jpsi\to\mu^+\mu^-$ candidates 
    together with the projections of the 6D fit of $\Lambda^{*}\to K^-p$ and
    $P_c^+\to\jpsi p$ amplitudes.   
    The distribution for the $K^-p$ mass slice in which the $|Lambda^{*}$ contributions 
    are smaller is shown in the inset.     
  \label{fig:pcs}
  }
\end{figure}

\begin{figure}[bthp]
  \begin{center}
%  \ifthenelse{\boolean{pdflatex}}{ 
        \includegraphics*[width=\figsize]{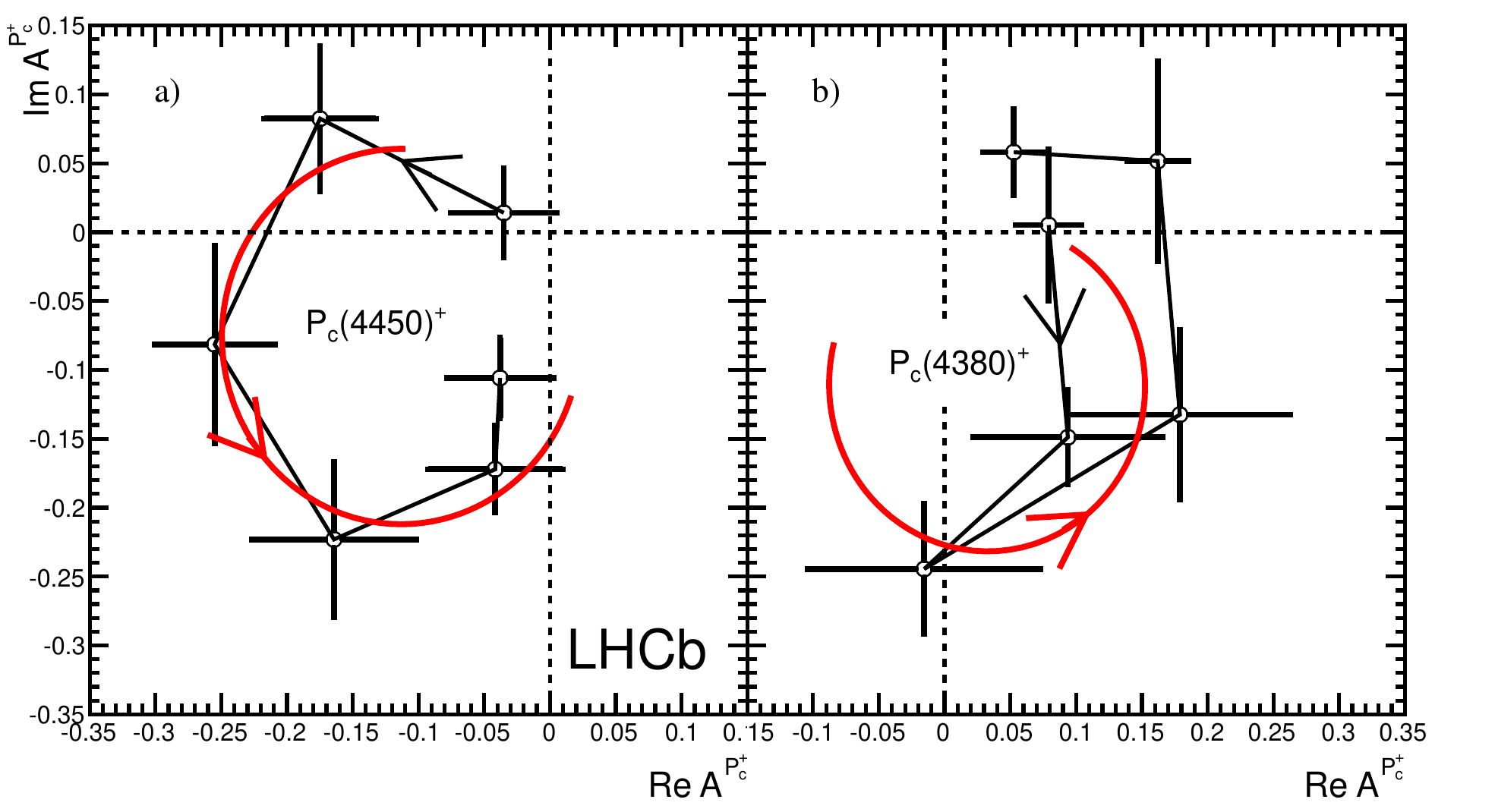}
%   }{
%        \includegraphics*[width=\figsize]{figs/DoubleArgand-final.pdf}
%   } 
  \end{center}
  \vskip-0.1cm\caption{\small 
Fitted values of the real and imaginary parts of the amplitudes for the baseline ($3/2^-$, $5/2^+$) fit for a) the $P_c(4450)^+$ state and b) the $P_c(4380)^+$ state, each divided into six $m_{\jpsi p}$ bins of equal width between $-\Gamma_0$ and $+\Gamma_0$ shown in the Argand diagrams as connected points with error bars ($m_{\jpsi p}$ increases counterclockwise).
The solid (red) curves are the predictions
from the Breit-Wigner formula for the same mass ranges
with  $M_0$ ($\Gamma_0$) of
 4450 (39) \mev and 4380 (205) \mev, respectively,
with the phases and magnitudes at the resonance masses set to the
average values between the two points around $M_0$.  
The phase convention assumes $L=0$, $S=\frac{1}{2}$ amplitude for $\Lambda(1520)$ to be real. 
Systematic uncertainties are not included.
  \label{fig:ArgandPc}
  }
\end{figure}

The 6D amplitude analysis of $\Lambda_b\to\jpsi p K^-$ decays \cite{LHCb-PAPER-2015-029}
followed in footsteps of the $B^0\to\psi'\pi^+K^-$ analysis.
The increased dimensionality is due to the $\Lambda_b$ spin. 
The observed $\jpsi p$ mass distribution, which contains a narrow peak at 4450 \mev,
could not be described with the excitations of the $\Lambda$ baryon. 
Two interfering $\jpsi p$ resonances of opposite parity with allowed spin combinations being either ($3/2$, $5/2$) 
or ($5/2$, $3/2$) are needed, in addition to the known $\Lambda^*$ resonances, for a 
successful description of the data (see Fig.~\ref{fig:pcs}). 
The two new states have been labeled $P_c(4380)^+$ and $P_c(4450)^+$ with the lighter one
having a larger width of $218\pm18\pm86$ \mev then the heavier one, which is very narrow for a resonance at such high mass,
$39\pm5\pm19$ \mev.
The Argand diagram for the narrower state is very consistent with the resonant behavior and is inconclusive for the broader
state (Fig.~\ref{fig:ArgandPc}). 
The pair can be accommodated in the diquark model of tightly bound pentaquarks, 
via a change of the orbital angular momentum
between the states, 
which explains the opposite parity and the prediction that the narrower state has a spin of $5/2$, which 
also would explain its smaller width due to the orbital angular momentum barrier 
factor in its decay \cite{Maiani:2015vwa,Lebed:2015tna,Anisovich:2015cia}. 
While the narrower state can be accommodated as a baryon-meson molecule, 
either $\Sigma D^{*0}$ \cite{Karliner:2015ina,Chen:2015loa,Roca:2015dva,He:2015cea}
or $p\chi_c(1P_1)$ \cite{Meissner:2015mza},
if its spin is $3/2^-$, the $5/2$ spin cannot be reached in the molecular model 
since $l=1$ molecular states are very unlikely to be bound.    
Rescattering models have been also proposed \cite{Guo:2015umn,Liu:2015fea,Mikhasenko:2015vca,Szczepaniak:2015hya}. 
While they can produce the peak at the $P_c(4450)^+$ mass, with the 
counterclockwise running of phase with the mass, it is not yet clear whether they can 
produce fits as good as those achieved from assuming bound state models in all 6 dimensions. 
Such mechanisms cannot produce a structure with a spin of $5/2$ at these masses.

While the $P_c^+$ states are the first plausible pentaquark candidates with heavy quarks inside, there are many other
observed neutral and charged tetraquark candidates with a $c\bar c$ pair inside (the neutral 
exotic candidates could also be hybrid states), but not necessarily confirmed, by the
other experiments (see e.g.~Ref.~\cite{Esposito:2014rxa}). 
Although many such states have been reported, including several $Z_b^+$ candidates, no clear 
theoretical model explaining all of them has emerged. 
The main competitors are tightly bound tetraquark and pentaquark models with diquarks inside, 
molecular models and rescattering processes. 
Once at work for one phenomenon 
(e.g. for the $P_c^+$ pentaquark candidates), 
the same mechanism should also be at work at other masses,
different multiquark structures (e.g. also for tetraquarks) 
and with different quark flavor structure (isospin and SU(3)$_f$ partners, $b\bar b$ and $c\bar c$ etc.).
In fact, different bound state models lead to different but always rich
``periodic tables'' of various expected states. 
Uncovering new entries to the experimentally observed tables is of crucial importance. 
However, it is also possible that more than one dynamics is at play for multiquark
systems, which would vastly complicate the interpretation of the data. 

Exotic bound states are likely to decay to more than one mode, and could be produced by more than one
process (e.g. prompt production, central-exclusive production, heavy ion collisions). Each observation
contributes a new insight into internal hadron structure. 
Rescattering processes are likely to produce structures which are specific for each 
final state and each production mechanism. 
Thus searches for the known exotic states in different decay modes or in different production 
environment are doubly important. 

The opportunities for experimental advancement of exotic hadron spectroscopy in LHCb are, therefore,
very rich and likely will be manpower limited for foreseeable future. 
Many channels known to contain important 
information, even easily accessible in LHCb like $B^0\to\jpsi \pi^+K^-$, await thorough amplitude analysis.
Other, harder to access channels, like $\chi_{cJ}\pi^+K^-$ will have lower statistics and worse purity but are still
usable for the important goal of verifying the states claimed in these channels by other collaborations.  
Survey of all accessible channels, especially in b-baryon decays, in a hunt for new entries to the pentaquark
candidate tables and for the existing ones in new decay modes or new production mechanisms, 
is an important responsibility for the LHCb collaboration given its unique experimental capabilities in this area.  
Search for even more complex multiquark structures, like hexaquarks (among them dibaryons) is also a unique window 
of opportunity. 
Theoretical guidance has been already provided by specific models, but open minded searches should be performed too.

Refining the analysis of the decays that already produced important results ($Z(4430)^+$ and 
$P_c^+$ states) should not be forgotten. Increased data statistics from the on-going Run II, and later from 
the upgraded LHCb, will help this goal. However, even the existing data samples may contain more information than
already extracted. For example, the exotic hadron structures cross many conventional resonances.
Careful investigation of their interference patterns in different parts of the Dalitz plane may provide
firmer evidence for their bound state nature or indicate rescattering processes. 

The LHCb detector is equipped with a trigger on purely hadronic final states with a detached secondary vertex. 
Such triggers have been successfully utilized to study $D(s)$ meson spectroscopy from  
$B(s)^0\to D K^+\pi^-$ and $D \pi^+\pi^-$ final states 
\cite{LHCb-PAPER-2014-035,lhcb3,lhcb5,lhcb4}.
Such studies are also important for interpretation of exotic hadron candidates. 
For example, the only remaining molecular explanation of the $Z(4430)^+$ state is a $D\,\Dbar(2600)$ hypothesis.
This calls for confirmation of the $D(2600)$ state and determination of its quantum numbers ($1^-$ has been assumed).  
With larger future data samples, a possibility of tetraquarks with one $c$ quark inside can perhaps be also explored.
Work on conventional states that constitute ``background'' components to the exotic candidates, like e.g. $K^*$ states 
in decays to $K\pi$ and $\phi K$, or $\Lambda^*$ states in decays to $Kp$ using the other channels than the ones 
known to contain strong exotic signals in the coupled decay mode is likely to be important for future precision amplitude fits. 

The theoretical community already provides plenty of guidance concerning possible interpretation of the existing
exotic hadron candidates, as well as prediction for other channels and states to look for.
It would be very useful to see more theoretical work on ways to distinguish rescattering signals from true bound 
states. At present the rescattering papers play the role of spoilers, 
providing post-dictions offering mundane explanation of the 
observed peaks. Promoting them to models with predictive power would be useful.
The sensitivity of high statistics amplitude analyses may become limited by theoretical limitations of the frameworks employed
in fitting the data. So far, the amplitude analyses in LHCb with exotic hadron components relied on isobar approximation
in spite of its known limitations. Theorists can help design variations of this framework 
to expose the limitations of such approach.
The other known problem is how to describe very broad, often referred to as ``non-resonant'', contributions.   
Since the amplitude analyses involving exotic candidates involve final particles with spin, theoretical works assuming
final particles with no spin are not terribly useful. 

The analysis of angular moments in the conventional hadron channel, reflected into mass distribution in the coupled exotic
channel, motivated by the past BaBar papers, has been also employed in LHCb and promoted to proper statistical analysis in
the $B^0\to\psi'\pi^+K^-$ channel \cite{Aaij:2015zxa}.
Such a method can also be employed in the analysis of other final states, though its
sensitivity is limited to prominent exotic peaks. At best it proves a need for exotic (or rescattering) contributions.
Negative outcomes do not carry useful information, as the sensitivity of this method cannot be evaluated without an amplitude model.
In fact, there is a danger of misinterpretation of null results, thus this method must be applied with proper awareness of
its limitation. When the outcome is positive, then the results are very valuable, but the amplitude analysis is still 
absolutely necessary for the interpretation of the exotic structures.
Suggestions for other alternative ways of analyzing multidimensional data with coupled conventional and exotic decay channels
may be helpful.

\newpage

\aSection{LHCb - Four body decays of $D$ mesons}{S. Harnew \&\ J. Rademacker, for LHCb}
\begin{center}
Sam Harnew \&\ Jonas Rademacker, for the LHCb collaboration \\
\vskip .2cm
{\small H. H. Wills Physics Laboratory, University of Bristol, \\
Bristol, United Kingdom}
\end{center}
\newcommand{\pspoint}{\ensuremath{\mathbf{x}}}
\newcommand{\pspointb}{\ensuremath{\mathbf{\bar{x}}}}
\newcommand{\dpspoint}{\ensuremath{\mathrm{d}^nx}}
\def\PD      {\ensuremath{D}}                 
\def\Dbar    {{\kern 0.2em\overline{\kern -0.2em \PD}{}}}
\def\D       {{\ensuremath{\PD}}}
\def\Db      {{\ensuremath{\Dbar}}}
\def\Dz      {{\ensuremath{\D^0}}}
\def\Dzb     {{\ensuremath{\Dbar{}^0}}}
\def\PK      {\ensuremath{K}}                 
\def\Kp      {{\ensuremath{K^+}}}
\def\Km      {{\ensuremath{K^-}}}
\def\pip    {{\ensuremath{\pi^+}}}
\def\pim    {{\ensuremath{\pi^-}}}
\def\DDb                        {{\ensuremath{\D\Dbar}}}
\def\Kbar    {{\kern 0.2em\overline{\kern -0.2em \PK}{}}}
\def\Kzb     {{\ensuremath{\Kbar{}^0}}}
\def\Dp      {{\ensuremath{\D^+}}}

\subsection{Formalism and differences to three body decays}
\label{sec:intro}
The differential decay rate of a $D$ meson into a final state $f$ of
$N_f$ pseudoscalars with four momenta $p_1, \ldots, p_{N_f}$ is given
by,
\begin{equation}
\mathrm{d} \Gamma[\D \to f(\pspoint)] = 
\left|\mathcal{M}(\pspoint)\right|^2 
\left | 
\frac{ \partial^n {\boldsymbol\phi(\pspoint)}
  }{ \partial (x_1, \ldots, x_n) }
\right | 
\dpspoint,
\end{equation}
where $\mathcal{M}$ is the matrix element describing the $\D \to f$
transition and $| \frac{ \partial^n {\boldsymbol\phi(\pspoint)}
  }{ \partial (x_1, \ldots, x_n) } |$ represents the density of
states at point $\pspoint = (x_1, \ldots, x_n)$ in $n$ dimensional
phase space, with $n= 3N_f -7$. For $N_f=3$, this leads the important
case of the two-dimensional Dalitz plot, discussed much in these
proceedings, and usually parametrised with $s_{12} = (p_1 + p_2)^{2}$
and $s_{23} = (p_2 + p_3)^{2}$. For $N_f=4$, phase space is five
dimensional. Five independent variables analogous to $s_{12}$ and
$s_{23}$ can be formed by combining and squaring different final state
4-momenta. However, such variables cannot fully describe four-body
decay kinematics. The reason for this is related to the 
%fundamentally different 
way in which three and four body decays transform under
parity. In the decay of a pseudoscalar to three pseudoscalars, the 
daughters' four-momenta form a plane in the mother's center-of-mass
frame; for this reason the
effect of the parity operation can also be achieved by a rotation.
Hence there can be no parity violating observables in
such decays and it is indeed possible to parametrise their kinematics
using the parity-invariant observables $s_{12}$ and $s_{23}$.  In
4-body decays, where the daughters are no longer constrained to a
plane, parity violating observables can be defined, and the decay
cannot be fully described using only parity-invariant
parameters. Although this might seem like an unwelcome complication,
it also means that four-body decays provide a new set of parity-odd
observables. These have unique sensitivity to $C\!P$ violation, as
discussed in Section~\ref{sec:podd}.

Usually, $\mathcal{M}$ is modelled using a coherent sum of decay
amplitudes, where each proceeds via a sequence of resonant two-body
decays. The theoretical limitations of this approach are discussed in
detail elsewhere in these proceedings. The amplitude structure of four
body decays is significantly more complicated than in three-body
decays, since there are usually two resonances in each decay sequence,
and there is a much larger variety of possible spin/helicity
structures. This complication also leads to another benefit of
four-body decays: while decays of a pseudoscalar to three
pseudoscalars can only contain intermediate resonances with natural
spin-parity ($J^{P} = 0^{+}, 1^{-}, 2^{+}, \ldots$), there is no such
restriction for four-body decays.

Four body amplitude analyses have for example been carried out in $\Dz
\to K^-\pi^+\pi^-\pi^+$, 
$\Dz\to \Kzb\pi^+\pi^-\pi^0$,
$\Dp \to K^-\pi^+\pi^+\pi^0$, 
$\Dp \to \Kzb\pi^+\pi^+\pi^-$, 
$\Dz \to K^-K^+\pi^-\pi^+$, 
$\Dz \to K^-K^-K^+\pi^+$, and 
$\Dz \to \pi^-\pi^+\pi^-\pi^+$ by the MARK~III,
FOCUS and CLEO collaborations~\cite{MarkIII_K3piModel, Link:2004wx,
  Link:2007fi,FOCUS3Kpi, KKpipiMint}. In this article, we focus on the
use of 4-body charm decays for $C\!P$ violation measurements, and the
role of amplitude models in this context.

\subsection{Measurement of the $C\!P$-violating phase $\gamma$}
\subsubsection{Formalism}
One of the key applications of an amplitude model is in the measurement of the
$C\!P$ violating phase $\gamma$ (or $\phi_3$) in $B^{-}\to\D_{B^{-}}K^{-}$ decays (and
similar) where $\D_{B^{-}}$ represents the following superposition of
\Dz and \Dzb:
\begin{equation}
\D_{B^{-}} \propto \Dz + r_{B}e^{i(\delta_B - \gamma)}\Dzb.
\end{equation}
Here $r_{B}$ is the magnitude of the ratio of the $B^{-}\to\Dz K^{-}$
to $B^{-}\to\Dzb K^{-}$ amplitude, and $\delta_B$ is the a
$C\!P$--conserving phase generated by the strong interaction. The
phase difference $(\delta_B - \gamma)$ between the two decay
amplitudes can be measured in decays of $\D_{B^{-}}$ to a final state
$f$ accessible from both \Dz\ and
\Dzb~\cite{GLW1,GLW2,ADS,DalitzGamma1,DalitzGamma2}. Performing the
measurment in both $B^{-}\to\D_{B^{-}}K^{-}$ and its $C\!P$-conjugate
decay mode allows to disentangle $\delta_B$ and $\gamma$.  The
$\D_{B^{-}}$ decay rate at phase space point \pspoint\ can be
expressed in terms of the \Dz\ and \Dzb\ amplitudes
$\mathcal{A}_{f}(\pspoint)$ and $\bar{\mathcal{A}}_{f}(\pspoint)$ as
\begin{equation}
\Gamma_{\D_{B^{-}}} \propto  \left| \mathcal{A}_{f}(\pspoint) \right|^2
+ r_B^2 \left| \bar{\mathcal{A}}_{f}(\pspoint)\right|^2
+ r_B \left|\mathcal{A}_{f}(\pspoint)
  \bar{\mathcal{A}}_{f}(\pspoint)\right| 
\cos( \delta_{B} - \gamma - \delta(\pspoint))
\end{equation}
where $\delta(\pspoint) \equiv \arg(\mathcal{A}_{f}(\pspoint)
\bar{\mathcal{A}}^{*}_{f}(\pspoint))$. To determine $\gamma$ requires
information on both the magnitude of $\mathcal{A}_{f}(\pspoint)$ and $\delta(\pspoint)$.
%(which implies that of $ \bar{\mathcal{A}}_{f}(\pspoint)$ in the
%absence of \$C\!P$\ violation in charm) . 
While $\left| \mathcal{A}_{f}(\pspoint) \right|^2$ can be taken directly
from $\Dz \to f$ data accessible at the B-factories or LHCb, obtaining
$\delta(\pspoint)$ is more challenging. One approach is to use an
amplitude model to obtain the phase information from a fit $\Dz \to f$ and $\Dzb \to f$
data. This model-dependent approach has been used successfully to
constrain gamma using three-body decays,
e.g.~\cite{Aubert:2005iz,Aubert:2008bd,Poluektov:2004mf,Poluektov:2006ia,LHCb-PAPER-2014-017}. 
With current datasets at LHCb, it should be
possible to reach a similar precision with four-body final states such as
$\Kp K^{-}\pip\pim$ and $\pip\pim\pip\pim$~\cite{Rademacker:2006zx}.

\subsubsection{Model-independent measurement of the $C\!P$-violating phase $\gamma$}

The theoretical limitations of the most widely-used amplitude models
are a key topic discussed in several contributions to these
proceedings. As a result of these limiations, the model-dependent
approach discussed above suffers from significant systematic
uncertainties. This motivates the use of model-independent methods.

The model-independent methods discussed here are based on data-driven
techniques that split phase space into one or more regions and
integrate over these regions. In phase space region $i$ one can define
the complex interference parameter,
\begin{equation}
Z^{f}_{i} =  
\frac{
\int _{\pspoint \in i} \mathcal{A}_{f}(\pspoint)
\bar{\mathcal{A}}^{*}_{f}(\pspoint) \left | \frac{ \partial^n
    {\boldsymbol \phi(\pspoint)} }{ \partial (x_1,\ldots,x_n) }   \right |
\dpspoint
}{
\sqrt{
\int _{\pspoint \in i} \left|\mathcal{A}_{f}(\pspoint)\right|
 \left | \frac{ \partial^n
    {\boldsymbol \phi(\pspoint)} }{ \partial (x_1,\ldots,x_n) }   \right |
\dpspoint
\;
\int _{\pspoint \in i}
\left| \bar{\mathcal{A}}_{f}(\pspoint)\right|^2 \left | \frac{ \partial^n
    {\boldsymbol \phi(\pspoint)} }{ \partial (x_1,\ldots,x_n) }   \right |
\dpspoint
}
}.
 \label{eqn:Zdef}
\end{equation}
 In this region of phase space the $\gamma$-sensitive
interference term in $B^{-}\to(\D\to f) K^{-}$ decays is proportional to $
|Z^{f}_{i}| \cos( \delta_{B} - \gamma - \arg(Z^{f}_{i} ) )$. 
The parameter $Z^f_i$ is equivalent to the coherence factor $R_i^f$
and average strong phase difference $\delta_i^f$ introduced
in~\cite{Atwood:coherenceFactor}, and the $c_i$ and $s_i$ parameters
introduced in~\cite{DalitzGamma1} through $Z^f_i = R_i^f
e^{-i{\delta_i^f}} = c_i + s_i$ (where we follow~\cite{Bondar:2005ki} for
the normalisation of $c_i, s_i$).
The phase of $Z^{f}_{i}$ is a weighted average of
$\delta(\pspoint)$ over $i$, and $|Z^{f}_{i}| \in [0,1]$
describes the dilution of the interference term from integrating over
$i$.
The advantage of this binned approach is that $Z^{f}_{i}$ can be
measured amplitude-model-independently, using decays of well-defined
superpositions of $\Dz$ and $\Dzb$, accessible at the charm threshold
and in charm mixing
\cite{DalitzGamma1,Bondar:2005ki,Atwood:coherenceFactor,coherenceFromMixing,coherenceFromMixing2}.

Somewhat paradoxically, input from an amplitude model is still
important to maximise the sensitivity of model-independent
analyses. With a model it is possible to select regions of phase space
such that $|Z^{f}_{i}|$, which multiplies the $\gamma$-sensitive term,
is as large as possible~\cite{Bondar:2007ir}. Any inaccuracies in the
model will not lead to a systematic bias, but just a reduction in
$|Z^{f}_{i}|$, and a corresponding reduction in sensitivity, which
will be apparent in the statistical uncertainty.
The same model-informed, model-unbiased constraints on $Z^f_i$ provide
valuable input to time-dependent mixing and $C\!P$ violation measurements in
the $\Dz - \Dzb$ system~\cite{Bondar:CharmMixingCP, Malde:2011mk,
  LHCb-PAPER-2015-042}.

\subsubsection{Quantum correlated $\psi(3770)\to \DDb$ decays}

Quantum correlated $\psi(3770)\to \DDb$ decays provide the
well-defined $\Dz - \Dzb$ superpositions needed to measure
$Z^f_i$. For example if
one $D$ meson (the `tag') decays to a $C\!P$ $\pm$ eigenstate, then the
other must be in the following superposition (using the convention
that $C\!P \Dz = + \Dzb$):
\begin{equation}
\D \propto \Dz \mp \Dzb.
\end{equation}
This method has been applied at CLEO-c to measure $Z_i^f$ for a
variety of different final states~\cite{Briere:2009aa, Insler:2012pm,
  Libby:2010nu, Lowery:2009id, Libby:2014rea,
  Evans:2016tlp,Malde:2015mha} including four-body decays $\Dz\ \to \Kp\pim\pip\pim$ (see
Figure~\ref{fig:Zcon}) and $\Dz \to \pip\pim\pip\pim$. In the
latter case, the measurement can be interpreted in terms of the
$C\!P$-even fraction of the $\D^0 \to \pip\pim\pip\pim$
decay~\cite{Nayak:2014tea,Malde:2015mha}.  The constraints have been
used in many $\gamma$ measurements,
including~\cite{BaBar_uses_us:2011up,
  Belle_uses_CLEO_2011,LHCb2012DalitzGamma, LHCb2013GammaCombination,
  LHCb-PAPER-2014-041, LHCb-PAPER-2016-003,LHCb-CONF-2016-001}.

\subsubsection{D-mixing as input for $\gamma$}

Another method to constrain $Z^{f}_{i}$ is by measuring \D-mixing in
large samples of $\Dz\to f$ decays, which are tagged as a \Dz\ at
production~\cite{coherenceFromMixing,coherenceFromMixing2}. At
decay-time $t$, the \D\ meson is in the following superposition:
\begin{equation}
\Dz(t) \propto  g_{+}(t) \Dz + g_{-}(t) \Dzb
\end{equation}
where $g_{+}$ and $g_{-}$ evolve slowly with decay-time and $g_{+}(0) = 1$ and $g_{+}(0) = 0$. 
The exact form of $g_{+}$ and $g_{-}$ depends on the dimensionless
mixing parameters $x$ and $y$, which are proportional to the mass and
width difference between the \D mass eigenstates, respectively.
% and have been measured to a good precision~\cite{several papers including hfag}.

\begin{figure}
\includegraphics[width=0.32\linewidth]{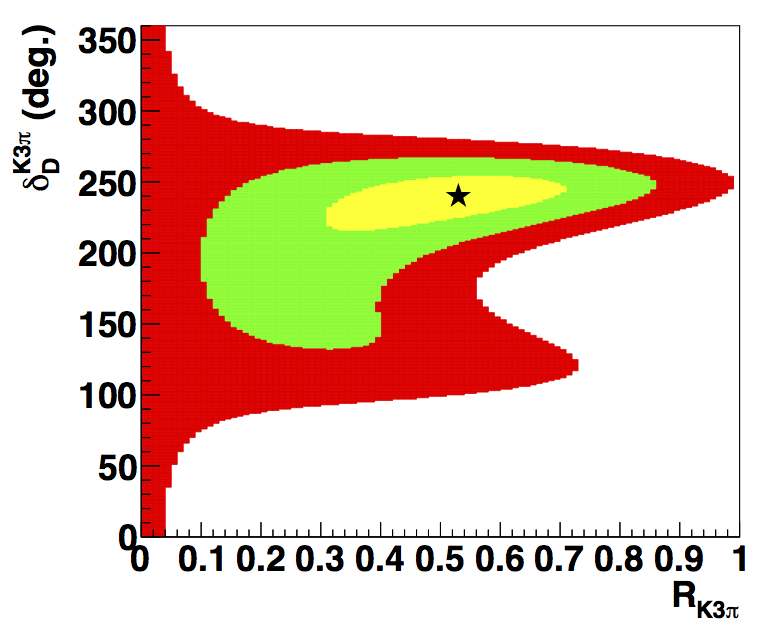}
\includegraphics[width=0.32\linewidth]{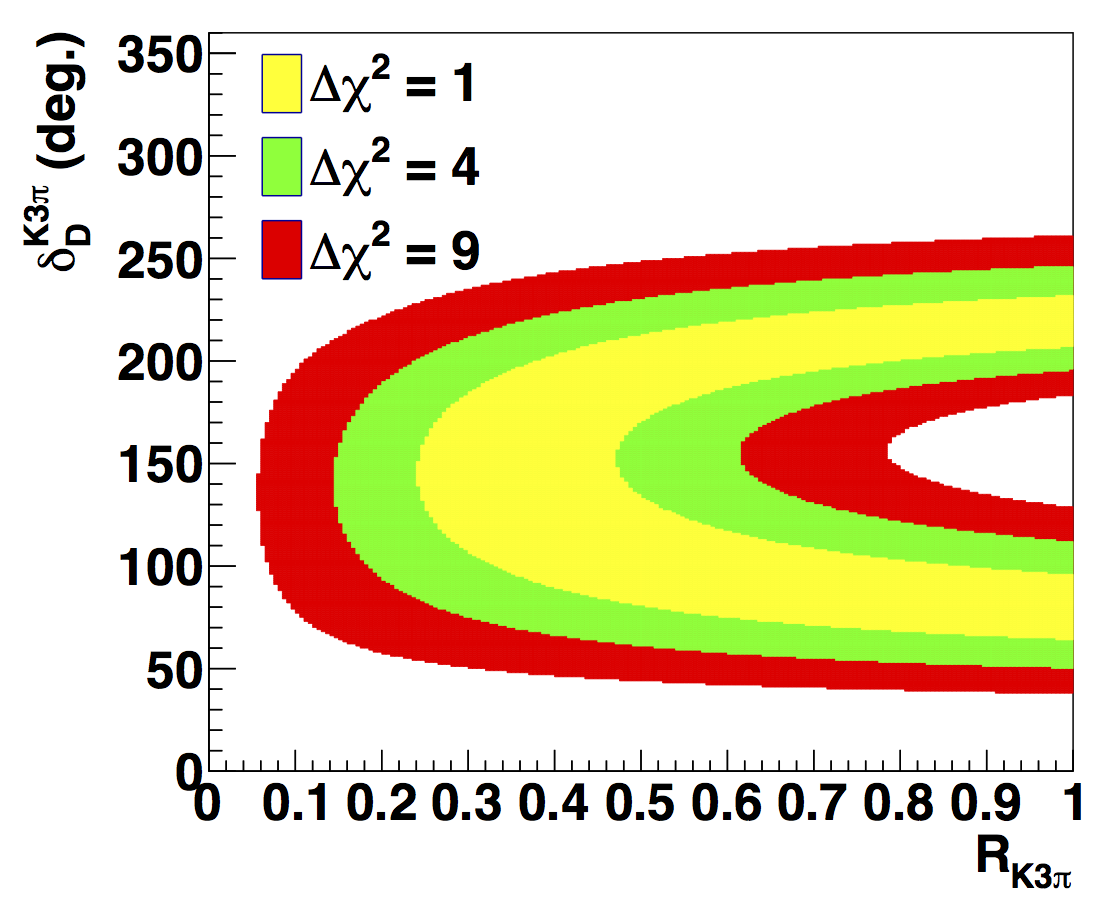}
\includegraphics[width=0.32\linewidth]{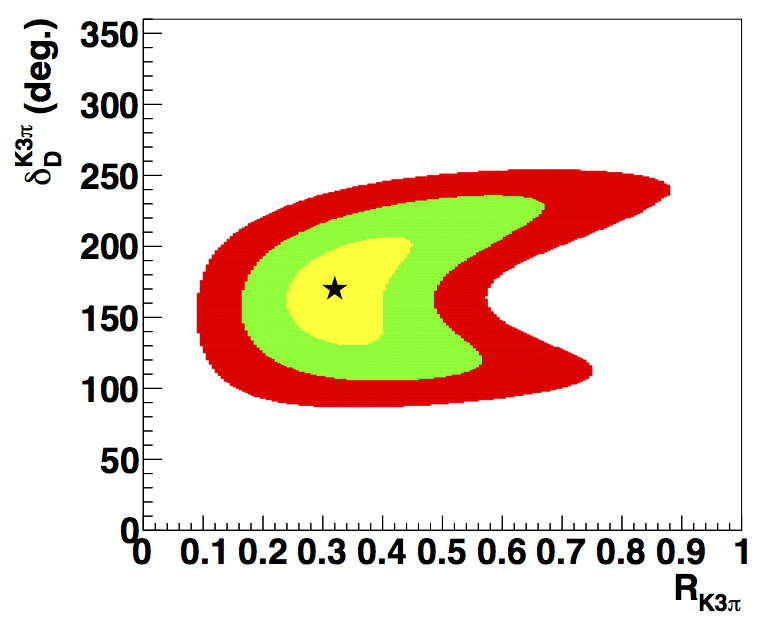}
\caption{Constraints on $Z^{K3\pi} = R_{K3\pi}e^{-i\delta^{K3\pi}_{D}}$ from (left) CLEO-c~\cite{Evans:2016tlp}, (center) LHCb~\cite{LHCb-PAPER-2015-057}, (right) CLEO-c and LHCb combined~\cite{Evans:2016tlp}. \label{fig:Zcon}  }
\end{figure}

Constraints on $Z^f$ have been obtained from \D\ mixing measurements
by LHCb for the final state
$f=\Kp\pim\pip\pim$~\cite{LHCb-PAPER-2015-057}, shown in
Figure~\ref{fig:Zcon}. Here, the omission of the subscript $i$
represents the fact that the integral defining $Z^f$ is performed over
the entire phase space.  The combined constraints on $Z^{K3\pi}$ from
LHCb and CLEO-c are also shown in
Figure~\ref{fig:Zcon}~\cite{Evans:2016tlp}. A binned measurement of
$Z^{K3\pi}_i$ is expected to significantly improve the mode's
sensitivity to $\gamma$~\cite{coherenceFromMixing2}. To obtain good
sensitivity with this approach requires a model for both the \Dz\ and
the \Dzb\ decay amplitude to $ K^+\pi^-\pi^+\pi^-$ to inform the
binning; at the time of writing, no such model exists for the
doubly-Cabibbo suppressed decay amplitude $\Dz \to
K^+\pi^-\pi^+\pi^-$.

\subsection{$C\!P$ violation in four body Charm Decays}
\subsubsection{Local $C\!P$ violation in four body decays}
Comparing $C\!P$-conjugate regions in phase space provides a measure of
$C\!P$ violation. Direct $C\!P$ violation results from interference
effects between two decay paths that have both a strong and a weak
phase difference. In multibody decays, there is a rich interference structure
with many different strong phase differences, so one can expect to find
locally enhanced $C\!P$ violating effects that would be `washed out' if one
integrated over the entire phase space i.e. considered the total decay
rates, only. The even richer structure of four-body decays, compared
to three-body decays, could lead to further enhanced sensitivity, at
the price of a substantially more complex analysis. Comparing the
results of amplitude fits for $C\!P$-conjugate decay modes provides a
measure of $C\!P$ violation. Such a model-dependent search for direct
$C\!P$ violation was performed by CLEO in $\Dz \to K^+ K^- \pi^+
\pi^-$~\cite{KKpipiMint}. Most model-independent searches for local $C\!P$
violation are based on performing a $\chi^2$ comparison of the number
of events in the bins of $C\!P$--conjugate Dalitz plots. This method was
pioneered by BaBar~\cite{Aubert:2008yd} and developed further
in~\cite{miranda1,miranda2}, and has been generalised
to four-body decays in LHCb's analysis of $\Dz \to \Kp\Km\pip\pim$ and
$\Dz \to \pip\pim\pip\pim$~\cite{LHCb-PAPER-2013-041}. All results have been
compatible with $C\!P$ conservation.

\subsubsection{$C\!P$ violation in P-odd observables}
\label{sec:podd}

While the above examples of $C\!P$ violation searches in four body
decays use generalisations of methods used in three-body analyses,
there is class of model-independent $C\!P$ violating parameters in
four-body decays that is inaccessible to Dalitz plot analyses. These
are based on $\hat{T}$ odd
observables~\cite{Golowich:1988ig,Valencia:1988it, Bensalem:2000hq,
  Bigi:2001sg, Bensalem:2002ys, Bensalem:2002pz,
  Datta:2003mj,Gronau:2011cf,Durieux:2015zwa}, where $\hat{T}$
represents motion reversal, also referred to as ``na\"ive time
reversal''~\cite{Durieux:2015zwa}. For the decay of a pseudoscalar to
pseudoscalars, $\hat{T}$ is equivalent to parity, so here we refer to
them as $P$-odd.  As mentioned in Section.\ref{sec:intro}, parity violation
is only possible in decays where $N_f > 3$. In such a case one can
define any parity-odd function $w(\pspoint) = -w(-\pspoint)$, then use
this to define the parity sensitive variable,
\begin{equation}
A_{w} = 
\frac{\int  w(\pspoint)  | \mathcal{A}_{f}(\pspoint) |^{2}  \left |
    \frac{ \partial  {\boldsymbol\phi(\pspoint)} }{ \partial (x_1,
      \ldots, x_n) }   \right | \dpspoint}
{\int  |w(\pspoint)|  | \mathcal{A}_{f}(\pspoint) |^{2}  \left |
    \frac{ \partial  {\boldsymbol\phi(\pspoint)} }{ \partial (x_1,
      \ldots, x_n) }   \right | \dpspoint},
\end{equation}
where a non-zero $A^{f}_{w}$ indicates parity-violation. The
same quantity can also be calculated for the $C\!P$ conjugate decay to
obtain $\bar{A}^{f}_{w}$. Any difference in parity violation between
the two decays i.e. $\mathcal{A}^{C\!P}_{w} = {A}_{w} - \bar{A}_{w} \neq
0$ would then indicate $C\!P$ violation.  This may just seem like a
convoluted method to find $C\!P$-violation, but it is a truly independent
observable; usually a prerequisite of $C\!P$-violation is a non-zero
strong-phase difference $\delta$ i.e. the magnitude of
$C\!P$-violation is proportional to $\sin \delta$.  $C\!P$-violation in
the observable $\mathcal{A}^{C\!P}_{w}$, however, is proportional to
$\cos(\delta_{+} - \delta_{-})$, where $\delta_{+}$ ($\delta_{-}$) is
the strong phase associated to a parity-even (parity-odd) part of the
amplitude~\cite{Durieux:2015zwa}.  Such measurements have been made in
$\Dz \to \Kp\Km\pip\pim$ and $\D_{s}^{+} \to K^+K_S\pip\pim$ using $w(p) =
\mathrm{sign}(p_1\cdot( p_2 \times p_3)
)$~\cite{FOCUSTodd2010,BaBarTodd1,BaBarTodd2,LHCB-PAPER-2014-046}. In
addition to a phase--space integrated result, LHCb's
analysis~\cite{LHCB-PAPER-2014-046} is also carried out locally in
sub--regions of phase space to enhance the sensitivity of the
method. All results so far have been consistent with $C\!P$
conservation.

\subsection{Summary}

Four body decays have unique sensitivity to $P$ and $C\!P$ violating
variables in charm and beauty decay spectroscopy. The vast, clean
samples of such decays available at LHCb and the B-factories, and
their upgrades, will lead to new opportunities as well as new
challenges. The unprecidented statistical precision will require
excellent control of systematic and theoretical
uncertainties. Theoretically well-motivated, practically implementable
amplitude models for four-body decays matching the superb quality of the
data would clearly be highly desirable. At this point, they seem
significantly further away than for the three body case.

In the meantime, the focus in controlling theory uncertainties will
remain on data-driven, model-independent approaches. Such approaches
have been developed, and are continuously being improved, for key
measurements like the precision determinations of $\gamma$ or $C\!P$
violation in charm decays. Many of these methods rely on input from
the charm threshold. For the measurement of $\gamma$, the precision
can be increased further through the use of phase information obtained
from charm mixing. However, even model-independent approaches rely on
input from amplitude models to optimise their precision. Since these
approaches are robust against model-induced biases, even a
`traditional' isobar model with Breit Wigner resonances and the much
vilified non-resonant term is expected to be very useful. Of course,
better descriptions of four body decay amplitudes would be even more
valuable.

\subsection{Acknowledgements}
We gratefully acknowledge funding from the European Research Council
under FP7 / ERC Grant Agreement number 307737.

\newpage

\newpage
\aSection{IIB's View of  Theoretical Landscape for LHCb Experimentalists}{I. I. Bigi}
\begin{center}
Ikaros I. Bigi \\
\vskip .2cm
{\small Department of Physics, University of Notre Dame du Lac\\
Notre Dame, Indiana, USA}
\end{center}

\vspace{3mm}

We have to deal with three cultures: Hadrodynamics (HD), HEP \& lattice QCD (LQCD). 
The first one deals with hadrons \& their diagrams, while the second \& third one with quarks \& gluons in different situations. 
The general challenge is how one can combine them. Obviously I am seen is somewhat 
`biased', but I think \& work in HEP, while I know about the strengths (\& weak points) of LQCD. 
There are sizable differences, as we have seen during the Rio WS. It is not trivial at all to combine them; 
in my view there is no other option to make progress in understanding underlying dynamics. 
We are not in a situation where we are looking for a culprit with a smoking gun at the scene of the crime. 
It is more complex to make the case. 

It is usually said that `model independent' analyses are the best information we can get from the data. To say it with different words: 
the `best fitted' analyses in the future with more data is the final step to understand the underlying dynamics. 
I agree that is very important. However in my view it is {\em not} the final step even with more data. I might been seen as biased 
as a theorist -- however it is crucial to enhance the collaboration between experimenters \& theorists on a long time scale. 
It is crucial to discuss which refined theoretical tools \& thinking are best in complex situations; it depends. One has only to look at the history 
in HEP: the best fits do not always give the best informations.  

In this case we cannot just `trust' diagrams; we are in a different era, where we have to go for real precision. My main points: 
\begin{itemize}
\item
`Popular' candidates are (broken) $SU(3)_{Fl}$ and in particular  U-spin symmetries. They are fine for spectroscopy, 
but not, when one includes weak transitions. The difference between U- vs. V-spin symmetries are `fuzzy' for well-know 
examples like: $D^0 \to K^+K^-$ vs. $D^0 \to \pi^+\pi^-$ and compare with $D^0 \to K^+K^-\pi^+\pi^-$ vs. 
$D^0 \to 2\pi^+2\pi^-$ etc. etc. It was discussed in some details in \cite{CICERONE1}.

Another example: the lifetimes of $B^0$ and $B^0_s$ are the same within 2\%; however we find 
BR$(B^0 \to K^+\pi^-) = (1.96 \pm 0.05)\cdot 10^{-5}$ vs. BR$(B^0_s \to K^+\pi^-) = (0.55 \pm 0.06)\cdot 10^{-5}$. 
There are sizable differences between "inclusive" vs. "exclusive" ones.

\item 
In general re-scattering/final state interactions \cite{1988BOOK,wolfenstein,CICERONE2} happen all the time in two-body final states (FS), 
but even more crucial for many-body FS. 
For the future we have to work on the connection of the `cultures' of HD 
and HEP. Here I use the language of quarks: $\bar q_i q_i \to \bar q_j q_j + \bar q_k q_k \bar q_l q_l +...$; 
it crucially depends on strong forces. However, it does not mean that it is under quantitative control of our understanding of QCD -- unless we get "help" with other tools.

\item 
The impacts of penguin diagrams were first shown about  
$\Delta I=3/2$ $\ll$ $\Delta I=1/2$ in kaon decays and direct CPV in 
$\epsilon^{\prime}/\epsilon_K \neq 0$. They are based on {\em local} operators. The expected SM 
predictions are effected by large uncertainties.  We have also sizable experimental uncertainties:
${\rm Re}(\epsilon^{\prime}/\epsilon_K)_{\rm exp.} = (1.66 \pm 0.23)\cdot 10^{-3}$.   
We have the first result from the LQCD culture \cite{LATTICE1}: 
${\rm Re}(\epsilon^{\prime}/\epsilon_K)_{\rm LQCD} = (0.138 \pm 0.515 \pm 0.443)\cdot 10^{-3}$.  
Obviously our community needs more lattice `data' \& more analyses. Of course the LHCb collaboration will not go after these transitions; my 
point is: when somebody goes after accuracy or even precision, one has be careful to use just diagrams. 

We can discuss the impact of penguin diagrams on inclusive lifetimes of beauty (\& charm) hadrons and in CKM 
suppressed decays of $B_{(s)}$ for $b\to q$ with $q=s,d$. Their impact is described with 
{\em short distance} QCD like for $\Delta \Gamma (B_q)$. 

We know about the impact of penguins in those items in at least semi-quantitatively. 
However the situations are much more `complex' for exclusive decays of heavy flavor hadrons. 
Penguin diagrams might show us the `road', 
but we need much more thinking, working and apply other tools like chiral symmetry, dispersion relations etc.; it needs time.  
\end{itemize}

Now I talk specifically for beauty and charm hadrons: 
\begin{itemize}
\item 
We know that the SM produces at least the leading source of CPV. Therefore the goal is to establish New Dynamics (ND) and maybe even its features. 
It is crucial to measure "regional" CP asymmetries in $B \to K\pi \pi$ vs. $B\to 3K$ and 
$B \to 3 \pi$ vs. $B \to \pi 2K$ with more accuracies. We have learnt that "regional" CP asymmetries  are large. 
The definition of "regional" asymmetries is important; furthermore it depends on the differences between narrow resonances vs. broad ones vs. non-resonances. 
Even CPT invariance seems to be `usable'. 
\item 
We cannot stop at three-body FS; we have to go to four-body FS and probe semi-regional asymmetries. 

\item 
We have to probe CPV in the CKM suppressed decays of $\Lambda_b$. I think we can deal with production  asymmetries 
in $pp$ collisions.  Finally it seems that even Belle II will not compete there.

\item 
Obviously we have to probe CPV both in SCS and DCS in the decays of charm mesons \& baryons. 
The SM gives basically zero 
CPV in DCS ones. Most decays of those give mostly three- and four-body FS. 

Again it is important to probe 
regional asymmetries; we have to go beyond moments with four-body FS and compare $D \to 4 \pi$ vs. 
$D \to \bar KK 2 \pi$. 

\item
However, there is a question: how do you "define regional" asymmetries. In particular it needs some "intelligent" judgment 
about four-body FS; probing phase spaces is not. Furthermore re-scattering connects 
$4\pi$ FS with $\bar KK 2\pi$ one. In general we have to compare $D^0$, $D^+$ \& $D^+_s$ transitions. 

Finally in the future we have to probe DCS decays like: $D^+ \to K^+\pi^+\pi^-/K^+K^+K^0$, 
$D^0 \to K^+\pi^+\pi^-\pi^-/K^+K^+K^-\pi^-$ and $D_s^+ \to K^+K^+\pi^-$. 

\item
We have to probe CPV in the decays of $\Lambda_c^+$ at least. 
\item 
Dynamics close to thresholds are very complex. Probing CP asymmetries give us much more information, where it comes 
from resonances or not, shows our limit so far about understanding strong forces - or the impact of 
New Dynamics \cite{DR_a,*DR_b,*DR_c,KUBIS_a, *KUBIS_b, *KUBIS_c}.

\end{itemize}
My main points are: we have to go beyond diagrams \& U-spin symmetry; they are connected with V-spin symmetry 
based on CPT invariance. 
We have to probe CP asymmetries in baryons, 
non-leading sources of CPV for beauty transitions \& in general for charm ones and the impact of ND \& its features. Again, it is crucial to combine 
two cultures in fundamental dynamics. Correlation, correlations, correlations!

In the real world there is a good reason to give short contributions
in Proceedings. I explain my points in detail in~\cite{BIGIRIO_a,
  BIGIRIO_b}.

\vspace{0.5cm}

{\bf Acknowledgments:} This work was supported by the NSF under the grant numbers PHY-1215979 
\& PHY-1520966. 

\vspace{4mm}

\newpage

\aSection{Parametrization of three-body hadronic $D$ decay amplitudes}{D. Boito \&\ B. Loiseau}
\begin{center}
Diogo Boito$^{a,b}$ and Beno\^it Loiseau$^c$ \\
\vskip .2cm
{\small $^a$Instituto de F\'isica de S\~ao Carlos, Universidade de S\~ao Paulo,\\
CP 369, 13560-970,  S\~ao Carlos, SP, Brazil}\\ \vskip .3cm
{\small $^b$Instituto de F\'isica, Universidade de S\~ao Paulo,\\  
S\~ao Paulo, SP, Brazil}\\ \vskip .3cm
\vskip .2cm
{\small $^c$Sorbonne Universit\'es, Pierre \& Marie Curie et Paris Diderot, IN2P3-CNRS,\\
Laboratoire de Physique Nucl\'eaire et de Hautes Energies, Groupe Phenomenologie, 4 place Jussieu, 75252 Paris, France}
\end{center}
\vskip .3cm

In order to constrain the Dalitz-plot analyses of hadronic
three-body $D$ decays we suggest here parametrizations of $D^+ \to K^-
\pi^+ \pi^+$ and $D^0 \to K^0_S \pi^- \pi^+$  amplitudes that can be
readily implemented in  experimental analyses. 
 Our parametrizations are derived in the
 % derive from the use of the 
quasi-two-body
factorization approach where the two-body final state interactions are
fully taken into account by unitary hadronic form factors.  In
particular, we tackle here the cases of the $S$- and $P$-wave   $K^-\pi^+$ pairs
in $D^+ \to K^- \pi^+ \pi^+$ and $K^0_S \pi^-$ in $D^0 \to K^0_S \pi^- \pi^+$, where the scalar and vector  $K\pi$ form factors play a decisive role.
Parametrizations of other  three-body hadronic $D$ and
$B$ decay amplitudes are in progress and will be presented elsewhere.

\subsection{Introduction: quasi two-body factorization approach}
\noindent
The two-body QCD factorization, as a
leading order approximation in an expansion in $\alpha_s$ and
inverse powers of the $b$ quark mass, has been applied with success to
charmless nonleptonic $B$ decays~(see e.g.
Ref.~\cite{Beneke2003}).  The charm quark mass,
$m_c\sim 1.2$~GeV, is smaller than the bottom quark mass which
  enhances corrections to the factorized results.  %Therefore,
  Factorization for $D$ decays is less predictive %, since 
inasmuch as it does not allow for systematic improvement.
  Nevertheless, as a phenomenological approach, since the initial
articles of Bauer, Stech and Wirbel~\cite{Bauer1987}, the
factorization hypothesis has been applied successfully to $D$ decays,
provided one treats Wilson coefficients as phenomenological parameters to account for possibly important non-factorizable corrections~\cite{Abbasetal}.

So far there is no factorization scheme for three-body decays.
However, three-body decays of $D$ and $B$ mesons have important
contributions from intermediate resonances --- such as those of the
$\rho(770)$ and $K^*(892)$ --- and can therefore be considered as {\it
  quasi two-body} decays.  One assumes that two of the three
final-state mesons form a single state originating from a
quark-antiquark pair, which leads to quasi two-body final states so
that the factorization procedure can be applied.  Then, the three-body
final state is reconstructed with the use of two-body mesonic form
factors to account for the important hadronic final state interactions. 
 For instance, for the $D^0 \to K^0_S \pi^- \pi^+ $ decays,
the three-meson final state $K^0_S \pi^+ \pi^-$ is supposed to be
formed by the quasi two-body pairs, $[K^0_S \pi^+]_{L} \ \pi^-$,
$[K^0_S \pi^-]_{L} \ \pi^+$ and $K^0_S \ [\pi^+{\pi^-}]_{L}$, where
two of the three mesons form a state in $L= S, P$ or $D$ wave.
This framework has been successfully employed to several
three-body $B$ and $D$
decays~\cite{Boitoetal09,BE2009,JPD_PRD89,ElBennichetal09}.

As a concrete example of this procedure let us apply it to the  $\DKpp$ case. The amplitude given by $\sand{K^-\pi^+\pi^+}{{\cal H}_{\rm eff}}{D^+}$ receives contributions
from two topologies and factorizes as
\begin{align}
\label{amplitude}
&{\cal A}(D^+\to K^-\pi^+\pi^+)
=
\frac{G_F}{\sqrt{2}}\cos^2\theta_C
[a_1\langle K^-\pi^+_1|\bar s\gamma^\mu(1-\gamma_5)c|D^+\rangle
        \langle\pi^+_2|\bar u\gamma_\mu(1-\gamma_5)d|0\rangle \nn \\
   & +a_2\langle K^-\pi^+_1|\bar s\gamma^\mu(1-\gamma_5)d|0\rangle
        \langle\pi^+_2|\bar u\gamma_\mu(1-\gamma_5)c|D^+\rangle] + (\pi_1^+ \leftrightarrow \pi_2^+).
\end{align}
In the above expression $G_F$ is the Fermi constant, $\theta_C$ the
Cabbibo angle and $a_{1,2}$~\cite{BurasNPB434_606}  are combinations of Wilson
coefficients of the effective weak Hamiltonian. %, ${\cal H}_{\rm eff}$.
In the expression factorized as above, the $K\pi$ form factors appear
explicitly in the matrix element $\langle K^-\pi^+_1|\bar s\gamma^\mu(1-\gamma_5)d|0\rangle$.
The evaluation of  $\langle K^-\pi^+_1|\bar s\gamma^\mu(1-\gamma_5)c|D^+\rangle$
is less straightforward.
However, assuming this transition to proceed through the dominant intermediate resonances, this
matrix element can also be written in terms of the $K\pi$ form factors~\cite{MG02}. The other matrix elements of Eq.~(\ref{amplitude}) can be written in terms of decay constants or transitions form factors that can be extracted from semi-leptonic decays.

In the next section we cast
%Next section expresses 
the amplitudes for the decays $D^+ \to
[K^- \pi^+]_{S,P}\ \pi^+ $ and $D^0 \to [K^0_S \pi^-]_{S,P,D}\ \pi^+ $
in  forms that can be readily used in  data analyses.

\subsection{Examples of amplitude parametrizations}
\subsubsection*{Parametrization of the  $D^+ \to [K^- \pi^+]_{S,P}\ \pi^+ $ amplitudes}

\noindent
We use the following notation for the invariant masses of the final state $s_1=(p_{K^-}+p_{\pi^+_1})^2$, $s_2=(p_{K^-}+p_{\pi^+_2})^2$  and $s_{12}=(p_{\pi^+_1}+p_{\pi^+_2})^2$.  From the results of Ref.~\cite{BE2009}, obtained  within the quasi two-body factorization approach, one can  parametrize the $S$- and $P$-wave 
amplitudes for $D^+ \to [K^- \pi^+]_{S,P}\ \pi^+ $ as
\begin{align}
\label{ASPD+}
\A^{D^+}_S(s_1,s_2,s_{12})&=  \left[ c_0^{S}(m_D^2 -s_1) +c_1^{S}  \frac{F_0^{D\pi}(s_1)}{s_1}   \right] F_0^{K\pi}(s_1),  \nn\\
\A^{D^+}_P(s_1,s_2,s_{12})&=\left[c_0^P \Omega(s_1,s_2,s_{12}) + c_1^P\left( \frac{s_2-s_{12}}{\Delta^2}    - \frac{1}{s_1}\right)  F_1^{D\pi} (s_1) \right] F_1^{K\pi}(s_1),
\end{align}
with $\Omega(s_1,s_2,s_{12}) = s_2-s_{12}-\Delta^2/s_1$ and $\Delta^2=(m_K^2 -m_\pi^2)(m^2_D-m^2_\pi)$.
 In Eq.~(\ref{ASPD+}), $ c_{0,1}^{S,P}$ are free complex parameters to be fitted.
We have assumed that the Wilson coefficients of the effective weak
Hamiltonian get wave dependent non-factorizable corrections.  The $S$- plus $P$-wave
 amplitude has then 7 (real) free parameters, since one global  phase cannot
be observed.  They can be related to those of
Ref.~\cite{BE2009} through a comparison with Eqs.~(3),~(8)
and~(35) of that work.  If we drop the assumption that the corrections
to Wilson coefficients depend on the angular momentum of the $K\pi$
pair, the number of free parameters is reduced to 4. In Eq.~(\ref{ASPD+}) $F_{0,1}^{D\pi}(s_1)$ are the scalar and vector $D$ to $\pi$ transition form factors and  $F_{0,1}^{K\pi}(s_1)$  the scalar and vector $K\pi$ form factors.

A parametrization of the type of Eq.~(\ref{ASPD+}) has been introduced
%employed successfully in Ref.~\cite{BE2009} in the description of the
$S$- and $P$-wave $K\pi$ pairs.  The parameters in that work were
obtained from integrated branching ratios, and not from a Dalitz plot
analysis. Nevertheless, the results were satisfactory for
$\sqrt{s_{1,2}}\lesssim $1.55~GeV. Additional contributions, e.g. with
higher angular momentum or the isospin-2 $\pi^+\pi^+$ interactions,
are very small in this process.  In a realistic Dalitz plot analysis,
however, they may be required and have to be included in the signal
function through usual isobar model expressions, for example.

\subsubsection*{Parametrization of the  $D^0 \to [K^0_S \pi^-]_{S,P,D}\ \pi^+ $ amplitudes}

In an amplitude analysis, using quasi-two body factorization, Ref.~\cite{JPD_PRD89} has obtained good fits to the Belle Dalitz plot density distribution and to that of a BABAR model for the $D^0 \to K^0_S \pi^- \pi^+ $ decays.
 The following parametrization, for the $D^0 \to [K^0_S \pi^-]_{S,P,D}\ \pi^+ $ amplitudes, is based on the amplitudes derived from Eqs.~(66)  and (68) in Ref.~\cite{JPD_PRD89}.
 One has, with $s_-=(p_{K^0_S}+p_{\pi^-})^2$, $s_+=(p_{K^0_S}+p_{\pi^+})^2$  and $s_{0}=(p_{\pi^-}+p_{\pi^+})^2$,
\bea
\label{ASPD0}
\A^{D^0}_S(s_0,s_-,s_{+})&=& \left( d_0^S+d_1^S s_-\right) F_0^{K\pi}(s_-),  \nn\\
\A^{D^0}_P(s_0,s_-,s_{+})&=&\left( s_0-s_{+} +\frac{\Delta^2}{s_-}\right)   d_0^P F_1^{K\pi}(s_-).
\eea
Here the $S$- plus $P$-wave amplitude  $\mathcal{M}^{D^0}(s_0,s_-,s_{+})=\A^{D^0}_S(s_0,s_-,s_{+})+\A^{D^0}_P(s_0,s_-,s_{+})$ depends on 3 free complex parameters, $d_{0,1}^S$ and $d_0^P$ which can also be related to the parameters appearing in the Eqs.~(66)  and (68) of Ref.~\cite{JPD_PRD89}.

 Despite its small fit fraction, the $D^0 \to [K^0_S \pi^-]_{D}\ \pi^+ $ amplitude can play an important role through interference and from Eq.~(76)  of Ref.~\cite{JPD_PRD89},  it can be parametrized  as
\beq
\label{ADD0}
\A^{D^0}_D(s_0,s_-,s_{+})= \left( d_0^D+d_1^D s_-\right) B_{K^*_2}(s_+,s_-),
\eeq
where  the contribution of the $K^{*-}_2(1430)$ resonance is described by the relativistic Breit-Wigner function $B_{K^*_2}(s_+,s_-)$ whose expression is given by Eq.~(73) of Ref.~\cite{JPD_PRD89}.
Here again the two  complex parameters $d_{0,1}^D$ are related to those entering Eqs.~(76) to (78) of  Ref.~\cite{JPD_PRD89}.

Finally, there will be analogous parametrizations for the $D^0 \to [K^0_S \pi^+]_{S,P,D}\ \pi^- $ amplitudes and the contribution from the $\pi^+\pi^-$  final state interactions can also be parametrized in a similar way by introducing $\pi\pi$ form factors~\cite{JPD_PRD89}.

\noindent
\subsubsection*{Relative phase}

It is important to emphasize that in the total amplitude for the decay
rates one cannot have access to a global phase. Therefore, one of the
global phases of the partial wave amplitudes can be set to zero. This
reduces the number of free (real) parameters in Eqs.~(\ref{ASPD+}) and
(\ref{ASPD0}) by one.  Experimental groups often take the $P$-wave,
dominated by the $\rho(770)$ or $K^*(892)$, as the reference wave and
hence impose, by convention, the $P$-wave phase to be zero.  This
global phase difference between the $S$- and $P$-waves plays an
important role in generating the correct pattern of interferences
observed in the Dalitz plot analyses.  In the parametrization outlined
here this phase difference is obtained empirically from data.

\noindent
\subsubsection*{Form factors}

\noindent
In Eqs.~(\ref{ASPD+}) and (\ref{ASPD0}), the $K\pi$ form factors play
a crucial role. Those employed in our
works are obtained from dispersion relations accounting for
analyticity and unitarity constraints~\cite{ElBennichetal09,FFs1,FFs2,FFs3}.  The scalar form factor includes 
 the contribution of the $K^*_0(800)$ and $K^*_0(1430)$ and the
vector form factor those of $K^*(892)$ and $K^*(1410)$.  The
non-resonant background is automatically included. 
The scalar and vector $D$ to $\pi$ transition form factors
can be parametrized following Ref.~\cite{Melikhov} (see, e.g., Eqs.~(116) and (117) of
Ref.~\cite{JPD_PRD89}).

\subsection{Concluding remarks and outlook}
In this note we suggest the replacement of the sums of Breit-Wigner
expressions, often used in Dalitz plot analyses, by amplitudes
parametrized in terms of unitary form factors. Those are determined by
coupled channel equations based on %using 
experimental meson-meson $T$-matrix
elements together with chiral symmetry and asymptotic QCD
constraints. Our amplitudes also carry information about the weak
vertex, described within the quasi-two body factorization approach.  In
collaboration with J.~P.~Dedonder, B.~El-Bennich, R.~Escribano,
R.~Kami\'nski, L.~Le\'sniak, we are working on
parametrizations for other  three-body hadronic $D$ and
$B$ decays, that will be presented elsewhere.  We will, upon request,
provide C++ functions ready for use, e.g., in CERN/Root.

\vspace{0.2cm}

\ni\textbf{Acknowledgements} It is a pleasure to thank Alberto dos Reis and the organizers of the workshop as well as  the Centro Brasileiro de Pesquisas F\'isicas (CBPF) for hosting
this fruitful meeting. Jean-Pierre Dedonder and Bruno El-Bennich must be thanked for very helpful discussions.  The work of DB was supported by the S\~ao Paulo Research Foundation (FAPESP) grant 2014/50683-0. BL's work 
received support from FAPESP (grant 2015/04772-6) and from LFTC, UNICSUL.

%\end{document}

\newpage

\aSection{Testing the SM with 3-body $B$ Decays}{B. Bhattacharya \&\ D. London}
\begin{center}
Bhubanjyoti Bhattacharya\footnote{bhujyo@lps.umontreal.ca}
and David London\footnote{london@lps.umontreal.ca}\\
\vskip .2cm
{\small {\it Physique des Particules, Universit\'e de Montr\'eal,}\\
{\it C.P. 6128, succ. centre-ville, Montr\'eal, QC,
Canada H3C 3J7}}
\end{center}
\vskip .3cm
% \documentclass[12pt]{article}

% \usepackage{graphicx}
% \usepackage{hyperref}
% \textheight 8.25in
% \textwidth 6in
% \voffset -0.5in
% \hoffset -0.25in

% \def \babar{B{\sc a}B{\sc ar}}
% \def\beq{\begin{equation}}
% \def\eeq{\end{equation}}
% \def\bea{\begin{eqnarray}}
% \def\eea{\end{eqnarray}}
% \def\nn{\nonumber}
% \def\roughly#1{\mathrel{\raise.3ex\hbox
% {$#1$\kern-.75em\lower1ex\hbox{$\sim$}}}}
% \def\lsim{\roughly<}
% \def\gsim{\roughly>}
% \def\ks{K_S}
% \def\kbar{{\bar K}^0}
% \def\bd{B^0}
% \def\bdbar{{\bar B}^0}
% \def\btod{{\bar b} \to {\bar d}}
% \def\btos{{\bar b} \to {\bar s}}
% \def\btokpi{B \to K \pi}
% \def\btokpipi{B \to K \pi \pi}
% \def\btokkk{B \to KK{\bar K}}
% \def\bs{B_s^0}
% \def\bsbar{{\bar B}_s^0}

% \def\bscc{{\bar b} \to {\bar s} c {\bar c}}
% \def\bsss{{\bar b} \to {\bar s} s {\bar s}}
% % Start of document
% % -----------------
% \pagestyle{plain}

% \begin{document}

%\begin{flushright}
%UdeM-GPP-TH-15-xxx \\
%\end{flushright}

% \begin{center}
% \bigskip
% {\Large \bf \boldmath Testing the SM with 3-body $B$ Decays} \\
% \bigskip
% \bigskip
% {\large
% Bhubanjyoti Bhattacharya\footnote{bhujyo@lps.umontreal.ca}
% and David London\footnote{london@lps.umontreal.ca}
% }
% \end{center}

% \begin{flushleft}
% ~~~~~~~~~~~~~~ {\it Physique des Particules, Universit\'e
% de Montr\'eal,}\\
% ~~~~~~~~~~~~~~~{\it C.P. 6128, succ. centre-ville, Montr\'eal, QC,
% Canada H3C 3J7}
% \end{flushleft}

\begin{center}
%\bigskip (\today)
\vskip0.5cm {\Large Abstract\\} \vskip3truemm
\parbox[t]{\textwidth}{An amplitude analysis can be used to extract
  the amplitudes of charmless $B \to PPP$ decays from their Dalitz
  plots. Here we describe two methods that use such an amplitude
  analysis to perform clean tests of the standard model (SM). Both
  methods use flavor SU(3) symmetry. We argue that SU(3)-breaking
  effects, which could be responsible for discrepancies with the SM
  predictions, may be reduced by averaging over the Dalitz plot. We
  show how this conjecture can be tested experimentally. We also
  address some questions, related to rescattering, that arose at the
  LHCb workshop.}

\end{center}

%\thispagestyle{empty}
%\newpage
%\setcounter{page}{1}
% Decrease texheight (for preprint numbers) again
%\textheight 23.0 true cm
%\baselineskip=14pt

\subsection{Introduction}

One of the main purposes of the LHCb workshop on multi-body decays of
$B$ and $D$ mesons was to re-examine the analysis used to reconstruct
the amplitudes of 3-body $B$ decays from their Dalitz plots. Here we
describe two methods that use such an amplitude analysis to perform
clean tests of the standard model (SM).

Under flavor SU(3) symmetry, there are three identical final-state
particles in charmless $B \to PPP$ decays, so that the six
permutations of these particles must be taken into account. This leads
to six final-state symmetries: there are one totally symmetric, one
totally antisymmetric, two mixed-symmetric, and two
mixed-antisymmetric states. In $B \to P_1 P_2 P_3$ decays, the
relative angular momenta of the $P_i$ are not fixed, so all six
symmetry states are possible. The physical $P_1 P_2 P_3$ state is a
linear combination of all six states.

The symmetry states can be isolated with an amplitude analysis. The $B
\to P_1 P_2 P_3$ amplitude is written as
\beq
{\cal M} (s_{12}, s_{13}) = {\cal N}_{\rm DP}\sum\limits_j c_j e^{i\theta_j} F_j
(s_{12}, s_{13})~.
\eeq
The form of the $F_j$ depends on the particular contribution to the
amplitude (resonant or non-resonant), and the $c_j$ and $\theta_j$ can
be extracted from a fit to the Dalitz-plot event distribution. Given
these quantities, ${\cal M}(s_{12},s_{13})$ can be reconstructed. The
amplitude for a state with a given symmetry is then found by applying
this symmetry to ${\cal M}(s_{12},s_{13})$. For example, the
fully-symmetric (FS) final-state amplitude is
\bea
{\cal M}_{\rm FS} & = & 
\frac{1}{\sqrt{6}} \left[ {\cal M}(s_{12},s_{13}) + {\cal M}(s_{13},s_{12}) +
  {\cal M}(s_{12},s_{23}) \right. \nn\\
&& \hskip1.5truecm \left. +~{\cal M}(s_{23},s_{12}) + {\cal M}(s_{23},s_{13}) + {\cal
    M}(s_{13},s_{23}) \right] ~.
\eea

The FS state is particularly interesting because it receives no
contributions from spin-1 resonances. Furthermore, the FS state can be
used to test the SM. Methods are described in the following two
sections.

\subsection{Amplitude Equalities}

It was shown in Ref.~\cite{BPPP_FS} that two amplitude equalities
follow from U-spin transformations. They are
\bea
\sqrt{2} {\cal A}(B^+\to K^+\pi^+\pi^-)_{\rm FS} & = & {\cal A}(B^+\to K^+K^+K^-)_{\rm FS} ~, \nn\\
\sqrt{2} {\cal A}(B^+\to \pi^+K^+K^-)_{\rm FS} &=& {\cal A}(B^+\to \pi^+\pi^+\pi^-)_{\rm FS} ~,
\eea
which we refer to as the $K\pi\pi$-$KK{\bar K}$ and $\pi K{\bar
  K}$-$\pi\pi\pi$ relations, respectively. These relations can be
probed experimentally, providing clean tests of the SM.

{\bf Question:} {\it Flavor SU(3) (or U spin) is only an approximate
  symmetry. Can it really be used in tests of the SM?} {\bf Answer:}
{\it Absolutely, but SU(3) breaking must be kept in mind. In
  particular, if either of the above amplitude equalities is found not
  to hold, it could simply be due to SU(3)-breaking effects.  In all
  of these tests, the possibility of SU(3) breaking must be
  addressed.}

{\bf Question:} {\it Could rescattering break the amplitude
  equalities?}  {\bf Answer:} {\it Only if it includes an
  SU(3)-breaking effect.  Rescattering that respects SU(3) is included
  in the amplitude equalities.}

Consider the $K\pi\pi$-$KK{\bar K}$ relation. It can be written as
\beq
\frac{|{\cal A}(B^+\to K^+K^+K^-)_{\rm FS}|}{\sqrt{2} |{\cal A}(B^+\to K^+\pi^+\pi^-)_{\rm FS}|} = 1 ~.
\eeq
(This also holds for $B^-$ decays.) The $K\pi\pi$-$KK{\bar K}$
relation is momentum dependent, and holds at every point in the Dalitz
plot.  Thus, the above ratio should be measured for each Dalitz-plot
point, and then one should average over all points. This will reduce
the statistical error. Also: recall that the FS amplitudes are
reconstructed from a fit to the full Dalitz plot. The errors on the
ratio at different points are therefore correlated.  These
correlations must be taken into account in calculating the total error
resulting from averaging.

A similar analysis can be done for the $\pi K{\bar K}$-$\pi\pi\pi$
relation.

\subsubsection{SU(3) breaking}

What about SU(3) breaking? In the presence of SU(3) breaking, we have
\beq
\frac{|{\cal A}(B^+\to K^+K^+K^-)_{\rm FS}|}{\sqrt{2} |{\cal A}(B^+\to K^+\pi^+\pi^-)_{\rm FS}|} = |X| ~,
\label{SU3test1}
\eeq
where $X$ is the SU(3)-breaking factor. It is a complex number that
depends in a complicated way on all the group-theoretical
SU(3)-breaking terms, and can take different values at different
points in the Dalitz plot.  Its size is unknown, though, based on
SU(3)-breaking effects in two-body decays, one might naively estimate
its deviation from 1 to be $O(30\%)$. Now, it seems reasonable to
expect that $|X| - 1 > 0$ at some points of the Dalitz plot, and $|X|
- 1 < 0$ at others.  In this situation, averaging over all Dalitz-plot
points will {\it reduce} the effect of SU(3) breaking. If so, the
deviation of $X$ from 1 will be found to be much smaller than
$O(30\%)$.

Of course, there is no guarantee that this occurs for the
$K\pi\pi$-$KK{\bar K}$ relation (and SU(3) breaking could be smaller
simply due to the fact that all spin-1 resonances, and the associated
SU(3) breaking, are absent from the FS amplitudes). However, the key
point is that one can experimentally test whether the theoretical
error is indeed reduced when one averages over the Dalitz plot. 

Now consider a pair of $B$ decays -- one $\btod$, the other $\btos$ --
that are related by U-spin reflection ($d \leftrightarrow s$). It has
been shown \cite{MGUspin} that, in the U-spin limit, the four
observables $B_{d,s}$ (branching ratios) and $A^{CP}_{d,s}$ (direct CP
asymmetries) are not independent, but obey
\beq
-\frac{A^{CP}_s}{A^{CP}_d} \, \frac{\tau(\bd) B_s}{\tau(\bs) B_d} = 1 ~.
\label{Uspinrel}
\eeq
One 3-body decay pair to which this relation applies is $B^+ \to \pi^+
K^+ K^-$ and $B^+ \to K^+ \pi^+ \pi^-$. The relation holds at each
point on the Dalitz plot.

But what about U-spin breaking? Using the same logic as above, we
expect that any U-spin breaking is reduced when one averages over all
Dalitz-plot points. That is, we write
\beq
-\frac{A^{CP}_s}{A^{CP}_d} \, \frac{\tau(\bd) B_s}{\tau(\bs) B_d} = N ~.
\label{SU3test2}
\eeq
In the U-spin limit, $N=1$. By measuring $B_{d,s}$ and $A^{CP}_{d,s}$,
and constructing the above ratio, it is possible to experimentally
determine if an average over all Dalitz-plot points leads to $N \to
1$. If so, this supports the conjecture that SU(3)-breaking effects
are also reduced by averaging. (Note that no amplitude analysis is
required for this test.)

\subsection{Extraction of $\gamma$ using an amplitude analysis}

In this section we describe the method for extracting $\gamma$ from $B
\to PPP$ decays using an amplitude analysis
\cite{BPPPgamma1,BPPPgamma2}.

It was shown in Ref.~\cite{BPPPdiagrams} that the amplitudes for each
$B \to PPP$ symmetry state can be expressed in terms of diagrams.
These are similar to those of two-body $B$ decays ($T$, $C$, etc.),
except that (i) they are momentum dependent, and (ii) for $B \to PPP$
decays one has to ``pop'' a quark pair from the vacuum. Diagrams have
the subscript ``1'' (``2'') if the popped quark pair is between two
non-spectator final-state quarks (two final-state quarks including the
spectator).

{\bf Question:} {\it Doesn't rescattering ruin the diagrammatic
  description of amplitudes? For example, what do penguin diagrams
  even mean in the presence of rescattering?} {\bf Answer:} {\it There
  was a great deal of confusion at the workshop about diagrams and
  rescattering. Indeed, the simple use of penguin diagrams was called
  into question. However, this is all a red herring. The diagrams
  {\underline{include}} rescattering. For example, consider the decay
  $B^+ \to \pi^+ K^0$, which is ${\bar b} \to {\bar s} d {\bar d}$ at
  the quark level. One of the diagrams contributing to this decay is a
  gluonic penguin: ${\bar b} \to {\bar s} g^* (\to d {\bar d})$. Now,
  one of the gluonic penguin diagrams has an internal $c$ quark, and
  is proportional to $V_{cb}^* V_{cs}$.  It represents the tree-level
  decay ${\bar b} \to {\bar c} c {\bar s}$, with the $c {\bar c}$ pair
  rescattering to $d {\bar d}$. If the momentum transfer is large,
  this is short-distance scattering, and can be thought of as $c {\bar
    c} \to d {\bar d}$ via the exchange of a single gluon. However, if
  the momentum transfer is small, this becomes a long-distance effect
  involving multiple gluon exchange, or taking place at the hadronic
  level. Regardless, this penguin diagram {\underline{includes}} such
  rescattering processes as $B^+ \to D_s^+ {\bar D}^0 \to \pi^+ K^0$.
  The point is that the diagrammatic decomposition of a decay
  amplitude takes rescattering into account, and this holds for $B \to
  PPP$ decays.}

Now consider the five decays $\bd \to K^+\pi^0\pi^-$, $\bd \to
K^0\pi^+\pi^-$, $B^+ \to K^+\pi^+\pi^-$, $\bd \to K^+ K^0 K^-$, and
$\bd \to K^0 K^0 \kbar$. The $\btokpipi$ diagrams involve a popped
$u{\bar u}$ or $d{\bar d}$ quark pair, while those of $\btokkk$ have a
popped $s{\bar s}$ pair. But flavor SU(3) symmetry makes no
distinction between $u{\bar u}$, $d{\bar d}$ or $s{\bar s}$, so that
the five amplitudes are written in terms of the same diagrams.

The decay amplitudes for the FS states are given by
\bea
\label{effamps}
2 A(\bd \to K^+\pi^0\pi^-)_{\rm FS} &=& b e^{i\gamma} - \kappa c ~, \nn\\
\sqrt{2} A(\bd \to K^0\pi^+\pi^-)_{\rm FS} &=& -d e^{i\gamma} - {\tilde P}'_{uc} e^{i\gamma} - a + \kappa d ~, \nn\\
\sqrt{2} A(B^+ \to K^+ \pi^+ \pi^-)_{\rm FS} &=& -c e^{i\gamma} -{\tilde P}'_{uc} e^{i\gamma} - a + \kappa b ~, \nn\\
\sqrt{2} A(\bd \to K^+ K^0 K^-)_{\rm FS} &=& \alpha_{SU(3)} (-c e^{i\gamma} -{\tilde P}'_{uc} e^{i\gamma} - a + \kappa b ) ~, \nn\\
A(\bd \to K^0 K^0 \kbar)_{\rm FS} &=& \alpha_{SU(3)} ({\tilde P}'_{uc} e^{i\gamma} + a ) ~,
\eea
where ${\tilde P}'_{uc}$ is a penguin diagram, $a$-$d$ are linear
combinations of diagrams (see Ref.~\cite{BPPPgamma2}), and 
\beq
\kappa \equiv - \frac{3}{2} \frac{|V^*_{tb} V_{ts}|}{|V^*_{ub} V_{us}|} \frac{c_9+c_{10}}{c_1+c_2} ~.
\eeq
There are many possible SU(3)-breaking effects between the $\btokpipi$
and $\btokkk$ diagrams, but they cannot all be included in a
fit. Instead, we assume that there is only a single SU(3)-breaking
parameter, $\alpha_{SU(3)}$.

As written above, the five FS amplitudes depend on 11 theoretical
parameters: the magnitudes of ${\tilde P}'_{uc}$ and $a$-$d$ (5),
their relative strong phases (4), $\gamma$, and $|\alpha_{SU(3)}|$.
But there are 13 experimental observables: the decay rates and direct
CP asymmetries of each of the five processes, and the indirect CP
asymmetries of $\bd \to K^0\pi^+\pi^-$, $\bd \to K^+ K^0 K^-$ and $\bd
\to K^0 K^0 \kbar$.  With more observables than theoretical
parameters, $\gamma$ can be extracted from a fit, even with the
inclusion of $|\alpha_{SU(3)}|$ as a fit parameter.

In addition, since the SU(3)-breaking effects are momentum dependent,
their effect on $\gamma$ will vary from point to point on the Dalitz
plot. Thus, as before, we expect that averaging over all Dalitz-plot
points will reduce the effect of SU(3) breaking, so that the extracted
value of $\gamma$ will be close to its true value. This also holds for
$\alpha_{SU(3)}$ alone.

In order to perform the fit, one must first extract the FS amplitudes
${\cal M}_{\rm FS}$ and ${\overline{\cal M}}_{\rm FS}$ for each of the
five processes and their CP conjugates using an amplitude analysis of
their Dalitz plots. Then, using these FS amplitudes, one generates the
observables for each decay. The effective CP-averaged branching ratio
($X$), direct CP asymmetry ($Y$), and indirect CP asymmetry ($Z$) are
obtained at each Dalitz-plot point as follows:
%BB X, Y, Z definition
\bea 
\label{XYZdef}
X(s_{12}, s_{13}) &=& |{\cal M}_{\rm FS}(s_{12}, s_{13})|^2 + |\overline
{{\cal M}}_{\rm FS}(s_{12}, s_{13})|^2~, \nn \\
Y(s_{12}, s_{13}) &=& |{\cal M}_{\rm FS}(s_{12}, s_{13})|^2 - |\overline
{{\cal M}}_{\rm FS}(s_{12}, s_{13})|^2~, \nn \\
Z(s_{12}, s_{13}) &=& {\rm Im}\left[{\cal M}^*_{\rm FS}(s_{12}, s_{13})
~\overline{{\cal M}}_{\rm FS}(s_{12}, s_{13})\right]~.
\eea

Given the theoretical expressions for the FS amplitudes
[Eq.~(\ref{effamps})], we know how $X$, $Y$ and $Z$ depend on the
theoretical unknowns. And given the values of these observables, one
can perform a fit at each Dalitz-plot point. Of all the theoretical
parameters, only $\gamma$ is the same at each point (i.e., it is
momentum-independent). One can therefore combine the results of all
the fits to extract the preferred value of $\gamma$. Note that the
errors on $X$, $Y$ and $Z$ at different points are correlated. These
correlations must be taken into account in calculating the total error
on $\gamma$.

In fact, \babar\ has measured the Dalitz plots for each of the five
decays of interest. In Ref.~\cite{BPPPgamma2}, we performed a
preliminary fit using this data and found the following intriguing
result. There are four preferred values for $\gamma$:
\beq
(31^{+2}_{-3})^\circ ~~,~~~~ (77 \pm 3)^\circ ~~,~~~~ (258^{+4}_{-3})^\circ ~~,~~~~ (315^{+3}_{-2})^\circ ~.
\eeq
Three of these indicate new physics (is this a ``$K \pi \pi$-$KK{\bar
  K}$ puzzle''?). But one solution -- $(77 \pm 3)^\circ$ -- is
consistent with the standard model.

Furthermore, we found that, when averaged over the points,
$|\alpha_{SU(3)}| = 0.97 \pm 0.05$ is found, i.e., this SU(3) breaking
is small. This supports the conjecture that the full SU(3) breaking is
small when averaged over the Dalitz plot. Of course, as noted above,
the full effect of SU(3) breaking is not simply encoded in
$\alpha_{SU(3)}$. However, an estimate of this theoretical error can
be obtained from the experimental measurements of SU(3) breaking in
Eqs.~(\ref{SU3test1}) and (\ref{SU3test2}).

\subsection{Conclusions}

An amplitude analysis of charmless $B \to PPP$ decays can be used to
construct the fully-symmetric (FS) amplitude of the $PPP$ final state.
We have described two methods that use the FS amplitudes of 3-body $B$
decays to perform clean tests of the SM. One of these uses the fact
that, within flavor SU(3), the SM predicts $\sqrt{2} {\cal A}(B^+\to
K^+\pi^+\pi^-)_{\rm FS} = {\cal A}(B^+\to K^+K^+K^-)_{\rm FS}$ and
$\sqrt{2} {\cal A}(B^+\to \pi^+K^+K^-)_{\rm FS} = {\cal A}(B^+\to
\pi^+\pi^+\pi^-)_{\rm FS}$. The other is a method, also using SU(3),
for extracting the weak phase $\gamma$ from the FS states of
$\btokpipi$ and $\btokkk$ decays. This value of $\gamma$ can be
compared with that found in independent measurements. In both of these
methods, SU(3)-breaking effects could lead to discrepancies with the
SM predictions.  We argue that such effects may be reduced by
averaging over the Dalitz plot, and show how this conjecture can be
tested experimentally.

At the LHCb workshop there was much discussion about rescattering and
its effect on the diagrammatic description of $B$ decay amplitudes,
particularly penguins. In fact, this concern is misplaced -- the
diagrams {\it include} rescattering. As regards the above methods for
testing the SM, rescattering is important only insofar as it breaks
SU(3). Rescattering that respects SU(3) will not lead to discrepancies
with the SM.

\bigskip
\noindent
{\bf Acknowledgments}:
This work was financially supported by the IPP (BB), and by NSERC of
Canada.

% \begin{thebibliography}{99}

% \bibitem{BPPP_FS}
% B.~Bhattacharya, M.~Gronau, M.~Imbeault, D.~London and J.~L.~Rosner,
%   ``Charmless $B\to PPP$ decays: The fully-symmetric final state,''
%   Phys.\ Rev.\ D {\bf 89}, no. 7, 074043 (2014)
%   [arXiv:1402.2909 [hep-ph]].
%   %%CITATION = ARXIV:1402.2909;%%

% \bibitem{MGUspin} 
% M.~Gronau,
%   ``U spin symmetry in charmless B decays,''
%   Phys.\ Lett.\ B {\bf 492}, 297 (2000)
%   [hep-ph/0008292].
%   %%CITATION = HEP-PH/0008292;%%

% \bibitem{BPPPgamma1}
% N.~Rey-Le Lorier and D.~London,
%   ``Measuring gamma with $B \to K \pi \pi$ and $B \to K K \bar{k}$ Decays,''
%   Phys.\ Rev.\ D {\bf 85}, 016010 (2012)
%   [arXiv:1109.0881 [hep-ph]].
%   %%CITATION = ARXIV:1109.0881;%%

% \bibitem{BPPPgamma2}
% B.~Bhattacharya, M.~Imbeault and D.~London,
%   ``Extraction of the CP-violating phase $\gamma$ using $B \to K \pi \pi$ and $B \to K K {\bar K}$ decays,''
%   Phys.\ Lett.\ B {\bf 728}, 206 (2014)
%   [arXiv:1303.0846 [hep-ph]].
%   %%CITATION = ARXIV:1303.0846;%%

% \bibitem{BPPPdiagrams} 
% N.~Rey-Le Lorier, M.~Imbeault and D.~London,
%   ``Diagrammatic Analysis of Charmless Three-Body B Decays,''
%   Phys.\ Rev.\ D {\bf 84}, 034040 (2011)
%   [arXiv:1011.4972 [hep-ph]].
%   %%CITATION = ARXIV:1011.4972;%%

% \end{thebibliography}

% \end{document}

\newpage

\aSection{Thoughts on B, D and $\Lambda_b$ decays}{E. Oset}
\begin{center}
Eulogio Oset
\vskip .2cm
{\small Departamento de
F\'{\i}sica Te\'orica and IFIC, Centro Mixto Universidad de
Valencia-CSIC Institutos de Investigaci\'on de Paterna, Aptdo.
22085, 46071 Valencia, Spain}
\end{center}
\subsection{Introduction}
Rather than making a review of my talk, and following the
recommendation of the organizers, here go some thoughts stimulated by
the discussions in the sessions and my view of these current issues.

\subsection{Three mesons in the final state}

One must differentiate two cases: 
\begin{enumerate}[a)]
\item One meson interacts weakly with the others.
\item The three mesons interact strongly among themselves. \\
\end{enumerate}

Case a) is for instance $B^0_s$ decay into $J/\psi$ and $\pi^+
\pi^-$. The $J/\psi$ and $\pi^+$ interaction can be neglected (even if
some resonance is seen in this channel because other more important
channels are not open). Then one concentrates on the $\pi \pi$
interaction. Recall that in this decay one has $s \bar s$ at the end,
which after hadronization gives $K \bar K$. Then how does one get $\pi
\pi$ at the end? The answer is re-scattering. This is the first lesson
in all these processes: Considering the interaction of the mesons is
essential to correctly interpreting the data. Second thought: Better
use a scheme with coupled channels. You produce $K \bar K$ at the
first step and measure $\pi \pi$ at the end. It is the re-scattering
in coupled channels that produces the pions at the end.

  When taking into account meson meson interaction, like in other cases, unitarity, analyticity, chiral symmetry, etc., should be implemented in the amplitudes. Although many approaches are suited, the chiral unitary approach conjugates all these things in a technically easy way \cite{npa,ramonet}. \\

Case b) is more difficult. In principle one should make approaches resembling Faddeev equations. This is very difficult to implement for experimental analysis. The minimum requirement is the addition of the proper interacting amplitudes of each pair to guarantee that all possible resonances will appear. \\

Note also that in these processes, like in the $\Lambda_b \to K^- p
J/\psi$, there is also a non resonant term from tree level production,
meaning the production of the particles prior to their final state
interaction. For this it is important to look at the process at the
quark level and perform the hadronization to obtain final mesons. This
is important because in some cases, like $J/\psi$ and $\pi^+ \pi^-$ in
the $B_s$ decay, there is no tree level, there is tree level for $K
\bar K$ but not for $\pi \pi$, see \cite{liang}. This is also a point
where the theoreticians can help, evaluating or estimating the
dependence of the tree level in the invariant masses.

\subsection{D with high statistics} 
There is nothing wrong about having such fantastic statistics in $D$
decays. In spite of it, there are still many observables that are
relevant and have not been properly exploited. One such example is the
mass distributions of $\pi \pi$, $\pi \eta$, $\pi K$ in $D^0$ decays
into $K^0_S$ and $f_0(500)$, $f_0(980)$, $a_0(980)$ which can help in
our understanding of the nature of the scalar resonances \cite{dai}.

\subsection{Quarks versus mesons}
The description of the weak processes in terms of quarks is realistic and most useful, so let us stick to it. However, if we observe mesons in the final states these quarks have to hadronize. From this moment on we have entered the arena of meson interactions, which thanks to Weinberg \cite{Weinberg:1978kz,Gasser:1983yg}, we know can be described more efficiently in terms of effective theories, chiral Lagrangians, local hidden gauge approach \cite{bando}, etc. 
  
Concerning the discussion of the $f_0(500)$ and $f_0(980)$ described in terms of quark components in some experimental papers, one should recall that these resonances are well described in terms of meson interactions in the chiral unitary approach. This does not mean that quarks are not there, simply that they are not the efficient degrees of freedom to account for these resonances. There is another way to look at this. In works where one starts with a $q \bar q$ seed to represents the scalars and then unitarizes the models to account for the inevitable coupling of these quarks to the meson meson components, it turns out that the meson meson components "eat up" the seed and remain as the only relevant component of the wave function 
\cite{vanBeveren:1986ea,Tornqvist:1995ay,Fariborz:2009cq,Fariborz:2009wf}.

\subsection{Threshold effects and new resonances}
This is an interesting topic not yet exploited and that should prove fruitful in the future. There are some resonances that appear as bound states of pairs of mesons (sometimes several channels collaborate) and are not trivially identified as such. One such example is the $D_{s0}^*(2317)$ which in most works appears roughly as a bound state of $DK$ \cite{gamermann}.  Another one is a predicted $D \bar D $ bound state around 3720 MeV  \cite{gamermann} (the equivalent to the $f_0(980)$ which stands as mostly a bound state of $K \bar K$). This state is difficult to identify because it is below the $D \bar D$ threshold, although the analysis of the $D \bar D$ spectrum close to threshold in the $e^+ e^- \to J/\psi D \bar D$ reaction supports the existence of this state \cite{danires}. 

In some papers we have stressed the relevant information to this respect that one finds by measuring simultaneously the $K D$ mass distribution close to threshold and the strength of the  $D_{s0}^*(2317)$ formation which will be seen in the $D_s \pi $ channel \cite{miguel,sekihara}. Similar cases for the production of X,Y,Z resonances are discussed in \cite{liangxyz}.

\subsection{When can we be sure to have a new resonance and not a threshold effect or reflection of another resonance?}

   This is not trivial but has a solution. Reflection of a resonance in one invariant mass for the other invariant mass of the Dalitz plot should come automatically in the present partial wave analysis that LHCb is doing. For instance in the analysis of the $\Lambda_b \to J/\psi p K^-$ \cite{penta} one can see in fig. 3 that the $\Lambda(1405)$ and $\Lambda(1520)$ resonances in the $K^- p $ distribution give rise to some broad peaks in the $J/\pi ~p $ distribution. 

  Detecting threshold effects from other channels requires to use a
  coupled channel analysis with a proper unitarization in coupled
  channels. This would require a different approach to the present
  partial wave analysis, feasible in principle.  In between, we will
  have to rely on theoretical calculations for such possible
  effects. Yet, we should be warned that not every possible threshold
  effect will revert into a visible peak, or account for the observed
  peaks. In this respect, the possible explanations for the narrow
  peak observed in the pentaquarks search of~\cite{penta}, all rely on
  a production amplitude involving the scattering amplitude of $J/\psi
  N$ or similar channels, which are necessarily small due to the OZI
  rule, and hence do not seem too likely to explain the big narrow
  peak observed.

\newpage

\aSection{A fully unitary and analytic description  
of  multichannel mesonic scattering and 
production processes}{G. Rupp \&\ E. van Beveren}
\begin{center}
George Rupp$^a$\footnote{george@ist.utl.pt} and Eef van Beveren$^b$\footnote{eef@teor.fis.uc.pt}\\
\vskip .3cm
{\small $^a$Centro de F\'{\i}sica e Engenharia de Materiais Avan\c{c}ados\\
{\it Instituto Superior T\'{e}cnico, Universidade de Lisboa}\\
{\it Edif\'{\i}cio Ci\^{e}ncia, P-1049-001 Lisboa, Portugal}}\\ \vskip .3cm
{\small $^b$Centro de F\'{\i}sica Computacional\\
{\it Departamento de F\'{\i}sica, Universidade de Coimbra}\\
{\it P-3004-516 Coimbra, Portugal}}
\end{center}

Hadron spectroscopy is incomparably more involved than atomic spectroscopy,
because of the multitude of strongly coupling decay channels and the usually
large resonance widths, which can even be of the same order of magnitude as the 
level splittings. This reality is hardly compatible with the continued
use of Breit-Wigner (BW) parametrisations of hadronic resonances, as BW shapes
can only be accurate for relatively narrow resonances, far from any threshold.
But also the often used {\it Flatt\'{e}} \cite{PLB63p224} form to parametrise
two-channel resonances near an important inelastic threshold can lead to
pathological effects, as in the often cited LASS \cite{NPB296p493} analysis
of $S$-wave $K\pi$ scattering, due to an unphysically large effect of the
subthreshold $K\eta^\prime$ channel a low energies.

Clearly, it would be very desirable for data analysts to have at their disposal
a simple-to-use expression for the fitting of multichannel hadronic data that
manifestly satisfies unitarity and analyticity, besides a proper handling of
overlapping resonances and also accounting for subthreshold contributions.
Now, such a framework has been proposed by us for mesons many years ago, on the
basis of the empirically observed dominance of OZI-allowed decays, triggered by
quark-pair creation from the vacuum and so with $^{3\!}P_0$ quantum numbers.
Of course, experimentalists need model-independent formulae to analyse their
data. However, as we shall show right away, our expressions for the amplitudes
can be stripped of most or even all of their model dependence, without
affecting the nice properties of the formalism.

In the year 2001, Jeffrey Appel, co-spokesperson of the E791 Collaboration,
contacted us by email \cite{Appel01} in connection with a Comment we had
published \cite{EPJC10p469} on a paper by T\"{o}rnqvist \cite{ZPC68p647}, in
which we criticised the failure to find a scalar $\kappa$ meson (nowadays
called $K_0^*(800)$), besides the $\sigma$ meson (now $f_0(500)$
``rehabilitated'' by T\"{o}rnqvist \& Roos in Ref.~\cite{Tornqvist:1995ay}. Appel
informed us that E791 had been encouraged by our Comment and indeed observed
preliminary evidence of a $\kappa$ resonance. In the ensuing email exchange, he
also challenged us \cite{Appel01} to provide easy-to-use formulae to fit the
awkward light scalars, as in our original work predicting a complete
light scalar nonet \cite{ZPC30p615} a very complicated multichannel quark
model had been employed, which was indeed impracticable for experimental
anaysis. Thus, we developed a very simple yet fully unitary and analytic ansatz
for non-exotic mesonic resonances without mixing in the quark sector and only
one open decay channel, hence particularly suited to describe $S$-wave $K\pi$
scattering below the $K\eta^\prime$ threshold. (Note that the $K\eta$ channel
almost entirely decouples for a realistic pseudoscalar mixing angle). The
resulting expression for the cotangent of the phase shift reads
\cite{EPJC22p493}
\begin{equation}
\cot(\delta_L) \; = \; \frac{n_L(ka)}{j_L(ka)} \; - \; \left\{2\lambda^2\mu\,ka
\,j_L^2(ka)\sum_{i=0}^N\,\frac{B_\ell^{(i)}}{E-E_\ell^{(i)}}\right\}^{-1} \; ,
\label{single}
\end{equation}
where $j_L$ and $n_L$ are spherical Bessel and Neumann functions, respectively,
with $L$ the orbital angular momentum in the meson-meson decay channel, $k$ and
$\mu$ are the relativistically defined momentum and reduced mass in that
channel, $a$ is the decay
radius of string breaking for OZI-allowed decay, $\lambda$ is an overall
coupling constant, $N$ is the number of included bare (``quenched'')
$q_1\bar{q_2}$ states, $\ell$ is the orbital angular momentum of these states,
$B_\ell^{(i)}$ are adjustable constants, $E_\ell^{(i)}$ are the real energies
of the bare $q_1\bar{q_2}$ states, and $E$ is the total energy of the
system, which becomes complex for physical mesonic resonances. The expression
in Eq.~(\ref{single}) was then fitted \cite{EPJC22p493} to experimental
$S$-wave $K\pi$ phase shifts, with only two bare strange scalar quark-antiquark
states included having fixed energies as predicted by the model of
Ref.~\cite{ZPC30p615}, and the rest of the infinite sum approximated by a
negative constant, while absorbing $\lambda$ in the constants $B_0^{(i)}$. Not
only was the resulting four-parameter fit of excellent quality, up to
1.45~GeV, but it also yielded automatically very reasonable values for the
$\kappa$ and $K_0^\star(1430)$ resonance pole positions \cite{EPJC22p493},
which can be obtained from Eq.~(\ref{single}) by finding the complex 
energies $E$ that satisfy $\cot(\delta_0)=i$. In Ref.~\cite{IJTPGTNLO11p179}
we developed a very detailed coupled-channel formalism, called
Resonance-Spectrum Expansion (RSE), to show the relation between our previous
coordinate-space model as employed e.g.\ in Ref.~\cite{ZPC30p615} and the
simple, more empirical expression in Eq.~(\ref{single}).

In Ref.~\cite{PRL91p012003} we demonstrated the predictive power of
Eq.~(\ref{single}), by applying exactly the same expression to the then
just observed enigmatic and very narrow $D_{sJ}(2317)$ meson, with unaltered
parameters, the only difference being the substitution of the mass of a strange
quark mass by that of a charmed quark. This allowed to explain the low mass of
what is now established as the scalar $c\bar{s}$ meson $D_{s0}^\star(2317)$,
while also predicting a broad $D_0^\star$(2100--2300) resonance, now listed in
the PDG meson summary table as $D_0^\star(2400)^0$, with a mass of
$(2318\pm29)$~MeV and a width of $(267\pm40)$~MeV.

Now, from an experimental point of view one may argue that the formula in 
Eq.~(\ref{single}) is still too model dependent. However, this dependence can
be almost completely eliminated. For instance, the bare energies
$E_\ell^{(i)}$ can be taken as free fit parameters instead of resulting from an
unquenched quark model like that of Ref.~\cite{ZPC30p615}. This would be
similar to fitting resonance masses and widths in a BW or Flatt\'{e}
parametrisation, with the additional advantage of needing to fit only one real
quantity for each resonance, as the pole positions follow straightforwardly
from Eq.~(\ref{single}). Moreover, the employed spherical Bessel and Neumann
functions may even be replaced by more general functions. We shall come back to
this point below. 

Clearly, most mesonic resonances have several inelastic decays, so that the
one-channel formula in Eq.~(\ref{single}) is inadequate in those cases. Thus,
we have also developed more general expressions, which can not only account
for multiple decays, but also for more than one $q\bar{q}$ channel. This is
important, for instance, when different $q\bar{q}$ angular-momentum states
having the same quantum numbers can mix, e.g.\ $^{3\!}S_1$ and $^{3\!}D_1$, or
when one is dealing with light isoscalar mesons, which are in principle
mixtures of $n\bar{n}$ ($n=u,d$) and $s\bar{s}$. Together with David Bugg and
Frieder Kleefeld, we applied such an extension to the light scalar nonet 
\cite{PLB641p265}, achieving a detailed description of phase shifts,
inelasticities, and line shapes, besides realistic pole positions. Another
application was a more detailed calculation \cite{PRL97p202001_a,*PRL97p202001_b} of the charmed
scalars $D_{s0}^\star(2317)$ and $D_0^\star(2400)$, also predicting their
first radial excitations, the former of which is a candidate for the
$D_{sJ}(2860)$ resonance.

The most general RSE framework for non-exotic meson-meson scattering, with an
arbitrary number of quark-antiquark and two-meson channels of different
types, we published in Ref.~\cite{AP324p1620}. This formalism has been very
successfully applied to e.g.\ $J^P=1^+$ open-charm mesons \cite{PRD84p094020}
and the mysterious axial-vector charmonium-like state $X(3872)$
\cite{EPJC71p1762}. The general expression for the fully off-energy-shell
multichannel $\mathcal{T}$-matrix reads
\begin{equation}
\tmat{i}{j}(p_i,p'_j;E) \; = \;
-2\lambda^2a\,\sqrt{\mu_ip_i\mu'_jp'_j}\:\bes{i}(p_ia)\sum_{m=1}^{N}
\rse_{im}(E)\,\left\{[\One-\Omega\,\mathcal{R}]^{-1}\right\}_{\!mj}\,
\bes{j}(p'_ja)\;, 
\label{tfinal}
\end{equation}
where \vspace*{-7mm}
\begin{eqnarray}
\mathcal{R}_{ij}(E) \; = \; 
\sum_{\alpha=1}^{N_{q_1\bar{q_2}}}\sum_{n=0}^{\infty}
\displaystyle\frac{g_i^{(\alpha)}(n) g_j^{(\alpha)}(n)}{E-E_n^{(\alpha)}}
\label{rse}
\end{eqnarray}
\vspace*{-4mm}
and
\begin{equation}
\Omega_{jr}(E) \; = \; -2i\lambda^2a\,\mu_jk_j\,\bes{j}(k_ja)\,
\han{j}(k_ja)\,\delta_{jr} \; . \\
\label{loop}
\end{equation}
The fully on-shell and unitary $\mathcal{S}$-matrix is then given by
\begin{equation}
\mathcal{S}_{ij}^{(L_i,L_j)}(k_i,k_j;E)\;=\;\delta_{ij}\:+\:2i
\tmat{i}{j}(k_i,k_j;E)\;.
\label{smatrix}
\end{equation}
Besides the quantities already defined in Eq.~(\ref{single}), we have in
Eqs.~(\ref{tfinal}--\ref{smatrix}) the orbital angular momenta in channels
$i,j$, $L_i$ resp.\ $L_j$, the total number of (open or closed) two-meson
channels, $N$, the number of quark-antiquark channels with the same quantum
numbers, $N_{q_1\bar{q}_2}$, the spherical Hankel function of the first kind
for two-meson channel $j$, $h_{L_j}^{(1)j}$, and the coupling of the
$(n\!-\!1)$th $q_1\bar{q}_2$ recurrence in channel $\alpha$
to meson-meson channel $i$, $g_i^{(\alpha)}(n)$. The latter couplings can be
straightforwardly computed in the general scheme developed many years ago by
one of us (EvB) \cite{ZPC21p291}, which is based on overlaps of normalised
harmonic-oscillator wave functions for the quark-antiquark pair of the
decaying meson,  the one created out of the vacuum with $^{3\!}P_0$ quantum
numbers, and the ones for the two outgoing mesons. Note that for the
quark-antiquark ground states ($n=0$), these coupling constants are usually
identical to those resulting from standard recouplings for spin, orbital
angular momentum, and isospin in point-particle approaches. As for the
$\mathcal{S}$-matrix in Eq.~(\ref{smatrix}), it is restricted to the open
meson-meson channels, with on-shell momenta $k_i,k_j$, contrary to
$\mathcal{T}$, which has been given fully off-shell, for the corresponding
momenta $p_i,p_j$. However, closed channels do contribute to $\mathcal{S}$ as
well, since the RSE matrix $\mathcal{R}$ in the denominator of
Eq.~(\ref{tfinal}) is always $N\times N$. Nevertheless, for phenomenological
reasons it may turn out to be necessary to further reduce the influence of
closed channels far below their thresholds, which can be achieved by using a
form factor \cite{PLB641p265,PRL97p202001_a,*PRL97p202001_b}.

For data-analysis purposes, Eqs.~(\ref{tfinal}--\ref{smatrix}) can be made 
less model dependent by truncating the infinite sum in Eq.~(\ref{rse}) to 
only a few terms and taking the corresponding real energies $E_n^{(\alpha)}$
as fit parameters, just like in Eq.~(\ref{single}). Furthermore, the $n$
dependence of the couplings $g_i^{(\alpha)}(n)$ can also be made adjustable,
since the specific fall-off for higher recurrences as derived in
Ref.~\cite{ZPC21p291} is indeed model dependent and will be different for other
types of wave functions. Nevertheless, the couplings for the same value of $n$
should not be allowed to vary, because they are essentially just determined by
Clebsch-Gordan coefficients. Finally, the employed spherical Bessel and Hankel
functions can also be generalised, by assuming a smoother string-breaking
potential than the spherical delta-shell mostly employed by us. Although the
latter singular function is a reasonable approximation to the peaked transition
rate between $q\bar{q}$ and meson-meson states as determined on the lattice
\cite{PRD71p114513}, more detailed shapes have in fact been derived in
Ref.~\cite{ZPC21p291} for the model of Ref.~\cite{ZPC30p615}. The corresponding
coordinate-space expressions can then be Fourier-transformed
so as to obtain alternative functions for the three-meson vertices, other than
the usual spherical Bessel function. One might even employ more empirical
functions to model those vertices in the effective separable RSE meson-meson
interaction \cite{AP324p1620}. The only necessary step is then to evaluate
a one-loop (``bubble'') integral to obtain the diagonal matrix $\Omega$ in
Eq.~(\ref{tfinal}). The separability of the meson-meson interaction always
allows to obtain an exact, closed-form expression for the $\mathcal{S}$-matrix
that is manifestly unitary and analytic. The latter can even be achieved if
one includes resonances in the two-meson final states, in order to mimic the
many multiparticle decays of excited mesonic resonances observed in experiment
\cite{PDG2014}. Now, the corresponding complex masses in asymptotic states
in principle destroy unitarity of the $\mathcal{S}$-matrix. However, by making
use of the unaltered symmetry of $\mathcal{S}$, an empirical algebraic
procedure \cite{EPJC71p1762} allows to redefine $\mathcal{S}$ so as to
restore exact unitarity. This has been applied with remarkable success to the
OZI-forbidden $\rho^0J\!/\!\psi$ and $\omega J\!/\!\psi$ decays of $X(3872)$
in Ref.~\cite{EPJC71p1762}.

To conclude the present brief review, we now focus on production processes.
Because direct meson-meson scattering is not feasible experimentally, mesonic
resonances initially used to be produced in meson scattering off nucleons, from
which meson-meson data were then extracted. For example, the famous LASS 
\cite{NPB296p493} data on $K^-\pi^+$ scattering were obtained from the reaction
$K^-p\to K^-\pi^+n$. However, nowadays mesonic resonances are either observed
in $e^+e^-$ collisions and decays of the resulting vector states, or in 
multiparticle decays of open-bottom and open-charm mesons. One may argue that,
from a theoretical point of view, there is no fundamental difference between
the various mechanisms to produce meson resonances, as resonance pole positions
are generally accepted to be universal and so independent of the process.
However, resonance line shapes may very well depend on the production
mechanism, which makes it all the more important to have an adequate formalism
at hand for a reliable analysis of the experimental data. But still more
importantly, the common feature of production processes is an initial state
with only one quark-antiquark pair and not a two-meson system. This yields
an extra term in the production amplitude as compared to meson-meson
scattering, which is even of leading order in the coupling constant for
$q\bar{q}$ creation \cite{AP323p1215}.

The formalism for production processes in Ref.~\cite{AP323p1215} was directly
derived from the RSE model \cite{AP324p1620}. Thus, the relation
between the production amplitude for an initial quark-antiquark state 
labelled $\alpha$ to a final meson-meson state labelled $i$ reads
\begin{equation}
\mathcal{P}_i^{(\alpha)} \; \propto \; \lambda\sum_{L,M}\,(-i)^{L}\,
Y^{(L)}_{M}(p_{i})\,
Q^{(\alpha )}_{\ell_{q\bar{q}}}(E)
\left\{ g_{\alpha i}\, j_{L}(p_{i}a)
-i\sum_{\nu}\,\mu_{\nu}\,p_{\nu}\,h^{(1)}_{L}(p_{\nu}a)\,
g_{\alpha\nu}\,\mathcal{T}^{(L)}_{i\nu}
\right\} \; ,
\label{prodscat}
\end{equation}
where $\mathcal{T}^{(\ell)}_{i\nu}$ is a partial-wave $\mathcal{T}$-matrix
element for transitions between the two-meson channels $i$ and $\nu$. Besides
the quantities already defined in Eqs.~(\ref{single}--\ref{smatrix}), we have
here the largely unknown function $Q^{(\alpha)}_{\ell_{q\bar{q}}}(E)$ to 
describe the initial quark-antiquark state $\alpha$ with energy E, which
should be determined by the precise production mechanism. Note that
$\mathcal{P}$ and $\mathcal{T}$ share the same resonance poles, as the
remaining functions in Eq.~(\ref{prodscat}) are all smooth. Furthermore, the
production amplitude $\mathcal{P}$ manifestly satisfies
\cite{AP323p1215,EPL81p61002} the so-called extended-unitarity relation
\begin{equation}
\Im{\mathcal{P}}\; =\; \mathcal{T}^{\ast}\, \mathcal{P} \; ,
\label{extended}
\end{equation}
despite the fact that the coefficients in front of the $\mathcal{T}$-matrix
elements in Eq.~(\ref{prodscat}) are complex \cite{EPL81p61002,EPL84p51002}
due to the spherical Hankel function $h^{(1)}_L$. Pennington \& Wilson argued
\cite{EPL84p51001} that these coefficients should be real, but we showed
\cite{EPL84p51002} this to imply that the coefficients themselves must contain
$\mathcal{T}$-matrix elements, whereas our definition in Eq.~(\ref{prodscat})
involves purely kinematical coefficients, proportional to $h^{(1)}_L$. Besides
this clear advantage, Eq.~(\ref{prodscat}) explicitly displays the mentioned
lead term in the production amplitude, which is linearly dependent on the
coupling $g_{\alpha\nu}$ between the initial $q\bar{q}$ state and meson-meson
channel $\nu$, and has the shape of a spherical Bessel function. This term will
generally result in an enhancement in the cross section right above a threshold
opening, which may be mistaken for a genuine resonance or distort the signal
when indeed a true resonance pole is nearby \cite{EPL85p61002,PRD80p074001}.
In Ref.~\cite{AP323p1215} we also showed the connection between our production
formalism and the BW approximation as well as the $\mathcal{K}$-matrix
approach.

Just as in the case of multichannel scattering above, $j_L$ and $h^{(1)}_L$
in Eq.~(\ref{prodscat}) may be substituted by more detailed functions to
describe string breaking or by purely phenomenological ones, in the context of
the RSE formalism. This will not affect the nice properties of the production
amplitude.

\newpage

\aSection{Questions/Requirements to Theory}{S. Paul}
\begin{center}
Stephan Paul \\
\vskip .2cm
{\small Physik Department\\Technische Universit\"at M\"unchen}
\end{center}

\subsection{Introduction}
With the advent of large data sets from diffractive production of  exclusive $(3\pi)^{-}$ ($50$ to $100$ Mevents), it has become possible to extract the structure of the underlying $(2\pi)^0$ isobars ($J^{PC}$: $0^{++}, 1^{--}, 2^{++}, 3^{--}, 4^{++}$) over a wide mass range of the $2\pi$-system using highly-developed partial wave analysis tools~\cite{Adolph:2015tqa}.
%\footnote{C. Adolph et al. [COMPASS Collaboration], arXiv:1509.00992 [hep-ex]}.
This extraction is being performed for different values of $J^{PC}$ and mass of
the 3-body system. Although some ambiguities still exist on what
concerns the independent extraction of all isobars in a fully model-independent way, we expect this information to be available soon, unless we encounter
unexpected fundamental problems with this approach.  Thus, we can study
the amplitudes of the
isobaric system and their dependence on the $(3\pi)^{-}$ mass
for various values of the angular momentum relative to the ``spectator'' pion. We expect this
information being valuable input to describe these systems using a set
of universal amplitudes, which could also incorporate final state
interaction (FSI) with the third pion. In addition, one can vary the
momentum of the spectator pion in the isobar rest frame by varying the
mass of the $(3\pi)^{-}$ system.

\subsubsection{$J^{PC}$-dependent description of the isobar in a 3-body decay}
\begin{figure}[tbhp]
  \begin{center}
%  \ifthenelse{\boolean{pdflatex}}{ 
        \includegraphics*[width=0.5\figsize]{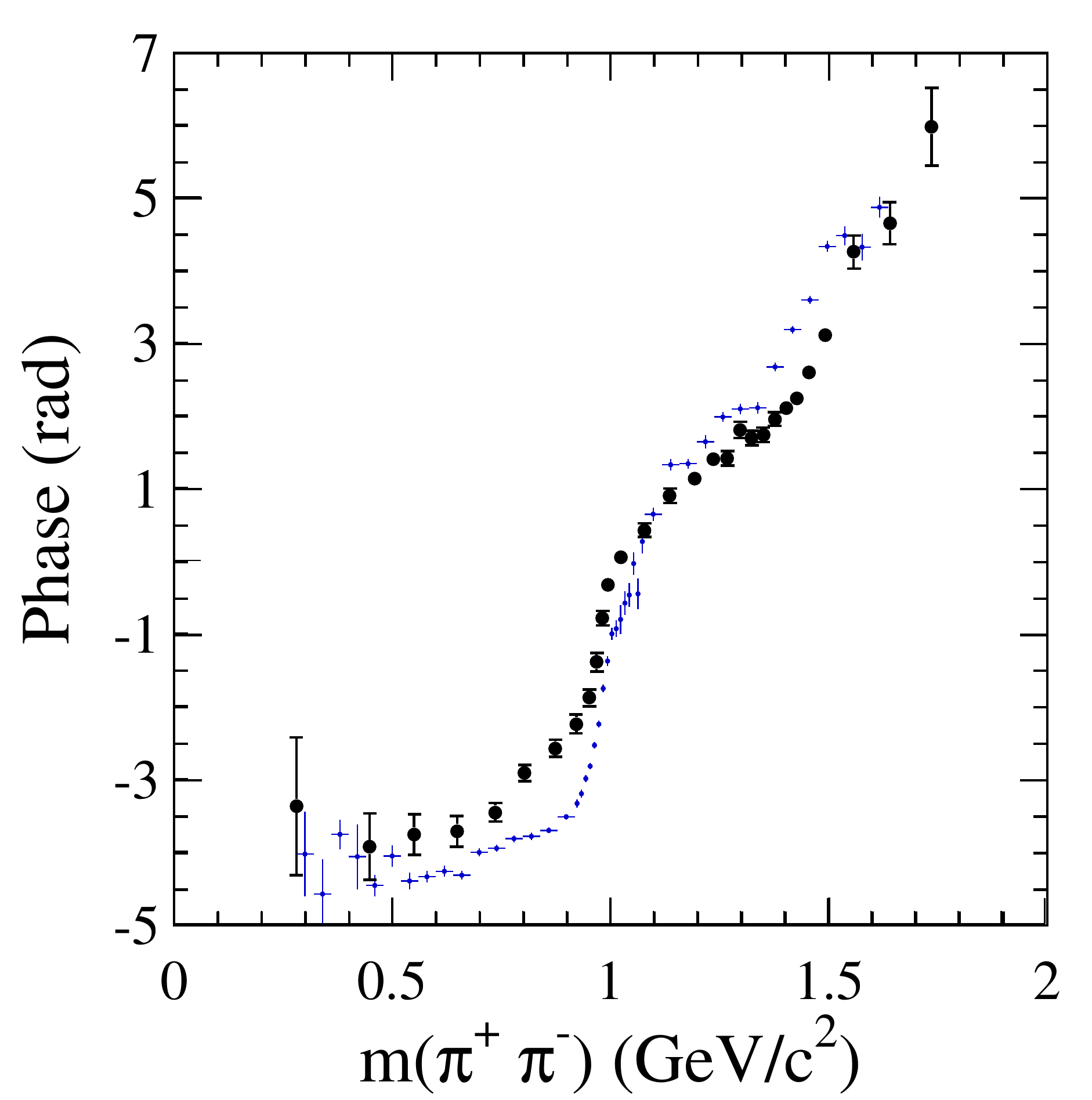}
%   }{
%        \includegraphics*[width=\figsize]{figs/c3_Z_1070_inset.pdf}
%   } 
  \end{center}
  \vskip-0.1cm\caption{\small 
    Comparison of the $\pi\pi$ S-wave phase as extracted from decays of $D_s$-mesons (BaBar \cite{antimo} - solid dots) and from a first freed-isobar analysis of the $(3\pi)^{-}$ system forming a $\pi(1800)$ (blue crosses) derived from~\cite{Adolph:2015tqa}. An arbitrary phase offset was added to the COMPASS data to roughly match the phase variation around $f_0(980)$
  \label{fig:pipi_s_compass}.
  }
\end{figure}

A first analysis of the $\pi\pi$ S-wave in COMPASS data~\cite{Adolph:2015tqa} has shown that the same
general features as seen in $\pi-\pi$ scattering data or in a
model--independent analysis of \textrm{$D_s$} decays~\cite{antimo}
%\footnote{B. Aubert et al. [BaBar collaboration], Phys. Rev. D 79, 032003 (2009)} 
(see LHCb, Belle, Belle II, BaBar, BESIII and others), can be
found in $(3\pi)^{-}$ decays (see fig. \ref{fig:pipi_s_compass}). It is also known from the analysis of \textrm{$\eta_c$}
decays, recently presented by the BaBar collaboration (see
hadron2015)~\cite{AntimoPalanoHadron2015},
%\footnote{https://www.jlab.org/conferences/hadron2015/talks/thursday/parallel/session3/\\3C1\_AntimoPalano.pdf},
that the moduli of the $\pi\pi$ amplitudes from various processes vary
considerably. However, the mass dependence of the $\pi\pi$ phase shows
little dependence on the system studied (except for an
overall phase offset).  In order to extract
information on more general system-independent amplitudes, theory input is required to give guidance on
the functional form. Chiral Perturbation Theory
 (\textrm{$\chi PT$}) can be used to calculate these amplitudes up to
$\pi\pi$ masses of about 400~\textrm{$MeV/c^2$}. The extrapolation to
the threshold for inelastic processes can be performed using
dispersion relations. However, no such recipe exists to continue the
$\pi\pi$ amplitude beyond this mass, nor how to include
well-known high-mass scalar resonances
such as $f_0(1500)$ (or above). This information, however, is
crucial for several physics goals:
\begin{itemize} 
\item to perform a quasi model-independent Partial-Wave
    Analysis (PWA) of the $(3\pi)^{-}$ system or even a
  multi-dimensional PWA, which includes the $(3\pi)^{-}$ and
  $(2\pi)^{0}$ amplitudes as well as their $t'$-dependence, in order to extract information on
  resonances in both the $(3\pi)^{-}$ and $(2\pi)^{0}$ system and
  their interrelation.
\item to perform a $C\!P$ violation analysis of \textrm{$D$}- and
  \textrm{$B$}-meson decays using purely hadronic final states on the level of physical
  amplitudes, instead of in bins within a Dalitz plot. Such
  a $C\!P$ violation analysis would also
  be based on the evolution of the amplitudes with decay
  time and thus requires the correct description of the real and
    imaginary parts of the amplitudes, not just their magnitude.
\item to extend all analyses to (true) 4-body and possibly 5-body final states
\end{itemize}

\subsubsection{Description of a resonance - poles}
Until recent, almost all determinations of meson and heavy-baryon resonances
and their parameters have
been performed using ``simple'' mass fits based on Breit-Wigner
amplitudes together with a coherent background amplitude (the
parametrization of the latter is mostly empirical).
Only elastic-scattering data have been used to extract pole information
from the S-matrix. While experimentalists are exposed to much
criticism for using Breit-Wigner amplitudes, few other methods have
been on the market and are easily available. Although BaBar and
COMPASS have started to use polynomials and \textrm{$K$}-matrix parametrizations for their
fits, this work has not been concluded and 
%there is a
%strong\change{}{ly} \change{theorist}{model}-dependent formulation of these analytical
%functions. \remark{is this `theorist-dependent' deliberate? If so,
%  fine. Alternatively, you could write: ``There is significant model
%  dependence in the formulation of these analytic functions''}
there is a significant model dependence in the formulation of these analytic functions.
Some
more guidance and a list of ``pros'' and ``cons'' of different
recipes, to which one can refer, would be very helpful. We see this as
the only way to obtain process independent information, also referred
to by the PDG.

\subsubsection{Threshold phenomena}
With the appearance of $X, Y, Z_{c,b}$ and $P_c$-states and also recently, the $a_1(1420)$, the question of their nature
dominates the discussion in hadron physics. This is particularly true for resonance structures
appearing close to thresholds of pairs of much lighter mesons. Here,
triangular or loop diagrams can be calculated (probably up to some
overall amplitude factor) resulting in analytical amplitudes,
which could be used by experimentalists in their mass-dependent fits,
replacing the above mentioned pole descriptions. This would complete 
all analyses of new states in terms of well-motivated mass-dependent amplitudes.
%\remark{I am not entirely sure I understand what's meant with the last sentence.}

\subsubsection{Non-resonant contributions in diffraction}

Pion diffraction off nucleons always results in a superposition of
production processes, e.g. Pomeron exchange and ``Deck'' effect,
where the latter is modeled to proceed via pion-nucleon elastic
scattering with Reggeon exchange at the
nucleon-vertex and  by the exchange of a pion to the ``upper''
pion vertex. The ``Deck'' process is thought to be the most prominent non-resonant
contribution in diffraction.  These production processes have
different dependences on the four-momentum transfer $t'$ to the target
proton and thus may be disentangled by
performing the PWA at various values of $t'$.
This may provide clearer access to resonances without
the use of simplistic
parametrizations of the non--resonant contribution, which is known to distort
resonance signals. Here, a better model for the non--resonant
production is required, which can be developed only if detailed
information from ``inside'' the production amplitude is known. This
task should be addressed jointly by theorists and experimentalists.

\subsection{Conclusion}
Partial-wave analyses of various multi-body
systems have reached a new level of accuracy and
complexity. With much closer links to theoretical
calculations, the extraction of physics information could be raised
to a new level, namely (nearly) ``universal'' amplitudes. Extraction
of pole parameters, the role of underlying dynamics and \textrm{$CP$} violation
based on calculable strong amplitudes could be much rewarding
results. We are aware that substantial effort is ongoing already (see e.g. ``white
paper''~\cite{Battaglieri:2014gca}),
%\footnote{M. Battaglieri et al.,``Analysis Tools for Next-Generation Hadron Spectroscopy Experiments'', Acta Phys.Polon. B46 (2015) 257,  arXiv:1412.6393 [hep-ph]}) 
but a worldwide coherent program, which includes heavy- and light-quark
physics processes, could accelerate the development, for which theory
is an absolutely necessary and crucial ingredient.

\newpage
\aSection{CPT, Final State Interaction and  CP Violation}{T.~Frederico, J.~H.~Alvarenga~Nogueira, I.~Bediaga, A.~B.~R.~Cavalcante, 
and O.~Louren\c{c}o}
\begin{center}
T.~Frederico$^a$, J.~H.~Alvarenga~Nogueira$^a$, I.~Bediaga$^b$, A.~B.~R.~Cavalcante$^b$, 
and O.~Louren\c{c}o$^c$\\
\vspace{.5cm}
$^a$Instituto Tecnol\'ogico de Aeron\'autica\\
 DCTA, 12228-900, S\~ao Jos\'e 
dos Campos, SP, Brazil 
\vspace{.5cm}\\
$^b$Centro Brasileiro de Pesquisas F\'isicas 
\\LAFEX, 22290-180, Rio de Janeiro, RJ, Brazil 
\vspace{.5cm}\\
$^c$Departamento de Ci\^encias da Natureza, Matem\'atica e Educa\c c\~ao 
\\CCA, Universidade Federal de S\~ao Carlos, 13600-970, Araras, SP, Brazil

\end{center}
\vskip .3cm
\definecolor{navyblue}{rgb}{0.3,0.3,1}
\definecolor{purple}{rgb}{0.6,0,0.5}

The CPT invariance of the fundamental theory implies identical particle and antiparticle  lifetimes. 
The partial decay widths of CP conjugate channels can be different due to CP violation (CPV), 
but in order to fulfill the CPT constraint their sum, namely the total width for particle and antiparticle, should be equal. 
Therefore, the amount of CPV in one channel should be compensated in other ones. The mechanism that
allows the flow of CPV  between different channels is the final state interaction (FSI), which couples 
different states in the decay channel. The discussion given in the following presents a CP asymmetry 
formula including resonances and FSI based on the CPT constraint. 

The CP asymmetry formula is discussed in the lowest order in the strong 
interaction and decomposed in  the two-body channels angular momentum. 
At the end of the section an expression for the CP 
asymmetry is  provided for the decays channels $B^{\pm}\to 
\pi^{\pm} \pi^+\pi^-$ and $B^{\pm}\to K^{\pm} \pi^+\pi^-$. 
For this example, we parametrise the Dalitz plot with an isobar model, with Breit Wigner lineshapes for resonances like the $\rho$ and the $f_0(980)$, plus a non-resonant background. In addition, the contributions from the $\pi \pi \leftrightarrow K K$ coupled channel amplitude is taken into account.

We follow closely  refs.~\cite{Marshak,Branco,bigibook,BedPRD14,NogPRD15} to introduce our notation 
and the  CPT constraint in $B$ meson decays.  
We start with the CPT invariance of the weak Hamiltonian $H_w$ and the strong Hamiltonian $H_s$, namely $(\mathcal{CPT})^{-1} H_{w, s} \mathcal{CPT} = H_{w, s}$. 
The hadron weak decay amplitude is obtained  as the matrix element  $\langle 
\lambda_{out}|H_w|h\rangle$, where the state $|\lambda_{out}\rangle$ 
carries the effect of the strong interaction between the hadrons in the final state. This is our main assumption.

The requirement of CPT invariance on the decay 
amplitude is
$\langle \lambda_{out}|H_w|h\rangle
= \chi_h\chi_\lambda\langle  \overline \lambda_{in} |H_w| \overline h  \rangle^*\, ,$
where we used that the hadron state $|h\rangle$ transforms under CPT as ${\mathcal {CPT} 
}\,|h\rangle=\chi\langle \overline h|$, where $\overline h$ is the charge conjugate 
state and $\chi$ a phase.  
The completeness relation for the strongly interacting states in terms of the eigenstates of $H_s$, $|\lambda_{out}\rangle$, given by $\sum_{\bar{\lambda}_{out}'}|\bar{\lambda}_{out} '\rangle \langle \bar{\lambda}_{out}'|$, and the hermiticity of $H_w$ implies that
\begin{eqnarray}
\langle \lambda_{out}|H_w|h\rangle
=\chi_h\chi_\lambda\sum_{\overline\lambda^\prime} 
S_{\overline\lambda^\prime,\overline\lambda}
\langle \overline\lambda^{\prime}_{out} |H_w| \overline h  \rangle^* \, , \label{cpt4}
\end{eqnarray} 
where  $S_{\overline\lambda^\prime,\overline\lambda}=
\langle \overline \lambda^\prime_{out}|\overline 
\lambda_{in}\rangle=S_{\lambda^\prime,\lambda} \ $
defines the matrix elements of the S-matrix. 
These matrix elements contain the information about the distortion of the final state due to the strong interaction, related to the strong phase. Note that, the final state in general can have more than two particles, and therefore the strong phase in the general case 
is not restricted to be directly associated with only the two-body phase-shift.

The full widths of the decaying particle and its charge 
conjugate are identical, i.e., the mean life of the particle and its conjugate are equal. From the CPT invariance 
relation (\ref{cpt4}) and the hermiticity of $H_w$, one can easily prove that the sum of partial widths of the hadron and its charge conjugate 
are indeed equal (see for details ref.~\cite{BedPRD14}). 

The expression (\ref{cpt4}) can be turned practical by introducing the expansion of the strongly interacting state as the free state
$(\lambda_0 \rangle)$ 
plus scattering corrections: $|\lambda_{out} \rangle=|\lambda_0 \rangle+\dots$. The free state is the mesonic noninteracting 
state, which does not carry the distortion due to the strong force. Thus, we write that:
\begin{eqnarray}
\langle \lambda_{out}|H_w|h\rangle
=\chi_h\chi_\lambda\sum_{\overline\lambda^\prime} 
S_{\overline\lambda^\prime,\overline\lambda}
\langle \overline\lambda^{\prime}_{0} |H_w| \overline h  \rangle^*+ \cdots 
=\sum_{\lambda^\prime} 
\left(\delta_{\lambda^\prime,\lambda}+ i   \, t_{\lambda^\prime,\lambda}\right)
\langle \lambda^{\prime}_{0} |H_w|  h  \rangle + \cdots\ ,
\label{cpt4a}
\end{eqnarray} 
where we have used the CPT invariance 
of the weak Hamiltonian, i.e., $\langle \lambda_{0}|H_w|h\rangle
= \chi_h\chi_\lambda\langle  \overline \lambda_{0} |H_w| \overline h  \rangle^*\, $, the CP invariance of the strong S-matrix, namely 
$S_{\lambda^\prime,\lambda}=S_{\overline\lambda^\prime,\overline\lambda}$ and we have written the S-matrix in terms of the scattering 
amplitude $t _{\lambda^\prime,\lambda}$, as 
$S_{\lambda^\prime,\lambda}=\delta_{\lambda^\prime,\lambda}+i \, t _{\lambda^\prime,\lambda}$.
The decay amplitudes computed with the non interacting terms in (\ref{cpt4a}) 
have the effect of CP violation  and are written in general as a sum of two terms:
\begin{eqnarray}
\langle \lambda_0|H_w|h\rangle=A_{0\lambda} + e^{-i\gamma}B_{0\lambda} , ~~~\text{   
and   }~~~~ \langle \overline \lambda_0|H_w|\overline h\rangle=A_{0\lambda} + 
e^{+i\gamma}B_{0\lambda} ,
\end{eqnarray} 
where $A_{0\lambda}$ and $B_{0\lambda} $ can be in principle associated with the tree 
and penguin amplitudes in the Bander-Silverman-Soni (BSS) model~\cite{BSS}.
The BSS mechanism is the direct
CPV due to the interference of the tree (weak phase) and penguin (strong phase)
amplitudes leading to the same final state. 
Within the present framework we name the amplitudes 
$A_{0\lambda}$ and $B_{0\lambda}$, as partonic or source amplitudes. In the region of the Dalitz plot where 
the effects of the final state interaction and formation of resonances are small, the source 
amplitudes are directly accessible and
are associated with the usual non-resonant background, although the source amplitudes, in principle, cover all the 
available phase-space. Substituting the source terms in (\ref{cpt4a}) one has that:
\begin{eqnarray}
\langle \lambda_{out}|H_w|h\rangle
=\sum_{\lambda^\prime} 
\left(\delta_{\lambda^\prime,\lambda}+ i   \, t_{\lambda^\prime,\lambda}\right)\left(
A_{0\lambda} + e^{-i\gamma}B_{0\lambda} \right)+ \cdots\ ,
\label{cpt4b}
\end{eqnarray} 
which gives for the CP asymmetry , $\Delta\Gamma_\lambda=|\langle \lambda_{out}|H_w|h\rangle|^2 - |\langle \overline \lambda_{out}|H_w|\overline h\rangle|^2 $, computed to leading order in the hadronic interaction, the following expression:
\begin{equation}
\Delta \Gamma_\lambda
=4(\sin\gamma) \, \mbox{Im}\Bigg\{ 
\left(B_{0\lambda}\right)^*A_{0\lambda} 
+i\sum_{\lambda^\prime}\left[\left(B_{0\lambda}\right)^*t_{\lambda^\prime, 
\lambda}\,A
_{0\lambda^\prime} - \left(B_{0\lambda^\prime}\, t_{\lambda^\prime,\lambda}\right)^*
A_{0\lambda}\right]\Bigg\},
\label{cp26-a}
\end{equation}
where $\lambda^\prime$ labels each  state coupled to the  decay channel $\lambda$ 
by the strong interaction. 
The last two terms in the 
right-hand side of eq.~(\ref{cp26-a}) can be interpreted as  the ``compound'' CP 
asymmetry~\cite{Soni2005_a,*Soni2005_b}. These two contributions cancel each other when summed in  
$\lambda$ (note that an integration over the phase-space is also implicit) resulting in the CPT condition, $\sum_\lambda\Delta\Gamma_
\lambda=0$, if  the source amplitudes  satisfy:
$\sum_{\lambda }\mbox{Im} \left[\left(B_{0\lambda}\right)^*A_{0\lambda}\right] = 0,$
which is a consequence of the CPT constraint at the microscopic level, e.g., as 
expressed by the tree and penguin amplitudes in the BSS model, that  should be valid 
in the absence of final state interactions. This term was neglected by Wolfenstein~\cite{wolfenstein}, which 
corresponds to the trivial solution of this equality, assuming that the phase
difference between the two CP-conserving  amplitudes is zero for all decay channels. 
Therefore, if we have only two decay
channels, namely, $\lambda$ and $\lambda^\prime$, which are coupled by the strong interaction, we can write: 
\begin{equation}
\Delta \Gamma_\lambda
=4(\sin\gamma) \, \mbox{Re}\Big\{ 
\left(B_{0\lambda}\right)^*t_{\lambda^\prime, 
\lambda}\,A_{0\lambda^\prime} - \left(B_{0\lambda^\prime}\, t_{\lambda^\prime,\lambda}\right)^*
A_{0\lambda}\Big\}=-\Delta \Gamma_{\lambda^\prime} \, ,
\label{cp26-b}
\end{equation}
To derive the above expression we used $ t_{\lambda^\prime,\lambda}= t_{\lambda,\lambda^\prime}$.
The relation (\ref{cp26-b}) has been used in Ref.~\cite{BedPRD14} for a qualitative analysis of the large CP violation effects observed by the LHCb collaboration~ in charmless three-body $B^\pm$ decays~\cite{LHCbPRL13}.
The focus of the study is the decay channels $B^\pm\to K^\pm K^+ K^-$ and $B^\pm\to K^\pm \pi^+ \pi^-$ in the low 
$K^+ K^-$ and $\pi^+\pi^-$ mass regions between 1 and 1.6 GeV, 
where these channels are coupled by the strong interaction~\cite{cern-munich2,Cohen1980}.%\cite{CERN-Munich,Cohen1980}.
Expression (\ref{cp26-b}) shows that apart from a phase-space factor, the asymmetries in these channels are proportional
to $\sqrt{1-\eta^2}\cos\left(\delta_{KK}+\delta_{\pi\pi}+\Phi\right)$. 
This factor arises from the
magnitude and phase of the  $\pi\pi\to KK$ transition amplitude~\cite{pelaprd05_a, *pelaprd05_b}: $\eta$ is the inelasticity parameter 
and $\delta_{\pi\pi(KK)}$ are the isoscalar spin zero phase-shifts. A more detailed analysis is presented in~\cite{NogPRD15} based 
on the recent experimental findings of LHCb~\cite{expnew}.
Both analyses based on Eq. (\ref{cp26-b}) led to the conclusion that  the ``compoundĞĞ contributions are essential to form
the observed pattern of the CP asymmetry data. 

We now discuss equation (\ref{cp26-b}) for the particular case when $A_{0\lambda}=0$ in the two-channel case, which gives that:
$\Delta \Gamma_\lambda
=4(\sin\gamma) \, \mbox{Re}\{ 
\left(B_{0\lambda}\right)^*t_{\lambda^\prime, 
\lambda}\,A_{0\lambda^\prime} \}\, .$ In this configuration, still, a CP asymmetry can appear due to the interference of two terms, where one of them arises 
from the strong coupling between the channels $\lambda$ and  $\lambda^{\prime}$ and the other is the one related with the weak phase. 
Now consider a $B^{\pm}$ decay channel where the penguin (tree) contribution is negligible, while in the corresponding coupled channel it is the opposite, i.e., the tree (penguin) contribution can be neglected. Then, the weak CP violating phase comes only through one channel and for the CP asymmetry to be visible, the interference has to occur between amplitudes originating in different channels, coupled by the strong interaction. In this case, FSI are the dominant mechanism leading to CP violating interference effects.

For  more detailed applications, the expression (\ref{cpt4a}) has to include resonances  
formed in the partonic process, like for example $B\to \pi\rho$ (see~\cite{NogPRD15}).
To take into account resonances the amplitudes $A_{0\lambda}$ and $B_{0\lambda}$ 
are separated in two parts:
\begin{equation}
A^J_{0\lambda}=A^J_{0\lambda NR}+\sum_R A^J_{0\lambda R}\quad\mbox{and}\quad
B^J_{0\lambda}=B^J_{0\lambda NR}+\sum_R B^J_{0\lambda R} \, , \label{cp24-1}\end{equation}
where the subscripts $R$ and $\mathit{NR}$ mean resonant and nonresonant channels, respectively.
The label  $J$ indicates the spin of the two-body state.
Accounting for (\ref{cp24-1}) the decay amplitude in Eq.~(\ref{cpt4b}) can be generalized to:
\begin{align}
\mathcal{A}^\pm_{LO}&=\sum_J\left[\sum_R A^J_{0\lambda R}+A^J_{0\lambda NR}+ \right. 
\left. e^{\pm i\gamma}\left(\sum_R B^J_{0\lambda R}+B^J_{0\lambda NR}\right) \right]+ 
\nonumber\\ &
+i\sum_{\lambda^\prime,J}t^J_{\lambda^\prime,\lambda}
\left[\sum_R A^J_{0\lambda^\prime R}+A^J_{0\lambda^\prime NR} + \right.
 \left. e^{\pm i\gamma}\left(\sum_R B^J_{0\lambda^\prime R}+B^J_{0\lambda^\prime 
NR}\right) \right] ,
\label{cp24-2}
\end{align}
where in the expression above the strong scattering amplitude $t_{\lambda',\lambda}$ is considered to leading order (LO).
In addition, the source resonant terms $A^J_{0\lambda R}$ and $B^J_{0\lambda R}$ should be interpreted 
as bare amplitudes, which are formed directly at the partonic level  where the two-hadron rescattering 
process is not yet included. Therefore, the Breit-Wigner (BW) amplitudes in each term are identified with
\begin{eqnarray}
(1\,+\,i\, t^J_{\lambda\lambda})A^J_{0\lambda R} \to a_0^R 
F_{R\,\lambda}^{BW}P_J(\cos\theta)\quad\mbox{and}\quad
(1\,+\,i\, t^J_{\lambda\lambda})B^J_{0\lambda R}\to b_{0\lambda}^R 
F_{R\,\lambda}^{BW}P_J(\cos\theta)  \label{ress-1},\quad
\end{eqnarray}
where $J$ is the spin of the resonance decaying to two spin zero particles and 
$P_J(\cos\theta)$ is the $J^{th}$ order Legendre polynomial and $\theta$ is the helicity angle between 
the same-charge particles in the Gottfried-Jackson frame. 

After substituting  (\ref{ress-1}) in (\ref{cp24-2}), we get that
\begin{multline}
\mathcal{A}^\pm_{LO}=\sum_{J\, R}\left(a_{0\lambda}^R +e^{\pm i\gamma} b_0^R\right) 
F_{R\,\lambda}^{BW}P_J(\cos\theta) +\sum_J \left(A^J_{0\lambda NR}+e^{\pm 
i\gamma}B^J_{0\lambda NR}\right)
+i\sum_{\lambda^\prime,J}t^J_{\lambda^\prime,\lambda}
\left(A^J_{0\lambda^\prime NR} +e^{\pm i\gamma}B^J_{0\lambda^\prime NR} \right) ,
\label{cp24-3}
\end{multline}
where the first and second term in the right-hand side are associated with the isobar model. 
The second term is the source amplitude for the final state channel, and the third one 
includes the hadronic interaction among the two of the hadrons with angular momentum $J$. 
We should note that the third term in eq. (\ref{cp24-3}) could also include a resonance in channel $\lambda^\prime$, which 
can decay to channel $\lambda$ due to the coupling between these states by the strong interaction, although we have not included this
possibility in the expression of the decay amplitude.
The resulting CP asymmetry from eq.~(\ref{cp24-3}), considering terms only in the LO  on the strong  scattering amplitude, 
can be cast in the following form:
\begin{small}
\begin{multline}
\Delta \Gamma_\lambda =\Gamma\left(h\to \lambda\right)-\Gamma(\overline h\to 
\overline\lambda) =\\ 
= 4(\sin\gamma) \, \sum_{J\,J^\prime}\mbox{Im}\left\{  \left(
\sum_{R}b_{0\lambda}^R F_{R\lambda}^{BW}P_J(\cos\theta) 
+B^J_{0\lambda NR}\right)^* \left(
\sum_{ R^\prime}a_{0\lambda}^{R^\prime} F_{R^\prime\lambda}^{BW}P_{J^\prime}(\cos\theta) 
+A^{J^\prime}_{0\lambda NR}\right) + \right. \\ 
+ \,i \sum_{\lambda^\prime}\left(
\sum_{R}b_{0\lambda}^R F_{R\lambda}^{BW}P_J(\cos\theta) 
+B^J_{0\lambda NR}\right)^*\, 
t^{J^\prime}_{\lambda^\prime,\lambda}\, %A^{J^\prime}_{0\lambda^\prime NR} 
 \left(
\sum_{  
R^\prime}a_{0\lambda^\prime}^{R^\prime}F_{R^\prime\lambda^\prime}^{BW}P_{J^\prime}
(\cos\theta) 
+A^{J^\prime}_{0\lambda^\prime NR}\right)
+ \\ \left. 
 -\, i \sum_{\lambda^\prime}\left(\sum_{R^\prime}b_{0\lambda^\prime}^{R^\prime} 
F_{R^\prime\lambda^\prime}^{BW}P_{J^\prime}(\cos\theta) 
+B^{J^\prime}_{0\lambda^\prime NR}\right)^*\, 
\left[t^{J^\prime}_{\lambda^\prime,\lambda}\right]^*\, \left(
\sum_{ R}a_{0\lambda}^{R} F_{R\lambda}^{BW}P_{J}(\cos\theta) 
+A^{J}_{0\lambda NR}\right) \right\} .
\label{cp27b}
\end{multline}
\end{small}
The subindex $\lambda$ also includes  different kinematical 
regions of the three-body channel. Since we have introduced the Breit-Wigner amplitudes 
in the decay amplitude $ (F_{R^\prime\lambda^\prime}^{BW})$, the CPT constraint has to 
be checked in the actual fit, i.e., if  
$\sum_\lambda\Delta \Gamma_\lambda=0$ is satisfied when one takes into account the 
integration over the phase-space besides the sum over all decay channels in the sum of 
$\lambda$. In the fitting procedure performed in~\cite{NogPRD15}, we kept only terms that 
are consistent with the CPT constraint.

The non-resonant source terms, $A^{J^\prime}_{0\lambda^\prime NR}$ and 
$B^{J^\prime}_{0\lambda^\prime NR}$, can contain a form factor like 
$\left(1+\frac{s}{\Lambda^2_\lambda}\right)^{-1}$ or similar
to parametrize the dependence on the square mass of 
the two particle system (either $\pi\pi$ or $KK$) produced by the short-range dynamics. 
Such form factor is associated with the partonic decay amplitude that produces the 
three-meson final state. In the production process, the relative momentum between the pair of mesons 
is distributed among the quarks in the 
momentum loop within  the microscopic amplitude computed with the tree and penguin 
diagrams. Therefore, by changing the relative momentum between the mesons in the final state 
the internal structure of the mesons involved in the initial and final states are probed. 
In general, the non resonant amplitude should depend on the two kinematically independent 
Mandelstam variables
;for simplicity, we chose only one of them in the example given above. 
The suggestion of a power-law form just reflects the hard momentum structure of the
mesons involved in the decay~\cite{ref8}. 
For example, by looking at the tree diagram, one observes that the decay of the $b-$quark in 
in its rest frame, produces a fast light quark and a pion back to back. The light quark has 
to share its momentum with a slow antiquark, which is driven by the gluon exchange and 
is damped by a momentum power-law form. A similar reasoning can be applied to the penguin 
diagram.

Expression (\ref{cp27b}) for the CP asymmetry is worked out for the example of a charmless three-body $B$ decay.
It is expanded in a form 
where the unknown parameters are exposed
in such a way that they can be easily fit to experimental data, 
\begin{eqnarray}
&\Delta\Gamma_\lambda& = 
\frac{\mathcal{B}\cos[2\delta_{\pi\pi}(s)] \sqrt{1-\eta^2(s)}}{\left(1+\frac{s}{ 
\Lambda^2_{\lambda}}\right)\left(1+\frac{s}{ \Lambda^2_{\lambda^\prime}}\right)} 
+ |F^{\rm BW}_\rho(s)|^2k(s)\cos\theta 
\left\{\frac{\mathcal{D}(m^2_\rho-s)}{1+\frac{s} {\Lambda^2_{\lambda}}} +
\frac{\mathcal{E}m_\rho\Gamma_\rho(s)}{1+\frac{s} {\Lambda^2_{\lambda}}} \right\} 
\nonumber\\
&+& F^{\rm BW}_{\rho}(s)|^2|F^{\rm BW}_f(s)|^2k(s)\cos\theta\times 
\nonumber\\
&\times&\left\{\mathcal{F}[(m_\rho^2 -s)(m_f^2-s) + m_\rho\Gamma_\rho(s)m_f\Gamma_f(s)] \right.
+ \left.
\mathcal{G}[(m_\rho^2 -s)m_f\Gamma_f(s) - m_\rho\Gamma_\rho(s)(m_f^2-s)]\right\}.
\label{dgamacos} 
\end{eqnarray}
In the expression above, $\lambda$ is associated to $\pi^\pm\pi^+\pi^-$ or $K^\pm\pi^+\pi^-$, and 
$\lambda^\prime$ to
$\pi^\pm K^+K^-$ or $K^\pm K^+K^-$. 
The Mandelstam variable $s$ is the interacting pair square mass. All parameters $\mathcal{B},\,\mathcal{D},\,\mathcal{E},$
$\,\mathcal{F}$ and $\mathcal{G}$ are related only with the partonic amplitudes $A^J_{0 \lambda}$ and $B^J_{0 \lambda}$ 
and with the phase $\gamma$. These relations are discussed in more detail in Ref.~\cite{NogPRD15}. 
The $F^{\rm BW}(s)$ is the relativistic Breit-Wigner and $\Gamma(s)$ is the width, and
for $\pi^\pm K^+K^-$ or $K^\pm K^+K^-$ channels, only $\mathcal{B}$ is non zero. 
Eq.~(\ref{dgamacos}) contains interferences involving the non-resonant background, 
the resonances $\rho(770)$ and $f_0(980)$ and the S-wave $KK \rightarrow \pi\pi$ amplitude. The parameters $\eta(s)$ and $\delta_{\pi\pi}$ are the inelasticity parameter and isoscalar s-wave $\pi\pi$ phase-shift, 
respectively~\cite{pelaprd05_a, *pelaprd05_b}.
All terms that violate CPT 
when integrated over the Dalitz give non vanishing contribution, 
were completely ignored in the above formula.

In the future, the analysis of 
the CP asymmetries in charmless $B$ decays will be extended to include corrections 
induced by the three-body rescattering processes 
following the same approach that has been applied in the charm sector to $D\to K\pi\pi$ decay (see \cite{dkpipi1,dkpipi2,dkpipi3} ).

\subsection*{Acknowledgments}

JHAN acknowledges the support from Funda\c c\~ao de Amparo \`a Pesquisa do Estado de S\~ao Paulo (FAPESP), Grant No. 2014/19094-8. We also thank the partial support from Conselho Nacional de Desenvolvimento Cient\'ifico e Tecnol\'ogico (CNPq).

\newpage

\aSection{Analytical improvements to Breit-Wigner isobar models}{A. P. Szczepaniak}
\begin{center}
Adam~P.~Szczepaniak$^{a,b,c}$\\
\vskip .2cm
{\small $^a$Center for Exploration of Energy and Matter, Indiana University, Bloomington, IN 47403}
{\small $^b$Physics Department, Indiana University, Bloomington, IN 47405, USA }
{\small $^c$Thomas Jefferson National Accelerator Facility, 
Newport News, VA 23606, USA }
\end{center}
\vskip .3cm
 
We discuss the derivation and properties of the general representation of partial wave amplitudes in the context of improving the models currently used in analysis of three particle Dalitz distributions.

\subsection{introduction}
In this note, after a brief introduction to aspects of  $S$-matrix theory 
 relevant in the analysis of three particle Dalitz plots, I focus on properties of Breit-Wigner (BW) amplitudes and the isobar model in general.  I discuss the LHCb analysis model in the context of 
 a general  isobar-type approximation and  show, for example, 
   which features of the  BW amplitude, {\it e.g.} barrier factors,  Blatt-Weisskopf factors, {\it etc.} are universal    and which are not,  {\it i.e.}  are process dependent.  The possibility of extending the BW description in a way that is consistent with analyticity, unitarity, and even crossing 
    would allow to access systematic uncertainties in data analysis. I concentrate on spinless particles. 
     Spin introduces kinematical complexities but  
    does not affect how unitarity, analyticity, and crossing are implemented, at least for a finite set of partial waves. 

 \subsection{Kinematical vs Dynamical Singularities} 
 
 We are interested in amplitudes describing a decay of a  quasi-stable particle $D$ with mass $M$ to 
  three  distinguishable particles  $A,B,C$  
  
\begin{equation} 
D  \to A + B + C   \label{D} 
\end{equation} 
 The decay amplitude depends on particle helicities,  $\lambda_i$, 
 $i=A,B,C,D$,   and  three Mandelstam invariants, which we define as $s = (p_A + p_B)^2$, $t = (p_B + p_C)^2$ and $u=(p_A + p_C)^2$. The invariants  are kinematically constrained by $s + t + u = \sum_i m_i^2$.  Analytical $S$-matrix theory states that, besides the decay channel, the same amplitude describes each of the three  two-to-two scattering process, {\it i.e} the $s$-channel reaction $D + \overline{C} \to A + B$, (bar denotes an antiparticle) as well as the $t$ and $u$ channel scattering.  What this means in practice is the following. For each combination of helicities  there is an analytical function $A_{\lambda_i}(s,t,u)$ of the three complex Mandelstam variables and complex $M^2$, such that the three physical 
    scattering amplitudes and the decay channel amplitude correspond to the limit of $A_{\lambda_i}$ 
     when  $s,t,u,$ and $M$ approach the real axis in the physical domain of the corresponding  reaction. 
 This is the essence of crossing symmetry. 
 In general crossing mixes helicity amplitudes and leads  
  to complicated relations for helicity amplitudes. Furthermore,  
     helicity amplitudes have kinematical singularities in the Mandelstam variables. 
% I don't understand this sentence: These are enforced by absence of such singularities in a covariant 
%representation.
 Despite such complexities it is possible to come up with parameterizations of helicity amplitudes that take into account both  kinematical and dynamical constraints~\cite{MartinSpearman}. 
   On the other hand it is also useful to consider 
    the  covariant from  {\it i.e.} a representation of helicity amplitudes in terms of  Lorentz-Dirac factors 
 that describe wave functions of the free particles participating in the reaction. The advantage of the covariant representation is that the scalar functions multiplying all independent covariants are simply related by crossing and are free from kinematical singularities. At the end of the day one still needs the 
     helicity amplitudes for partial wave analysis. In ~\cite{Danilkin:2014cra} the reaction $\omega \to \pi^+\pi^-\pi^0$ was  studied and this example provides a good illustration of the issues discussed above.

The main postulate of relativistic reaction theory is that reaction amplitudes are analytical functions of kinematical variables. It follows from Cauchy's theorem  that an analytical function is fully determined by its domain of analyticity, {\it i.e.} location of singularities. Thus knowing amplitude singularities allows 
to determine the amplitude elsewhere, including the various physical regions. 
     In S-matrix theory it is assumed that all singularities 
   can be traced to unitarity. In the absence of an explicit solution to the scattering problem in QCD, 
   analyticity and unitarity  provide the least model dependent description  of hadron scattering. 
    
  Unitarity operates in any of the Mandelstam variables. 
 In the  $s$-channel the physical domain is located on the positive real axis in the $s$-plane above the elastic threshold.    Unitarity makes amplitudes singular at each open channel threshold, the singularity being of the square-root type. The same happens in variables $u$ and $t$.  The amplitude  also has singularities in the variable $M$ since it represents an unstable channel. In studying a particular decay process, {\it e.g.} $J/\psi \to 3\pi$,  $M$ is fixed in a very narrow range (within the width of the $J/\psi$),  and dependence on $M$ is effectively fixed and its singularity structure irrelevant.

In general,  it is not known how to write an amplitude that has correct unitarity 
 constraints in two or more overlapping channels. The reason being that it is only simple to implement 
  unitarity on partial waves where it is an algebraic constraint. A single partial wave in one channel corresponds to an infinite number of partial  waves in another one thus imposing unitarity on say both $s$ and $t$-channel partial waves simultaneously requires an infinite number of partial waves. Regge theory  extends the 
  concept of analytical continuation to the angular momentum variable of partial waves. Singularities in the angular  momentum plane  {\it e.g.} Regge poles, determine the behavior of infinite sums of partial waves, thus Regge theory is used to implement  cross channel unitarity.

 In general, amplitude singularities are known only in a limited domain, but as long as they dominate in 
  a kinematical region of interest one may be able to construct a realistic amplitude model.  Amplitude models fall into two main categories. One is that of dual models, {\it e.g.} the Veneziano model and  the other is the isobar  model category,  {\it e.g.} the Khuri-Treiman model. Dual models attempt to incorporate $S$-matrix constraints directly on the full amplitude that depends on the Mandelstam variables. 
  Since isobar is synonymous with a partial wave, isobar models are models based on a (truncated) partial wave expansion. 

   In the following I will discus the isobar model in some detail since this is almost exclusively the model used at present in analyses of three particle Dalitz distributions.

 \subsection{The Breit-Wigner amplitude} 
 
 In the  Breit-Wigner formula, a partial wave, $f_l(s)$ is approximated by a pole in $s$ 
  located in the complex energy plane at $s_P = Re s_P - i \mathit{Im} s_P$, 
\begin{equation} 
f^{p}_l(s) \propto \frac{1}{s_P - s}  \label{p} 
\end{equation} 
 The real and imaginary parts are related to the mass,  $M = \sqrt{Re\, s_P}$  and the width, $\Gamma = Im\, s_P/M$  of a resonance. For comparison with experiment a reaction amplitude 
  is evaluated at physical, real values of  kinematical variables {\it e.g} when $s$ approaches
    the real axis from above. The contribution form a  BW  pole to a partial cross section, $\sigma_l$ is  proportional to 

\begin{equation} 
\sigma_l \propto \lim_{\epsilon \to 0} |f^p_l(s + i\epsilon) |^2 \propto \frac{1}{(s - M^2)^2 + (M \Gamma)^2}.  
\end{equation} 

Since the resonance pole is located in the complex s-plane, on the real s-axis,  where experimental data is taken, it produces a smooth variation in the cross section. The closer the pole is to  the real axis, {\it i.e.} the smaller the resonances width, the more rapid is the variation in cross section or event distribution. 
It is worth keeping in mind that once energy  is considered as a complex variable any variation of the reaction amplitude in the physical region can be traced to the existence of singularities in the complex energy plane 
  {\it e.g.} a pole as in the BW formula.

   \subsection{ Unitarity and the Breit-Wigner amplitude} 
   
 The BW formula of Eq.~(\ref{p}) is an analytical function of $s$ except for a single pole at $s=s_P$. How does unitarity constrain the BW pole? Resonance decay is possible because of open channels and it is unitarity that controls the distribution of probability across decays.  It thus follows that unitarity must 
  constrain resonance decay widths and thus the imaginary part of the BW pole. But Eq.~(\ref{p})  is only an approximation to the ``true", unitary amplitude valid for $s$ near the position of the complex pole. 
 Since  unitarity operates in the physical domain, {\it i.e.} 
       on the real axis, the constraint of unitarity on the ``true" amplitude is lost in the pole approximation, {\it i.e.} at a finite distance from the real axis. In this case implementing unitarity is related to using 
       energy dependent widths. 

      Suppose the lowest mass open channel is a state of  particles $A$ and $B$ with threshold at $s=s_{th} = (m_A + m_B)^2$. In the mass range between $s_{th}$  and the first threshold, 
    unitarity constrains the ``true" amplitude to satisfy
   
   \begin{equation} 
   Im \hat f_l(s) = \hat t^*_l(s) \rho_l(s) \hat f_l(s).  \label{f} 
  \end{equation} 
  Here $\hat f_l(s)  =  f_l(s)/(a q)^{l}$  and $\hat t_l(s) = t_l(s)/(a q)^{2l}$ are the $s$-channel 
   reduced partial waves 
    representing production  of $AB$ in $D + \overline{C} \to A + B$ and elastic $A + B \to A + B$ scattering, respectively. Near threshold $q\to0$, with  $q$  being the relative momentum between $A$ and $B$ in the 
     $s$-channel center of mass frame,  partial waves vanish as $(aq)^l$,  where $a$ is given 
      by the position of the lowest mass singularity in the crossed channels, {\it i.e.} the range of interaction.  $\rho_l(s)$ is a known kinematical function describing the two-body phase space. It has 
      a  square-root  branch point at $s = s_{tr}$. As $s$ increases beyond the first inelastic threshold, $Im \hat f_l(s)$ receives a  ``kick" from another square-root type singularity form channel openings  
       and the {\it r.h.s} of Eq.~(\ref{f}) needs to be modified.  Eventually three and more particle channels open. 
        In practice unitarity is a useful constraint in a limited energy range that covers a small number of open channels  {\it e.g.} close to the elastic threshold. 
              Replacing, in Eq.~(\ref{f}),  $\hat f_l$  by $\hat t$ one obtains the unitarity relation for the elastic 
         $A + B \to A + B$ partial waves in the elastic region, 
   \begin{equation} 
   Im \hat t_l(s) = |\hat t(s)|^2 \rho_l(s). \label{t} 
  \end{equation} 
 Above the inelastic threshold the {\it r.h.s} of Eq.~(\ref{t}) should be modified as discussed above. 
  In case there is a finite number $N$  of  relevant inelastic channels, the unitarity condition can be expressed in a matrix form 
\begin{equation} 
Im \hat t_{l,ij}(s) = \sum_k  \hat t_{l,ik}^*(s) \rho_{l,k}(s) \hat t_{kj}(s) 
\end{equation} 
 where $\hat t_{l,ij}(s) = t_l(s)/(aq)^{l_i}(aq)^{l_j}$  and 
 $\rho_{l,k}(s)$ is the appropriate, reduced phase space in channel $k$. 
 The unitarity relation for $\hat f_{l,i}(s)$ takes on a similar form, 
\begin{equation} 
Im \hat f_{l,i}(s) = \sum_k  \hat t_{l,ik}^*(s) \rho_{l,k}(s) \hat f_{k}(s). 
\end{equation} 

For a given $\hat t_l(s)$,  the analytical amplitude $\hat f_l(s)$ that satisfies Eq.~(\ref{f}) can be written as  

    \begin{equation} 
  \hat f_l(s) = \hat t_l(s) G_l(s)  \label{sf} 
  \end{equation} 
and sometimes the so-called Muskhelishvili-Omnes function is used instead of $\hat t_l(s)$ on the {\it r.h.s} of Eq.~(\ref{sf})   \cite{Pham:1976yi}.  The function $G_l(s)$ is an analytical function of $s$ with cuts 
 except on the real axis in the elastic region. The latter are accounted for by the elastic amplitude $t_l(s)$. 
  The singularities  of $G_l(s)$ correspond to the (often unknown) contributions from the unitarity-imposed 
   left hand cuts cuts that exist in the crossed, $t$ and $u$ channels. 
 The inelastic contributions {\it i.e.}  right cuts in $\hat f_l$ are related to the inelastic contributions to the amplitude    $\hat t_l(s)$ which is easy to show if the matrix representation is used, 
   \begin{equation} 
   \hat f_{l,i} = \sum_k \hat t_{l,ik} G_{l,k}(s) \label{sf2}
   \end{equation} 
 and with the functions $G_{l,k}(s)$ bearing only left hand cuts.    
     Now we go  back to the pole formula. From Eq.~(\ref{f})  it follows that poles of the production amplitude  
     $\hat f_l$ are also poles of $\hat t_l$. This is because no resonance poles appear on sheets connected 
   to the left hand cuts~\cite{gribov}.  Since $\hat t_l(s) $  satisfies  Eq.~(\ref{t}) it can be shown that the most general parametrization has the form, 
  \begin{equation} 
  t_l(s) = \frac{1 }{ C_l(s)  - I_l(s)  } \label{st} 
  \end{equation}
  which in the inelastic case generalizes to the matrix form 
  \begin{equation} 
  t_{l,ij}(s) = [ C_{l}(s) - I_l(s) ]^{-1}_{ij} 
  \end{equation}
 with $C_l$ and $I_l$ becoming   $N\times N$ matrices in the channel space.   
     The function $C_l(s)$ ($C_{l,ij}(s)$)   has  similar properties to the function $G_l(s)$  ($G_{l,i}(s)$)  in Eq.~(\ref{sf}),     {\it i.e.} it is real for real $s$ in the elastic region and have only left hand cuts. 
   The function $I_l(s)$ is a known analytical function {\it i.e.} the Chew-Mandelstam function, 
  %given by 
%\begin{equation}  
 %I(s) =   \frac{s}{\pi} \int ds' \frac{\rho(s')}{s' (s' - s)} 
 %\end{equation} 
with its imaginary part for real $s$ given by $Im I_l(s) = \rho_l(s)$. 
The reason why the analytical solution of Eq.~(\ref{t}) is more complicated than that of Eq.~(\ref{f}), is that the former is a non-linear relation for the amplitude. It is easy to check that this equation becomes a linear condition for the inverse of $\hat t_l$  
 and this is the reason why dependence on phase space appears in the denominator in Eq.~(\ref{st}). 
 It is straightforward  to check that Eq.~(\ref{st}) satisfies Eq.~(\ref{f}), or in the inelastic case,  its matrix generalization.  The presence of  ``denominators" in amplitudes are a direct consequence of unitarity and so are the resonance  poles, which correspond to zeros of the denominators.

  One  immediately recognizes that the $K$-matrix, or the $K$ function in the elastic case,  
   corresponds to 
 \begin{equation} 
 K^{-1}_l(s) = C_l(s) - Re I_l(s). 
 \end{equation} 
 Since the real  part of a function is not an analytical function, an analytical approximation to the $K$ matrix 
  violates analyticity of the amplitude and may lead to spurious ``kinks" from square-root unitarity branch points in the physical region. 
 It  is much better to use Eq.~(\ref{st}), {\it aka} the Chew-Mandelstam representation, with the analytical ($K$-matrix type) parametrization reserved for the analytical  function $C_l(s)$. 
       
  Combining Eq.~(\ref{st}) with Eq.~(\ref{sf}) one obtains 
  \begin{equation} 
  \hat f_l(s) = \frac{G_l(s)}{C_l(s) - I_l(s)}  \label{solf1} 
  \end{equation} 
  or in the matrix form for the inelastic case 
  \begin{equation} 
  \hat f_{l,i}(s) = \sum_k [C_l(s) - I_l(s)]^{-1}_{i,k} G_{l,k}.  \label{solf2} 
  \end{equation}

  Now we can finally see how  the BW  pole formula of Eq.~(\ref{p})  emerges. Suppose in Eq.~(\ref{solf1}) the denominator vanishes  at some complex $s= s_p$. Near the pole  of  $\hat f_l$ 

\begin{equation} 
 \hat  f_l(s) \sim \frac{\beta_l}{s - s_p}  
  \end{equation} 
  where 
  \begin{equation} 
  \beta_l = G_l(s_p)/(C'_l(s_p) - I'_l(s_p)).
  \end{equation} 
  In the inelastic case the role of the denominator in the {\it r.h.s} of  Eq.~(\ref{solf1}) is  played by the  determinant of the $N \times N$ matrix $[C_l(s) - I_l(s)]$. 
  Even though the residue of the pole $\beta_l$  is in general  a complex number, it can be shown that only its magnitude is to be related with  a coupling of a resonance to a decay channel~\cite{gribov}. The residue $\beta_l$ should be distinguished from the numerator $G_l(s)$. The latter is energy dependent and represents the production amplitude of the final state  $A + B$ given the initial state $D + \overline{C}$. The former is a number representing the  product of couplings of the resonance to the  initial and final states. 
     
 Finally we  ``derive'' the more familiar  BW formula, with energy dependent widths. In  the ``LHCb'' notation 
 \begin{equation} 
 f^{LHCb}_{l_s}(s) = F_{l_s}(q) R_{l_s}(s) F_{l_t}(p)  
 \end{equation} 
 so that the reduced amplitude is given by 
 \begin{equation} 
 \hat f^{LHCb}_{l_s}(s) = \frac{ F_{l_s}(q)}{(aq)^{l_s}} R_{l_s}(s) F_{l_t}(p)
 \end{equation} 
 where $q = q(s)$ is the decay channel relative momentum between $A$ and $B$ 
  and $p = p(s)$ is the decay channel relative momentum between the $(AB)$ pair and the spectator particle $C$. For comparison with the analysis given above all that is needed is to replace the decay channel expressions for $q$ and $p$ by the $s$-channel ones. 
 The function $F_l(x)$ is a product of an angular momentum barrier factor  $x^{l}$ and a Blatt-Weisskopf  factor   
 \begin{equation} 
 F_l(x) = (ax)^l F'_l(x)
 \end{equation} 
where, for example, 
 \begin{equation} 
 F'_2(q) =  \sqrt{\frac{13}{((aq)-3)^2 + 9(aq)}}  
 \end{equation} 
 The propagator $R_{l_s}(s)$ is given by 
 \begin{eqnarray} 
& &  R_{l_s}(s) = \frac{1}{ m_r^2 - s - i \rho_{l_s}(s) X(p) } \nonumber \\
 & &  = 
  \frac{1}{  X(p) (  C^{LHCb}(s) -  i \rho_{l_s}(s) ) }, \, C^{LHCb}(s)   \equiv \frac{m_r^2 - s}{X(p)}  \nonumber \\
 \end{eqnarray} 
 so that one can rewrite the LHCb amplitude model as 
 \begin{equation} 
 \hat f^{LHCb}_{l_s}(s) = \frac{G^{LHCb}_{l_s}(s)}{ C^{LHcB}(s) - i\rho_{l_s}(s) }  \label{BWlhcb} 
 \end{equation} 
 where 
 \begin{equation} 
 G^{LHCb}_{l_s}(s)   =  \frac{F'_{l_s}(q) F_{l_t}(p)}{X(s)}. 
 \end{equation} 
 Eq.~(\ref{BWlhcb}) is a specific case of Eq.~(\ref{solf1}). It is in fact the $K$ matrix (function) approximation 
 since only the imaginary part, $\rho_{l_s}(s)$ of the dispersive integral $I_l(s)$ is used.
   The functions $C_l$ and $G_l$ containing, through left hand cuts,  physics of elastic production of $AB$ in  $A + B \to A + B$ and  in decay $D  \to A + B + C$, respectively, 
    have been replaced by a specific product of Blatt-Wisskopf factors. The latter originate 
     from a potential model in non-relativistic scattering and at best can be considered as a crude  approximation. In precision data analysis they should be replaced by a more flexible 
       parametrization.  In LHCb, analysis contributions from several poles are included by adding 
        BW amplitudes. Eq.~(\ref{solf1}) shows how all such poles need to appear as zeros of the common denominator. Finally the matrix representation of Eq.~(\ref{solf2})  is  the correct formula for dealing with multiple channels.

   \subsection{ Combining $s$, $t$, and $u$, channel isobars and corrections to the isobar model} 
 
In the previous section we discussed  how the energy dependent  BW amplitude is related to the general 
 expression for the partial wave. Here we discuss how the partial wave amplitudes build the full amplitude in the decay channel. 
 
In the scattering domain of the $s$-channel, the  partial wave expansion of the full amplitude 
 converges and a finite number of partial waves waves may give a good approximation to the whole sum.  Energy dependence of individual 
  partial waves can be represented using expressions like the one in Eq.~(\ref{sf}) or  Eq.~(\ref{sf2}). 
  In a decay channel, the partial wave series also  converges, but one cannot simply replace the one sum but the other. This is because  the decay channel partial waves have extra ''complexity'' compared to the scattering channel, due to $t$ and $u$ channel singularities, {\it i.e.} resonances begin in the physical region. 
     Ignoring spins of external particles,    the $s$-channel partial wave series is given by 
  \begin{equation} 
  A(s,t,u) = \frac{1}{4\pi} \sum_{l=0}^\infty (2l+1) f_l(s) P_l(z_s). \label{spw} 
  \end{equation} 
On the {\it r.h.s} the dependence on $t$ and $u$ is algebraic, through $z_s$ and the rotational functions. Thus in the physical domain of the decay,  the $s$-channel series diverges because the {\it l.h.s} has singularities in $t$ and $u$. Furthermore, it follows that any truncated, finite set of $s$-channel partial waves cannot reproduce $t$ or $u$-channel singularities, {\it e.g.} resonance which appear inside the Dalitz plot. These issues are resolved in the isobar model by replacing the infinite number of $s$-channel partial waves by a  finite set and  adding  a (finite)  set of $t$ and $u$ channel partial waves. Thus the amplitude has a mixed form that includes partial waves (isobars) in the three channels simultaneously, 
 
 \begin{equation}
 A(s,t,u) = \frac{1}{4\pi} \sum_{l=0}^{L_{max}} (2l+1) a^{(s)}_l(s) P_l(z_s)  + (s \to t) + (s \to u) \label{im} 
 \end{equation} 
 We refer to $a^{(s)}_l(s)$ as the isobaric amplitude in the $s$-channel and analogously, 
   $a^{(t)}_l(t)$ and $a^{(u)}_l(u)$ are the isobaric amplitudes in the $t$ and $u$-channels, respectively. 
 In a typical Dalitz plot analysis, the isobaric amplitudes are parametrized using the energy dependent BW amplitudes discussed in the previous sections.  
 
 Projecting the {\it r.h.s} of Eq.~(\ref{im}) into the $s$-channel gives the $s$-channel partial waves, which 
   we denoted by $f_l(s)$,  
 \begin{eqnarray} 
& &  f_l(s)  = a^{(s)}_l(s) \nonumber \\
& &  +\frac{1}{2} \int_{-1}^1  dz_s P_l(z_s) \sum_{l'=0}^{L_{max}} (2l' + 1) a^{(t)}_{l'}(t) P_{l'}(z_t) + (t \to u) \nonumber \\  & & \equiv a^{(s)}_l(s) + b^{(s)}_l(s). 
 \end{eqnarray} 
 Under the integral, $t$ and $u$ are to be considered as function of $s$ and $z_s$, the cosine of the $s$-channel scattering angle.  Since the integral contributes to partial waves with arbitrary $l$, Eq.~(\ref{im}) defines a model   for an infinite number of partial waves $f_l(s)$ and gives the result of the  analytical continuation of the   series in Eq.~(\ref{spw}).  
Application of unitarity in the $s$-channel leads to a relation between the isobaric  amplitudes 
\begin{equation} 
a^{(s)}_l(s) = t_l(s) \left[ \frac{1}{\pi} \int_{s_{tr}} ds' \frac{ \rho_l(s') b^{(s)}_l(s')}{s' - s} \right]. \label{a} 
\end{equation} 
The amplitude  $b^{(s)}_l(s)$ is  the $s$-channel projection of the $t$ and $u$ channel exchanges. 
 As a function of $s$ it has a left hand cut but no right hand, unitary  cut. The 
dispersive integral in Eq.~(\ref{a})  has the $s$-channel unitary cut. 
 Thus $s$-channel unitarity demands that the $s$-channel isobaric amplitude $a^{(s)}_l(s)$ has the right cut 
  coming not only form the elastic $A + B \to A + B$ amplitude, $t_l(s)$, but also from the dispersion of $s$-channel projections of the  $t$ and $u$-channel amplitudes. It is important to note this difference between the partial wave amplitudes $f_l(s)$ and the isobaric amplitudes $a^{(s)}_l(s)$. In case of the former the right hand cut discontinuity comes entirely  from the elastic scattering, {\it c.f.} Eq.~(\ref{sf}),~(\ref{sf2}). In the isobar model, the partial wave amplitudes are given by, 
\begin{eqnarray}  
& &  f_l(s) = t_l(s) \left[ \frac{ b^{(s)}_l(s) }{t_l(s)}+  \frac{1}{\pi} \int_{s_{tr}} ds' \frac{ \rho(s') b^{(s)}_l(s')}{s' - s} \right]  \nonumber \\
& & \equiv  t_l(s) G_l(s)  
\end{eqnarray} 
which are indeed of the form given by Eq.~(\ref{sf}), since it can be shown that the right hand cuts cancel between the two terms in the square bracket so that  $G_l(s)$ has only left hand cuts,

\begin{equation}  
  G_l(s) =  \frac{1}{\pi} \int_{s_{tr}} ds' \rho(s') \frac{  b^{(s)}_l(s') - b^{(s)}_l(s)}{s' - s}.
\end{equation}

 In the inelastic case this generalizes to the form given by Eq.~(\ref{sf2}). 
 The left hand cuts, as expected, originate from exchanges in the crossed channels determined by the amplitude $b^{(s)}_l(s)$. 
  
   One goes further and also  impose $t$ and $u$-channel unitarity, thereby correlating isobar expansions in the three channels. This is the analytical, unitary description of ``final state interaction"   that relies on the model independent features of the amplitudes only. 
 
  The implications for event distributions  in the Dalitz plot, given the left hand cut singularities of 
  $b^{(s)}_l(s)$,  were studied in ~\cite{schmid}. In the next section we take a  look at the left hand cuts of the amplitude $G_l(s)$.

  \subsection{Watson's theorem} 
  Since the imaginary part of a complex function is itself a real function, it follows from Eq.~(\ref{f}) that in the elastic region, the phase of $\hat f_l(s)$ equals that of the elastic scattering amplitude $\hat t_l(s)$. 
It is often forgotten to be mentioned, however, that in many cases the {\it r.h.s}  in 
 Eq.~(\ref{f}) does not saturate the imaginary part even in the elastic regime. When this happens,  Watson theorem is violated even in the elastic regime. And this is the case for a decay process. In this case there is nevertheless a relation between $\hat f_l$ and $\hat t_l$, or a generalized Watson relation. 
 
 In the decay kinematics, even if  the energy of the $A B$ state is below the inelastic threshold, the imaginary part of the $s$-channel partial amplitude, $\hat f_l(s)$,   is not in given by Eq.~(\ref{f}). 
  In this case because  in the decay kinematics,  cross channel singularities (from thresholds/unitarity) {\it e.g.} isobar exchanges in $t$ or $u$ channel,  are located in the physical region of the $s$-channel and contribute to the imaginary part of the  $s$-channel partial wave. Another way of saying this, is that in the decay kinematics singularities of $\hat f_l(s)$, which otherwise are to the left of the  elastic unitarity cut,  move to the right of the elastic unitary branch point.  
 
In the previous section we argued that  Eq.~(\ref{sf}) follows from the assumption that $\hat f_l(s)$ is an analytical function.  The relation between analyticity and casualty applies to full amplitudes, while partial waves are related to full amplitudes in a complicated way. Analytical partial waves are  obtained by a continuation of the unitary relation to the complex energy plane. It can be shown  (see {\it e.g.} \cite{gribov}) 
  that when left and right hand cuts are separated $Im \hat f_(s)   = \Delta \hat f(s_+) \equiv  (\hat f_l(s_+) - \hat f_l(s_-) )/2i $, where $s_\pm = s \pm  i \epsilon $. That is,  the imaginary part of the amplitude as measured in the experiment, which by itself is a real function, is  equal to the discontinuity across the real axis of the unique extension of $\hat f_l(s)$ to the complex energy plane. It turns out,  however,  that in decay kinematics, the proper extension of $\hat f_l(s)$ to the complex energy plane is such that in 
  Eq.~(\ref{f}) the {\it l.h.s} should be replaced by 
 $\Delta f(s_+)$ and the {\it r.h.s} should be replaced by the product 
 $t_l(s_-) \rho_l(s) f_(s_+)$.  Thus, elastic unitarity still determines the discontinuity across the right hand cut but it is no longer a real function. This is the generalized Watson or discontinuity relation.

 As a result the representation given by Eqs.~(\ref{sf}),(\ref{sf2}) is still valid 
with  the exception that the numerators, $G_l(s)$ become complex in the elastic region. They do not have the unitarity cut though. Just like in the ``standard" Watson's theorem the right hand cut discontinuity comes entirely from the elastic amplitude $t_l(s)$.  The  complexity of $G_l(s)$ is still of the left hand cut nature, except that left hand cuts have moved onto the right hand side under the right hand cut and into the second sheet. 
\cite{Szczepaniak:2015eza}.

\subsection*{Acknowledgments}
I would like to thank the organizers and participants of the LHCb workshop on multi-body decays of B and D mesons for stimulating discussions. 
 This  work is  supported by the U.S. Department of Energy, Office of Science, Office of Nuclear Physics under contract DE-AC05-06OR23177. It is  also supported in part by the U.S. Department of Energy under Grant No. DE-FG0287ER40365.

% \begin{thebibliography}{40}

% \bibitem{MartinSpearman} 
% A.D. Martin and T.D. Spearman, {\it Elementary Particle Theory}, (North-Holland, Amsterdam, and Elsevier, New York, 1970.) 

% \bibitem{Danilkin:2014cra} 
%   I.~V.~Danilkin, C.~Fern·ndez-RamÌrez, P.~Guo, V.~Mathieu, D.~Schott, M.~Shi and A.~P.~Szczepaniak,
%   %``Dispersive analysis of ?/??3?,??*,''
%   Phys.\ Rev.\ D {\bf 91}, no. 9, 094029 (2015)
%   [arXiv:1409.7708 [hep-ph]].
  
%  \bibitem{Pham:1976yi} 
%   T.~N.~Pham and T.~N.~Truong,
%   %``Muskhelishvili-Omnes Integral Equation with Inelastic Unitarity: Single and Coupled Channel Equations,''
%   Phys.\ Rev.\ D {\bf 16}, 896 (1977).
%   %%CITATION = PHRVA,D16,896;%%
 
%  \bibitem{gribov} 
% V.N.~Gribov, J.~Nyiri, Y.~Dokshitzer, {\it  Strong interactions of hadrons at high energies : Gribov lectures on theoretical physics}, (Cambridge University Press Cambridge, UK, New York 2009). 

% \bibitem{Szczepaniak:2015eza} 
%   A.~P.~Szczepaniak,
%   %``Triangle Singularities and XYZ Quarkonium Peaks,''
%   Phys.\ Lett.\ B {\bf 747}, 410 (2015)
%   [arXiv:1501.01691 [hep-ph]].

%  \bibitem{schmid} 
%  C.Schmid, Phys. Rev. 154, 1363 (1964). 
% \end{thebibliography}

%   \end{document} 

\newpage

\aSection{Comments on the LHCb determination of the $\sigma/f_0(500)$-meson composition from 
$B$ decays interpreted within constituent quark models}{J.R.~Pelaez}
\begin{center}
J.R.~Pelaez
\vskip .2cm
{\small Departamento de F\'isica Te\'orica II, Universidad Complutense de Madrid, 28040 Madrid, Spain}
\end{center}

We elaborate here on the discussion about the LHCb determination of the  tetraquark
versus quarkonium composition of the $\sigma/f_0(500)$ meson. We are not discussing the 
experimental measurements of $\bar B^0\rightarrow J/\psi \pi\pi$ decays, but the interpretation in terms of $f_0(500)$ and $f_0(980)$ resonance production and the use of 
decay ratios in order to extract the alleged quark composition of these states.
It is argued that the model used by LHCb relies on strong assumptions which are hard to justify, in view of the huge $\sigma$ decay width into two pions.

Although in this meeting I talked about the status of the $\sigma$ or $f_0(500)$ meson, how it has evolved from a confusing situation to a relatively precise status, and how there are strong indications of its non-ordinary $\bar qq$ nature, these considerations can be found in other conference proceedings of mine \cite{Pelaez:2015qba}, or, even better, in an extensive recent review \cite{Pelaez:2015qba}. Thus, instead of repeating myself
in yet another proceeding contribution, it might be more interesting to
summarize a follow-up discussion that was triggered in this Rio LHCb meeting but that was developed over some mail exchanges with members of the collaboration during the following couple of months. The discussion, triggered by Tim Gershon's talk in this workshop,  concerns 
the $\sigma/f_0(500)$ and $f_0(980)$ composition in terms of 
simple models and the use by LHCb of $B$ meson decay ratios to determine the structure of these states in terms of quarks. For brevity I will use $\sigma$ and $f_0$, respectively.
The relevant LHCb publications are \cite{Aaij:2013zpt} and \cite{Aaij:2014siy}, which are based on the phenomenological work by Stone and Zhang \cite{Stone} (which is built along the lines of \cite{Fleischer:2011au}).

Of course, the main issue with the $\sigma$ is that it has a width as large as its mass.
Recall the Review of Particle Physics (RPP) \cite{PDG} quotes a pole mass of 400-550 MeV and pole width (twice the imaginary part of the pole position) of 400-700 MeV.
It is a resonance that barely propagates and disintegrates to two-pions extremely fast. 
It is therefore very different from other resonances and the approximations 
that we take for granted for 
most resonances
do not always make sense for it.
Let us discuss some details on the caveats the theorists at Rio\footnote{Let me remark that, among others, E.Oset stated right there that one should avoid such a formulation} had on the equation shown in the LHCb papers and
\cite{Stone}, for
the decomposition of the sigma. The equations in question were these:
\begin{equation}
\ket{f_0}=\cos \phi \ket{s\bar s} +\sin \phi\ket {n\bar n}; \quad \ket{\sigma}=-\sin \phi \ket{s \bar s} +\cos \phi\ket {n \bar n};\label{naivedecomp}.
\end{equation}
and the decay ratios that followed to determine that angle. Here $\ket {n \bar n}=(\ket {u \bar u}+\ket {d \bar d})/\sqrt{2}$.

1)  The above equations  are just a change of basis between normalized states.
A first caveat about the use of Eq.\ref{naivedecomp} above is that
resonances are not bound states and their spacial wave function is not normalizable (which very naively and intuitively means that the wave function of two pions resonating as a sigma extends to infinity).
Of course, it was immediately pointed out that a similar mixing treatment is used for other pairs of resonances with 
the same quantum numbers,
like $\eta-\eta'$ and $\phi-\omega$. This is correct, but one has to keep in mind that these four resonances have minute widths (the largest being that of the $\omega$ which is 8.5 MeV to be compared with its mass of 782 MeV). When mixing equations similar to Eq.\ref{naivedecomp} are written for these very narrow resonances their width is being ignored. I, for instance, have used  some similar ``Fock decomposition'' of the $\sigma$ \cite{Cohen:2014vta} (although considering even more quark-level states), but only in the large $N_c$ limit, where the relevant widths tend to zero. 
In particular, for the $\eta-\eta'$ and $\omega-\phi$ systems one neglects their widths 
{\it assuming} they are so small that the resonance formation dynamics and the description of their ``structure'' or whatever is to be described about them is barely changed within such an approximation. But that assumption is hard to sustain for the $\sigma$, 
whose mass is as large as its width. Moreover, it is hard to understand  
why hadronization effects should be the same, or even similar, for the $\sigma$
 and the $f_0$, which are so different. Just recall that the $f_0$ pole mass is $990\pm20\,$MeV
and its with is 40-100 MeV, according to the RPP\cite{PDG2014}. Furthermore, given the energy spread of the sigma meson, 
one could even wonder why its composition should be the same at all energies, or why it would have the same mixing 
with the $f_0$ at different energies over its mass distribution.

2) Even with the very strong assumption that Eq.\ref{naivedecomp} may describe the mixing of the $f_0(980)$ with such a wide state as the $\sigma$ without taking into account the two-meson continuum,
one could think that since we do not know how to describe hadronization well, 
we could get rid of hadronization factors in ratios of observables. 
This is the idea in the LHCb papers above. 
This could be more or less justified for cases like $\phi/\omega$ mixing
since they are extremely narrow, well established $\bar qq$ states\footnote{In the naive, intuitive, quark-model sense. The real theory is QCD, of course. But lets keep it simple.}, 
with relatively similar masses, and one could expect that, 
apart from kinematic factors due to their different mass, the dynamics 
that bind quarks or whatever inside them are very similar. 
Therefore if one was to form ratios of decays for these narrow resonances as done in the above papers, 
one may not complain much if their mixing angle was extracted just considering ratios and taking
into account kinematic factors like those called $\Phi(M_1)/\Phi(M_2)$ in those papers. 
In other words, we expect the unknown non-perturbative QCD dynamics that bind 
two-quarks inside the $\phi$ and the $\omega$ to roughly cancel in the ratios.
But let me remark that this is an assumption, although it may look reasonable. 
We cannot calculate the deviations from this assumption, but we expect them to be small in cases like $\phi/\omega$ mixing.

However, if for the $\sigma$ and $f_0$ one tries to use  Eq.\ref{naivedecomp} above and form ratios like 
\begin{equation}
\tan^2\phi=\frac{{\cal B} (\bar B^0\rightarrow J/\Psi f_0(980))}{{\cal B} (\bar B^0\rightarrow J/\Psi f_0(500))}
\frac{\Phi(500)}{\Phi(980)}
\label{ec:ratio}\end{equation}
i.e. the ones in the LHCb references we are discussing, it is implicitly assumed not only that Eq.\ref{naivedecomp}
makes sense, but also that the dynamics to create a $\sigma$ and an $f_0$ are the same and cancel in the ratios. But 
looking at the $\sigma$`s huge width of hundreds of MeV against 
the small tens of MeV width of the $f_0$, and also considering the difference in masses, 
this assumption seems rather strong, if not just unjustified.

Using this kind of ratios to extract mixing angles or other meson properties implies therefore two radical assumptions:
I) that neglecting the huge $\sigma$ width does not change its nature and the dynamics of its formation,
the internal structure and distribution of its components
and II) that these dynamics, structure, distribution of components, etc... are exactly the same for the $f_0$. In the paper of Stone and Zhang \cite{Stone},
this is done in two steps. First, in their perturbative diagrams made with quarks, all the complications of hadronization and the formation of mesons are included
in 
some $F_B^f$ form factors and some $\cal Z$ constants (given in their Table I) 
that "represent the coupling amplitude that depends on the quark configuration after the $\bar B$ meson decay and the quark content of the light meson in either the $\bar qq$ or tetraquark model". Second, it is claimed that all
these form factors and amplitudes cancel in ratios in such a way  that only the kinematics and the
simple flavor factors remain in their ratios of amplitudes.
But for instance, decays and formation processes 
might depend on the value of the quark wave functions at the origin 
(or their derivatives, or many other things) and there is no 
reason why this effect should be the same for the $\sigma$ and $f_0$, 
not to talk once again about the effects of pion/kaon loops, which can 
contribute differently to the $\sigma$ and $f_0$ and, at least for the sigma, are huge.
Something similar might happen in order to form the $\sigma$ and the $f_0$.
Therefore, there should be a ratio
$F_B^{f_0}{\cal Z}_{f_0}/F_B^\sigma {\cal Z}_\sigma$, 
which {\it for each particular  model} may contain flavor factors 
in the form of sines or cosines, but apart from that, 
nothing is known of its value. 
Yes, there are some estimates in the literature, but these implicitely assume a narrow
state and only a tree level calculation {\it without meson loops} that can provide the decay and therefore with a stable sigma. With such an unrealistic assumption, it is 
not strange that this ratio is predictied to be one, and this is assumed in the LHCb works. In real life, however, the width of the $\sigma$ and the $f_0$ are so different that
I would not even dare to say that the numerator and denominator in the ratio 
 should be of the same order of magnitude.
Of course, phase space kinematic factors are also there, although they do not take into account the hadronization dynamics.

Neglecting the width, pions, etc... is a very crude and unlikely model for the sigma.. 
Once again, it is the same problem... pions!! 
and two pions are essential to describe not only the huge width and the coupling to two pions, 
 but also the mass (through loops). 
The formation of the sigma and the position of its pole are
actually strongly dependent on the pion loops which provide its huge width (see the review \cite{Pelaez:2015qba}).

3) One might also be confused because it is true that equations like Eq.\ref{naivedecomp} are being ``abused'' in the literature, and people talk about the ``mixing angle''. But it is not necessarily the mixing angle above.
This is because
the spacial wave function may not be normalizable, but still other parts of the ``state''  might be of interest.
For instance, some people talk about the $\sigma/f_0$ ``mixing angle'', but this usually refers to the 
``singlet/octet'' mixing. In purity, people mean something like this:
\begin{equation}
\ket{f_0}=\cos \phi \ket{8} +\sin \phi\ket{1}; \quad \ket{\sigma}=-\sin \phi \ket{8} +\cos \phi\ket{1}.
\end{equation}
But very often people (most likely including myself), instead of writing $\ket{8}$ and $\ket{1}$, write the equivalent $q\bar q$ singlet and octet decomposition, because we are used to it since 
we {\it intuitively} understand 
normal mesons as $\bar qq$ states within a naive quark-model.
Moreover, even more frequently, people use the $\ket{n\bar n}$ and $\ket{s\bar s}$ basis as in Eq.\ref{naivedecomp}, since the $\omega/\phi$ system 
is very close to ``ideal'' mixing in terms of SU(3) factors
because the $\phi$ is very close to the $\ket{s\bar s}$ state.

But strictly speaking this is only correct for the $SU(3)$ flavor structure part of the ``state'', 
which is useful for instance 
to obtain the Clebsch-Gordan factors between different reactions where such resonances appear. 
Note ``state'' has been written between inverted commas, 
when referring to the resonance, because due to points 1) and 2) above,
resonance ``states'' are not well defined if one talks about their spacial wave function. 
The well-defined states with those flavor quantum numbers are 
the two-meson states where the resonance appears, since in practice these 
are the stable states that propagate to infinity as asymptotic states
\footnote{We are neglecting here the pion and kaon decays, i.e., we are the neglecting electroweak interaction. But this is a good approximation since their widths
are tiny compared to strong interaction scales. We do see pions and kaons propagating over macroscopic distances.}.
Then, as long as you do not make further assumptions on the  
dynamical or spacial description, this ``abuse'' of notation is fine. But one must keep in mind that it refers only to the flavor structure of the channel where you see the resonances. One should be very careful
to use such a decomposition to calculate or describe other features of  resonances.

4) Why consider only two quarks? In the paper by Stone and Zhang \cite{Stone} the authors also consider the alternative
four quark scenario... but these are particular cases. Actually, as we discussed during the conference, there are at least two independent tetraquark configurations (due to color index contractions), and only one is being considered. In a particular basis
one tetraquark configuration is the unbound two-meson state.

But if all these states are possible, and once there is strong evidence that
the $\sigma$ dynamics is dominated by pions  (because it has such a huge width)
as well as strong hints that it might be a non-ordinary meson
one might think about adding the $\pi\pi$ state to the equation, even though the  $\pi\pi$ system would be a scattering non-normalizable state. Note, however, that the effect of kaon and pion loops, while non-negligible, is very different, at least in size, for the $f_0$ than for the $\sigma$, just look at the much smaller $f_0(980)$ width.

There are nice ways to study this, as in 
the chiral unitary approach (see \cite{Liang:2014tia} and references therein) as E. Oset suggested during the workshop.
There is also a classic study \cite{weinbergdeuterium_a,*weinbergdeuterium_b,*weinbergdeuterium_c} 
that S. Weinberg introduced for deuterium, considering simultaneously a 
bound state and a proton-neutron state below the scattering threshold.
That argument was restricted to bound states, but it has been 
reanalysed for the $f_0$ case in \cite{Baru:2003qq}.

4) There is also a relatively minor issue of the phase space factor $\Phi(500)$ with the arbitrary choice of argument at 500 MeV, when
the sigma effect is sizable from the $\pi\pi$ threshold up to 1 GeV. Why 500 MeV? The PDG gives 400-550 MeV,
the precise and rigorous dispersive approaches give more like 450 MeV. Recall also that expressing the pole as $s_p=(M-i\Gamma/2)^2$
or as $s=M^2-iM \Gamma$, which is just a change in {\it definition}, the value of $M$ changes. 
At LHCb some kind of convolution or weighted mass distribution is assumed, which is
fine for the kinematic functions but, as mentiones earlier, in Eq. \ref{ec:ratio} there should be a ratio 
of the sort $F_B^{f_0}{\cal Z}_{f_0}/F_B^\sigma {\cal Z}_\sigma$,
for the different formation dynamics of the $\sigma$ and the $f_0$, 
and this ratio is not necessarily a constant but most likely would also have an energy dependence following that
of the convolution with phase space 
which is, once again, neglected. 

In addition,  there is the issue on how to assign a decay to 
the $\sigma$  or the $f_0$ decay channel, if they have the same quantum numbers and a huge overlap. 
It seems that this separation
necessarily depends on the specific choice of resonance parameterizations of the amplitude, 
which for LHCb includes some kind of Breit-Wigner parameterization of the $\sigma$ meson 
(or slight variations). 
Thus the separation of $\pi\pi$ final states between those coming from a $\sigma$ or an $f_0$
must be very model dependent, and, as many theoreticians emphasized during the workshop, 
a Breit-Wigner model (or slight variations), as used by the LHCb, is a bad choice for the $\sigma$ meson.

5)  Finally, in the discussion after the workshop, the theoretical references 
to the work in \cite{Stone} were also commented upon.
The vast majority of theoretical citations come from E.Oset, who, as already pointed out, made 
it clear during the discussion session that this formalism should be abandoned. 

There are two other citations from Christoph Hanhart and collaborators 
(one is the PDG ``Note on light scalars'' \cite{PDG2014}), who have recently reexamined  these decays within 
 a dispersive formalism \cite{Daub:2015xja}, which I think is a correct and model-independent
 way of dealing with these decays, i.e. through the Omn\'es formalism. 
The above mixing equation is not even mentioned. Moreover, the fact that \cite{Daub:2015xja} can describe the dta shows that there is no more information on the $f_0$ and $\sigma$ structure in B decays than in $\pi\pi/\bar K$ scattering \footnote{I thank C. Hanhart for this comment.}.

 The other theoreticians who quote the paper are:

 i) Stan Brodsky and collaborators, who quote the paper as their reference [33] 
 as follows when talking about the $a_0$ and $f_0(980)$ {\it ``... (note also recent
 work [33, 34] to measure their tetraquark content), it
 is not clear that their  diquark and antidiquark  components are sufficiently
 compact that they do not overlap and instantly mix with
 a meson-meson configuration''}. 
 Once again this is the very same point that has been emphasized here: 
 Meson-meson components are unavoidable in a description  of light scalars. 
They do not get into the pros and cons or even discuss the approach in any detail, but just make a clear warning about meson-meson contributions.

 ii) Frank Close and collaborators \cite{Close}. 
They redo the model of Stone and Zhang \cite{Stone} including isospin violation and therefore $a_0/f_0$ mixing.
 Surprisingly they reach OPPOSITE conclusions to those obtained from the LHCb works.
 Quoting literally: {\it ``In conclusion: the LHCb data appear to be consistent 
 with the picture of scalar mesons below 1 GeV being tetraquark
 states, and those above 1 GeV being a canonical nonet mixed with a scalar glueball.''}. 
Remember that in the LHCb paper \cite{Aaij:2013zpt} 
 it is claimed that  the result {\it ``favors somewhat a quark-antiquark interpretation''}.
And in \cite{Aaij:2014siy} LHCb makes a strong claim, already in the abstract, that 
the observed rates for the sigma {\it ``...is inconsistent with a model where these scalar mesons are formed from
two quarks and two antiquarks (tetraquarks) at the eight standard deviation level''}.
 This Close paper is at odds with that of Stone and Zhang \cite{Stone} even when using the same LHCb data.

Incidentally, note that Close and Tornqvist \cite{Close:2002zu} 
have a model where they suggest the very same picture of scalars they defend on the previous paragraph, but
made of ``layers'': an outer layer is made of mesons and would provide the meson-meson component and an
inner layer made of compact quark-antiquark or tetraquark structure. However, they rarely include the meson 
cloud in their own calculations. 

In any case, from what has been explained above, 
the picture in both the Stone and Zhang \cite{Stone} and Close et al. \cite{Close} papers
neglects the $\sigma$ width, the meson-meson state
and assumes that the  $\sigma$ and $f_0$ formation dynamics cancel in the ratio. 
I do not see how these approximations and assumptions are justified.
At the very least the analysis would 
be strongly model dependent on unjustified assumptions.

iii) Finally, a recent lattice paper on the sigma meson \cite{Howarth:2015caa}, also quotes 
the  \cite{Aaij:2014siy} LHCb paper as reporting that the sigma  {\it ``is not a mesonic bound state according to their models''} \footnote{LHCb says a tetraquark, not a mesonic state, so I think they are
making an additional identification.}. Then the authors
go on to make several brief remarks about the model dependence of LHCb's approach, which coincide with those
already discussed above in some points... i) the mass and widths of the sigma are comparable, ii) there could be more states beyond $\bar qq$ (although they think of glueballs, which are certainly suppressed at that mass) and iii) The use of a Breit Wigner parameterization.

\vskip .5 cm

It is hard to tell the physical significance of being eight deviations away from a model that does not take into account the pion continuum state and does not describe the most salient feature of the $\sigma$, which is its huge width to two pions. But being in an LHCb abstract, this might be erroneously interpreted as the end of the discussion on the sigma composition for readers who might assume that LHCb is using the currently most accepted model in the literature. Although, of course, LHCb is not making that claim about the model.  Unfortunately, I have already met several people who made that interpretation.

In conclusion, the main concern, as discussed in Rio, 
is that  LHCb is  making  a very strong claim on the composition using a very particular model based
on assumptions which are hard to reconcile, if not just inconsistent,
 with the observation of the huge $\sigma$ width.

I hope the above comments are useful and I really appreciate very much the efforts of the LHCb Collaboration. 
Let me be clear, I think LHCb can be very helpful for our understanding of the light 
and not so light scalars. Those data are great for us theoreticians!! 
We wait for them eagerly. Please keep up the good work analyzing
the decays with two-mesons in the scalar sector.
However, for the ``composition'' kind of theoretical or phenomenological analyses one really needs to use  model independent formalisms and only make very robust assumptions, which may not be the case when trying to make a quark-level description neglecting the sigma width and the two-pion state. The theory for such wide states is more complicated than the simple approximations valid
for other narrow and more conventional states, 
but to my view that makes this problem even more challenging and interesting.

\vskip -.5 cm
\subsection*{Acknowledgements}
I want to thank Alberto Reis and Ignacio Bediaga for their kind invitation to participate in this workshop and for creating such a nice environment for Physics discussions. I also want to thank Tim Gershon for his questions, his interests and for suggesting me to write this comment. I have profited from the discussions with the three of them together with Eulogio Oset and Jonas Rademacker in Rio, as well as with F. J. Llanes Estrada later on. I would also like to thnank Jonas Rademacker and the LHCb "Amplitude Analysis" sub-group for their  kind invitation to give a seminar at CERN to discuss this issue. Let me also thank Sheldon Stone for his comments on an earlier version of this manuscript and during the talk at CERN. I also thank C. Hanhart for some corrections and comments.
 
\vskip -.5 cm

%\begin{thebibliography}{99}

\newpage

\aSection{Final state interaction in $\bppp$}{I. Bediaga \&\ P. C. Magalh\~{a}es}
\begin{center}
\vskip .2cm
I. Bediaga and P. C. Magalh\~{a}es\\
{\small Centro Brasileiro de Pesquisas F\'isicas  --  CBPF\\
Rio de Janeiro - Brazil}
\end{center}

In this note  we address the issue of re-scattering effects in the charmless three-body decays
of $B$ mesons. In particular, we study the  case of  the $\bppp$ decay and show that the presence of 
hadronic loops shifts the P-wave phase near threshold to below zero, and 
modify the position of the $\rho$-meson peak in the Dalitz plot.

\subsection{Weak $\times$ strong interaction}
One can think of heavy meson  hadronic decays as a sequence of two processes. The first stage
consists of the heavy quark weak transition, a short-distance process that is usually described by 
factorization methods. We refer to this as the weak vertex. This short-distance transition
is followed by the hadronization and the long-distance process of
final state interactions (FSI), where the light mesons   interact in all possible ways before reaching 
the detector. These two processes involve different scales and in general are
treated separately with different theoretical frameworks.
\vskip .4cm
{\bf{Weak sector}}

The most common  way to treat B and D  weak decays  is using factorization techniques, as in Refs.~\cite{BenoitBppp, JPD_PRD89}, where the decay is divided in combinations of quark currents that do not interact strongly with each other. An effective Hamiltonian describes the transitions between the quark currents with the correct CKM matrix elements and the Wilson coefficients. The latter describe perturbatively the quark-gluon interaction and are the direct link to short-distance physics. 
An alternative way is the heavy mesons ChPT, developed by Burdman and Donoghue~\cite{BD} and  Wise~\cite{wise}. In this approach, the charm and bottom mesons are coupled to the light SU(3) (ChPT) by means of SU(3) operators. In the  effective Lagrangian, the quark gluon interactions are hidden in coupling constants that need to be determined by experiments. 
\vskip .4cm
{ \bf Final State Interactions}\\
It is very important to clarify that in three-body decays  there are two distinct FSI mechanisms. The first one is related to the two-body  scattering and includes all possible interactions between the two particles, changing resonances and coupled channels. This pure two-body interaction  have been extensively studied and models based on ChPT~\cite{GL_a, *Gasser:1983yg, EGPR, Bernard, CGL} and dispersion relations~\cite{Pelaez, Moussalam, Pelaez:2015qba}, applying  the constraints of analyticity and unitarity 
of the S-matrix. These models are able to describe the scattering data up to $\sim 2$ GeV.  
Approaches that consider only this kind of FSI in three-body decays  are called (2+1), or quasi-two-body, because interactions involving  the third particle, usually referred to as ``bachelor", are ignored. %This is also the case of the isobar model, build up by adding two-body resonances.  
However, re-scattering between the bachelor and the other mesons
can  also happen and br relevant. In these three-body FSI, involving  hadronic loops, there is a
momentum  sharing between the three particles in the final state. The exhaustive consideration of those effects in the amplitude is possible only by means of three-body calculations such as the solution of the Faddeev equation~\cite{lc09}. But the theoretical treatment 
of these effects  is also possible using approximate methods~\cite{dkpipi1}. 

  \subsection{ $B\to 3\p$}
 The main topologies that contribute to the process $\bppp$ are described in the diagrams shown in
 Fig.~10. One can see that the   $v.1$ diagram does not contribute at  tree level, since it leads
 to the $\p^+\p^+\p^0$ final state. The  $v.2$ diagram, the internal $W$ emission is colour suppressed. Experimental data~\cite{expnew}  indicate that the tree process involving $\r^0$ resonance seems to be 
dominant in this decay. For these reasons we choose to start our study investigating the contribution of diagram 
$a.1$.  
  \begin{figure}[ht]
\begin{center}
\includegraphics[width=.4\columnwidth,angle=0]{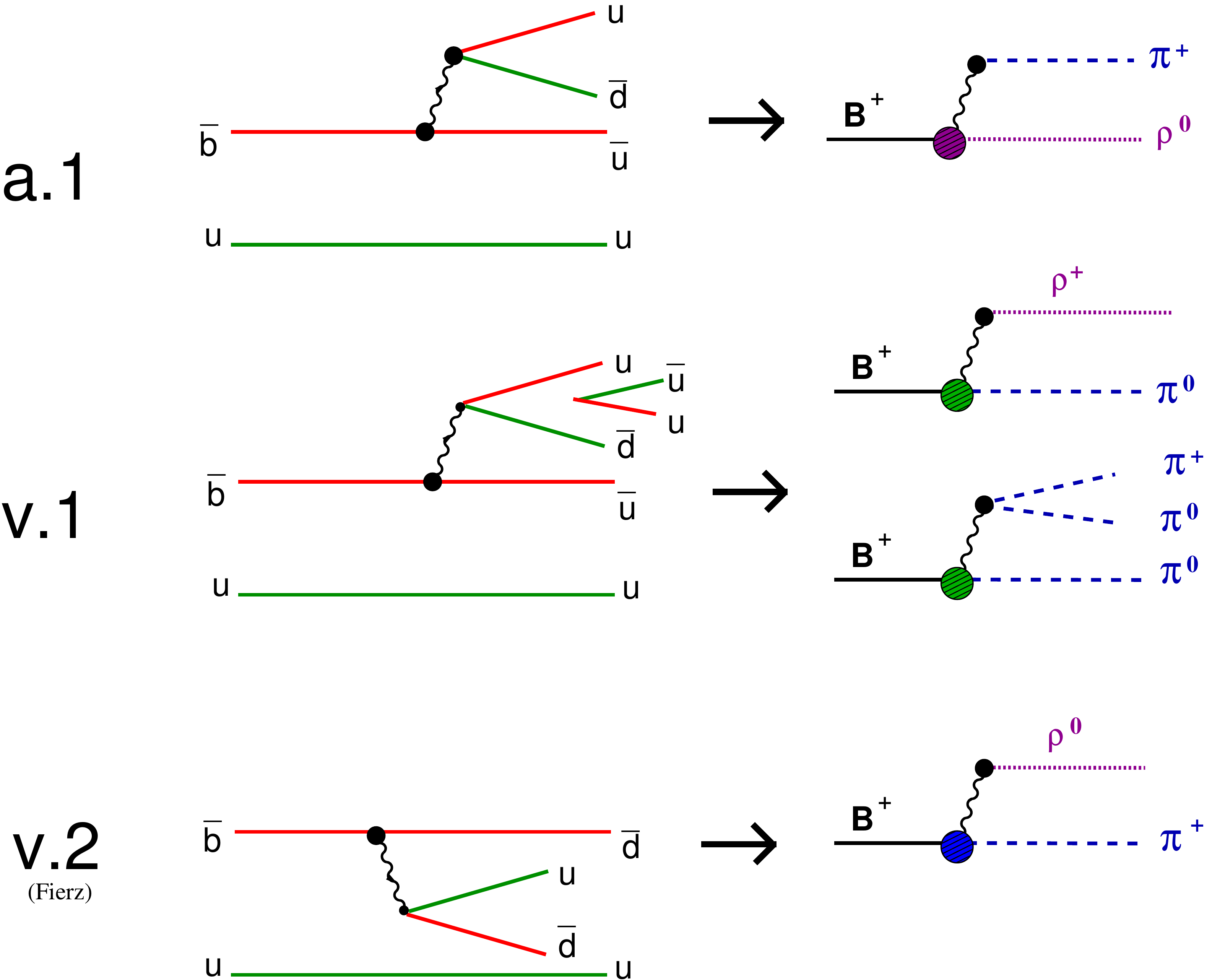}
\caption{From quarks to hadronized Feynman diagrams contribution to $\bppp$ decay.\label{fig:Bppp:F1}}
\end{center}
\end{figure}
  
In addition to the tree diagram,  the decay amplitude includes a contribution from  
re-scattering. By re-scattering, here, we mean the interaction between the pion produced  from the vacuum with the one produced from the $\rho$- meson. Therefore, the amplitude for $\bppp$ decay is represented  by the series of diagrams in Fig.~11.
\begin{figure}[ht]
\begin{center}
\includegraphics[width=.49\columnwidth,angle=0]{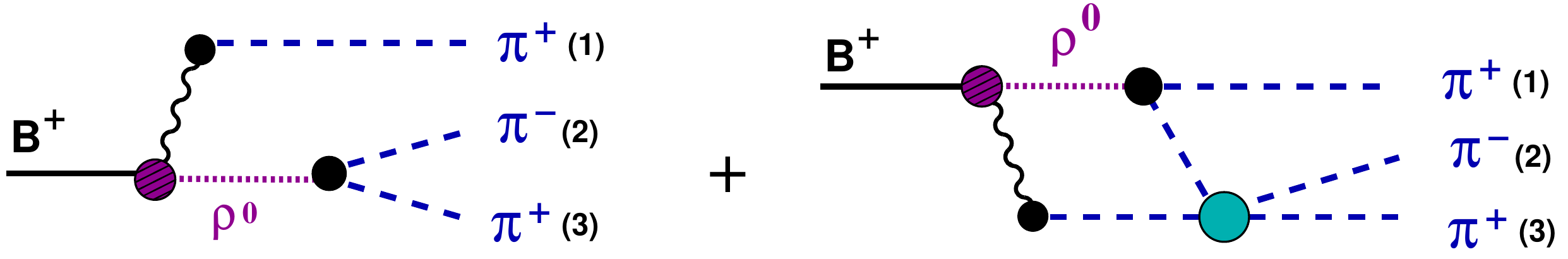}
\caption{ Diagrams contributing to $\bppp$ decay amplitude.}
\end{center}
\label{fig:Bppp:F2}
\end{figure}
In this qualitative study, we focus on the role of the  second diagram in Fig.~11 in the $\bppp$  decay 
amplitude. For this reason, the contribution from the weak vertex is not included, and this will be studied in the near future.  It is important to note that neglecting this contribution will not affect the 
conclusions of this study, since both diagrams in Fig.~11 have the same weak vertex.

\subsection*{Tree amplitude - $A_0$ }
The tree level contribution to the $\bppp$ decay amplitude, the first diagram in Fig. 11, denoted here as $A_0$, is given by:

 \bea
 A_0 &=& C_0\,[-p_1\,\cdot(p_2-p_3)]\,F^1_{\p\p}(s_{23});\label{A0}
 \eea
 where $C_0$ is a constant and $F_0^{B \rho}(p_1^2)$ is the form factor related to the $B\to \r$ amplitude. The product $-p_1\,\cdot(p_2-p_3) \equiv (s_{13} - s_{12})$  accounts for the angular distribution of the P-wave. 
 
 \subsection*{Production amplitude: $\rho \to \p\p$ }
The  $\rho$ production amplitude is a crucial element in the description of the $\bppp$ decay  for both tree and re-scattering  amplitudes. 

\begin{figure}[ht]
\begin{center}
\includegraphics[width=.59\columnwidth,angle=0]{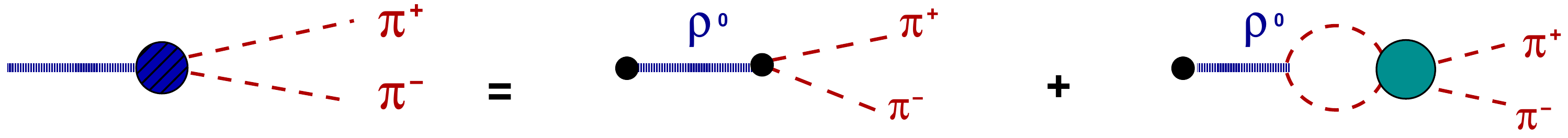}
\caption{Diagrams for process $\r\to\p\p$.\label{fig1}
}
\end{center}
\end{figure}
 The corresponding amplitude of this process, shown in Fig. 12, can be expressed terms  of the 
 $\p\p$ scattering amplitude  using the K-matrix approximation, as in Ref.~\cite{Diogo},
\beq
F^1_{\p\p}(s) = G_\r \frac{\cos\d(s)}{m_\rho^2-s}\, e^{i\d(s)}
\label{2.15}
\eeq
where $\d(s)$ is the $\p\p$ phase shift. In Eq. (\ref{2.15}) we use the  CERN-Munich 
parameterization~\cite{cern-munich1, *cern-munich2}  $\p\p$ phase shift,
instead of using a model for the  $\p\p$ scattering amplitude.

The production amplitude, Eq. (\ref{2.15}), is inside the hadronic loop in the re-scattering calculation, therefore, the amplitude needs to be integrated with the loop. 
We parametrize the experimental data by a sum of poles with a complex constant in the numerator~\cite{pat}, and write
 \bea
F^1_{\p\p}(s)= \sum_{i=1}^3 \frac{N_{\rho i}}{s - \Theta_i}\,;
\label{rhop}
\eea
where the values of $N_{\rho i} $ and $\Theta_i = a + i b$ are available in table I of Ref.~\cite{pat}. The advantage of this form is to allow us to use Feynman rules to integrate the loop, as will become clear in the following.

\subsection*{Re-scattering amplitude - A1}

The re-scattering contribution, second diagram in Fig. 11, can be written  in terms of the tree amplitude as:
\bea
A_1= -i \int \frac{d^4 \ell}{(2\p)^4} \; 
\frac{T_{\p\p} \, A_0}{\D_\p^+ \,\D_\p^- }\;,
\label{A1}
\eea
where $T_{\p\p}$ is the $\p\p$ scattering amplitude represented by the green bubble and $\D_\p^+$,  $\,\D_\p^- $ are the pion propagators inside the loop, %given by:
%\bea
$\D_\p^+ = (P_B-l)^2 - M^2_\p$ and $\D_\p^- = (l-p_3)^2 - M^2_\p$.
%\eea
It is important to note that the $B\to\r$ form factor is not a constant any more because the transferred momentum is integrated over in the loop. In this case, we use a monopole approximation and 
\bea
F_0^{B \rho}(p_1^2)   = -\frac{F_{B\r}(0)\,m^{2}_{B*}}{\,\D_{B*}} ; \,\,\,\,\D_{B*} = (P_B-l)^2 -  m^{2}_{B*} \,.
\eea

Finally, the re-scattering amplitude for $\bppp$ is given by:
\bea
A_1 &=& i \frac{ C_O \,m^{2}_{B*}}{2\, }\,T^P_{\p\p}(s_{23}) \,N_{\rho}\,\lc\, I_1 - I^t_2 \,\, \rc,
\label{A1.3}
\eea
  where $I_1$ and $I^t_2$  are  functions of  scalar and tensorial loop integrals, respectively,
\bea
 I_1 = (s_{12} - s_{13})\,&\frac{i}{16\p^2}&\,\lc \lp  M^2_B - \,2\,s_{23} + 3\,M^2_\p + \T_i\rp \Pi_{\p^+\p^- B^*\rho_i} + \Pi_{\p^- B^*\rho_i} \right. \nn\\&&+ \left.  2\,\Pi_{\p^+ B^*\rho_i} -\,\Pi_{\p^+\p^- B*} \rc \,,
 \label{I1} \\[2mm]
 I^t_2 = 2\,(p_2-p_3)_\m\,&\frac{i}{16\p^2}&\,\lc \lp  M^2_B - \,2\,s_{23} + 3\,M^2_\p + \T_i\rp \Pi^\m_{\p^+\p^- B*\rho_i} + \Pi^\m_{\p^- B^*\rho_i} \right. \nn\\&&+\left. 2\,\Pi^\m_{\p^+ B^*\rho_i} -\,\Pi^\m_{\p^+\p^- B^*} \rc \, ,\label{I2}
\eea
\ni  and the indices on functions type $\Pi_{xyz}$ refer to the particles that are taking part in the loop. 
All these loop integrals can be solved using Feynman integral technique. A guide for 
all calculations can be find in Ref.~\cite{PatThesis}.
 
\subsection{ Results}
We compare numerically the individual contributions from the tree and re-scattering 
amplitudes to the total $\bppp$ amplitude.

In Fig. 13  individual contributions to the  P-wave phase are shown as a function of 
the $\p\p$ invariant mass. Although they look similar, at
threshold the   re-scattering contribution, Eq. (\ref{A1.3}), starts  below $-60^o$  
and crosses  $90^o$ in a different position than the tree contribution. Even if one shift up the re-scattering contribution, they have a clear different energy dependence on the lower sector.
\begin{figure}[h]
\begin{center}
%\hspace*{-10mm}
\includegraphics[width=.5\columnwidth,angle=0]{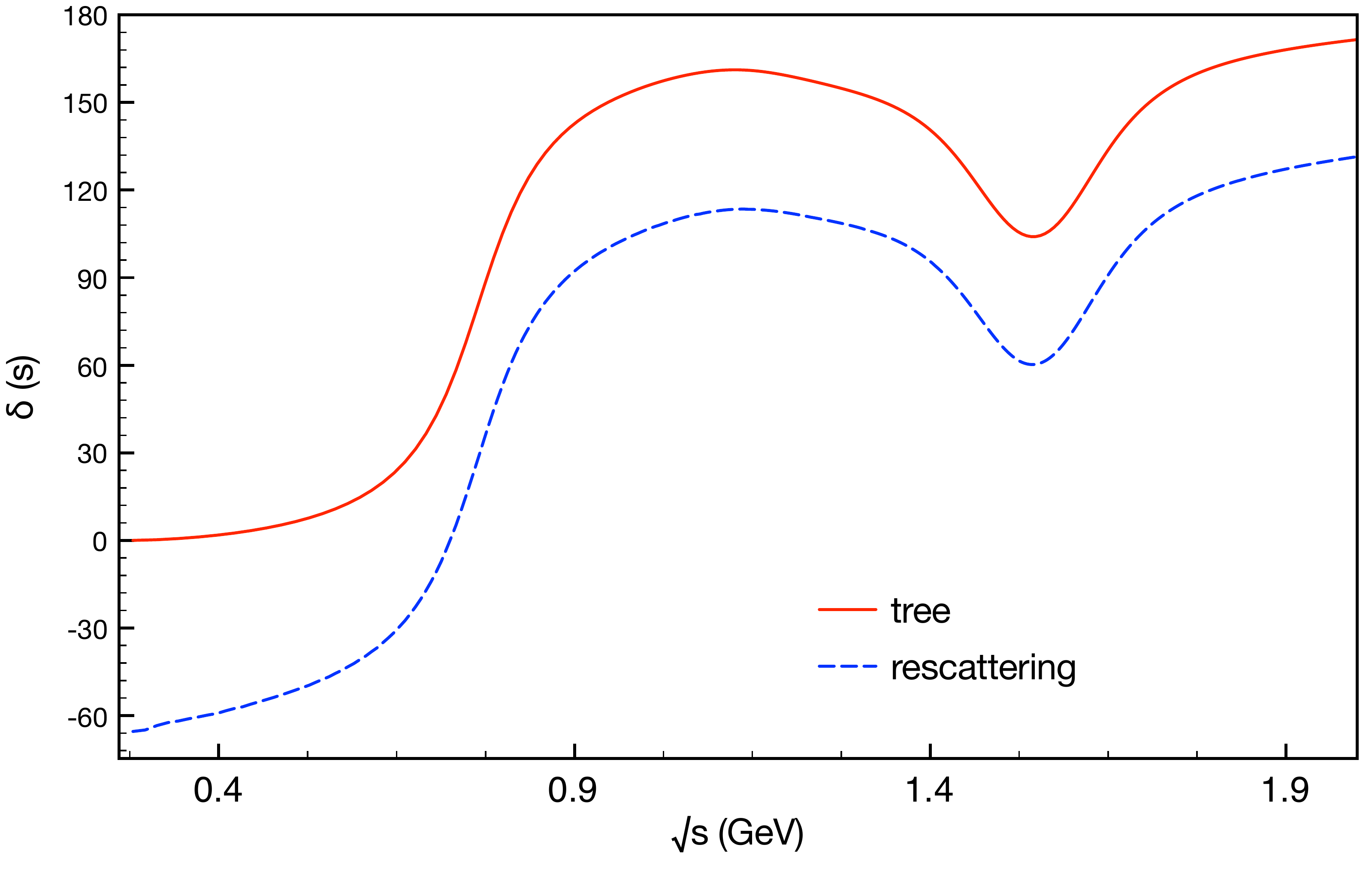}
\caption{Amplitude phase for tree, Eq.(\ref{A0}), and re-scattering, Eq.(\ref{A1.3}),  contributions.}
\label{phase}
\end{center}
\end{figure} 

In Fig. 14 we show the results for the modulus. In the plot on the left  one can see that the 
two amplitudes interfere destructively with the dominance of the former. Is possible to note a presence of a smaller peak around $1.7$ GeV, which is the contribution of the $\r(1450)$ coming from the $T_{\p\p}$ scattering amplitude. 
In the plot on the right, the two contributions are superimposed, with the re-scattering contribution
rescaled. The two line shapes are very different. Not only is the position of the peak  different, 
but also the two curves are asymmetric in opposite directions, with the re-scattering curve
being wider than the curve from the tree amplitude.

\begin{figure}[h]
%\hspace*{-10mm}
\hspace*{-2mm}\includegraphics[width=.5\columnwidth,angle=0]{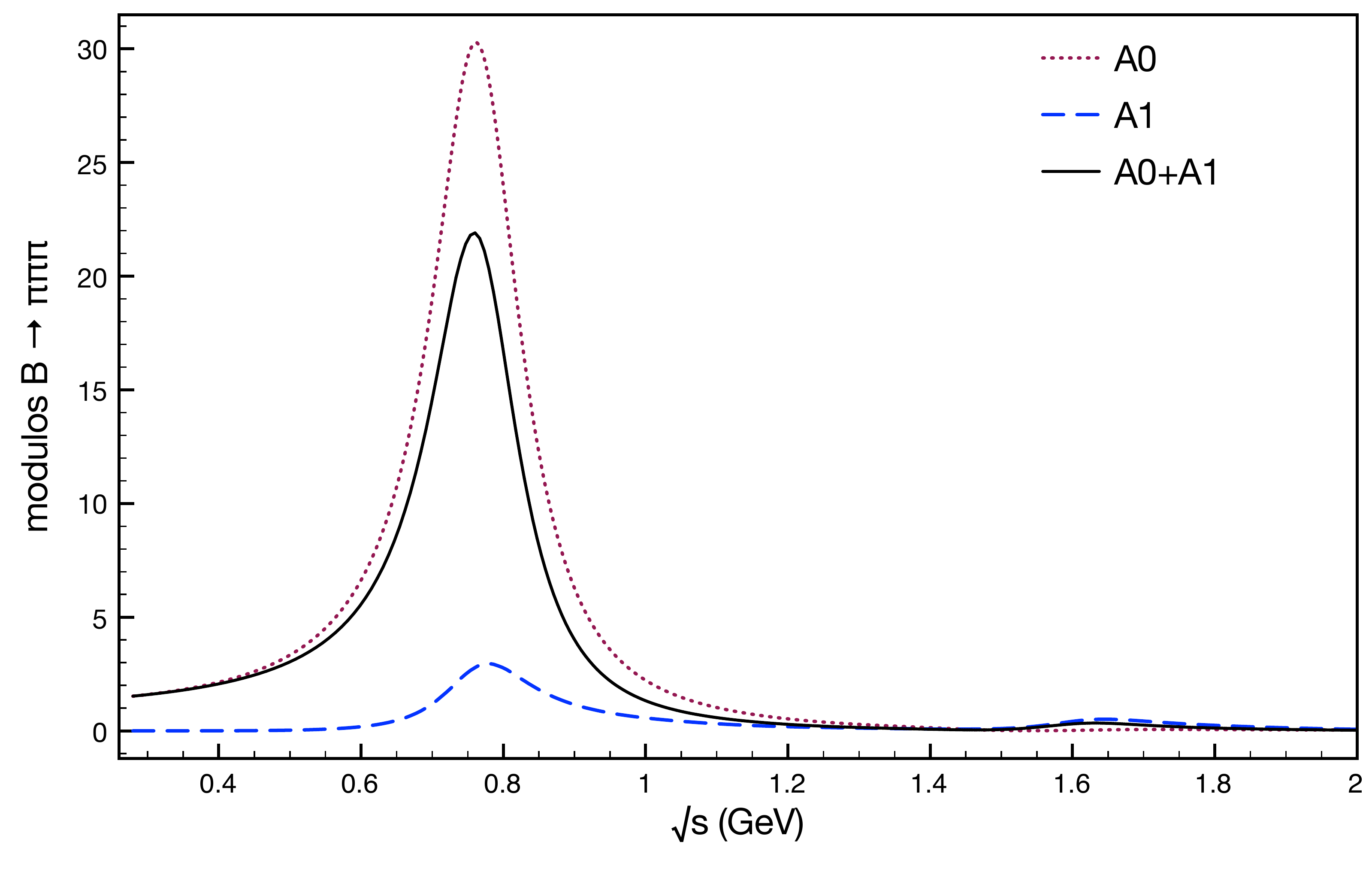}
\includegraphics[width=.5\columnwidth,angle=0]{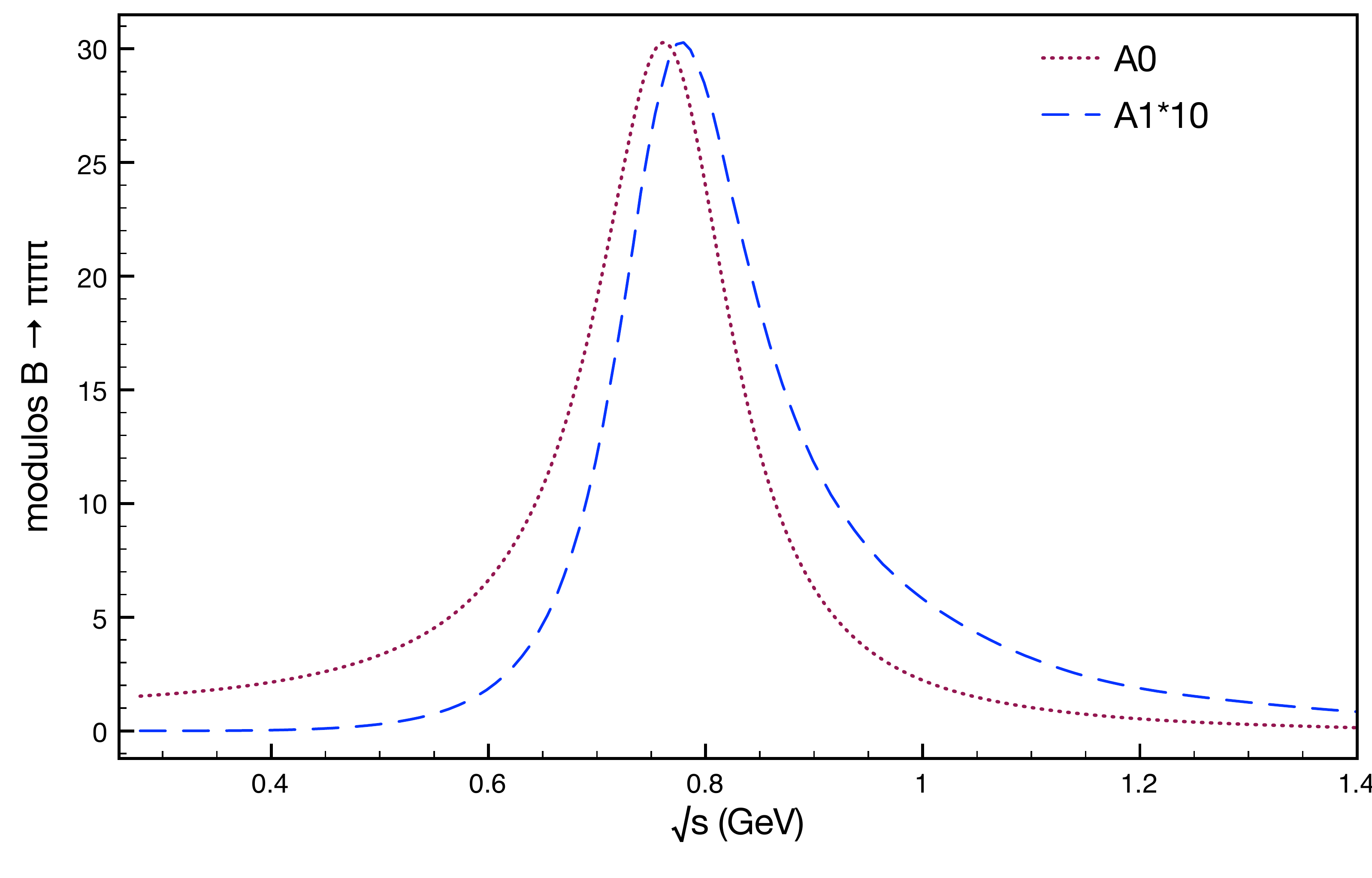}
\caption{Magnitude square of the  tree Eq.(\ref{A0}), re-scattering Eq.(\ref{A1.3}) and final amplitude.
The plot on the right has the two contributions superimposed with a scale factor applied to the
re-scattering curve.}
\label{moduloZ}
\end{figure}

\subsection*{Data analysis}

Two samples of Toy Monte-Carlo with 10,000 events were generated to $B\to\rho\pi$ decays using the amplitude derived in the previous 
section. A first sample is generated with only the tree amplitude. A second sample is generated
with the full amplitude, including the re-scattering contribution. The samples are then fitted with
 isobar model tools, such as Relativistic Breit-Wigner functions with no barriers factors. 
 A good fit is obtained for the tree-only sample. This is expected, since the only ingredient 
 is the $\rho$ production amplitude of Eq.(\ref{2.15}). The results for the  resonance parameters are: 
$m_\rho = 0.775 \pm 0.001 $GeV, $\Gamma_\rho = 0.148 \pm 0.001$GeV, where the  uncertainties  are due to the sample size. Comparing to the  PDG\cite{PDG} values:  $m_\rho =0.77526 \pm 0.00025$ GeV and $\Gamma_\rho = 0.1491 \pm 0.0008$ GeV, one can see that they nicely agree. 

%$m_\rho =0.776420(99) $, $\Gamma_\rho =0.149492(237)$. 

When we add the re-scattering contribution to the tree one, defining the complete amplitude to $\bppp$ decay, there is a presence of a second bump in the amplitude due to the $\rho(1450)$ present on the re-scattering amplitude. Therefore the sample simulated with the full amplitude can  be fitted by a sum of  two   Breit-Wigner functions, yielding  the following parameters: 
 $m_\rho = 0.756  \pm 0.001$GeV, $\Gamma_\rho =0.163 \pm 0.003 $GeV and $m_{\rho2} =1.50 \pm 0.01  $GeV, $\Gamma_{\rho2} =0.194  \pm 0.033 $GeV; where the uncertainties  are statistical. 
%$m_\rho = 0.774186 (1424) $, $\Gamma_\rho =0.171478(3454) $ and $m_{\rho2} =1.52967 (1297) $, $\Gamma_{\rho2} =0.18734 (3096) $. 
The first $\rho$ contribution is the result of the tree and re-scattering interference  and the second $\rho$ came only from the rescattering (Fig.~14, left). 
Comparing  the results of the two fits, one can see that the inclusion of the re-scattering 
amplitude changes slightly the $\rho$ mass with a significant increase in the width.
The re-scattering contribution to the amplitude, function $A_1$, is not fitted with two Breit-Wigner functions. However, we manage to parametrized this function using one Landau and two Gaussian functions, the result is shown in Fig. 15.
 \begin{figure}[h]
%\hspace*{-10mm}
\includegraphics[width=.5\columnwidth,angle=0]{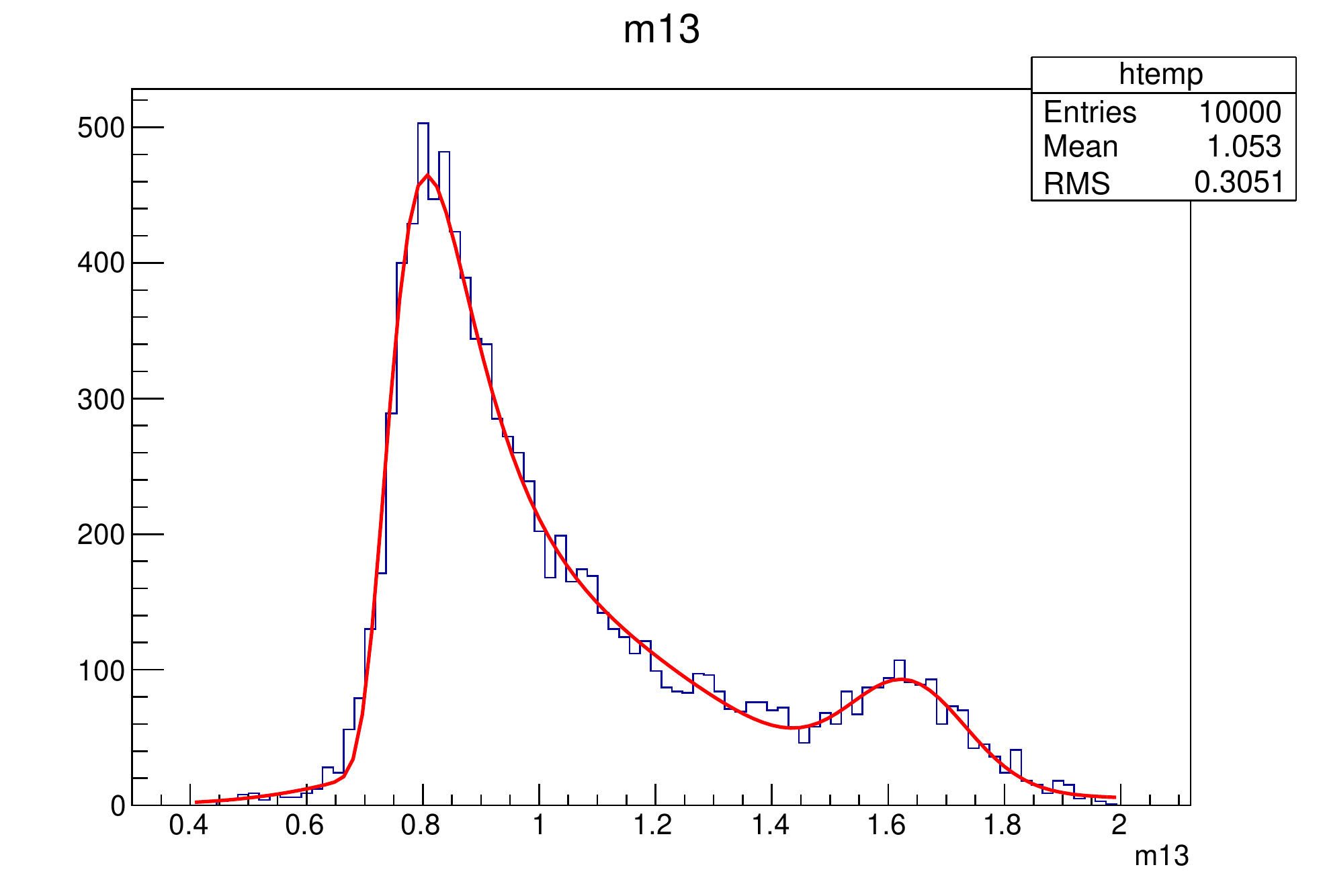}
\caption{ Fit to re-scattering amplitude, Eq.(\ref{A1.3})}
\label{landauGauss}
\end{figure}

\subsection{Final remarks}
The inclusion of re-scattering introduce new complexities into the game, it shifts the phase at threshold to bellow zero  and change the shape of the $\rho$ meson in the final amplitude. Inspecting the diagram in fig.\ref{fig:Bppp:F1}, we can conclude that these effect came from the hadronic loop.  It is important to note that the $\rho$ meson we seeing in the Dalitz plot, or data, is an overlap of all $\rho$'s: the one produced directly from B decays and the one produced from the re-scattering. Therefore, we need to take  this in the account when performing fits. 

\subsection{Acknowledgements} 
We would like to thank the workshop organizer for the pleasant environment and friendly reception. The work of PCM was supported by CNPq (Conselho Nacional de Desenvolvimento Cient\'{i}fico e Tecnol\'{o}gico).

\newpage

\aSection{Breit-Wigner, resonance propagator, scattering amplitudes, and production amplitudes: 
Who is who?}{P. C. Magalh\~{a}es \&\ M.R. Robilotta}
\begin{center}
P. C. Magalh\~{a}es$^a$ and M.R. Robilotta$^b$
\vskip .2cm
{\small $^a$Centro Brasileiro de Pesquisas F\'isicas  --  CBPF\\
Rio de Janeiro, Brazil}
\vskip .2cm
{\small$^b$Instituto de F\'isica, Universidade de S\~ao Paulo,\\  
 S\~ao Paulo, Brazil }
\end{center}
%\input{define.tex}
%\newcommand{\rpp}{\rho^0 \to \pi^+\pi^-}

%\begin{document}

The LHCb workshop on amplitude analyses, held in CBPF, raised many interesting discussions. 
Among them, the role of Breit-Wigner functions embodied into two-body scattering amplitudes 
used in experimental analyses. 
It became clear that there is considerable confusion on this topic,
associated mostly with the use of the name Breit-Wigner in different contexts. 
In this contribution, we aim to clarify this issue, 
by discussing the differences between resonance propagators, scattering amplitudes and 
production amplitudes involving resonances, 
emphasizing their connection with the different Breit-Wigner functions.  
This paper is a short version of a more detailed work to be presented in a near future.

\subsection{Introduction}

In their original 1936 paper, Breit and Wigner \cite{BW} described the capture of slow 
neutrons by nuclei by means of cross sections, written generically as
\bea
\s= \a \; \frac{1}{(\n - \n_0)^2 + \G^2}\;,
\label{1}
\eea
where  $\a$ was a suitably chosen function of energy, $\n$  and $\n_0$ were energies, 
and $\G$ was related with the  {\em half width breadth}. 
As time went by, several other functions, sharing the same kind of denominator, became also known
as Breit-Wigner functions (BW). 
As these various functions describe different objects, such as propagators, two-body scattering 
amplitudes or many-body production amplitudes, which occur in different situations, 
the result was  a big mess of names and meanings. 

As the denominators we are interested in originate from resonance propagators, 
we begin by sketching their main features and discuss their insertion into amplitudes 
afterwards. 
While doing this, it is useful to distinguish {\em theory} from {\em models}.
As one is dealing with relativistic particles, the natural framework for the former
is quantum field theory (QFT).
Models, on the other hand, depart from theoretical results and introduce 
reasonable  {\em ad hoc} modifications, aimed at improving their efficacy
in describing data. 

%PPPPPPPPPPPPPPPPPPPPPPPPPPPPPPPPPPPPPPPPPPPPPPPPPPPPPPPPPPPPPPPPPpp
\subsection{Propagators - theory}

\subsubsection*{Bare propagator}
This discussion is standard and can be found, for instance, in the book by
Bjorken and Drell \cite{BjorkenDrell}.
In QFT, one is entitled to imagine a 
world in which a particle, such as the $\r$-meson, lives in complete isolation. 
The propagator of this particle, $\D_0\,$,
can be derived from a Lagrangian involving its bare mass $m_0$, and is written as  
\bea
i\D_0(s) &\!=\!& i \, \frac{\Pi_J}{D_0(s)} \;,
%\label{2}\\[2mm]
%D_0(s) &\!=\!& s - m_0^2 \;,
\hspace*{10mm}
D_0(s)\!=\! s - m_0^2 \;,
\label{3}
\eea
where the numerator $\Pi_J$ is a projector, associated with the spin $J$.
For instance, $J=0 \rar \Pi_S=1;  J=1 \rar \Pi_P=g^{\m\n}- p^\m p^\n/m^2  ...$ . 
In this work, we concentrate just on denominators and do not care about numerators. 

\subsubsection*{Dressed propagator}
If the propagating particle can interact with two other particles, which we will call $A$ and $B$, its energy changes as does its mass.
In the case of the $\r$, the other two particles could be
$[A\,B] \rar [\g\,\r], [\p\;\p], [K\,\Kb], [p \, \pb], \cdots$, and
systems involving more than two particles are also possible. 

%F   ^^^^^^^^^^^^^^^^^^^^^^^^^^^^^^^^^^^^^^^^^^^^^^^^^^^^^^^^^^^^^^^^^^^^
\begin{figure}[h] 
\begin{center}
\includegraphics[width=.55\columnwidth,angle=0]{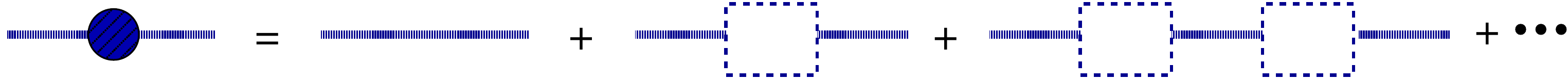}
\caption{The dressing of a bare propagator.}
\label{F1}
\end{center}
\end{figure}
%^^^^^^^^^^^^^^^^^^^^^^^^^^^^^^^^^^^^^^^^^^^^^^^^^^^^^^^^^^^^^^^^^^^^^^^^
 
The possibility of interactions means that the original propagating particle can transform itself into 
particles $A$ and $B$ which, in turn,  can yield back the propagating particle.
In the case of the $\r$, one could have
$\r \rar \p\,\p$ and, afterwards $\p\,\p \rar \r$.        
This yields a characteristic self-energy, usually denoted by $\S$,
which, by means of the geometric series indicated in eq. (\ref{4}) and Fig. \ref{F1}, give rise to the dressed propagator $\D$, 
\bea
i\D = i \D_0 + i \D_0 \, \lb - i\,\S\, \rb\, i\D_0 
+ i \D_0 \, \lb - i\,\S\, \rb\, i\D_0 \, \lb - i\,\S\, \rb\, i\D_0 \, 
+ \cdots \;,
\label{4}
\eea 
This sum can be expressed in a more compact form as
\bea
i\D(s) &\!=\!& i\, \frac{\Pi_J}{D(s)} \;,
%\label{5}\\[2mm]
%D(s) &\!=\!& s - m_0^2 - \S(s)\;.
\hspace*{10mm}
 D(s) \!=\! s - m_0^2 - \S(s)\;.
\label{6}
\eea
The important feature to be noted is that the availability of different intermediate
states for the propagating particle modifies the function $D(s)$.
This denominator is a Lorentz scalar so as the function $\S$.  
This result is model independent and  specific choices of the function $\S$ eventually yield a BW functions. 

In QFT, the function $\S$ is formally
associated with two effects, namely the 
{\em wave function renormalization} $Z$ and  the {\em modification of the particle mass}.
These features are formally introduced by noting that the physical mass $m$ is fixed as the solution 
of the equation
\bea 
m_0^2 + \S(m^2) = m^2 \;.
\label{7}
\eea
After fixing $m$, one expands the self-energy as \cite{Scherer}
\bea
&& \S(s) = \S(m^2) + (s - m^2)\, \S'(m^2)+ \tilde{\S}(s) \;,
\label{8}
\eea
and rewrites the denominator as 
\bea
&& D(s) = \lb 1 - \S'(m^2) \rb \lb (s-m^2) - \frac{\tilde{\S}(s)}{1 - \S'(m^2)} \rb  \;.
\label{9}
\eea
The first factor in this result is absorbed into a redefinition of the fields
and the renormalized propagator, indicated by a bar, is given by
\bea
i\bar{\D}(s) &\!=\! & i\, \frac{\Pi_J}{\bar{D}(s)} \;,\hspace*{10mm}
\bar{D}(s) \!=\!  s - m^2 -  \frac{\tilde{\S}(s)}{1 - \S'(m^2)}  \;.
\label{11}
\eea

If the only effect of the self-energy were the shift in the mass, 
the denominator $D(s)$ would have a simple pole at $s=m^2$. 
However, the term proportional to $\tilde{\S}(s)$ cannot be neglected and may give rise
to important consequences, such as bound states or resonance widths. 

\subsubsection*{The self-energy $\S$}
The properties of the self-energy $\S$ can be discussed in general terms
and one finds that:
\\
- $\S$ can be either a real function or a complex function, depending
on how the mass of the propagating particle relates with the masses present in the intermediate system.
\\
- the form of $\S$ depends on the angular momentum of the propagating particle.

As the general  discussion  is well beyond our limited scope, 
here  we introduce these features by means of  specific examples.
\vskip .2cm
\begin{center} 
\underline{Scalar resonances} 
\end{center} 
\vskip .2cm
%F   ^^^^^^^^^^^^^^^^^^^^^^^^^^^^^^^^^^^^^^^^^^^^^^^^^^^^^^^^^^^^^^^^^^^^
\begin{figure}[h] 
\begin{center}
\includegraphics[width=.35\columnwidth,angle=0]{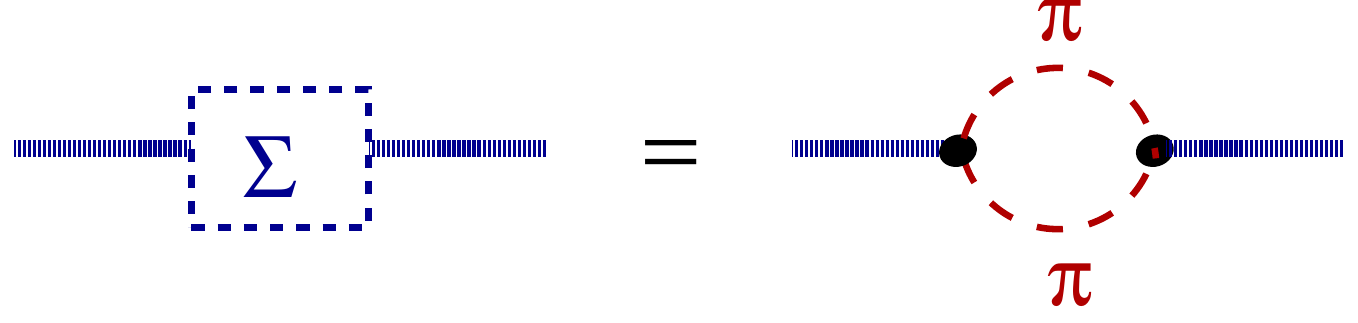}
\caption{The self-energy mechanism.}
\label{F2}
\end{center}
\end{figure}

In the case of a scalar resonance, such as the $f_0$, which can decay into two pions,
the expression for the self energy involves twice the factor $G_S$,
associated with the transitions $f_0 \rar \p\p$ and $\p\p \rar f_0$,
together with the two-pion propagator.
%In chiral models, $G_S$ is a polynomial in $s$ and $M_\p$, but this is not important for this discussion.
The important point is that $G_S$ is a real function and
the scalar self-energy reads
\bea
\S_S = G_S \, \Ob_S \, G_S \;,
\label{12}
\eea
where $\Ob_S$ is a function related with the two-pion propagator and written as 
{\small
\bea
\Omega_S(s) &\!=\!& i\, \int \frac{d^4\ell}{(2\p)^4} \; 
\frac{1}{[(\ell \sp p/2)^2 - M_\p^2+i\,\e\,]\;[(\ell \sm p/2)^2 - M_\p^2 +i\,\e\,]} \;,
\label{13}
\eea}
with $p^2=s$.
This integral is divergent and the strategy for dealing with this unwanted feature
is to write 
$\Omega_S(s)= \Ob_S(s)+\Omega_S(0)$,
where $\Ob_S(s)$ is a regular function and the divergence is kept in $\Omega_S(0)$.
The integral is then regularized and the net effect is the replacement
$\Omega_S(0) \rar C$, where $C$  is a finite but unknown constant, which must 
be fixed by data.
The regular part $\Ob_S(s)$, can be evaluated explicitly along different intervals of the variable 
$\l^2 = (s- 4M_\p^2)/s $ and one has 
%
%Thus
%
%
{\footnotesize
\bea
 s \!<\! 0  
&\rar & \Ob_S = - \, \frac{1}{32 \p^2} \,
\lb  2 - 2 \; \frac{\sqrt{|s|\sp 4M_\p^2}}{\sqrt{|s|}} \, 
\ln \frac{\sqrt{|s|+4M_\p^2} + \sqrt{|s|}}{2\,M_\p } \rb \;,
\label{15}\\[2mm]
 0 \! \le \! s \! < \! 4\,M_\p^2 
&\rar & \Ob_S = - \, \frac{1}{32 \p^2} \,
\lb  2 -\; 2\, \frac{\sqrt{4M_\p^2-s}}{\sqrt{s}} \; \tan^{-1} \lp \frac{\sqrt{s}}{\sqrt{4M_\p^2-s}} \rp \rb \;,
\label{16}\\[2mm]
s \! \geq  4\, M_\p^2 
&\rar & \Ob_S = - \, \frac{1}{32 \p^2} \,
\lb  2 - 2 \,\frac{\sqrt{s - 4M_\p^2}}{\sqrt{s}} \; 
\ln \frac{\sqrt{s-4M_\p^2}+\sqrt{s}}{2\,M_\p}
+ \,i\,  \p\;\frac{\sqrt{s-4M_\p^2}}{\sqrt{s}} \; \rb \;.
\label{17}
\eea}
 
On quite general grounds, the imaginary part of a self-energy can be associated with $\tau^{-1}$, 
where $\tau$ is the proper lifetime of the propagating particle \cite{LSB} .
In the present example, the resonance can decay only when $s \geq 4\,M_\p^2 $
and this is the reason why only Eq.(\ref{17}) contains an imaginary component. 
This idea can be reinforced by recalling that the very concept of lifetime does 
not apply to virtual states.

Assembling all these pieces into Eq.(\ref{8}) one obtains the self-energy for a scalar resonance, where the  numerical value for $\S'(m^2)$ is small and the factor $\frac{1}{1- \S'(m^2)}$ can be set to one in Eq(\ref{9}). Then, the renormalized resonance propagator, Eq.(\ref{11}), becomes

{\small
\bea
&& D_S(s) = s - \cM_S^2(s)  + \,i \; \Theta(s-4 M_\p^2) \; G_S^2 \,
\frac{\sqrt{s-4M_\p^2}}{32\p\,\sqrt{s}}
\label{19}
\eea}
where
{\small
\bea
&& \cM_S^2(s) =  m_0^2 + G_S^2 \, \lc \lb \Ob_S \rb_R + C \rc \;
\label{20}
\eea}
\ni is the running mass. The physical mass $m_S$ of the resonance is the solution 
of the equation $\cM_S^2(s=m_S^2) =  m_S^2 \;.
$
Once the value of $m$ is fixed, the resonance propagator is conventionally written as

{\small
\bea
&&  i\D_S(s) = \frac{i}{D_S(s)} \;,
\hspace*{10mm}
 D_S(s) = s - \cM_S^2(s)  + \,i \; \Theta(s-4 M_\p^2) \; m_S \;  \G_S(s) \;,
\label{23}\\%[2mm]
&& \G_S(s) = \frac{G_S^2}{32\p \,m_S} \; \frac{\sqrt{s - 4 M_\p^2}} {\sqrt{s}} \;.
\label{24}
\eea
}
Comparing the final expression for the propagator in Eq.(\ref{23}) with the one derived from QFT, Eq.(\ref{11}), one can see the relevance of the factor $\tilde{\S}(s)$ for the width, Eq.(\ref{24}), and the running mass, Eq.(\ref{20}).
This final form of the propagator also indicates that the expressions that look like BWs arise naturally in the framework of QFT. 

One notes that the  functions $D(s)$, $\cM(s)$ and $\G(s)$ entering the propagator
are Lorentz scalars
and the width is linear in $q$, the three-momentum of the intermediate pions in their CM.
\vskip .2cm
\begin{center}
\underline{Vector resonances}  
\end{center}
\vskip .2cm
As discussed above, the bare propagator of a vector resonance involves a characteristic numerator.
In the dressing process, this factor also influences the self-energy $\S_P$ and, therefore, 
the denominator.
The net effect corresponds to the inclusion of an extra kinematic factor. % into the 

{\small
\bea
&&  \Ob_S \rar \Ob_P = \frac{1}{3} \, [s \sm 4\,M_\p^2] \; \Ob_S \;,
\label{25} 
\eea
}
and this leads us to 
{\small
\bea
&& i\D_P(s) = \frac{i}{D_P(s)} \;,
\hspace*{10mm} 
 D_P(s) = s - \cM_P^2(s)  + \,i \; \Theta(s-4 M_\p^2) \; m_P \;  \G_P(s) \;,
\label{28}\\[1mm]
&& \cM_P^2(s) = m_0^2 + G_P^2 \, \lb \Ob_P + C_P \,\rb \;,
\hspace*{8mm}
 \G_P(s) = \frac{G_P^2}{96\p\,m_P} \; \frac{[s \sm 4\,M_\p^2]  \, \sqrt{s \sm  4 M_\p^2}} {\sqrt{s}} \;.
\label{30}
\eea}
In the rest frame  of the vector resonance, the width is proportional to $q^3$. Once again, the vector resonance propagator shows that the contribution from $\tilde{\S}(s)$ to the self-energy, eq.(\ref{8}), is very significant to both the resonance width and running mass. It is very interesting to point-out that the expression to the vector propagator, Eq.(\ref{28}), is the same as the one proposed by Gounaris-Sakurai\cite{GS} as we show below, in Eq.(\ref{35}).

\subsubsection*{Poles}
Poles of propagators are important and can be determined from denominators such as 
those given by eqs. (\ref{23}) and (\ref{28}).
Schematically, they are obtained by writing $s = (a - i\, b)^2$ and 
then determining the values of $a$ and $b$ by means of the condition 
{\small
\bea
D_J[(a-i\,b)^2]=0.
\label{31}
\eea}
Calling $a=m_{pole}$ and $b=\G_{pole}/2$, one finds representations for the propagator
of the form
{\small
\bea
i \, \D_J(s) = \frac{\mathrm{residue}}{ s - (m_{J\,pole} - i \, \G_{J\,pole}/2)^2 }
+ \mathrm{non\!-\!pole} \;.
\label{32}
\eea}
One notes that:
\\
-  The values of $m_{J\,pole}$ and $\G_{J\,pole}$ need not to be close to the values
of $m_J$ and $\G_J$ in eqs.(\ref{23}) and (\ref{28}).
\\
-  There may exist more than one solution to eq. (\ref{31}), representing dynamically generated states.
%
%
%
%\subsubsection{two-resonances}
%
%In case two resonances, with different masses, can coexist in the same channel ...
%{\bf \large ??????}
%
%
%
%
%mmmmmmmmmmmmmmmmmmmmmmmmmmmmmmmmmmmmmmmmmmmmmmmmmmmmmmmmmmm
%
\subsection{Models}
The description of systems involving resonances may become cumbersome in QFT, 
when several different degrees of freedom are involved.
In this case, it is usual that results from QFT are extended by means of 
{\em ad hoc} modifications, based on educated guesses.  
Some popular choices for the case of the $\r$-meson are presented below.

\subsubsection*{$K$-matrix}
In the so-called $K$-matrix approximation, the running masses $\cM^2(s)$ in eqs. (\ref{23})
and (\ref{28}) are replaced with the fixed empirical resonance masses, and one 
employs denominators of the form.
{\small
\bea
&& D_J(s) = s - m_J^2  + \,i \; \Theta(s-4 M_\p^2) \; m_J \;  \G_J(s) \;.
\label{33}
\eea}
This expression is the same as that used in the isobar model without barriers factors.
\subsubsection*{Gounaris-Sakurai}
The form of the propagator known as Gounaris-Sakurai (GS) was developed in 1968 \cite{GS} and
expresses the physics just described.
Using their notation,  we have
{\small
\bea
i\D_{GS}(s) &\!=\! & \frac{i}{D_{GS}(s)}\,,
%\label{34}\\[2mm]
%
%D_{GS}(s) &\!=\! & s - [m_\rho^2  + f(s)]  + i  m_\rho \Gamma(s)\,,
\hspace*{10mm}
D_{GS}(s) \!=\!  s - [m_\rho^2  + f(s)]  + i  m_\rho \Gamma(s)\,,
 \label{35}\\[3mm]
%\eea
%
%where
%\bea
%
&&\Gamma(s) = \Gamma_\r \;\frac{m_\rho}{\sqrt{s}}\;\lb \frac{k(s)}{k(m_\rho^2)}\rb^3  \,;
\hspace*{10mm}
 k(s)= \sqrt{\frac{s-4M_\p^2}{4}} \,;
\label{36}\\[1mm]
f(s) &=& \Gamma_\r \;\frac{m_\rho^2}{k(m_\rho^2)^3} \; 
\lb k(s)^2\, \lp h(s) - h(m_\rho^2)\rp  + (m_\rho^2 - s) \, h' \, k(m_\rho^2)^2  \rb \,; 
\label{37}\\[1mm]
&& h(s)= \frac{2}{\pi}  \,\frac{k(s)}{\sqrt{s}} \;\ln \lb\frac{\sqrt{s} + 2 k(s)}{2 M_\pi}\rb\,;
%\label{39}\\[2mm]
\hspace*{8mm}
h'  = h(m_\rho^2) \; \lb \frac{1}{8 k(m_\rho^2)^2} -  \frac{1}{2 m_\rho^2}\rb  +  \frac{1}{2 \pi m_\rho^2}.
\label{40}
\eea}

The Gounaris-Sakurai  expression for the width is identical to that of eq. (\ref{30}),
provided one defines $\G_\r=\G(s=m_\r^2)$  and then  uses this to replace  the factor 
$G_P^2/(96 \p \, m_\r)$ present there.
Eq. (\ref{23})  yieds $h(s)=16\p\,\Ob_S -1/\p$
and the function $f(s)$ amounts to the expansion of the real part of the self-enegy around $s=m_\r^2$,
given by eq. (\ref{8}).

\subsubsection*{Isobar model}

In the isobar model the following expression is widely used in fits to data
{\small
\bea
i\D_{IM}^{\rho}(p) &\!=\!& \frac{i}{ s - m_\rho^2 + i  m_\rho\, \G_\rho^{IM}(s)} \;,
\hspace*{10mm}
\Gamma_\r^{IM}(p) = \Gamma_\r^{GS} 
\lb \frac{ F_r(s)}{F_r(m_{\rho}^2)}\rb ^2 \;,
\eea}
where $\Gamma_\r^{GS}$ is given by eq.(\ref{36}) and $ F_r(s) = 1/\sqrt{1+[r_r q_r(s)]^2} $. 
This function becomes identical to the K-Matrix, eq. (\ref{33}), if we take out the barrier factors.

\subsection{Scattering amplitudes}
%F   ^^^^^^^^^^^^^^^^^^^^^^^^^^^^^^^^^^^^^^^^^^^^^^^^^^^^^^^^^^^^^^^^^^^^
\begin{figure}[h] 
\begin{center}
\includegraphics[width=.35\columnwidth,angle=0]{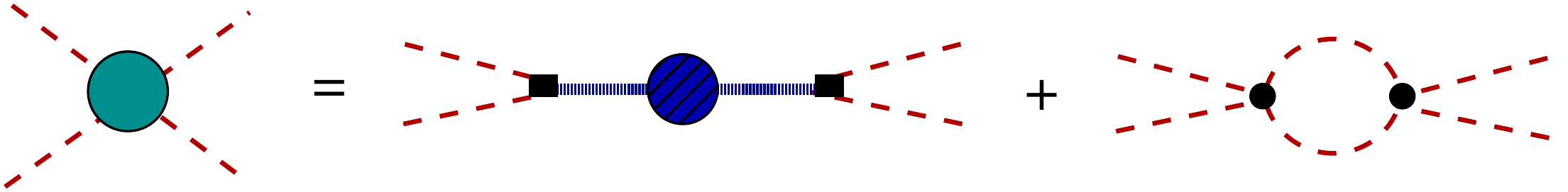}
\caption{$\p\p$ scattering amplitude: resonances exchange and meson-meson interaction.}
\label{F3}
\end{center}
\end{figure}

Scattering amplitudes are the usual  sources of resonance information. 
However, scattering amplitudes may receive contributions from many kinds of processes,
not all of them associated with resonances, as in Fig. \ref{F3}. 
This gives rise to important differences between denominators entering propagators and amplitudes.
We discuss this feature by means of an example concerning the $P$-wave $\p\p$
elastic amplitude.
One assumes a simple $\p\p$ interaction, based on just an $s$-channel $\rho$-pole 
and a background $B$, as in Fig. \ref{F3}.

Using a chiral model \cite{pat}  as a guide, the construction of the amplitude begins 
by describing the tree diagrams, which yield
{\small \bea 
&& \bar{T^1} = (t-u)  
\lb B - G_\r^2\, \frac{s}{s-m_\rho^2}\rb \,, 
\label{B.1} 
\eea} 
where $s$, $t$ and $u$ are the  Mandelstam variables and $B=1/F^2$. For free particles in the center of mass frame,  
$(t-u)=(s - 4M_\p^2)\, \cos\, \theta$ and the $P$-wave 
projection gives rise to the kernel 
{\small \bea 
\cK^{P1} &=& -\, \frac{(s-4 M_\p^2)}{3\,} \;  
\lb B + G_\r^2 \frac{s}{s - m_\rho^2} \rb  %\\[2mm]
= -\, \frac{(s-4 M_\p^2)}{3\,} \;  
\lb \frac{ B (s- m_\rho^2) + G_\r^2\, s}{s - m_\rho^2} \rb  
\label{B.3} 
\eea }
which can be used to construct a 
unitarized scattering amplitude \cite{OO},  given by
{\small \bea 
T_{\p\p}^{P1} = \frac{\cK^{P1}}{1 + \cK^{P1}\,(\Ob_{\p\p} + C_{\p\p})} \; 
 = \frac{ \, (4 M_\p^2 - s)/3\;  
\lb  B (s- m_\rho^2) + G_\r^2\, s \rb }{   s - m_\rho^2  -     
\lb  B (s- m_\rho^2) + G_\r^2\, s \rb \,(\Ob_P + C_P)} \;. 
\label{B.7} 
\eea }
This gives rise to a denominator which is different from all the previous ones discussed 
and, in particular, to a width of the form
\bea
 \G_\r^{T}(s) = \frac{\lb  B (s- m_\rho^2) + G_\r^2\, s \rb }{96\p} \; \frac{[s \sm 4\,M_\p^2]  \, \sqrt{s \sm  4 M_\p^2}} {\sqrt{s}} \;. 
\label{B.8} 
\eea

The presence of the background $B$ in this result is a clear indication that it does influence the width profile
for $s\neq m_\r^2$.

%xxxxxxxxxxxxxxxxxxxxxxxxxxxxxxxxxxxxxxxxxxxxxxxxxxxxxxxxxxxxxxxxxxxxx
\subsection{Production amplitude: $\rpp$ } 

The $\r$ meson can be produced directly in  B or  D  decays  before it  decays into two mesons.
%Once we are interest in the compare the modifications on the propagation of the resonance, we can neglect  the source of the $\r$.  
The same model used to derive the scattering amplitude, eq. (\ref{B.7}), 
yields the production amplitude 
 {\small \bea
\Pi(s) &=& A \;\frac{G_\r}{m_\rho^2-s} 
\lb 1 + (-\bar\Omega) \frac{\cK}{1 + \bar\Omega\cK} \rb
= A\;\frac{G_\r}{m_\rho^2-s} 
\lb \frac{1}{1 + \bar\Omega\cK} \rb 
\\[2mm]& =& A\; \frac{G_\r}{m_\rho^2-s +
\lb  B (s- m_\rho^2) + G_\r^2\, s \rb \,(\Ob_P + C_P)}  .
\label{2.12}
\eea}
where $A$ is a factor describing the transition of the heavy mesons into $\r+$other mesons.  
Comparing eq. (\ref{2.12}) with the resonance  propagator, eq. (\ref{28}), 
one  can see that, again, both the width and the running mass are different due to the presence of
a background in the kernel. 

\subsection{Final remarks}

We inspect the general ground of the structures used to describe resonance propagators in different context. It became clear that the functional form of the resonance propagation depend  strongly on the environment and the dynamical model considered. We show that availability of different intermediate states can give an important contributions  and change both the mass and the width of the resonances propagator. This difference is also manifest on the extraction of the pole position, once the source propagation is not universal so as the associated pole position. 

%The numerator can change the behave of the functions,and we are not  with this at the moment.

\subsection{Acknowledgements}
We would like to thank the workshop organizer for the pleasant environment and friendly reception. The work of PCM was supported by CNPq (Conselho Nacional de Desenvolvimento Cient\'{i}fico e Tecnol\'{o}gico).

%\end{document}
\newpage
\newpage

%\section{Conclusion}
%\begin{center}
%Jonas Rademacker$^a$ and Alberto C. dos Reis$^b$ \\
%\vskip .2cm
%{\small $^a$H. H. Wills Physics Laboratory, University of Bristol, \\
%Bristol, United Kingdom}
%\vskip .2cm
%{\small $^b$Centro Brasileiro de Pesquisas F\'isicas  --  CBPF\\
%Rio de Janeiro, Brazil}
%\end{center}
%\input{conclusion}
\newpage
%\section*{References}
%\input{bib}

%\bibliographystyle{plainnat}
\bibliography{main}

\begin{thebibliography}{240}%
\makeatletter
\providecommand \@ifxundefined [1]{%
 \@ifx{#1\undefined}
}%
\providecommand \@ifnum [1]{%
 \ifnum #1\expandafter \@firstoftwo
 \else \expandafter \@secondoftwo
 \fi
}%
\providecommand \@ifx [1]{%
 \ifx #1\expandafter \@firstoftwo
 \else \expandafter \@secondoftwo
 \fi
}%
\providecommand \natexlab [1]{#1}%
\providecommand \enquote  [1]{``#1''}%
\providecommand \bibnamefont  [1]{#1}%
\providecommand \bibfnamefont [1]{#1}%
\providecommand \citenamefont [1]{#1}%
\providecommand \href@noop [0]{\@secondoftwo}%
\providecommand \href [0]{\begingroup \@sanitize@url \@href}%
\providecommand \@href[1]{\@@startlink{#1}\@@href}%
\providecommand \@@href[1]{\endgroup#1\@@endlink}%
\providecommand \@sanitize@url [0]{\catcode `\\12\catcode `\$12\catcode
  `\&12\catcode `\#12\catcode `\^12\catcode `\_12\catcode `\%12\relax}%
\providecommand \@@startlink[1]{}%
\providecommand \@@endlink[0]{}%
\providecommand \url  [0]{\begingroup\@sanitize@url \@url }%
\providecommand \@url [1]{\endgroup\@href {#1}{\urlprefix }}%
\providecommand \urlprefix  [0]{URL }%
\providecommand \Eprint [0]{\href }%
\providecommand \doibase [0]{http://dx.doi.org/}%
\providecommand \selectlanguage [0]{\@gobble}%
\providecommand \bibinfo  [0]{\@secondoftwo}%
\providecommand \bibfield  [0]{\@secondoftwo}%
\providecommand \translation [1]{[#1]}%
\providecommand \BibitemOpen [0]{}%
\providecommand \bibitemStop [0]{}%
\providecommand \bibitemNoStop [0]{.\EOS\space}%
\providecommand \EOS [0]{\spacefactor3000\relax}%
\providecommand \BibitemShut  [1]{\csname bibitem#1\endcsname}%
\let\auto@bib@innerbib\@empty
%</preamble>
\bibitem [{\citenamefont {Aaij}\ \emph
  {et~al.}(2014{\natexlab{a}})\citenamefont {Aaij} \emph {et~al.}}]{lhcb3}%
  \BibitemOpen
  \bibfield  {author} {\bibinfo {author} {\bibfnamefont {R.}~\bibnamefont
  {Aaij}} \emph {et~al.} (\bibinfo {collaboration} {LHCb collaboration}),\
  }\href {\doibase 10.1103/PhysRevD.90.072003} {\bibfield  {journal} {\bibinfo
  {journal} {Phys. Rev.}\ }\textbf {\bibinfo {volume} {D90}},\ \bibinfo {pages}
  {072003} (\bibinfo {year} {2014}{\natexlab{a}})},\ \Eprint
  {http://arxiv.org/abs/1407.7712} {arXiv:1407.7712 [hep-ex]} \BibitemShut
  {NoStop}%
%%CITATION = ARXIV:1407.7712;%%
\bibitem [{\citenamefont {Aaij}\ \emph
  {et~al.}(2015{\natexlab{a}})\citenamefont {Aaij} \emph {et~al.}}]{lhcb4}%
  \BibitemOpen
  \bibfield  {author} {\bibinfo {author} {\bibfnamefont {R.}~\bibnamefont
  {Aaij}} \emph {et~al.} (\bibinfo {collaboration} {LHCb collaboration}),\
  }\href {\doibase 10.1103/PhysRevD.91.092002} {\bibfield  {journal} {\bibinfo
  {journal} {Phys. Rev.}\ }\textbf {\bibinfo {volume} {D91}},\ \bibinfo {pages}
  {092002} (\bibinfo {year} {2015}{\natexlab{a}})},\ \Eprint
  {http://arxiv.org/abs/1503.02995} {arXiv:1503.02995 [hep-ex]} \BibitemShut
  {NoStop}%
%%CITATION = ARXIV:1503.02995;%%
\bibitem [{\citenamefont {Aaij}\ \emph
  {et~al.}(2015{\natexlab{b}})\citenamefont {Aaij} \emph {et~al.}}]{lhcb5}%
  \BibitemOpen
  \bibfield  {author} {\bibinfo {author} {\bibfnamefont {R.}~\bibnamefont
  {Aaij}} \emph {et~al.} (\bibinfo {collaboration} {LHCb collaboration}),\
  }\href {\doibase 10.1103/PhysRevD.92.032002} {\bibfield  {journal} {\bibinfo
  {journal} {Phys. Rev.}\ }\textbf {\bibinfo {volume} {D92}},\ \bibinfo {pages}
  {032002} (\bibinfo {year} {2015}{\natexlab{b}})},\ \Eprint
  {http://arxiv.org/abs/1505.01710} {arXiv:1505.01710 [hep-ex]} \BibitemShut
  {NoStop}%
%%CITATION = ARXIV:1505.01710;%%
\bibitem [{\citenamefont {Aaij}\ \emph
  {et~al.}(2015{\natexlab{c}})\citenamefont {Aaij} \emph {et~al.}}]{lhcb6}%
  \BibitemOpen
  \bibfield  {author} {\bibinfo {author} {\bibfnamefont {R.}~\bibnamefont
  {Aaij}} \emph {et~al.} (\bibinfo {collaboration} {LHCb collaboration}),\
  }\href {\doibase 10.1103/PhysRevD.92.012012} {\bibfield  {journal} {\bibinfo
  {journal} {Phys. Rev.}\ }\textbf {\bibinfo {volume} {D92}},\ \bibinfo {pages}
  {012012} (\bibinfo {year} {2015}{\natexlab{c}})},\ \Eprint
  {http://arxiv.org/abs/1505.01505} {arXiv:1505.01505 [hep-ex]} \BibitemShut
  {NoStop}%
%%CITATION = ARXIV:1505.01505;%%
\bibitem [{\citenamefont {Gershon}(2009)}]{tim1}%
  \BibitemOpen
  \bibfield  {author} {\bibinfo {author} {\bibfnamefont {T.}~\bibnamefont
  {Gershon}},\ }\href {\doibase 10.1103/PhysRevD.79.051301} {\bibfield
  {journal} {\bibinfo  {journal} {Phys. Rev.}\ }\textbf {\bibinfo {volume}
  {D79}},\ \bibinfo {pages} {051301} (\bibinfo {year} {2009})},\ \Eprint
  {http://arxiv.org/abs/0810.2706} {arXiv:0810.2706 [hep-ph]} \BibitemShut
  {NoStop}%
%%CITATION = ARXIV:0810.2706;%%
\bibitem [{\citenamefont {Gershon}\ and\ \citenamefont
  {Williams}(2009)}]{tim2}%
  \BibitemOpen
  \bibfield  {author} {\bibinfo {author} {\bibfnamefont {T.}~\bibnamefont
  {Gershon}}\ and\ \bibinfo {author} {\bibfnamefont {M.}~\bibnamefont
  {Williams}},\ }\href {\doibase 10.1103/PhysRevD.80.092002} {\bibfield
  {journal} {\bibinfo  {journal} {Phys. Rev.}\ }\textbf {\bibinfo {volume}
  {D80}},\ \bibinfo {pages} {092002} (\bibinfo {year} {2009})},\ \Eprint
  {http://arxiv.org/abs/0909.1495} {arXiv:0909.1495 [hep-ph]} \BibitemShut
  {NoStop}%
%%CITATION = ARXIV:0909.1495;%%
\bibitem [{\citenamefont {Latham}\ and\ \citenamefont {Gershon}(2009)}]{tom}%
  \BibitemOpen
  \bibfield  {author} {\bibinfo {author} {\bibfnamefont {T.}~\bibnamefont
  {Latham}}\ and\ \bibinfo {author} {\bibfnamefont {T.}~\bibnamefont
  {Gershon}},\ }\href {\doibase 10.1088/0954-3899/36/2/025006} {\bibfield
  {journal} {\bibinfo  {journal} {J. Phys.}\ }\textbf {\bibinfo {volume}
  {G36}},\ \bibinfo {pages} {025006} (\bibinfo {year} {2009})},\ \Eprint
  {http://arxiv.org/abs/0809.0872} {arXiv:0809.0872 [hep-ph]} \BibitemShut
  {NoStop}%
%%CITATION = ARXIV:0809.0872;%%
\bibitem [{\citenamefont {Charles}\ \emph {et~al.}(1998)\citenamefont
  {Charles}, \citenamefont {Le~Yaouanc}, \citenamefont {Oliver}, \citenamefont
  {Pene},\ and\ \citenamefont {Raynal}}]{beta}%
  \BibitemOpen
  \bibfield  {author} {\bibinfo {author} {\bibfnamefont {J.}~\bibnamefont
  {Charles}}, \bibinfo {author} {\bibfnamefont {A.}~\bibnamefont {Le~Yaouanc}},
  \bibinfo {author} {\bibfnamefont {L.}~\bibnamefont {Oliver}}, \bibinfo
  {author} {\bibfnamefont {O.}~\bibnamefont {Pene}}, \ and\ \bibinfo {author}
  {\bibfnamefont {J.~C.}\ \bibnamefont {Raynal}},\ }\href {\doibase
  10.1016/S0370-2693(98)00250-0} {\bibfield  {journal} {\bibinfo  {journal}
  {Phys. Lett.}\ }\textbf {\bibinfo {volume} {B425}},\ \bibinfo {pages} {375}
  (\bibinfo {year} {1998})},\ \bibinfo {note} {[Erratum: Phys.
  Lett.B433,441(1998)]},\ \Eprint {http://arxiv.org/abs/hep-ph/9801363}
  {arXiv:hep-ph/9801363 [hep-ph]} \BibitemShut {NoStop}%
%%CITATION = HEP-PH/9801363;%%
\bibitem [{\citenamefont {Bugg}(2009)}]{dabba}%
  \BibitemOpen
  \bibfield  {author} {\bibinfo {author} {\bibfnamefont {D.~V.}\ \bibnamefont
  {Bugg}},\ }\href {\doibase 10.1088/0954-3899/36/7/075003} {\bibfield
  {journal} {\bibinfo  {journal} {J. Phys.}\ }\textbf {\bibinfo {volume}
  {G36}},\ \bibinfo {pages} {075003} (\bibinfo {year} {2009})},\ \Eprint
  {http://arxiv.org/abs/0901.2217} {arXiv:0901.2217 [hep-ph]} \BibitemShut
  {NoStop}%
%%CITATION = ARXIV:0901.2217;%%
\bibitem [{\citenamefont {Veneziano}(1968)}]{ven}%
  \BibitemOpen
  \bibfield  {author} {\bibinfo {author} {\bibfnamefont {G.}~\bibnamefont
  {Veneziano}},\ }\href {\doibase 10.1007/BF02824451} {\bibfield  {journal}
  {\bibinfo  {journal} {Nuovo Cim.}\ }\textbf {\bibinfo {volume} {A57}},\
  \bibinfo {pages} {190} (\bibinfo {year} {1968})}\BibitemShut {NoStop}%
%%CITATION = NUCIA,A57,190;%%
\bibitem [{\citenamefont {Aaij}\ \emph
  {et~al.}(2014{\natexlab{b}})\citenamefont {Aaij} \emph {et~al.}}]{expnew}%
  \BibitemOpen
  \bibfield  {author} {\bibinfo {author} {\bibfnamefont {R.}~\bibnamefont
  {Aaij}} \emph {et~al.} (\bibinfo {collaboration} {LHCb collaboration}),\
  }\href {\doibase 10.1103/PhysRevD.90.112004} {\bibfield  {journal} {\bibinfo
  {journal} {Phys. Rev.}\ }\textbf {\bibinfo {volume} {D90}},\ \bibinfo {pages}
  {112004} (\bibinfo {year} {2014}{\natexlab{b}})},\ \Eprint
  {http://arxiv.org/abs/1408.5373} {arXiv:1408.5373 [hep-ex]} \BibitemShut
  {NoStop}%
%%CITATION = ARXIV:1408.5373;%%
\bibitem [{\citenamefont {Aubert}\ \emph
  {et~al.}(2008{\natexlab{a}})\citenamefont {Aubert} \emph {et~al.}}]{babar}%
  \BibitemOpen
  \bibfield  {author} {\bibinfo {author} {\bibfnamefont {B.}~\bibnamefont
  {Aubert}} \emph {et~al.} (\bibinfo {collaboration} {BaBar}),\ }\href
  {\doibase 10.1103/PhysRevD.78.051102} {\bibfield  {journal} {\bibinfo
  {journal} {Phys. Rev.}\ }\textbf {\bibinfo {volume} {D78}},\ \bibinfo {pages}
  {051102} (\bibinfo {year} {2008}{\natexlab{a}})},\ \Eprint
  {http://arxiv.org/abs/0802.4035} {arXiv:0802.4035 [hep-ex]} \BibitemShut
  {NoStop}%
%%CITATION = ARXIV:0802.4035;%%
\bibitem [{\citenamefont {Bediaga}\ \emph {et~al.}(2009)\citenamefont
  {Bediaga}, \citenamefont {Bigi}, \citenamefont {Gomes}, \citenamefont
  {Guerrer}, \citenamefont {Miranda},\ and\ \citenamefont {Reis}}]{miranda1}%
  \BibitemOpen
  \bibfield  {author} {\bibinfo {author} {\bibfnamefont {I.}~\bibnamefont
  {Bediaga}}, \bibinfo {author} {\bibfnamefont {I.~I.}\ \bibnamefont {Bigi}},
  \bibinfo {author} {\bibfnamefont {A.}~\bibnamefont {Gomes}}, \bibinfo
  {author} {\bibfnamefont {G.}~\bibnamefont {Guerrer}}, \bibinfo {author}
  {\bibfnamefont {J.}~\bibnamefont {Miranda}}, \ and\ \bibinfo {author}
  {\bibfnamefont {A.~C.~d.}\ \bibnamefont {Reis}},\ }\href {\doibase
  10.1103/PhysRevD.80.096006} {\bibfield  {journal} {\bibinfo  {journal} {Phys.
  Rev.}\ }\textbf {\bibinfo {volume} {D80}},\ \bibinfo {pages} {096006}
  (\bibinfo {year} {2009})},\ \Eprint {http://arxiv.org/abs/0905.4233}
  {arXiv:0905.4233 [hep-ph]} \BibitemShut {NoStop}%
%%CITATION = ARXIV:0905.4233;%%
\bibitem [{\citenamefont {Bediaga}\ \emph {et~al.}(2012)\citenamefont
  {Bediaga}, \citenamefont {Miranda}, \citenamefont {dos Reis}, \citenamefont
  {Bigi}, \citenamefont {Gomes}, \citenamefont {Otalora~Goicochea},\ and\
  \citenamefont {Veiga}}]{miranda2}%
  \BibitemOpen
  \bibfield  {author} {\bibinfo {author} {\bibfnamefont {I.}~\bibnamefont
  {Bediaga}}, \bibinfo {author} {\bibfnamefont {J.}~\bibnamefont {Miranda}},
  \bibinfo {author} {\bibfnamefont {A.~C.}\ \bibnamefont {dos Reis}}, \bibinfo
  {author} {\bibfnamefont {I.~I.}\ \bibnamefont {Bigi}}, \bibinfo {author}
  {\bibfnamefont {A.}~\bibnamefont {Gomes}}, \bibinfo {author} {\bibfnamefont
  {J.~M.}\ \bibnamefont {Otalora~Goicochea}}, \ and\ \bibinfo {author}
  {\bibfnamefont {A.}~\bibnamefont {Veiga}},\ }\href {\doibase
  10.1103/PhysRevD.86.036005} {\bibfield  {journal} {\bibinfo  {journal} {Phys.
  Rev.}\ }\textbf {\bibinfo {volume} {D86}},\ \bibinfo {pages} {036005}
  (\bibinfo {year} {2012})},\ \Eprint {http://arxiv.org/abs/1205.3036}
  {arXiv:1205.3036 [hep-ph]} \BibitemShut {NoStop}%
%%CITATION = ARXIV:1205.3036;%%
\bibitem [{\citenamefont {Bediaga}\ \emph {et~al.}(2014)\citenamefont
  {Bediaga}, \citenamefont {Frederico},\ and\ \citenamefont
  {Lourenço}}]{BedPRD14}%
  \BibitemOpen
  \bibfield  {author} {\bibinfo {author} {\bibfnamefont {I.}~\bibnamefont
  {Bediaga}}, \bibinfo {author} {\bibfnamefont {T.}~\bibnamefont {Frederico}},
  \ and\ \bibinfo {author} {\bibfnamefont {O.}~\bibnamefont {Lourenço}},\
  }\href {\doibase 10.1103/PhysRevD.89.094013} {\bibfield  {journal} {\bibinfo
  {journal} {Phys. Rev.}\ }\textbf {\bibinfo {volume} {D89}},\ \bibinfo {pages}
  {094013} (\bibinfo {year} {2014})},\ \Eprint {http://arxiv.org/abs/1307.8164}
  {arXiv:1307.8164 [hep-ph]} \BibitemShut {NoStop}%
%%CITATION = ARXIV:1307.8164;%%
\bibitem [{\citenamefont {Alvarenga~Nogueira}\ \emph
  {et~al.}(2015)\citenamefont {Alvarenga~Nogueira}, \citenamefont {Bediaga},
  \citenamefont {Cavalcante}, \citenamefont {Frederico},\ and\ \citenamefont
  {Lourenço}}]{NogPRD15}%
  \BibitemOpen
  \bibfield  {author} {\bibinfo {author} {\bibfnamefont {J.~H.}\ \bibnamefont
  {Alvarenga~Nogueira}}, \bibinfo {author} {\bibfnamefont {I.}~\bibnamefont
  {Bediaga}}, \bibinfo {author} {\bibfnamefont {A.~B.~R.}\ \bibnamefont
  {Cavalcante}}, \bibinfo {author} {\bibfnamefont {T.}~\bibnamefont
  {Frederico}}, \ and\ \bibinfo {author} {\bibfnamefont {O.}~\bibnamefont
  {Lourenço}},\ }\href {\doibase 10.1103/PhysRevD.92.054010} {\bibfield
  {journal} {\bibinfo  {journal} {Phys. Rev.}\ }\textbf {\bibinfo {volume}
  {D92}},\ \bibinfo {pages} {054010} (\bibinfo {year} {2015})},\ \Eprint
  {http://arxiv.org/abs/1506.08332} {arXiv:1506.08332 [hep-ph]} \BibitemShut
  {NoStop}%
%%CITATION = ARXIV:1506.08332;%%
\bibitem [{\citenamefont {Magalh\~aes}\ \emph {et~al.}(2011)\citenamefont
  {Magalh\~aes}, \citenamefont {Robilotta}, \citenamefont {Guimar\~aes},
  \citenamefont {Frederico}, \citenamefont {de~Paula}, \citenamefont {Bediaga},
  \citenamefont {Reis}, \citenamefont {Maekawa},\ and\ \citenamefont
  {Zarnauskas}}]{dkpipi1}%
  \BibitemOpen
  \bibfield  {author} {\bibinfo {author} {\bibfnamefont {P.~C.}\ \bibnamefont
  {Magalh\~aes}}, \bibinfo {author} {\bibfnamefont {M.~R.}\ \bibnamefont
  {Robilotta}}, \bibinfo {author} {\bibfnamefont {K.~S. F.~F.}\ \bibnamefont
  {Guimar\~aes}}, \bibinfo {author} {\bibfnamefont {T.}~\bibnamefont
  {Frederico}}, \bibinfo {author} {\bibfnamefont {W.}~\bibnamefont {de~Paula}},
  \bibinfo {author} {\bibfnamefont {I.}~\bibnamefont {Bediaga}}, \bibinfo
  {author} {\bibfnamefont {A.~C.~d.}\ \bibnamefont {Reis}}, \bibinfo {author}
  {\bibfnamefont {C.~M.}\ \bibnamefont {Maekawa}}, \ and\ \bibinfo {author}
  {\bibfnamefont {G.~R.~S.}\ \bibnamefont {Zarnauskas}},\ }\href {\doibase
  10.1103/PhysRevD.84.094001} {\bibfield  {journal} {\bibinfo  {journal} {Phys.
  Rev.}\ }\textbf {\bibinfo {volume} {D84}},\ \bibinfo {pages} {094001}
  (\bibinfo {year} {2011})},\ \Eprint {http://arxiv.org/abs/1105.5120}
  {arXiv:1105.5120 [hep-ph]} \BibitemShut {NoStop}%
%%CITATION = ARXIV:1105.5120;%%
\bibitem [{\citenamefont {Guimar\~aes}\ \emph {et~al.}(2014)\citenamefont
  {Guimar\~aes}, \citenamefont {Lourenço}, \citenamefont {de~Paula},
  \citenamefont {Frederico},\ and\ \citenamefont {dos Reis}}]{dkpipi2}%
  \BibitemOpen
  \bibfield  {author} {\bibinfo {author} {\bibfnamefont {K.~S. F.~F.}\
  \bibnamefont {Guimar\~aes}}, \bibinfo {author} {\bibfnamefont
  {O.}~\bibnamefont {Lourenço}}, \bibinfo {author} {\bibfnamefont
  {W.}~\bibnamefont {de~Paula}}, \bibinfo {author} {\bibfnamefont
  {T.}~\bibnamefont {Frederico}}, \ and\ \bibinfo {author} {\bibfnamefont
  {A.~C.}\ \bibnamefont {dos Reis}},\ }\href {\doibase 10.1007/JHEP08(2014)135}
  {\bibfield  {journal} {\bibinfo  {journal} {JHEP}\ }\textbf {\bibinfo
  {volume} {08}},\ \bibinfo {pages} {135} (\bibinfo {year} {2014})},\ \Eprint
  {http://arxiv.org/abs/1404.3797} {arXiv:1404.3797 [hep-ph]} \BibitemShut
  {NoStop}%
%%CITATION = ARXIV:1404.3797;%%
\bibitem [{\citenamefont {Frederico}\ \emph {et~al.}(2014)\citenamefont
  {Frederico}, \citenamefont {Guimar\~aes}, \citenamefont {Louren\c{c}o},
  \citenamefont {de~Paula}, \citenamefont {Bediaga},\ and\ \citenamefont {dos
  Reis}}]{dkpipi3}%
  \BibitemOpen
  \bibfield  {author} {\bibinfo {author} {\bibfnamefont {T.}~\bibnamefont
  {Frederico}}, \bibinfo {author} {\bibfnamefont {K.~S. F.~F.}\ \bibnamefont
  {Guimar\~aes}}, \bibinfo {author} {\bibfnamefont {O.}~\bibnamefont
  {Louren\c{c}o}}, \bibinfo {author} {\bibfnamefont {W.}~\bibnamefont
  {de~Paula}}, \bibinfo {author} {\bibfnamefont {I.}~\bibnamefont {Bediaga}}, \
  and\ \bibinfo {author} {\bibfnamefont {A.~C.}\ \bibnamefont {dos Reis}},\
  }\bibfield  {booktitle} {\emph {\bibinfo {booktitle} {{Proceedings, Venturing
  off the lightcone - local versus global features (Light Cone 2013)}}},\
  }\href {\doibase 10.1007/s00601-014-0861-z} {\bibfield  {journal} {\bibinfo
  {journal} {Few Body Syst.}\ }\textbf {\bibinfo {volume} {55}},\ \bibinfo
  {pages} {441} (\bibinfo {year} {2014})},\ \Eprint
  {http://arxiv.org/abs/1402.6975} {arXiv:1402.6975 [hep-ph]} \BibitemShut
  {NoStop}%
%%CITATION = ARXIV:1402.6975;%%
\bibitem [{\citenamefont {Aitala}\ \emph {et~al.}(2002)\citenamefont {Aitala}
  \emph {et~al.}}]{kppe791}%
  \BibitemOpen
  \bibfield  {author} {\bibinfo {author} {\bibfnamefont {E.~M.}\ \bibnamefont
  {Aitala}} \emph {et~al.} (\bibinfo {collaboration} {E791}),\ }\href {\doibase
  10.1103/PhysRevLett.89.121801} {\bibfield  {journal} {\bibinfo  {journal}
  {Phys. Rev. Lett.}\ }\textbf {\bibinfo {volume} {89}},\ \bibinfo {pages}
  {121801} (\bibinfo {year} {2002})},\ \Eprint
  {http://arxiv.org/abs/hep-ex/0204018} {arXiv:hep-ex/0204018 [hep-ex]}
  \BibitemShut {NoStop}%
%%CITATION = HEP-EX/0204018;%%
\bibitem [{\citenamefont {Aitala}\ \emph
  {et~al.}(2001{\natexlab{a}})\citenamefont {Aitala} \emph
  {et~al.}}]{d3pi-e791}%
  \BibitemOpen
  \bibfield  {author} {\bibinfo {author} {\bibfnamefont {E.~M.}\ \bibnamefont
  {Aitala}} \emph {et~al.} (\bibinfo {collaboration} {E791}),\ }\href {\doibase
  10.1103/PhysRevLett.86.770} {\bibfield  {journal} {\bibinfo  {journal} {Phys.
  Rev. Lett.}\ }\textbf {\bibinfo {volume} {86}},\ \bibinfo {pages} {770}
  (\bibinfo {year} {2001}{\natexlab{a}})},\ \Eprint
  {http://arxiv.org/abs/hep-ex/0007028} {arXiv:hep-ex/0007028 [hep-ex]}
  \BibitemShut {NoStop}%
%%CITATION = HEP-EX/0007028;%%
\bibitem [{\citenamefont {Aitala}\ \emph
  {et~al.}(2001{\natexlab{b}})\citenamefont {Aitala} \emph
  {et~al.}}]{ds3pi-e791}%
  \BibitemOpen
  \bibfield  {author} {\bibinfo {author} {\bibfnamefont {E.~M.}\ \bibnamefont
  {Aitala}} \emph {et~al.} (\bibinfo {collaboration} {E791}),\ }\href {\doibase
  10.1103/PhysRevLett.86.765} {\bibfield  {journal} {\bibinfo  {journal} {Phys.
  Rev. Lett.}\ }\textbf {\bibinfo {volume} {86}},\ \bibinfo {pages} {765}
  (\bibinfo {year} {2001}{\natexlab{b}})},\ \Eprint
  {http://arxiv.org/abs/hep-ex/0007027} {arXiv:hep-ex/0007027 [hep-ex]}
  \BibitemShut {NoStop}%
%%CITATION = HEP-EX/0007027;%%
\bibitem [{\citenamefont {Bonvicini}\ \emph {et~al.}(2007)\citenamefont
  {Bonvicini} \emph {et~al.}}]{d3pi-cleo}%
  \BibitemOpen
  \bibfield  {author} {\bibinfo {author} {\bibfnamefont {G.}~\bibnamefont
  {Bonvicini}} \emph {et~al.} (\bibinfo {collaboration} {CLEO}),\ }\href
  {\doibase 10.1103/PhysRevD.76.012001} {\bibfield  {journal} {\bibinfo
  {journal} {Phys. Rev.}\ }\textbf {\bibinfo {volume} {D76}},\ \bibinfo {pages}
  {012001} (\bibinfo {year} {2007})},\ \Eprint {http://arxiv.org/abs/0704.3954}
  {arXiv:0704.3954 [hep-ex]} \BibitemShut {NoStop}%
%%CITATION = ARXIV:0704.3954;%%
\bibitem [{\citenamefont {Asner}(2003)}]{asner}%
  \BibitemOpen
  \bibfield  {author} {\bibinfo {author} {\bibfnamefont {D.}~\bibnamefont
  {Asner}},\ }\href@noop {} {\  (\bibinfo {year} {2003})},\ \Eprint
  {http://arxiv.org/abs/hep-ex/0410014} {arXiv:hep-ex/0410014 [hep-ex]}
  \BibitemShut {NoStop}%
%%CITATION = HEP-EX/0410014;%%
\bibitem [{\citenamefont {Link}\ \emph {et~al.}(2004)\citenamefont {Link} \emph
  {et~al.}}]{focus-3pi}%
  \BibitemOpen
  \bibfield  {author} {\bibinfo {author} {\bibfnamefont {J.~M.}\ \bibnamefont
  {Link}} \emph {et~al.} (\bibinfo {collaboration} {FOCUS}),\ }\href {\doibase
  10.1016/j.physletb.2004.01.065} {\bibfield  {journal} {\bibinfo  {journal}
  {Phys. Lett.}\ }\textbf {\bibinfo {volume} {B585}},\ \bibinfo {pages} {200}
  (\bibinfo {year} {2004})},\ \Eprint {http://arxiv.org/abs/hep-ex/0312040}
  {arXiv:hep-ex/0312040 [hep-ex]} \BibitemShut {NoStop}%
%%CITATION = HEP-EX/0312040;%%
\bibitem [{\citenamefont {Anisovich}\ and\ \citenamefont
  {Sarantsev}(2003)}]{sarantsev}%
  \BibitemOpen
  \bibfield  {author} {\bibinfo {author} {\bibfnamefont {V.~V.}\ \bibnamefont
  {Anisovich}}\ and\ \bibinfo {author} {\bibfnamefont {A.~V.}\ \bibnamefont
  {Sarantsev}},\ }\href {\doibase 10.1140/epja/i2002-10068-x} {\bibfield
  {journal} {\bibinfo  {journal} {Eur. Phys. J.}\ }\textbf {\bibinfo {volume}
  {A16}},\ \bibinfo {pages} {229} (\bibinfo {year} {2003})},\ \Eprint
  {http://arxiv.org/abs/hep-ph/0204328} {arXiv:hep-ph/0204328 [hep-ph]}
  \BibitemShut {NoStop}%
%%CITATION = HEP-PH/0204328;%%
\bibitem [{\citenamefont {Aitala}\ \emph {et~al.}(2006)\citenamefont {Aitala}
  \emph {et~al.}}]{brian}%
  \BibitemOpen
  \bibfield  {author} {\bibinfo {author} {\bibfnamefont {E.~M.}\ \bibnamefont
  {Aitala}} \emph {et~al.} (\bibinfo {collaboration} {E791}),\ }\href {\doibase
  10.1103/PhysRevD.73.032004, 10.1103/PhysRevD.74.059901} {\bibfield  {journal}
  {\bibinfo  {journal} {Phys. Rev.}\ }\textbf {\bibinfo {volume} {D73}},\
  \bibinfo {pages} {032004} (\bibinfo {year} {2006})},\ \bibinfo {note}
  {[Erratum: Phys. Rev.D74,059901(2006)]},\ \Eprint
  {http://arxiv.org/abs/hep-ex/0507099} {arXiv:hep-ex/0507099 [hep-ex]}
  \BibitemShut {NoStop}%
%%CITATION = HEP-EX/0507099;%%
\bibitem [{\citenamefont {Aubert}\ \emph
  {et~al.}(2009{\natexlab{a}})\citenamefont {Aubert} \emph {et~al.}}]{antimo}%
  \BibitemOpen
  \bibfield  {author} {\bibinfo {author} {\bibfnamefont {B.}~\bibnamefont
  {Aubert}} \emph {et~al.} (\bibinfo {collaboration} {BaBar}),\ }\bibfield
  {booktitle} {\emph {\bibinfo {booktitle} {{Proceedings, 34th International
  Conference on High Energy Physics (ICHEP 2008)}}},\ }\href {\doibase
  10.1103/PhysRevD.79.032003} {\bibfield  {journal} {\bibinfo  {journal} {Phys.
  Rev.}\ }\textbf {\bibinfo {volume} {D79}},\ \bibinfo {pages} {032003}
  (\bibinfo {year} {2009}{\natexlab{a}})},\ \Eprint
  {http://arxiv.org/abs/0808.0971} {arXiv:0808.0971 [hep-ex]} \BibitemShut
  {NoStop}%
%%CITATION = ARXIV:0808.0971;%%
\bibitem [{\citenamefont {Hyams}\ \emph {et~al.}(1973)\citenamefont {Hyams}
  \emph {et~al.}}]{cern-munich1}%
  \BibitemOpen
  \bibfield  {author} {\bibinfo {author} {\bibfnamefont {B.}~\bibnamefont
  {Hyams}} \emph {et~al.},\ }\href {\doibase 10.1016/0550-3213(73)90618-4}
  {\bibfield  {journal} {\bibinfo  {journal} {Nucl. Phys.}\ }\textbf {\bibinfo
  {volume} {B64}},\ \bibinfo {pages} {134} (\bibinfo {year}
  {1973})}\BibitemShut {NoStop}%
%%CITATION = NUPHA,B64,134;%%
\bibitem [{\citenamefont {Grayer}\ \emph {et~al.}(1974)\citenamefont {Grayer}
  \emph {et~al.}}]{cern-munich2}%
  \BibitemOpen
  \bibfield  {author} {\bibinfo {author} {\bibfnamefont {G.}~\bibnamefont
  {Grayer}} \emph {et~al.},\ }\href {\doibase 10.1016/0550-3213(74)90545-8}
  {\bibfield  {journal} {\bibinfo  {journal} {Nucl. Phys.}\ }\textbf {\bibinfo
  {volume} {B75}},\ \bibinfo {pages} {189} (\bibinfo {year}
  {1974})}\BibitemShut {NoStop}%
%%CITATION = NUPHA,B75,189;%%
\bibitem [{\citenamefont {Link}\ \emph {et~al.}(2009)\citenamefont {Link} \emph
  {et~al.}}]{focuspwa}%
  \BibitemOpen
  \bibfield  {author} {\bibinfo {author} {\bibfnamefont {J.~M.}\ \bibnamefont
  {Link}} \emph {et~al.} (\bibinfo {collaboration} {FOCUS}),\ }\href {\doibase
  10.1016/j.physletb.2009.09.057} {\bibfield  {journal} {\bibinfo  {journal}
  {Phys. Lett.}\ }\textbf {\bibinfo {volume} {B681}},\ \bibinfo {pages} {14}
  (\bibinfo {year} {2009})},\ \Eprint {http://arxiv.org/abs/0905.4846}
  {arXiv:0905.4846 [hep-ex]} \BibitemShut {NoStop}%
%%CITATION = ARXIV:0905.4846;%%
\bibitem [{\citenamefont {Aston}\ \emph
  {et~al.}(1988{\natexlab{a}})\citenamefont {Aston} \emph {et~al.}}]{lass}%
  \BibitemOpen
  \bibfield  {author} {\bibinfo {author} {\bibfnamefont {D.}~\bibnamefont
  {Aston}} \emph {et~al.},\ }\href {\doibase 10.1016/0550-3213(88)90028-4}
  {\bibfield  {journal} {\bibinfo  {journal} {Nucl. Phys.}\ }\textbf {\bibinfo
  {volume} {B296}},\ \bibinfo {pages} {493} (\bibinfo {year}
  {1988}{\natexlab{a}})}\BibitemShut {NoStop}%
%%CITATION = NUPHA,B296,493;%%
\bibitem [{\citenamefont {Watson}(1952)}]{watson}%
  \BibitemOpen
  \bibfield  {author} {\bibinfo {author} {\bibfnamefont {K.~M.}\ \bibnamefont
  {Watson}},\ }\href {\doibase 10.1103/PhysRev.88.1163} {\bibfield  {journal}
  {\bibinfo  {journal} {Phys. Rev.}\ }\textbf {\bibinfo {volume} {88}},\
  \bibinfo {pages} {1163} (\bibinfo {year} {1952})}\BibitemShut {NoStop}%
%%CITATION = PHRVA,88,1163;%%
\bibitem [{\citenamefont {Magalh\~aes}\ and\ \citenamefont
  {Robilotta}(2015)}]{pat}%
  \BibitemOpen
  \bibfield  {author} {\bibinfo {author} {\bibfnamefont {P.~C.}\ \bibnamefont
  {Magalh\~aes}}\ and\ \bibinfo {author} {\bibfnamefont {M.~R.}\ \bibnamefont
  {Robilotta}},\ }\href {\doibase 10.1103/PhysRevD.92.094005} {\bibfield
  {journal} {\bibinfo  {journal} {Phys. Rev.}\ }\textbf {\bibinfo {volume}
  {D92}},\ \bibinfo {pages} {094005} (\bibinfo {year} {2015})},\ \Eprint
  {http://arxiv.org/abs/1504.06346} {arXiv:1504.06346 [hep-ph]} \BibitemShut
  {NoStop}%
%%CITATION = ARXIV:1504.06346;%%
\bibitem [{\citenamefont {Achasov}\ \emph {et~al.}(1979)\citenamefont
  {Achasov}, \citenamefont {Devyanin},\ and\ \citenamefont
  {Shestakov}}]{achasov}%
  \BibitemOpen
  \bibfield  {author} {\bibinfo {author} {\bibfnamefont {N.~N.}\ \bibnamefont
  {Achasov}}, \bibinfo {author} {\bibfnamefont {S.~A.}\ \bibnamefont
  {Devyanin}}, \ and\ \bibinfo {author} {\bibfnamefont {G.~N.}\ \bibnamefont
  {Shestakov}},\ }\href {\doibase 10.1016/0370-2693(79)90488-X} {\bibfield
  {journal} {\bibinfo  {journal} {Phys. Lett.}\ }\textbf {\bibinfo {volume}
  {B88}},\ \bibinfo {pages} {367} (\bibinfo {year} {1979})}\BibitemShut
  {NoStop}%
%%CITATION = PHLTA,B88,367;%%
\bibitem [{\citenamefont {Aaij}\ \emph
  {et~al.}(2014{\natexlab{c}})\citenamefont {Aaij} \emph
  {et~al.}}]{Aaij:2014siy}%
  \BibitemOpen
  \bibfield  {author} {\bibinfo {author} {\bibfnamefont {R.}~\bibnamefont
  {Aaij}} \emph {et~al.} (\bibinfo {collaboration} {LHCb collaboration}),\
  }\href {\doibase 10.1103/PhysRevD.90.012003} {\bibfield  {journal} {\bibinfo
  {journal} {Phys. Rev.}\ }\textbf {\bibinfo {volume} {D90}},\ \bibinfo {pages}
  {012003} (\bibinfo {year} {2014}{\natexlab{c}})},\ \Eprint
  {http://arxiv.org/abs/1404.5673} {arXiv:1404.5673 [hep-ex]} \BibitemShut
  {NoStop}%
%%CITATION = ARXIV:1404.5673;%%
\bibitem [{\citenamefont {Aaij}\ \emph
  {et~al.}(2013{\natexlab{a}})\citenamefont {Aaij} \emph
  {et~al.}}]{LHCb-PAPER-2013-001}%
  \BibitemOpen
  \bibfield  {author} {\bibinfo {author} {\bibfnamefont {R.}~\bibnamefont
  {Aaij}} \emph {et~al.} (\bibinfo {collaboration} {LHCb collaboration}),\
  }\href {\doibase 10.1103/PhysRevLett.110.222001} {\bibfield  {journal}
  {\bibinfo  {journal} {Phys. Rev. Lett.}\ }\textbf {\bibinfo {volume} {110}},\
  \bibinfo {pages} {222001} (\bibinfo {year} {2013}{\natexlab{a}})},\ \Eprint
  {http://arxiv.org/abs/1302.6269} {arXiv:1302.6269 [hep-ex]} \BibitemShut
  {NoStop}%
%%CITATION = ARXIV:1302.6269;%%
\bibitem [{\citenamefont {Aaij}\ \emph
  {et~al.}(2015{\natexlab{d}})\citenamefont {Aaij} \emph
  {et~al.}}]{LHCb-PAPER-2015-015}%
  \BibitemOpen
  \bibfield  {author} {\bibinfo {author} {\bibfnamefont {R.}~\bibnamefont
  {Aaij}} \emph {et~al.} (\bibinfo {collaboration} {LHCb collaboration}),\
  }\href {\doibase 10.1103/PhysRevD.92.011102} {\bibfield  {journal} {\bibinfo
  {journal} {Phys. Rev.}\ }\textbf {\bibinfo {volume} {D92}},\ \bibinfo {pages}
  {011102(R)} (\bibinfo {year} {2015}{\natexlab{d}})},\ \Eprint
  {http://arxiv.org/abs/1504.06339} {arXiv:1504.06339 [hep-ex]} \BibitemShut
  {NoStop}%
%%CITATION = ARXIV:1504.06339;%%
\bibitem [{\citenamefont {Bugg}(2008)}]{Bugg:2008wu}%
  \BibitemOpen
  \bibfield  {author} {\bibinfo {author} {\bibfnamefont {D.~V.}\ \bibnamefont
  {Bugg}},\ }\href {\doibase 10.1088/0954-3899/35/7/075005} {\bibfield
  {journal} {\bibinfo  {journal} {J. Phys.}\ }\textbf {\bibinfo {volume}
  {G35}},\ \bibinfo {pages} {075005} (\bibinfo {year} {2008})},\ \Eprint
  {http://arxiv.org/abs/0802.0934} {arXiv:0802.0934 [hep-ph]} \BibitemShut
  {NoStop}%
%%CITATION = ARXIV:0802.0934;%%
\bibitem [{\citenamefont {Esposito}\ \emph {et~al.}(2014)\citenamefont
  {Esposito}, \citenamefont {Guerrieri}, \citenamefont {Piccinini},
  \citenamefont {Pilloni},\ and\ \citenamefont {Polosa}}]{Esposito:2014rxa}%
  \BibitemOpen
  \bibfield  {author} {\bibinfo {author} {\bibfnamefont {A.}~\bibnamefont
  {Esposito}}, \bibinfo {author} {\bibfnamefont {A.~L.}\ \bibnamefont
  {Guerrieri}}, \bibinfo {author} {\bibfnamefont {F.}~\bibnamefont
  {Piccinini}}, \bibinfo {author} {\bibfnamefont {A.}~\bibnamefont {Pilloni}},
  \ and\ \bibinfo {author} {\bibfnamefont {A.~D.}\ \bibnamefont {Polosa}},\
  }\href {\doibase 10.1142/S0217751X15300021} {\bibfield  {journal} {\bibinfo
  {journal} {Int. J. Mod. Phys.}\ }\textbf {\bibinfo {volume} {A30}},\ \bibinfo
  {pages} {1530002} (\bibinfo {year} {2014})},\ \Eprint
  {http://arxiv.org/abs/1411.5997} {arXiv:1411.5997 [hep-ph]} \BibitemShut
  {NoStop}%
%%CITATION = ARXIV:1411.5997;%%
\bibitem [{\citenamefont {Aaij}\ \emph
  {et~al.}(2014{\natexlab{d}})\citenamefont {Aaij} \emph
  {et~al.}}]{Aaij:2014jqa}%
  \BibitemOpen
  \bibfield  {author} {\bibinfo {author} {\bibfnamefont {R.}~\bibnamefont
  {Aaij}} \emph {et~al.} (\bibinfo {collaboration} {LHCb collaboration}),\
  }\href {\doibase 10.1103/PhysRevLett.112.222002} {\bibfield  {journal}
  {\bibinfo  {journal} {Phys.Rev.Lett.}\ }\textbf {\bibinfo {volume} {112}},\
  \bibinfo {pages} {222002} (\bibinfo {year} {2014}{\natexlab{d}})},\ \Eprint
  {http://arxiv.org/abs/1404.1903} {arXiv:1404.1903 [hep-ex]} \BibitemShut
  {NoStop}%
%%CITATION = ARXIV:1404.1903;%%
\bibitem [{\citenamefont {Choi}\ \emph {et~al.}(2008)\citenamefont {Choi} \emph
  {et~al.}}]{Choi:2007wga}%
  \BibitemOpen
  \bibfield  {author} {\bibinfo {author} {\bibfnamefont {S.}~\bibnamefont
  {Choi}} \emph {et~al.} (\bibinfo {collaboration} {Belle collaboration}),\
  }\href {\doibase 10.1103/PhysRevLett.100.142001} {\bibfield  {journal}
  {\bibinfo  {journal} {Phys.Rev.Lett.}\ }\textbf {\bibinfo {volume} {100}},\
  \bibinfo {pages} {142001} (\bibinfo {year} {2008})},\ \Eprint
  {http://arxiv.org/abs/0708.1790} {arXiv:0708.1790 [hep-ex]} \BibitemShut
  {NoStop}%
%%CITATION = ARXIV:0708.1790;%%
\bibitem [{\citenamefont {Aubert}\ \emph
  {et~al.}(2009{\natexlab{b}})\citenamefont {Aubert} \emph
  {et~al.}}]{Aubert:2008aa}%
  \BibitemOpen
  \bibfield  {author} {\bibinfo {author} {\bibfnamefont {B.}~\bibnamefont
  {Aubert}} \emph {et~al.} (\bibinfo {collaboration} {BaBar collaboration}),\
  }\href {\doibase 10.1103/PhysRevD.79.112001} {\bibfield  {journal} {\bibinfo
  {journal} {Phys.Rev.}\ }\textbf {\bibinfo {volume} {D79}},\ \bibinfo {pages}
  {112001} (\bibinfo {year} {2009}{\natexlab{b}})},\ \Eprint
  {http://arxiv.org/abs/0811.0564} {arXiv:0811.0564 [hep-ex]} \BibitemShut
  {NoStop}%
%%CITATION = ARXIV:0811.0564;%%
\bibitem [{\citenamefont {Chilikin}\ \emph {et~al.}(2013)\citenamefont
  {Chilikin} \emph {et~al.}}]{Chilikin:2013tch}%
  \BibitemOpen
  \bibfield  {author} {\bibinfo {author} {\bibfnamefont {K.}~\bibnamefont
  {Chilikin}} \emph {et~al.} (\bibinfo {collaboration} {Belle collaboration}),\
  }\href {\doibase 10.1103/PhysRevD.88.074026} {\bibfield  {journal} {\bibinfo
  {journal} {Phys.Rev.}\ }\textbf {\bibinfo {volume} {D88}},\ \bibinfo {pages}
  {074026} (\bibinfo {year} {2013})},\ \Eprint {http://arxiv.org/abs/1306.4894}
  {arXiv:1306.4894 [hep-ex]} \BibitemShut {NoStop}%
%%CITATION = ARXIV:1306.4894;%%
\bibitem [{\citenamefont {Pakhlov}\ and\ \citenamefont
  {Uglov}(2015)}]{Pakhlov:2014qva}%
  \BibitemOpen
  \bibfield  {author} {\bibinfo {author} {\bibfnamefont {P.}~\bibnamefont
  {Pakhlov}}\ and\ \bibinfo {author} {\bibfnamefont {T.}~\bibnamefont
  {Uglov}},\ }\href {\doibase 10.1016/j.physletb.2015.06.074} {\bibfield
  {journal} {\bibinfo  {journal} {Phys. Lett.}\ }\textbf {\bibinfo {volume}
  {B748}},\ \bibinfo {pages} {183} (\bibinfo {year} {2015})},\ \Eprint
  {http://arxiv.org/abs/1408.5295} {arXiv:1408.5295 [hep-ph]} \BibitemShut
  {NoStop}%
%%CITATION = ARXIV:1408.5295;%%
\bibitem [{\citenamefont {Aaij}\ \emph
  {et~al.}(2015{\natexlab{e}})\citenamefont {Aaij} \emph
  {et~al.}}]{LHCb-PAPER-2015-029}%
  \BibitemOpen
  \bibfield  {author} {\bibinfo {author} {\bibfnamefont {R.}~\bibnamefont
  {Aaij}} \emph {et~al.} (\bibinfo {collaboration} {LHCb}),\ }\href {\doibase
  10.1103/PhysRevLett.115.072001} {\bibfield  {journal} {\bibinfo  {journal}
  {Phys. Rev. Lett.}\ }\textbf {\bibinfo {volume} {115}},\ \bibinfo {pages}
  {072001} (\bibinfo {year} {2015}{\natexlab{e}})},\ \Eprint
  {http://arxiv.org/abs/1507.03414} {arXiv:1507.03414 [hep-ex]} \BibitemShut
  {NoStop}%
%%CITATION = ARXIV:1507.03414;%%
\bibitem [{\citenamefont {Maiani}\ \emph {et~al.}(2015)\citenamefont {Maiani},
  \citenamefont {Polosa},\ and\ \citenamefont {Riquer}}]{Maiani:2015vwa}%
  \BibitemOpen
  \bibfield  {author} {\bibinfo {author} {\bibfnamefont {L.}~\bibnamefont
  {Maiani}}, \bibinfo {author} {\bibfnamefont {A.~D.}\ \bibnamefont {Polosa}},
  \ and\ \bibinfo {author} {\bibfnamefont {V.}~\bibnamefont {Riquer}},\ }\href
  {\doibase 10.1016/j.physletb.2015.08.008} {\bibfield  {journal} {\bibinfo
  {journal} {Phys. Lett.}\ }\textbf {\bibinfo {volume} {B749}},\ \bibinfo
  {pages} {289} (\bibinfo {year} {2015})},\ \Eprint
  {http://arxiv.org/abs/1507.04980} {arXiv:1507.04980 [hep-ph]} \BibitemShut
  {NoStop}%
%%CITATION = ARXIV:1507.04980;%%
\bibitem [{\citenamefont {Lebed}(2015)}]{Lebed:2015tna}%
  \BibitemOpen
  \bibfield  {author} {\bibinfo {author} {\bibfnamefont {R.~F.}\ \bibnamefont
  {Lebed}},\ }\href {\doibase 10.1016/j.physletb.2015.08.032} {\bibfield
  {journal} {\bibinfo  {journal} {Phys. Lett.}\ }\textbf {\bibinfo {volume}
  {B749}},\ \bibinfo {pages} {454} (\bibinfo {year} {2015})},\ \Eprint
  {http://arxiv.org/abs/1507.05867} {arXiv:1507.05867 [hep-ph]} \BibitemShut
  {NoStop}%
%%CITATION = ARXIV:1507.05867;%%
\bibitem [{\citenamefont {Anisovich}\ \emph {et~al.}(2015)\citenamefont
  {Anisovich}, \citenamefont {Matveev}, \citenamefont {Nyiri}, \citenamefont
  {Sarantsev},\ and\ \citenamefont {Semenova}}]{Anisovich:2015cia}%
  \BibitemOpen
  \bibfield  {author} {\bibinfo {author} {\bibfnamefont {V.~V.}\ \bibnamefont
  {Anisovich}}, \bibinfo {author} {\bibfnamefont {M.~A.}\ \bibnamefont
  {Matveev}}, \bibinfo {author} {\bibfnamefont {J.}~\bibnamefont {Nyiri}},
  \bibinfo {author} {\bibfnamefont {A.~V.}\ \bibnamefont {Sarantsev}}, \ and\
  \bibinfo {author} {\bibfnamefont {A.~N.}\ \bibnamefont {Semenova}},\
  }\href@noop {} {\  (\bibinfo {year} {2015})},\ \Eprint
  {http://arxiv.org/abs/1507.07652} {arXiv:1507.07652 [hep-ph]} \BibitemShut
  {NoStop}%
%%CITATION = ARXIV:1507.07652;%%
\bibitem [{\citenamefont {Karliner}\ and\ \citenamefont
  {Rosner}(2015)}]{Karliner:2015ina}%
  \BibitemOpen
  \bibfield  {author} {\bibinfo {author} {\bibfnamefont {M.}~\bibnamefont
  {Karliner}}\ and\ \bibinfo {author} {\bibfnamefont {J.~L.}\ \bibnamefont
  {Rosner}},\ }\href {\doibase 10.1103/PhysRevLett.115.122001} {\bibfield
  {journal} {\bibinfo  {journal} {Phys. Rev. Lett.}\ }\textbf {\bibinfo
  {volume} {115}},\ \bibinfo {pages} {122001} (\bibinfo {year} {2015})},\
  \Eprint {http://arxiv.org/abs/1506.06386} {arXiv:1506.06386 [hep-ph]}
  \BibitemShut {NoStop}%
%%CITATION = ARXIV:1506.06386;%%
\bibitem [{\citenamefont {Chen}\ \emph {et~al.}(2015)\citenamefont {Chen},
  \citenamefont {Liu}, \citenamefont {Li},\ and\ \citenamefont
  {Zhu}}]{Chen:2015loa}%
  \BibitemOpen
  \bibfield  {author} {\bibinfo {author} {\bibfnamefont {R.}~\bibnamefont
  {Chen}}, \bibinfo {author} {\bibfnamefont {X.}~\bibnamefont {Liu}}, \bibinfo
  {author} {\bibfnamefont {X.-Q.}\ \bibnamefont {Li}}, \ and\ \bibinfo {author}
  {\bibfnamefont {S.-L.}\ \bibnamefont {Zhu}},\ }\href {\doibase
  10.1103/PhysRevLett.115.132002} {\bibfield  {journal} {\bibinfo  {journal}
  {Phys. Rev. Lett.}\ }\textbf {\bibinfo {volume} {115}},\ \bibinfo {pages}
  {132002} (\bibinfo {year} {2015})},\ \Eprint
  {http://arxiv.org/abs/1507.03704} {arXiv:1507.03704 [hep-ph]} \BibitemShut
  {NoStop}%
%%CITATION = ARXIV:1507.03704;%%
\bibitem [{\citenamefont {Roca}\ \emph {et~al.}(2015)\citenamefont {Roca},
  \citenamefont {Nieves},\ and\ \citenamefont {Oset}}]{Roca:2015dva}%
  \BibitemOpen
  \bibfield  {author} {\bibinfo {author} {\bibfnamefont {L.}~\bibnamefont
  {Roca}}, \bibinfo {author} {\bibfnamefont {J.}~\bibnamefont {Nieves}}, \ and\
  \bibinfo {author} {\bibfnamefont {E.}~\bibnamefont {Oset}},\ }\href@noop {}
  {\  (\bibinfo {year} {2015})},\ \Eprint {http://arxiv.org/abs/1507.04249}
  {arXiv:1507.04249 [hep-ph]} \BibitemShut {NoStop}%
%%CITATION = ARXIV:1507.04249;%%
\bibitem [{\citenamefont {He}(2015)}]{He:2015cea}%
  \BibitemOpen
  \bibfield  {author} {\bibinfo {author} {\bibfnamefont {J.}~\bibnamefont
  {He}},\ }\href@noop {} {\  (\bibinfo {year} {2015})},\ \Eprint
  {http://arxiv.org/abs/1507.05200} {arXiv:1507.05200 [hep-ph]} \BibitemShut
  {NoStop}%
%%CITATION = ARXIV:1507.05200;%%
\bibitem [{\citenamefont {Mei\ss{}ner}\ and\ \citenamefont
  {Oller}(2015)}]{Meissner:2015mza}%
  \BibitemOpen
  \bibfield  {author} {\bibinfo {author} {\bibfnamefont {U.-G.}\ \bibnamefont
  {Mei\ss{}ner}}\ and\ \bibinfo {author} {\bibfnamefont {J.~A.}\ \bibnamefont
  {Oller}},\ }\href@noop {} {\  (\bibinfo {year} {2015})},\ \Eprint
  {http://arxiv.org/abs/1507.07478} {arXiv:1507.07478 [hep-ph]} \BibitemShut
  {NoStop}%
%%CITATION = ARXIV:1507.07478;%%
\bibitem [{\citenamefont {Guo}\ \emph {et~al.}(2015)\citenamefont {Guo},
  \citenamefont {Mei{\ss}ner}, \citenamefont {Wang},\ and\ \citenamefont
  {Yang}}]{Guo:2015umn}%
  \BibitemOpen
  \bibfield  {author} {\bibinfo {author} {\bibfnamefont {F.-K.}\ \bibnamefont
  {Guo}}, \bibinfo {author} {\bibfnamefont {U.-G.}\ \bibnamefont
  {Mei{\ss}ner}}, \bibinfo {author} {\bibfnamefont {W.}~\bibnamefont {Wang}}, \
  and\ \bibinfo {author} {\bibfnamefont {Z.}~\bibnamefont {Yang}},\ }\href
  {\doibase 10.1103/PhysRevD.92.071502} {\bibfield  {journal} {\bibinfo
  {journal} {Phys. Rev.}\ }\textbf {\bibinfo {volume} {D92}},\ \bibinfo {pages}
  {071502} (\bibinfo {year} {2015})},\ \Eprint
  {http://arxiv.org/abs/1507.04950} {arXiv:1507.04950 [hep-ph]} \BibitemShut
  {NoStop}%
%%CITATION = ARXIV:1507.04950;%%
\bibitem [{\citenamefont {Liu}\ \emph {et~al.}(2015)\citenamefont {Liu},
  \citenamefont {Wang},\ and\ \citenamefont {Zhao}}]{Liu:2015fea}%
  \BibitemOpen
  \bibfield  {author} {\bibinfo {author} {\bibfnamefont {X.-H.}\ \bibnamefont
  {Liu}}, \bibinfo {author} {\bibfnamefont {Q.}~\bibnamefont {Wang}}, \ and\
  \bibinfo {author} {\bibfnamefont {Q.}~\bibnamefont {Zhao}},\ }\href@noop {}
  {\  (\bibinfo {year} {2015})},\ \Eprint {http://arxiv.org/abs/1507.05359}
  {arXiv:1507.05359 [hep-ph]} \BibitemShut {NoStop}%
%%CITATION = ARXIV:1507.05359;%%
\bibitem [{\citenamefont {Mikhasenko}(2015)}]{Mikhasenko:2015vca}%
  \BibitemOpen
  \bibfield  {author} {\bibinfo {author} {\bibfnamefont {M.}~\bibnamefont
  {Mikhasenko}},\ }\href@noop {} {\  (\bibinfo {year} {2015})},\ \Eprint
  {http://arxiv.org/abs/1507.06552} {arXiv:1507.06552 [hep-ph]} \BibitemShut
  {NoStop}%
%%CITATION = ARXIV:1507.06552;%%
\bibitem [{\citenamefont
  {Szczepaniak}(2015{\natexlab{a}})}]{Szczepaniak:2015hya}%
  \BibitemOpen
  \bibfield  {author} {\bibinfo {author} {\bibfnamefont {A.~P.}\ \bibnamefont
  {Szczepaniak}},\ }\href@noop {} {\  (\bibinfo {year} {2015}{\natexlab{a}})},\
  \Eprint {http://arxiv.org/abs/1510.01789} {arXiv:1510.01789 [hep-ph]}
  \BibitemShut {NoStop}%
%%CITATION = ARXIV:1510.01789;%%
\bibitem [{\citenamefont {Aaij}\ \emph
  {et~al.}(2014{\natexlab{e}})\citenamefont {Aaij} \emph
  {et~al.}}]{LHCb-PAPER-2014-035}%
  \BibitemOpen
  \bibfield  {author} {\bibinfo {author} {\bibfnamefont {R.}~\bibnamefont
  {Aaij}} \emph {et~al.} (\bibinfo {collaboration} {LHCb collaboration}),\
  }\href {\doibase 10.1103/PhysRevLett.113.162001} {\bibfield  {journal}
  {\bibinfo  {journal} {Phys. Rev. Lett.}\ }\textbf {\bibinfo {volume} {113}},\
  \bibinfo {pages} {162001} (\bibinfo {year} {2014}{\natexlab{e}})},\ \Eprint
  {http://arxiv.org/abs/1407.7574} {arXiv:1407.7574 [hep-ex]} \BibitemShut
  {NoStop}%
%%CITATION = ARXIV:1407.7574;%%
\bibitem [{\citenamefont {Aaij}\ \emph
  {et~al.}(2015{\natexlab{f}})\citenamefont {Aaij} \emph
  {et~al.}}]{Aaij:2015zxa}%
  \BibitemOpen
  \bibfield  {author} {\bibinfo {author} {\bibfnamefont {R.}~\bibnamefont
  {Aaij}} \emph {et~al.} (\bibinfo {collaboration} {LHCb}),\ }\href@noop {} {\
  (\bibinfo {year} {2015}{\natexlab{f}})},\ \Eprint
  {http://arxiv.org/abs/1510.01951} {arXiv:1510.01951 [hep-ex]} \BibitemShut
  {NoStop}%
%%CITATION = ARXIV:1510.01951;%%
\bibitem [{\citenamefont {Coffman}\ \emph {et~al.}(1992)\citenamefont {Coffman}
  \emph {et~al.}}]{MarkIII_K3piModel}%
  \BibitemOpen
  \bibfield  {author} {\bibinfo {author} {\bibfnamefont {D.}~\bibnamefont
  {Coffman}} \emph {et~al.} (\bibinfo {collaboration} {Mark III}),\ }\href
  {\doibase 10.1103/PhysRevD.45.2196} {\bibfield  {journal} {\bibinfo
  {journal} {Phys. Rev. D}\ }\textbf {\bibinfo {volume} {45}},\ \bibinfo
  {pages} {2196} (\bibinfo {year} {1992})}\BibitemShut {NoStop}%
\bibitem [{\citenamefont {Link}\ \emph
  {et~al.}(2005{\natexlab{a}})\citenamefont {Link} \emph
  {et~al.}}]{Link:2004wx}%
  \BibitemOpen
  \bibfield  {author} {\bibinfo {author} {\bibfnamefont {J.~M.}\ \bibnamefont
  {Link}} \emph {et~al.} (\bibinfo {collaboration} {FOCUS}),\ }\href {\doibase
  10.1016/j.physletb.2005.02.005} {\bibfield  {journal} {\bibinfo  {journal}
  {Phys. Lett.}\ }\textbf {\bibinfo {volume} {B610}},\ \bibinfo {pages} {225}
  (\bibinfo {year} {2005}{\natexlab{a}})},\ \Eprint
  {http://arxiv.org/abs/hep-ex/0411031} {arXiv:hep-ex/0411031 [hep-ex]}
  \BibitemShut {NoStop}%
%%CITATION = HEP-EX/0411031;%%
\bibitem [{\citenamefont {Link}\ \emph {et~al.}(2007)\citenamefont {Link} \emph
  {et~al.}}]{Link:2007fi}%
  \BibitemOpen
  \bibfield  {author} {\bibinfo {author} {\bibfnamefont {J.~M.}\ \bibnamefont
  {Link}} \emph {et~al.} (\bibinfo {collaboration} {FOCUS}),\ }\href {\doibase
  10.1103/PhysRevD.75.052003} {\bibfield  {journal} {\bibinfo  {journal} {Phys.
  Rev.}\ }\textbf {\bibinfo {volume} {D75}},\ \bibinfo {pages} {052003}
  (\bibinfo {year} {2007})},\ \Eprint {http://arxiv.org/abs/hep-ex/0701001}
  {arXiv:hep-ex/0701001 [hep-ex]} \BibitemShut {NoStop}%
%%CITATION = HEP-EX/0701001;%%
\bibitem [{\citenamefont {Link}\ \emph {et~al.}(2003)\citenamefont {Link} \emph
  {et~al.}}]{FOCUS3Kpi}%
  \BibitemOpen
  \bibfield  {author} {\bibinfo {author} {\bibfnamefont {J.~M.}\ \bibnamefont
  {Link}} \emph {et~al.} (\bibinfo {collaboration} {FOCUS}),\ }\href {\doibase
  10.1016/j.physletb.2003.09.054} {\bibfield  {journal} {\bibinfo  {journal}
  {Phys. Lett.}\ }\textbf {\bibinfo {volume} {B575}},\ \bibinfo {pages} {190}
  (\bibinfo {year} {2003})},\ \Eprint {http://arxiv.org/abs/hep-ex/0308054}
  {arXiv:hep-ex/0308054 [hep-ex]} \BibitemShut {NoStop}%
%%CITATION = HEP-EX/0308054;%%
\bibitem [{\citenamefont {Artuso}\ \emph {et~al.}(2012)\citenamefont {Artuso}
  \emph {et~al.}}]{KKpipiMint}%
  \BibitemOpen
  \bibfield  {author} {\bibinfo {author} {\bibfnamefont {M.}~\bibnamefont
  {Artuso}} \emph {et~al.} (\bibinfo {collaboration} {CLEO collaboration}),\
  }\href {\doibase 10.1103/PhysRevD.85.122002} {\bibfield  {journal} {\bibinfo
  {journal} {Phys. Rev.}\ }\textbf {\bibinfo {volume} {D85}},\ \bibinfo {pages}
  {122002} (\bibinfo {year} {2012})},\ \Eprint {http://arxiv.org/abs/1201.5716}
  {arXiv:1201.5716 [hep-ex]} \BibitemShut {NoStop}%
%%CITATION = ARXIV:1201.5716;%%
\bibitem [{\citenamefont {Gronau}\ and\ \citenamefont {Wyler}(1991)}]{GLW1}%
  \BibitemOpen
  \bibfield  {author} {\bibinfo {author} {\bibfnamefont {M.}~\bibnamefont
  {Gronau}}\ and\ \bibinfo {author} {\bibfnamefont {D.}~\bibnamefont {Wyler}},\
  }\href {\doibase 10.1016/0370-2693(91)90034-N} {\bibfield  {journal}
  {\bibinfo  {journal} {Phys.Lett.}\ }\textbf {\bibinfo {volume} {B265}},\
  \bibinfo {pages} {172} (\bibinfo {year} {1991})}\BibitemShut {NoStop}%
%%CITATION = PHLTA,B265,172;%%
\bibitem [{\citenamefont {Gronau}\ and\ \citenamefont {London}(1991)}]{GLW2}%
  \BibitemOpen
  \bibfield  {author} {\bibinfo {author} {\bibfnamefont {M.}~\bibnamefont
  {Gronau}}\ and\ \bibinfo {author} {\bibfnamefont {D.}~\bibnamefont
  {London}},\ }\href {\doibase 10.1016/0370-2693(91)91756-L} {\bibfield
  {journal} {\bibinfo  {journal} {Phys.Lett.}\ }\textbf {\bibinfo {volume}
  {B253}},\ \bibinfo {pages} {483} (\bibinfo {year} {1991})}\BibitemShut
  {NoStop}%
%%CITATION = PHLTA,B253,483;%%
\bibitem [{\citenamefont {Atwood}\ \emph {et~al.}(1997)\citenamefont {Atwood},
  \citenamefont {Dunietz},\ and\ \citenamefont {Soni}}]{ADS}%
  \BibitemOpen
  \bibfield  {author} {\bibinfo {author} {\bibfnamefont {D.}~\bibnamefont
  {Atwood}}, \bibinfo {author} {\bibfnamefont {I.}~\bibnamefont {Dunietz}}, \
  and\ \bibinfo {author} {\bibfnamefont {A.}~\bibnamefont {Soni}},\ }\href
  {\doibase 10.1103/PhysRevLett.78.3257} {\bibfield  {journal} {\bibinfo
  {journal} {Phys.Rev.Lett.}\ }\textbf {\bibinfo {volume} {78}},\ \bibinfo
  {pages} {3257} (\bibinfo {year} {1997})},\ \Eprint
  {http://arxiv.org/abs/hep-ph/9612433} {arXiv:hep-ph/9612433 [hep-ph]}
  \BibitemShut {NoStop}%
%%CITATION = HEP-PH/9612433;%%
\bibitem [{\citenamefont {Giri}\ \emph {et~al.}(2003)\citenamefont {Giri},
  \citenamefont {Grossman}, \citenamefont {Soffer},\ and\ \citenamefont
  {Zupan}}]{DalitzGamma1}%
  \BibitemOpen
  \bibfield  {author} {\bibinfo {author} {\bibfnamefont {A.}~\bibnamefont
  {Giri}}, \bibinfo {author} {\bibfnamefont {Y.}~\bibnamefont {Grossman}},
  \bibinfo {author} {\bibfnamefont {A.}~\bibnamefont {Soffer}}, \ and\ \bibinfo
  {author} {\bibfnamefont {J.}~\bibnamefont {Zupan}},\ }\href {\doibase
  10.1103/PhysRevD.68.054018} {\bibfield  {journal} {\bibinfo  {journal} {Phys.
  Rev. D}\ }\textbf {\bibinfo {volume} {68}},\ \bibinfo {pages} {054018}
  (\bibinfo {year} {2003})}\BibitemShut {NoStop}%
\bibitem [{\citenamefont {Poluektov}\ \emph
  {et~al.}(2004{\natexlab{a}})\citenamefont {Poluektov} \emph
  {et~al.}}]{DalitzGamma2}%
  \BibitemOpen
  \bibfield  {author} {\bibinfo {author} {\bibfnamefont {A.}~\bibnamefont
  {Poluektov}} \emph {et~al.} (\bibinfo {collaboration} {Belle}),\ }\href
  {\doibase 10.1103/PhysRevD.70.072003} {\bibfield  {journal} {\bibinfo
  {journal} {Phys. Rev. D}\ }\textbf {\bibinfo {volume} {70}},\ \bibinfo
  {pages} {072003} (\bibinfo {year} {2004}{\natexlab{a}})}\BibitemShut
  {NoStop}%
\bibitem [{\citenamefont {Aubert}\ \emph {et~al.}(2005)\citenamefont {Aubert}
  \emph {et~al.}}]{Aubert:2005iz}%
  \BibitemOpen
  \bibfield  {author} {\bibinfo {author} {\bibfnamefont {B.}~\bibnamefont
  {Aubert}} \emph {et~al.} (\bibinfo {collaboration} {BaBar}),\ }\href
  {\doibase 10.1103/PhysRevLett.95.121802} {\bibfield  {journal} {\bibinfo
  {journal} {Phys. Rev. Lett.}\ }\textbf {\bibinfo {volume} {95}},\ \bibinfo
  {pages} {121802} (\bibinfo {year} {2005})},\ \Eprint
  {http://arxiv.org/abs/hep-ex/0504039} {arXiv:hep-ex/0504039 [hep-ex]}
  \BibitemShut {NoStop}%
%%CITATION = HEP-EX/0504039;%%
\bibitem [{\citenamefont {Aubert}\ \emph
  {et~al.}(2008{\natexlab{b}})\citenamefont {Aubert} \emph
  {et~al.}}]{Aubert:2008bd}%
  \BibitemOpen
  \bibfield  {author} {\bibinfo {author} {\bibfnamefont {B.}~\bibnamefont
  {Aubert}} \emph {et~al.} (\bibinfo {collaboration} {BaBar}),\ }\href
  {\doibase 10.1103/PhysRevD.78.034023} {\bibfield  {journal} {\bibinfo
  {journal} {Phys. Rev.}\ }\textbf {\bibinfo {volume} {D78}},\ \bibinfo {pages}
  {034023} (\bibinfo {year} {2008}{\natexlab{b}})},\ \Eprint
  {http://arxiv.org/abs/0804.2089} {arXiv:0804.2089 [hep-ex]} \BibitemShut
  {NoStop}%
%%CITATION = ARXIV:0804.2089;%%
\bibitem [{\citenamefont {Poluektov}\ \emph
  {et~al.}(2004{\natexlab{b}})\citenamefont {Poluektov} \emph
  {et~al.}}]{Poluektov:2004mf}%
  \BibitemOpen
  \bibfield  {author} {\bibinfo {author} {\bibfnamefont {A.}~\bibnamefont
  {Poluektov}} \emph {et~al.} (\bibinfo {collaboration} {Belle}),\ }\href
  {\doibase 10.1103/PhysRevD.70.072003} {\bibfield  {journal} {\bibinfo
  {journal} {Phys. Rev.}\ }\textbf {\bibinfo {volume} {D70}},\ \bibinfo {pages}
  {072003} (\bibinfo {year} {2004}{\natexlab{b}})},\ \Eprint
  {http://arxiv.org/abs/hep-ex/0406067} {arXiv:hep-ex/0406067 [hep-ex]}
  \BibitemShut {NoStop}%
%%CITATION = HEP-EX/0406067;%%
\bibitem [{\citenamefont {Poluektov}\ \emph {et~al.}(2006)\citenamefont
  {Poluektov} \emph {et~al.}}]{Poluektov:2006ia}%
  \BibitemOpen
  \bibfield  {author} {\bibinfo {author} {\bibfnamefont {A.}~\bibnamefont
  {Poluektov}} \emph {et~al.} (\bibinfo {collaboration} {Belle}),\ }\href
  {\doibase 10.1103/PhysRevD.73.112009} {\bibfield  {journal} {\bibinfo
  {journal} {Phys. Rev.}\ }\textbf {\bibinfo {volume} {D73}},\ \bibinfo {pages}
  {112009} (\bibinfo {year} {2006})},\ \Eprint
  {http://arxiv.org/abs/hep-ex/0604054} {arXiv:hep-ex/0604054 [hep-ex]}
  \BibitemShut {NoStop}%
%%CITATION = HEP-EX/0604054;%%
\bibitem [{\citenamefont {Aaij}\ \emph
  {et~al.}(2014{\natexlab{f}})\citenamefont {Aaij} \emph
  {et~al.}}]{LHCb-PAPER-2014-017}%
  \BibitemOpen
  \bibfield  {author} {\bibinfo {author} {\bibfnamefont {R.}~\bibnamefont
  {Aaij}} \emph {et~al.} (\bibinfo {collaboration} {LHCb collaboration}),\
  }\href {\doibase 10.1016/j.nuclphysb.2014.09.015} {\bibfield  {journal}
  {\bibinfo  {journal} {Nucl. Phys.}\ }\textbf {\bibinfo {volume} {B888}},\
  \bibinfo {pages} {169} (\bibinfo {year} {2014}{\natexlab{f}})},\ \Eprint
  {http://arxiv.org/abs/1407.6211} {arXiv:1407.6211 [hep-ex]} \BibitemShut
  {NoStop}%
%%CITATION = ARXIV:1407.6211;%%
\bibitem [{\citenamefont {Rademacker}\ and\ \citenamefont
  {Wilkinson}(2007)}]{Rademacker:2006zx}%
  \BibitemOpen
  \bibfield  {author} {\bibinfo {author} {\bibfnamefont {J.}~\bibnamefont
  {Rademacker}}\ and\ \bibinfo {author} {\bibfnamefont {G.}~\bibnamefont
  {Wilkinson}},\ }\href {\doibase 10.1016/j.physletb.2007.01.071} {\bibfield
  {journal} {\bibinfo  {journal} {Phys.Lett.}\ }\textbf {\bibinfo {volume}
  {B647}},\ \bibinfo {pages} {400} (\bibinfo {year} {2007})},\ \Eprint
  {http://arxiv.org/abs/hep-ph/0611272} {arXiv:hep-ph/0611272 [hep-ph]}
  \BibitemShut {NoStop}%
%%CITATION = HEP-PH/0611272;%%
\bibitem [{\citenamefont {Atwood}\ and\ \citenamefont
  {Soni}(2003)}]{Atwood:coherenceFactor}%
  \BibitemOpen
  \bibfield  {author} {\bibinfo {author} {\bibfnamefont {D.}~\bibnamefont
  {Atwood}}\ and\ \bibinfo {author} {\bibfnamefont {A.}~\bibnamefont {Soni}},\
  }\href {\doibase 10.1103/PhysRevD.68.033003} {\bibfield  {journal} {\bibinfo
  {journal} {Phys. Rev.}\ }\textbf {\bibinfo {volume} {D68}},\ \bibinfo {pages}
  {033003} (\bibinfo {year} {2003})}\BibitemShut {NoStop}%
\bibitem [{\citenamefont {Bondar}\ and\ \citenamefont
  {Poluektov}(2006)}]{Bondar:2005ki}%
  \BibitemOpen
  \bibfield  {author} {\bibinfo {author} {\bibfnamefont {A.}~\bibnamefont
  {Bondar}}\ and\ \bibinfo {author} {\bibfnamefont {A.}~\bibnamefont
  {Poluektov}},\ }\href {\doibase 10.1140/epjc/s2006-02590-x} {\bibfield
  {journal} {\bibinfo  {journal} {Eur. Phys. J.}\ }\textbf {\bibinfo {volume}
  {C47}},\ \bibinfo {pages} {347} (\bibinfo {year} {2006})},\ \Eprint
  {http://arxiv.org/abs/hep-ph/0510246} {arXiv:hep-ph/0510246 [hep-ph]}
  \BibitemShut {NoStop}%
%%CITATION = HEP-PH/0510246;%%
\bibitem [{\citenamefont {Harnew}\ and\ \citenamefont
  {Rademacker}(2014)}]{coherenceFromMixing}%
  \BibitemOpen
  \bibfield  {author} {\bibinfo {author} {\bibfnamefont {S.}~\bibnamefont
  {Harnew}}\ and\ \bibinfo {author} {\bibfnamefont {J.}~\bibnamefont
  {Rademacker}},\ }\href {\doibase 10.1016/j.physletb.2013.11.065} {\bibfield
  {journal} {\bibinfo  {journal} {Phys.Lett.}\ }\textbf {\bibinfo {volume}
  {B728}},\ \bibinfo {pages} {296} (\bibinfo {year} {2014})},\ \Eprint
  {http://arxiv.org/abs/1309.0134} {arXiv:1309.0134 [hep-ph]} \BibitemShut
  {NoStop}%
%%CITATION = ARXIV:1309.0134;%%
\bibitem [{\citenamefont {Harnew}\ and\ \citenamefont
  {Rademacker}(2015)}]{coherenceFromMixing2}%
  \BibitemOpen
  \bibfield  {author} {\bibinfo {author} {\bibfnamefont {S.}~\bibnamefont
  {Harnew}}\ and\ \bibinfo {author} {\bibfnamefont {J.}~\bibnamefont
  {Rademacker}},\ }\href {\doibase 10.1007/JHEP03(2015)169} {\bibfield
  {journal} {\bibinfo  {journal} {JHEP}\ }\textbf {\bibinfo {volume} {03}},\
  \bibinfo {pages} {169} (\bibinfo {year} {2015})},\ \Eprint
  {http://arxiv.org/abs/1412.7254} {arXiv:1412.7254 [hep-ph]} \BibitemShut
  {NoStop}%
%%CITATION = ARXIV:1412.7254;%%
\bibitem [{\citenamefont {Bondar}\ and\ \citenamefont
  {Poluektov}(2007)}]{Bondar:2007ir}%
  \BibitemOpen
  \bibfield  {author} {\bibinfo {author} {\bibfnamefont {A.}~\bibnamefont
  {Bondar}}\ and\ \bibinfo {author} {\bibfnamefont {A.}~\bibnamefont
  {Poluektov}},\ }\href@noop {} {\  (\bibinfo {year} {2007})},\ \Eprint
  {http://arxiv.org/abs/hep-ph/0703267} {arXiv:hep-ph/0703267 [HEP-PH]}
  \BibitemShut {NoStop}%
%%CITATION = HEP-PH/0703267;%%
\bibitem [{\citenamefont {Bondar}\ \emph {et~al.}(2010)\citenamefont {Bondar},
  \citenamefont {Poluektov},\ and\ \citenamefont
  {Vorobiev}}]{Bondar:CharmMixingCP}%
  \BibitemOpen
  \bibfield  {author} {\bibinfo {author} {\bibfnamefont {A.}~\bibnamefont
  {Bondar}}, \bibinfo {author} {\bibfnamefont {A.}~\bibnamefont {Poluektov}}, \
  and\ \bibinfo {author} {\bibfnamefont {V.}~\bibnamefont {Vorobiev}},\ }\href
  {\doibase 10.1103/PhysRevD.82.034033} {\bibfield  {journal} {\bibinfo
  {journal} {Phys.Rev.}\ }\textbf {\bibinfo {volume} {D82}},\ \bibinfo {pages}
  {034033} (\bibinfo {year} {2010})},\ \Eprint {http://arxiv.org/abs/1004.2350}
  {arXiv:1004.2350 [hep-ph]} \BibitemShut {NoStop}%
%%CITATION = ARXIV:1004.2350;%%
\bibitem [{\citenamefont {Malde}\ and\ \citenamefont
  {Wilkinson}(2011)}]{Malde:2011mk}%
  \BibitemOpen
  \bibfield  {author} {\bibinfo {author} {\bibfnamefont {S.}~\bibnamefont
  {Malde}}\ and\ \bibinfo {author} {\bibfnamefont {G.}~\bibnamefont
  {Wilkinson}},\ }\href {\doibase 10.1016/j.physletb.2011.05.072} {\bibfield
  {journal} {\bibinfo  {journal} {Phys.Lett.}\ }\textbf {\bibinfo {volume}
  {B701}},\ \bibinfo {pages} {353} (\bibinfo {year} {2011})},\ \Eprint
  {http://arxiv.org/abs/1104.2731} {arXiv:1104.2731 [hep-ph]} \BibitemShut
  {NoStop}%
%%CITATION = ARXIV:1104.2731;%%
\bibitem [{\citenamefont {Aaij}\ \emph
  {et~al.}(2015{\natexlab{g}})\citenamefont {Aaij} \emph
  {et~al.}}]{LHCb-PAPER-2015-042}%
  \BibitemOpen
  \bibfield  {author} {\bibinfo {author} {\bibfnamefont {R.}~\bibnamefont
  {Aaij}} \emph {et~al.} (\bibinfo {collaboration} {LHCb collaboration}),\
  }\href@noop {} {\  (\bibinfo {year} {2015}{\natexlab{g}})},\ \bibinfo {note}
  {{to appear in JHEP}},\ \Eprint {http://arxiv.org/abs/1510.01664}
  {arXiv:1510.01664 [hep-ex]} \BibitemShut {NoStop}%
%%CITATION = ARXIV:1510.01664;%%
\bibitem [{\citenamefont {Briere}\ \emph {et~al.}(2009)\citenamefont {Briere}
  \emph {et~al.}}]{Briere:2009aa}%
  \BibitemOpen
  \bibfield  {author} {\bibinfo {author} {\bibfnamefont {R.~A.}\ \bibnamefont
  {Briere}} \emph {et~al.} (\bibinfo {collaboration} {CLEO}),\ }\href {\doibase
  10.1103/PhysRevD.80.032002} {\bibfield  {journal} {\bibinfo  {journal}
  {Phys.Rev.}\ }\textbf {\bibinfo {volume} {D80}},\ \bibinfo {pages} {032002}
  (\bibinfo {year} {2009})},\ \Eprint {http://arxiv.org/abs/0903.1681}
  {arXiv:0903.1681 [hep-ex]} \BibitemShut {NoStop}%
%%CITATION = ARXIV:0903.1681;%%
\bibitem [{\citenamefont {Insler}\ \emph {et~al.}(2012)\citenamefont {Insler}
  \emph {et~al.}}]{Insler:2012pm}%
  \BibitemOpen
  \bibfield  {author} {\bibinfo {author} {\bibfnamefont {J.}~\bibnamefont
  {Insler}} \emph {et~al.} (\bibinfo {collaboration} {CLEO}),\ }\href {\doibase
  10.1103/PhysRevD.85.092016} {\bibfield  {journal} {\bibinfo  {journal}
  {Phys.Rev.}\ }\textbf {\bibinfo {volume} {D85}},\ \bibinfo {pages} {092016}
  (\bibinfo {year} {2012})},\ \Eprint {http://arxiv.org/abs/1203.3804}
  {arXiv:1203.3804 [hep-ex]} \BibitemShut {NoStop}%
%%CITATION = ARXIV:1203.3804;%%
\bibitem [{\citenamefont {Libby}\ \emph {et~al.}(2010)\citenamefont {Libby}
  \emph {et~al.}}]{Libby:2010nu}%
  \BibitemOpen
  \bibfield  {author} {\bibinfo {author} {\bibfnamefont {J.}~\bibnamefont
  {Libby}} \emph {et~al.} (\bibinfo {collaboration} {CLEO}),\ }\href {\doibase
  10.1103/PhysRevD.82.112006} {\bibfield  {journal} {\bibinfo  {journal}
  {Phys.Rev.}\ }\textbf {\bibinfo {volume} {D82}},\ \bibinfo {pages} {112006}
  (\bibinfo {year} {2010})},\ \Eprint {http://arxiv.org/abs/1010.2817}
  {arXiv:1010.2817 [hep-ex]} \BibitemShut {NoStop}%
%%CITATION = ARXIV:1010.2817;%%
\bibitem [{\citenamefont {Lowrey}\ \emph {et~al.}(2009)\citenamefont {Lowrey}
  \emph {et~al.}}]{Lowery:2009id}%
  \BibitemOpen
  \bibfield  {author} {\bibinfo {author} {\bibfnamefont {N.}~\bibnamefont
  {Lowrey}} \emph {et~al.} (\bibinfo {collaboration} {CLEO}),\ }\href {\doibase
  10.1103/PhysRevD.80.031105} {\bibfield  {journal} {\bibinfo  {journal}
  {Phys.Rev.}\ }\textbf {\bibinfo {volume} {D80}},\ \bibinfo {pages} {031105}
  (\bibinfo {year} {2009})},\ \Eprint {http://arxiv.org/abs/0903.4853}
  {arXiv:0903.4853 [hep-ex]} \BibitemShut {NoStop}%
%%CITATION = ARXIV:0903.4853;%%
\bibitem [{\citenamefont {Libby}\ \emph {et~al.}(2014)\citenamefont {Libby},
  \citenamefont {Malde}, \citenamefont {Powell}, \citenamefont {Wilkinson},
  \citenamefont {Asner} \emph {et~al.}}]{Libby:2014rea}%
  \BibitemOpen
  \bibfield  {author} {\bibinfo {author} {\bibfnamefont {J.}~\bibnamefont
  {Libby}}, \bibinfo {author} {\bibfnamefont {S.}~\bibnamefont {Malde}},
  \bibinfo {author} {\bibfnamefont {A.}~\bibnamefont {Powell}}, \bibinfo
  {author} {\bibfnamefont {G.}~\bibnamefont {Wilkinson}}, \bibinfo {author}
  {\bibfnamefont {D.}~\bibnamefont {Asner}},  \emph {et~al.},\ }\href {\doibase
  10.1016/j.physletb.2014.02.032} {\bibfield  {journal} {\bibinfo  {journal}
  {Phys.Lett.}\ }\textbf {\bibinfo {volume} {B731}},\ \bibinfo {pages} {197}
  (\bibinfo {year} {2014})},\ \Eprint {http://arxiv.org/abs/1401.1904}
  {arXiv:1401.1904 [hep-ex]} \BibitemShut {NoStop}%
%%CITATION = ARXIV:1401.1904;%%
\bibitem [{\citenamefont {Evans}\ \emph {et~al.}(2016)\citenamefont {Evans},
  \citenamefont {Harnew}, \citenamefont {Libby}, \citenamefont {Malde},
  \citenamefont {Rademacker},\ and\ \citenamefont {Wilkinson}}]{Evans:2016tlp}%
  \BibitemOpen
  \bibfield  {author} {\bibinfo {author} {\bibfnamefont {T.}~\bibnamefont
  {Evans}}, \bibinfo {author} {\bibfnamefont {S.}~\bibnamefont {Harnew}},
  \bibinfo {author} {\bibfnamefont {J.}~\bibnamefont {Libby}}, \bibinfo
  {author} {\bibfnamefont {S.}~\bibnamefont {Malde}}, \bibinfo {author}
  {\bibfnamefont {J.}~\bibnamefont {Rademacker}}, \ and\ \bibinfo {author}
  {\bibfnamefont {G.}~\bibnamefont {Wilkinson}},\ }\href@noop {} {\  (\bibinfo
  {year} {2016})},\ \Eprint {http://arxiv.org/abs/1602.07430} {arXiv:1602.07430
  [hep-ex]} \BibitemShut {NoStop}%
%%CITATION = ARXIV:1602.07430;%%
\bibitem [{\citenamefont {Malde}\ \emph {et~al.}(2015)\citenamefont {Malde},
  \citenamefont {Thomas}, \citenamefont {Wilkinson}, \citenamefont {Naik},
  \citenamefont {Prouve}, \citenamefont {Rademacker}, \citenamefont {Libby},
  \citenamefont {Nayak}, \citenamefont {Gershon},\ and\ \citenamefont
  {Briere}}]{Malde:2015mha}%
  \BibitemOpen
  \bibfield  {author} {\bibinfo {author} {\bibfnamefont {S.}~\bibnamefont
  {Malde}}, \bibinfo {author} {\bibfnamefont {C.}~\bibnamefont {Thomas}},
  \bibinfo {author} {\bibfnamefont {G.}~\bibnamefont {Wilkinson}}, \bibinfo
  {author} {\bibfnamefont {P.}~\bibnamefont {Naik}}, \bibinfo {author}
  {\bibfnamefont {C.}~\bibnamefont {Prouve}}, \bibinfo {author} {\bibfnamefont
  {J.}~\bibnamefont {Rademacker}}, \bibinfo {author} {\bibfnamefont
  {J.}~\bibnamefont {Libby}}, \bibinfo {author} {\bibfnamefont
  {M.}~\bibnamefont {Nayak}}, \bibinfo {author} {\bibfnamefont
  {T.}~\bibnamefont {Gershon}}, \ and\ \bibinfo {author} {\bibfnamefont
  {R.~A.}\ \bibnamefont {Briere}},\ }\href {\doibase
  10.1016/j.physletb.2015.05.043} {\bibfield  {journal} {\bibinfo  {journal}
  {Phys. Lett.}\ }\textbf {\bibinfo {volume} {B747}},\ \bibinfo {pages} {9}
  (\bibinfo {year} {2015})},\ \Eprint {http://arxiv.org/abs/1504.05878}
  {arXiv:1504.05878 [hep-ex]} \BibitemShut {NoStop}%
%%CITATION = ARXIV:1504.05878;%%
\bibitem [{\citenamefont {Nayak}\ \emph {et~al.}(2015)\citenamefont {Nayak},
  \citenamefont {Libby}, \citenamefont {Malde}, \citenamefont {Thomas},
  \citenamefont {Wilkinson}, \citenamefont {Briere}, \citenamefont {Naik},
  \citenamefont {Gershon},\ and\ \citenamefont {Bonvicini}}]{Nayak:2014tea}%
  \BibitemOpen
  \bibfield  {author} {\bibinfo {author} {\bibfnamefont {M.}~\bibnamefont
  {Nayak}}, \bibinfo {author} {\bibfnamefont {J.}~\bibnamefont {Libby}},
  \bibinfo {author} {\bibfnamefont {S.}~\bibnamefont {Malde}}, \bibinfo
  {author} {\bibfnamefont {C.}~\bibnamefont {Thomas}}, \bibinfo {author}
  {\bibfnamefont {G.}~\bibnamefont {Wilkinson}}, \bibinfo {author}
  {\bibfnamefont {R.~A.}\ \bibnamefont {Briere}}, \bibinfo {author}
  {\bibfnamefont {P.}~\bibnamefont {Naik}}, \bibinfo {author} {\bibfnamefont
  {T.}~\bibnamefont {Gershon}}, \ and\ \bibinfo {author} {\bibfnamefont
  {G.}~\bibnamefont {Bonvicini}},\ }\href {\doibase
  10.1016/j.physletb.2014.11.022} {\bibfield  {journal} {\bibinfo  {journal}
  {Phys. Lett.}\ }\textbf {\bibinfo {volume} {B740}},\ \bibinfo {pages} {1}
  (\bibinfo {year} {2015})},\ \Eprint {http://arxiv.org/abs/1410.3964}
  {arXiv:1410.3964 [hep-ex]} \BibitemShut {NoStop}%
%%CITATION = ARXIV:1410.3964;%%
\bibitem [{\citenamefont {Lees}\ \emph
  {et~al.}(2011{\natexlab{a}})\citenamefont {Lees} \emph
  {et~al.}}]{BaBar_uses_us:2011up}%
  \BibitemOpen
  \bibfield  {author} {\bibinfo {author} {\bibfnamefont {J.}~\bibnamefont
  {Lees}} \emph {et~al.} (\bibinfo {collaboration} {BaBar}),\ }\href {\doibase
  10.1103/PhysRevD.84.012002} {\bibfield  {journal} {\bibinfo  {journal}
  {Phys.Rev.}\ }\textbf {\bibinfo {volume} {D84}},\ \bibinfo {pages} {012002}
  (\bibinfo {year} {2011}{\natexlab{a}})},\ \Eprint
  {http://arxiv.org/abs/1104.4472} {arXiv:1104.4472 [hep-ex]} \BibitemShut
  {NoStop}%
%%CITATION = ARXIV:1104.4472;%%
\bibitem [{\citenamefont {Adachi}\ \emph {et~al.}(2011)\citenamefont {Adachi}
  \emph {et~al.}}]{Belle_uses_CLEO_2011}%
  \BibitemOpen
  \bibfield  {author} {\bibinfo {author} {\bibfnamefont {I.}~\bibnamefont
  {Adachi}} \emph {et~al.} (\bibinfo {collaboration} {Belle}),\ }\href@noop {}
  {\  (\bibinfo {year} {2011})},\ \Eprint {http://arxiv.org/abs/1106.4046}
  {arXiv:1106.4046 [hep-ex]} \BibitemShut {NoStop}%
%%CITATION = ARXIV:1106.4046;%%
\bibitem [{\citenamefont {Aaij}\ \emph {et~al.}(2012)\citenamefont {Aaij} \emph
  {et~al.}}]{LHCb2012DalitzGamma}%
  \BibitemOpen
  \bibfield  {author} {\bibinfo {author} {\bibfnamefont {R.}~\bibnamefont
  {Aaij}} \emph {et~al.} (\bibinfo {collaboration} {LHCb}),\ }\href {\doibase
  10.1016/j.physletb.2012.10.020} {\bibfield  {journal} {\bibinfo  {journal}
  {Phys. Lett.}\ }\textbf {\bibinfo {volume} {B718}},\ \bibinfo {pages} {43}
  (\bibinfo {year} {2012})},\ \Eprint {http://arxiv.org/abs/1209.5869}
  {arXiv:1209.5869 [hep-ex]} \BibitemShut {NoStop}%
%%CITATION = ARXIV:1209.5869;%%
\bibitem [{\citenamefont {Aaij}\ \emph
  {et~al.}(2013{\natexlab{b}})\citenamefont {Aaij} \emph
  {et~al.}}]{LHCb2013GammaCombination}%
  \BibitemOpen
  \bibfield  {author} {\bibinfo {author} {\bibfnamefont {R.}~\bibnamefont
  {Aaij}} \emph {et~al.} (\bibinfo {collaboration} {LHCb}),\ }\href {\doibase
  10.1016/j.physletb.2013.08.020} {\bibfield  {journal} {\bibinfo  {journal}
  {Phys.Lett.}\ }\textbf {\bibinfo {volume} {B726}},\ \bibinfo {pages} {151}
  (\bibinfo {year} {2013}{\natexlab{b}})},\ \Eprint
  {http://arxiv.org/abs/1305.2050} {arXiv:1305.2050 [hep-ex]} \BibitemShut
  {NoStop}%
%%CITATION = ARXIV:1305.2050;%%
\bibitem [{\citenamefont {Aaij}\ \emph
  {et~al.}(2014{\natexlab{g}})\citenamefont {Aaij} \emph
  {et~al.}}]{LHCb-PAPER-2014-041}%
  \BibitemOpen
  \bibfield  {author} {\bibinfo {author} {\bibfnamefont {R.}~\bibnamefont
  {Aaij}} \emph {et~al.} (\bibinfo {collaboration} {LHCb collaboration}),\
  }\href {\doibase 10.1007/JHEP10(2014)097} {\bibfield  {journal} {\bibinfo
  {journal} {JHEP}\ }\textbf {\bibinfo {volume} {10}},\ \bibinfo {pages} {097}
  (\bibinfo {year} {2014}{\natexlab{g}})},\ \Eprint
  {http://arxiv.org/abs/1408.2748} {arXiv:1408.2748 [hep-ex]} \BibitemShut
  {NoStop}%
%%CITATION = ARXIV:1408.2748;%%
\bibitem [{\citenamefont {Aaij}\ \emph {et~al.}(2016)\citenamefont {Aaij} \emph
  {et~al.}}]{LHCb-PAPER-2016-003}%
  \BibitemOpen
  \bibfield  {author} {\bibinfo {author} {\bibfnamefont {R.}~\bibnamefont
  {Aaij}} \emph {et~al.} (\bibinfo {collaboration} {LHCb collaboration}),\
  }\href@noop {} {\  (\bibinfo {year} {2016})},\ \bibinfo {note} {{submitted to
  Phys. Lett. B}},\ \Eprint {http://arxiv.org/abs/1603.08993} {arXiv:1603.08993
  [hep-ex]} \BibitemShut {NoStop}%
%%CITATION = ARXIV:1603.08993;%%
\bibitem [{LHC(2016)}]{LHCb-CONF-2016-001}%
  \BibitemOpen
  \href@noop {} {\enquote {\bibinfo {title} {{LHCb $\gamma$ combination update
  from $B \to DKX$ decays}},}\ } (\bibinfo {year} {{2016}})\BibitemShut
  {NoStop}%
\bibitem [{\citenamefont {Aaij}\ \emph
  {et~al.}(2015{\natexlab{h}})\citenamefont {Aaij} \emph
  {et~al.}}]{LHCb-PAPER-2015-057}%
  \BibitemOpen
  \bibfield  {author} {\bibinfo {author} {\bibfnamefont {R.}~\bibnamefont
  {Aaij}} \emph {et~al.} (\bibinfo {collaboration} {LHCb collaboration}),\
  }\href@noop {} {\  (\bibinfo {year} {2015}{\natexlab{h}})},\ \bibinfo {note}
  {{submitted to Phys. Rev. Lett.}},\ \Eprint {http://arxiv.org/abs/1602.07224}
  {arXiv:1602.07224 [hep-ex]} \BibitemShut {NoStop}%
%%CITATION = ARXIV:1602.07224;%%
\bibitem [{\citenamefont {Aubert}\ \emph
  {et~al.}(2008{\natexlab{c}})\citenamefont {Aubert} \emph
  {et~al.}}]{Aubert:2008yd}%
  \BibitemOpen
  \bibfield  {author} {\bibinfo {author} {\bibfnamefont {B.}~\bibnamefont
  {Aubert}} \emph {et~al.} (\bibinfo {collaboration} {BaBar}),\ }\href
  {\doibase 10.1103/PhysRevD.78.051102} {\bibfield  {journal} {\bibinfo
  {journal} {Phys. Rev.}\ }\textbf {\bibinfo {volume} {D78}},\ \bibinfo {pages}
  {051102} (\bibinfo {year} {2008}{\natexlab{c}})},\ \Eprint
  {http://arxiv.org/abs/0802.4035} {arXiv:0802.4035 [hep-ex]} \BibitemShut
  {NoStop}%
%%CITATION = ARXIV:0802.4035;%%
\bibitem [{\citenamefont {Aaij}\ \emph
  {et~al.}(2013{\natexlab{c}})\citenamefont {Aaij} \emph
  {et~al.}}]{LHCb-PAPER-2013-041}%
  \BibitemOpen
  \bibfield  {author} {\bibinfo {author} {\bibfnamefont {R.}~\bibnamefont
  {Aaij}} \emph {et~al.} (\bibinfo {collaboration} {LHCb collaboration}),\
  }\href {\doibase 10.1016/j.physletb.2013.09.011} {\bibfield  {journal}
  {\bibinfo  {journal} {Phys. Lett.}\ }\textbf {\bibinfo {volume} {B726}},\
  \bibinfo {pages} {623} (\bibinfo {year} {2013}{\natexlab{c}})},\ \Eprint
  {http://arxiv.org/abs/1308.3189} {arXiv:1308.3189 [hep-ex]} \BibitemShut
  {NoStop}%
%%CITATION = ARXIV:1308.3189;%%
\bibitem [{\citenamefont {Golowich}\ and\ \citenamefont
  {Valencia}(1989)}]{Golowich:1988ig}%
  \BibitemOpen
  \bibfield  {author} {\bibinfo {author} {\bibfnamefont {E.}~\bibnamefont
  {Golowich}}\ and\ \bibinfo {author} {\bibfnamefont {G.}~\bibnamefont
  {Valencia}},\ }\href {\doibase 10.1103/PhysRevD.40.112} {\bibfield  {journal}
  {\bibinfo  {journal} {Phys. Rev.}\ }\textbf {\bibinfo {volume} {D40}},\
  \bibinfo {pages} {112} (\bibinfo {year} {1989})}\BibitemShut {NoStop}%
%%CITATION = PHRVA,D40,112;%%
\bibitem [{\citenamefont {Valencia}(1989)}]{Valencia:1988it}%
  \BibitemOpen
  \bibfield  {author} {\bibinfo {author} {\bibfnamefont {G.}~\bibnamefont
  {Valencia}},\ }\href {\doibase 10.1103/PhysRevD.39.3339} {\bibfield
  {journal} {\bibinfo  {journal} {Phys. Rev.}\ }\textbf {\bibinfo {volume}
  {D39}},\ \bibinfo {pages} {3339} (\bibinfo {year} {1989})}\BibitemShut
  {NoStop}%
%%CITATION = PHRVA,D39,3339;%%
\bibitem [{\citenamefont {Bensalem}\ and\ \citenamefont
  {London}(2001)}]{Bensalem:2000hq}%
  \BibitemOpen
  \bibfield  {author} {\bibinfo {author} {\bibfnamefont {W.}~\bibnamefont
  {Bensalem}}\ and\ \bibinfo {author} {\bibfnamefont {D.}~\bibnamefont
  {London}},\ }\href {\doibase 10.1103/PhysRevD.64.116003} {\bibfield
  {journal} {\bibinfo  {journal} {Phys. Rev.}\ }\textbf {\bibinfo {volume}
  {D64}},\ \bibinfo {pages} {116003} (\bibinfo {year} {2001})},\ \Eprint
  {http://arxiv.org/abs/hep-ph/0005018} {arXiv:hep-ph/0005018 [hep-ph]}
  \BibitemShut {NoStop}%
%%CITATION = HEP-PH/0005018;%%
\bibitem [{\citenamefont {Bigi}(2001)}]{Bigi:2001sg}%
  \BibitemOpen
  \bibfield  {author} {\bibinfo {author} {\bibfnamefont {I.~I.~Y.}\
  \bibnamefont {Bigi}},\ }in\ \href@noop {} {\emph {\bibinfo {booktitle}
  {{KAON2001: International Conference on CP Violation Pisa, Italy, June 12-17,
  2001}}}}\ (\bibinfo {year} {2001})\ \Eprint
  {http://arxiv.org/abs/hep-ph/0107102} {arXiv:hep-ph/0107102 [hep-ph]}
  \BibitemShut {NoStop}%
%%CITATION = HEP-PH/0107102;%%
\bibitem [{\citenamefont {Bensalem}\ \emph
  {et~al.}(2002{\natexlab{a}})\citenamefont {Bensalem}, \citenamefont {Datta},\
  and\ \citenamefont {London}}]{Bensalem:2002ys}%
  \BibitemOpen
  \bibfield  {author} {\bibinfo {author} {\bibfnamefont {W.}~\bibnamefont
  {Bensalem}}, \bibinfo {author} {\bibfnamefont {A.}~\bibnamefont {Datta}}, \
  and\ \bibinfo {author} {\bibfnamefont {D.}~\bibnamefont {London}},\ }\href
  {\doibase 10.1103/PhysRevD.66.094004} {\bibfield  {journal} {\bibinfo
  {journal} {Phys. Rev.}\ }\textbf {\bibinfo {volume} {D66}},\ \bibinfo {pages}
  {094004} (\bibinfo {year} {2002}{\natexlab{a}})},\ \Eprint
  {http://arxiv.org/abs/hep-ph/0208054} {arXiv:hep-ph/0208054 [hep-ph]}
  \BibitemShut {NoStop}%
%%CITATION = HEP-PH/0208054;%%
\bibitem [{\citenamefont {Bensalem}\ \emph
  {et~al.}(2002{\natexlab{b}})\citenamefont {Bensalem}, \citenamefont {Datta},\
  and\ \citenamefont {London}}]{Bensalem:2002pz}%
  \BibitemOpen
  \bibfield  {author} {\bibinfo {author} {\bibfnamefont {W.}~\bibnamefont
  {Bensalem}}, \bibinfo {author} {\bibfnamefont {A.}~\bibnamefont {Datta}}, \
  and\ \bibinfo {author} {\bibfnamefont {D.}~\bibnamefont {London}},\ }\href
  {\doibase 10.1016/S0370-2693(02)02028-2} {\bibfield  {journal} {\bibinfo
  {journal} {Phys. Lett.}\ }\textbf {\bibinfo {volume} {B538}},\ \bibinfo
  {pages} {309} (\bibinfo {year} {2002}{\natexlab{b}})},\ \Eprint
  {http://arxiv.org/abs/hep-ph/0205009} {arXiv:hep-ph/0205009 [hep-ph]}
  \BibitemShut {NoStop}%
%%CITATION = HEP-PH/0205009;%%
\bibitem [{\citenamefont {Datta}\ and\ \citenamefont
  {London}(2004)}]{Datta:2003mj}%
  \BibitemOpen
  \bibfield  {author} {\bibinfo {author} {\bibfnamefont {A.}~\bibnamefont
  {Datta}}\ and\ \bibinfo {author} {\bibfnamefont {D.}~\bibnamefont {London}},\
  }\href {\doibase 10.1142/S0217751X04018300} {\bibfield  {journal} {\bibinfo
  {journal} {Int. J. Mod. Phys.}\ }\textbf {\bibinfo {volume} {A19}},\ \bibinfo
  {pages} {2505} (\bibinfo {year} {2004})},\ \Eprint
  {http://arxiv.org/abs/hep-ph/0303159} {arXiv:hep-ph/0303159 [hep-ph]}
  \BibitemShut {NoStop}%
%%CITATION = HEP-PH/0303159;%%
\bibitem [{\citenamefont {Gronau}\ and\ \citenamefont
  {Rosner}(2011)}]{Gronau:2011cf}%
  \BibitemOpen
  \bibfield  {author} {\bibinfo {author} {\bibfnamefont {M.}~\bibnamefont
  {Gronau}}\ and\ \bibinfo {author} {\bibfnamefont {J.~L.}\ \bibnamefont
  {Rosner}},\ }\href {\doibase 10.1103/PhysRevD.84.096013} {\bibfield
  {journal} {\bibinfo  {journal} {Phys. Rev.}\ }\textbf {\bibinfo {volume}
  {D84}},\ \bibinfo {pages} {096013} (\bibinfo {year} {2011})},\ \Eprint
  {http://arxiv.org/abs/1107.1232} {arXiv:1107.1232 [hep-ph]} \BibitemShut
  {NoStop}%
%%CITATION = ARXIV:1107.1232;%%
\bibitem [{\citenamefont {Durieux}\ and\ \citenamefont
  {Grossman}(2015)}]{Durieux:2015zwa}%
  \BibitemOpen
  \bibfield  {author} {\bibinfo {author} {\bibfnamefont {G.}~\bibnamefont
  {Durieux}}\ and\ \bibinfo {author} {\bibfnamefont {Y.}~\bibnamefont
  {Grossman}},\ }\href {\doibase 10.1103/PhysRevD.92.076013} {\bibfield
  {journal} {\bibinfo  {journal} {Phys. Rev.}\ }\textbf {\bibinfo {volume}
  {D92}},\ \bibinfo {pages} {076013} (\bibinfo {year} {2015})},\ \Eprint
  {http://arxiv.org/abs/1508.03054} {arXiv:1508.03054 [hep-ph]} \BibitemShut
  {NoStop}%
%%CITATION = ARXIV:1508.03054;%%
\bibitem [{\citenamefont {Link}\ \emph
  {et~al.}(2005{\natexlab{b}})\citenamefont {Link} \emph
  {et~al.}}]{FOCUSTodd2010}%
  \BibitemOpen
  \bibfield  {author} {\bibinfo {author} {\bibfnamefont {J.~M.}\ \bibnamefont
  {Link}} \emph {et~al.} (\bibinfo {collaboration} {FOCUS}),\ }\href {\doibase
  10.1016/j.physletb.2005.07.024} {\bibfield  {journal} {\bibinfo  {journal}
  {Phys. Lett.}\ }\textbf {\bibinfo {volume} {B622}},\ \bibinfo {pages} {239}
  (\bibinfo {year} {2005}{\natexlab{b}})},\ \Eprint
  {http://arxiv.org/abs/hep-ex/0506012} {arXiv:hep-ex/0506012 [hep-ex]}
  \BibitemShut {NoStop}%
%%CITATION = HEP-EX/0506012;%%
\bibitem [{\citenamefont {del Amo~Sanchez}\ \emph {et~al.}(2010)\citenamefont
  {del Amo~Sanchez} \emph {et~al.}}]{BaBarTodd1}%
  \BibitemOpen
  \bibfield  {author} {\bibinfo {author} {\bibfnamefont {P.}~\bibnamefont {del
  Amo~Sanchez}} \emph {et~al.} (\bibinfo {collaboration} {BaBar}),\ }\href
  {\doibase 10.1103/PhysRevD.81.111103} {\bibfield  {journal} {\bibinfo
  {journal} {Phys. Rev.}\ }\textbf {\bibinfo {volume} {D81}},\ \bibinfo {pages}
  {111103} (\bibinfo {year} {2010})},\ \Eprint {http://arxiv.org/abs/1003.3397}
  {arXiv:1003.3397 [hep-ex]} \BibitemShut {NoStop}%
%%CITATION = ARXIV:1003.3397;%%
\bibitem [{\citenamefont {Lees}\ \emph
  {et~al.}(2011{\natexlab{b}})\citenamefont {Lees} \emph
  {et~al.}}]{BaBarTodd2}%
  \BibitemOpen
  \bibfield  {author} {\bibinfo {author} {\bibfnamefont {J.~P.}\ \bibnamefont
  {Lees}} \emph {et~al.} (\bibinfo {collaboration} {BaBar}),\ }\href {\doibase
  10.1103/PhysRevD.84.031103} {\bibfield  {journal} {\bibinfo  {journal} {Phys.
  Rev.}\ }\textbf {\bibinfo {volume} {D84}},\ \bibinfo {pages} {031103}
  (\bibinfo {year} {2011}{\natexlab{b}})},\ \Eprint
  {http://arxiv.org/abs/1105.4410} {arXiv:1105.4410 [hep-ex]} \BibitemShut
  {NoStop}%
%%CITATION = ARXIV:1105.4410;%%
\bibitem [{\citenamefont {Aaij}\ \emph
  {et~al.}(2014{\natexlab{h}})\citenamefont {Aaij} \emph
  {et~al.}}]{LHCB-PAPER-2014-046}%
  \BibitemOpen
  \bibfield  {author} {\bibinfo {author} {\bibfnamefont {R.}~\bibnamefont
  {Aaij}} \emph {et~al.} (\bibinfo {collaboration} {LHCb collaboration}),\
  }\href {\doibase 10.1007/JHEP10(2014)005} {\bibfield  {journal} {\bibinfo
  {journal} {JHEP}\ }\textbf {\bibinfo {volume} {10}},\ \bibinfo {pages} {005}
  (\bibinfo {year} {2014}{\natexlab{h}})},\ \Eprint
  {http://arxiv.org/abs/1408.1299} {arXiv:1408.1299 [hep-ex]} \BibitemShut
  {NoStop}%
%%CITATION = ARXIV:1408.1299;%%
\bibitem [{\citenamefont {Bianco}\ \emph {et~al.}(2003)\citenamefont {Bianco},
  \citenamefont {Fabbri}, \citenamefont {Benson},\ and\ \citenamefont
  {Bigi}}]{CICERONE1}%
  \BibitemOpen
  \bibfield  {author} {\bibinfo {author} {\bibfnamefont {S.}~\bibnamefont
  {Bianco}}, \bibinfo {author} {\bibfnamefont {F.~L.}\ \bibnamefont {Fabbri}},
  \bibinfo {author} {\bibfnamefont {D.}~\bibnamefont {Benson}}, \ and\ \bibinfo
  {author} {\bibfnamefont {I.}~\bibnamefont {Bigi}},\ }\href {\doibase
  10.1393/ncr/i2003-10003-1} {\bibfield  {journal} {\bibinfo  {journal} {Riv.
  Nuovo Cim.}\ }\textbf {\bibinfo {volume} {26N7}},\ \bibinfo {pages} {1}
  (\bibinfo {year} {2003})},\ \Eprint {http://arxiv.org/abs/hep-ex/0309021}
  {arXiv:hep-ex/0309021 [hep-ex]} \BibitemShut {NoStop}%
%%CITATION = HEP-EX/0309021;%%
\bibitem [{\citenamefont {Bigi}\ \emph {et~al.}(1988)\citenamefont {Bigi},
  \citenamefont {Khoze}, \citenamefont {Uraltsev},\ and\ \citenamefont
  {Sanda}}]{1988BOOK}%
  \BibitemOpen
  \bibfield  {author} {\bibinfo {author} {\bibfnamefont {I.}~\bibnamefont
  {Bigi}}, \bibinfo {author} {\bibfnamefont {V.}~\bibnamefont {Khoze}},
  \bibinfo {author} {\bibfnamefont {N.}~\bibnamefont {Uraltsev}}, \ and\
  \bibinfo {author} {\bibfnamefont {A.}~\bibnamefont {Sanda}},\ }\enquote
  {\bibinfo {title} {The question of {CP} noninvariance - as seen through the
  eyes of neutral beauty},}\ in\ \href@noop {} {\emph {\bibinfo {booktitle} {CP
  Violation}}}\ (\bibinfo  {publisher} {World Scientific},\ \bibinfo {year}
  {1988})\ \bibinfo {note} {in particular about the item of FSI in pps. 190 -
  194}\BibitemShut {NoStop}%
\bibitem [{\citenamefont {Wolfenstein}(1991)}]{wolfenstein}%
  \BibitemOpen
  \bibfield  {author} {\bibinfo {author} {\bibfnamefont {L.}~\bibnamefont
  {Wolfenstein}},\ }\href {\doibase 10.1103/PhysRevD.43.151} {\bibfield
  {journal} {\bibinfo  {journal} {Phys. Rev.}\ }\textbf {\bibinfo {volume}
  {D43}},\ \bibinfo {pages} {151} (\bibinfo {year} {1991})}\BibitemShut
  {NoStop}%
%%CITATION = PHRVA,D43,151;%%
\bibitem [{CIC()}]{CICERONE2}%
  \BibitemOpen
  \href@noop {} {}\bibinfo {note} {Sects. 11.1.1. , pps. 163 - 166 in
  \cite{CICERONE1}}\BibitemShut {NoStop}%
\bibitem [{\citenamefont {Garron}(2015)}]{LATTICE1}%
  \BibitemOpen
  \bibfield  {author} {\bibinfo {author} {\bibfnamefont {N.}~\bibnamefont
  {Garron}},\ }in\ \href@noop {} {\emph {\bibinfo {booktitle} {{8th
  International Workshop on Chiral Dynamics (CD 2015) Pisa, Italy, June 29-July
  3, 2015}}}}\ (\bibinfo {year} {2015})\ \Eprint
  {http://arxiv.org/abs/1512.02440} {arXiv:1512.02440 [hep-lat]} \BibitemShut
  {NoStop}%
%%CITATION = ARXIV:1512.02440;%%
\bibitem [{\citenamefont {Pennington}(2002)}]{DR_a}%
  \BibitemOpen
  \bibfield  {author} {\bibinfo {author} {\bibfnamefont {M.~R.}\ \bibnamefont
  {Pennington}},\ }in\ \href@noop {} {\emph {\bibinfo {booktitle} {{Production,
  properties and interaction of mesons. Proceedings, 7th International
  Workshop, Meson 2002, Krakow, Poland, May 24-28, 2002}}}}\ (\bibinfo {year}
  {2002})\ pp.\ \bibinfo {pages} {1--14},\ \Eprint
  {http://arxiv.org/abs/hep-ph/0207220} {arXiv:hep-ph/0207220 [hep-ph]}
  \BibitemShut {NoStop}%
%%CITATION = HEP-PH/0207220;%%
\bibitem [{\citenamefont {Pelaez}\ \emph {et~al.}(2011)\citenamefont {Pelaez},
  \citenamefont {Pennington}, \citenamefont {Ruiz~de Elvira},\ and\
  \citenamefont {Wilson}}]{DR_b}%
  \BibitemOpen
  \bibfield  {author} {\bibinfo {author} {\bibfnamefont {J.~R.}\ \bibnamefont
  {Pelaez}}, \bibinfo {author} {\bibfnamefont {M.~R.}\ \bibnamefont
  {Pennington}}, \bibinfo {author} {\bibfnamefont {J.}~\bibnamefont {Ruiz~de
  Elvira}}, \ and\ \bibinfo {author} {\bibfnamefont {D.~J.}\ \bibnamefont
  {Wilson}},\ }in\ \href@noop {} {\emph {\bibinfo {booktitle} {{Proceedings,
  14th International Conference on Hadron spectroscopy (Hadron 2011)}}}}\
  (\bibinfo {year} {2011})\ \Eprint {http://arxiv.org/abs/1109.2392}
  {arXiv:1109.2392 [hep-ph]} \BibitemShut {NoStop}%
%%CITATION = ARXIV:1109.2392;%%
\bibitem [{\citenamefont {Donoghue}(1996)}]{DR_c}%
  \BibitemOpen
  \bibfield  {author} {\bibinfo {author} {\bibfnamefont {J.~F.}\ \bibnamefont
  {Donoghue}},\ }in\ \href@noop {} {\emph {\bibinfo {booktitle} {{Advanced
  School on Effective Theories Almunecar, Spain, June 25-July 1, 1995}}}}\
  (\bibinfo {year} {1996})\ \Eprint {http://arxiv.org/abs/hep-ph/9607351}
  {arXiv:hep-ph/9607351 [hep-ph]} \BibitemShut {NoStop}%
%%CITATION = HEP-PH/9607351;%%
\bibitem [{\citenamefont {Hanhart}(2013)}]{KUBIS_a}%
  \BibitemOpen
  \bibfield  {author} {\bibinfo {author} {\bibfnamefont {C.}~\bibnamefont
  {Hanhart}},\ }in\ \href@noop {} {\emph {\bibinfo {booktitle} {{Proceedings,
  6th International Workshop on Charm Physics (Charm 2013)}}}}\ (\bibinfo
  {year} {2013})\ \Eprint {http://arxiv.org/abs/1311.6627} {arXiv:1311.6627
  [hep-ph]} \BibitemShut {NoStop}%
%%CITATION = ARXIV:1311.6627;%%
\bibitem [{\citenamefont {Kubis}\ \emph {et~al.}(2012)\citenamefont {Kubis},
  \citenamefont {Niecknig},\ and\ \citenamefont {Schneider}}]{KUBIS_b}%
  \BibitemOpen
  \bibfield  {author} {\bibinfo {author} {\bibfnamefont {B.}~\bibnamefont
  {Kubis}}, \bibinfo {author} {\bibfnamefont {F.}~\bibnamefont {Niecknig}}, \
  and\ \bibinfo {author} {\bibfnamefont {S.~P.}\ \bibnamefont {Schneider}},\
  }\bibfield  {booktitle} {\emph {\bibinfo {booktitle} {{Proceedings, 8th
  International Workshop on e+ e- Collisions from Phi to Psi (PHIPSI11)}}},\
  }\href {\doibase 10.1016/j.nuclphysbps.2012.02.017} {\bibfield  {journal}
  {\bibinfo  {journal} {Nucl. Phys. Proc. Suppl.}\ }\textbf {\bibinfo {volume}
  {225-227}},\ \bibinfo {pages} {75} (\bibinfo {year} {2012})},\ \Eprint
  {http://arxiv.org/abs/1111.6799} {arXiv:1111.6799 [hep-ph]} \BibitemShut
  {NoStop}%
%%CITATION = ARXIV:1111.6799;%%
\bibitem [{\citenamefont {Gardner}\ and\ \citenamefont
  {Meissner}(2002{\natexlab{a}})}]{KUBIS_c}%
  \BibitemOpen
  \bibfield  {author} {\bibinfo {author} {\bibfnamefont {S.}~\bibnamefont
  {Gardner}}\ and\ \bibinfo {author} {\bibfnamefont {U.-G.}\ \bibnamefont
  {Meissner}},\ }\href {\doibase 10.1103/PhysRevD.65.094004} {\bibfield
  {journal} {\bibinfo  {journal} {Phys. Rev.}\ }\textbf {\bibinfo {volume}
  {D65}},\ \bibinfo {pages} {094004} (\bibinfo {year} {2002}{\natexlab{a}})},\
  \Eprint {http://arxiv.org/abs/hep-ph/0112281} {arXiv:hep-ph/0112281 [hep-ph]}
  \BibitemShut {NoStop}%
%%CITATION = HEP-PH/0112281;%%
\bibitem [{\citenamefont {Bigi}(2015{\natexlab{a}})}]{BIGIRIO_a}%
  \BibitemOpen
  \bibfield  {author} {\bibinfo {author} {\bibfnamefont {I.~I.}\ \bibnamefont
  {Bigi}},\ }\href {\doibase 10.1007/s11467-015-0476-y} {\bibfield  {journal}
  {\bibinfo  {journal} {Front. Phys. China}\ }\textbf {\bibinfo {volume}
  {10}},\ \bibinfo {pages} {240} (\bibinfo {year} {2015}{\natexlab{a}})},\
  \Eprint {http://arxiv.org/abs/1503.07719} {arXiv:1503.07719 [hep-ph]}
  \BibitemShut {NoStop}%
%%CITATION = ARXIV:1503.07719;%%
\bibitem [{\citenamefont {Bigi}(2015{\natexlab{b}})}]{BIGIRIO_b}%
  \BibitemOpen
  \bibfield  {author} {\bibinfo {author} {\bibfnamefont {I.~I.}\ \bibnamefont
  {Bigi}},\ }\href@noop {} {\  (\bibinfo {year} {2015}{\natexlab{b}})},\
  \Eprint {http://arxiv.org/abs/1509.03899} {arXiv:1509.03899 [hep-ph]}
  \BibitemShut {NoStop}%
%%CITATION = ARXIV:1509.03899;%%
\bibitem [{\citenamefont {Beneke}\ and\ \citenamefont
  {Neubert}(2003)}]{Beneke2003}%
  \BibitemOpen
  \bibfield  {author} {\bibinfo {author} {\bibfnamefont {M.}~\bibnamefont
  {Beneke}}\ and\ \bibinfo {author} {\bibfnamefont {M.}~\bibnamefont
  {Neubert}},\ }\href {\doibase 10.1016/j.nuclphysb.2003.09.026} {\bibfield
  {journal} {\bibinfo  {journal} {Nucl. Phys.}\ }\textbf {\bibinfo {volume}
  {B675}},\ \bibinfo {pages} {333} (\bibinfo {year} {2003})},\ \Eprint
  {http://arxiv.org/abs/hep-ph/0308039} {arXiv:hep-ph/0308039 [hep-ph]}
  \BibitemShut {NoStop}%
%%CITATION = HEP-PH/0308039;%%
\bibitem [{\citenamefont {Bauer}\ \emph {et~al.}(1987)\citenamefont {Bauer},
  \citenamefont {Stech},\ and\ \citenamefont {Wirbel}}]{Bauer1987}%
  \BibitemOpen
  \bibfield  {author} {\bibinfo {author} {\bibfnamefont {M.}~\bibnamefont
  {Bauer}}, \bibinfo {author} {\bibfnamefont {B.}~\bibnamefont {Stech}}, \ and\
  \bibinfo {author} {\bibfnamefont {M.}~\bibnamefont {Wirbel}},\ }\href
  {\doibase 10.1007/BF01561122} {\bibfield  {journal} {\bibinfo  {journal} {Z.
  Phys.}\ }\textbf {\bibinfo {volume} {C34}},\ \bibinfo {pages} {103} (\bibinfo
  {year} {1987})}\BibitemShut {NoStop}%
%%CITATION = ZEPYA,C34,103;%%
\bibitem [{\citenamefont {Biswas}\ \emph {et~al.}(2015)\citenamefont {Biswas},
  \citenamefont {Sinha},\ and\ \citenamefont {Abbas}}]{Abbasetal}%
  \BibitemOpen
  \bibfield  {author} {\bibinfo {author} {\bibfnamefont {A.}~\bibnamefont
  {Biswas}}, \bibinfo {author} {\bibfnamefont {N.}~\bibnamefont {Sinha}}, \
  and\ \bibinfo {author} {\bibfnamefont {G.}~\bibnamefont {Abbas}},\ }\href
  {\doibase 10.1103/PhysRevD.92.014032} {\bibfield  {journal} {\bibinfo
  {journal} {Phys. Rev.}\ }\textbf {\bibinfo {volume} {D92}},\ \bibinfo {pages}
  {014032} (\bibinfo {year} {2015})},\ \Eprint
  {http://arxiv.org/abs/1503.08176} {arXiv:1503.08176 [hep-ph]} \BibitemShut
  {NoStop}%
%%CITATION = ARXIV:1503.08176;%%
\bibitem [{\citenamefont {Boito}\ \emph {et~al.}(2009)\citenamefont {Boito},
  \citenamefont {Dedonder}, \citenamefont {El-Bennich}, \citenamefont
  {Leitner},\ and\ \citenamefont {Loiseau}}]{Boitoetal09}%
  \BibitemOpen
  \bibfield  {author} {\bibinfo {author} {\bibfnamefont {D.~R.}\ \bibnamefont
  {Boito}}, \bibinfo {author} {\bibfnamefont {J.~P.}\ \bibnamefont {Dedonder}},
  \bibinfo {author} {\bibfnamefont {B.}~\bibnamefont {El-Bennich}}, \bibinfo
  {author} {\bibfnamefont {O.}~\bibnamefont {Leitner}}, \ and\ \bibinfo
  {author} {\bibfnamefont {B.}~\bibnamefont {Loiseau}},\ }\href {\doibase
  10.1103/PhysRevD.79.034020} {\bibfield  {journal} {\bibinfo  {journal} {Phys.
  Rev.}\ }\textbf {\bibinfo {volume} {D79}},\ \bibinfo {pages} {034020}
  (\bibinfo {year} {2009})},\ \Eprint {http://arxiv.org/abs/0812.3843}
  {arXiv:0812.3843 [hep-ph]} \BibitemShut {NoStop}%
%%CITATION = ARXIV:0812.3843;%%
\bibitem [{\citenamefont {Boito}\ and\ \citenamefont
  {Escribano}(2009)}]{BE2009}%
  \BibitemOpen
  \bibfield  {author} {\bibinfo {author} {\bibfnamefont {D.~R.}\ \bibnamefont
  {Boito}}\ and\ \bibinfo {author} {\bibfnamefont {R.}~\bibnamefont
  {Escribano}},\ }\href {\doibase 10.1103/PhysRevD.80.054007} {\bibfield
  {journal} {\bibinfo  {journal} {Phys. Rev.}\ }\textbf {\bibinfo {volume}
  {D80}},\ \bibinfo {pages} {054007} (\bibinfo {year} {2009})},\ \Eprint
  {http://arxiv.org/abs/0907.0189} {arXiv:0907.0189 [hep-ph]} \BibitemShut
  {NoStop}%
%%CITATION = ARXIV:0907.0189;%%
\bibitem [{\citenamefont {Dedonder}\ \emph {et~al.}(2014)\citenamefont
  {Dedonder}, \citenamefont {Kaminski}, \citenamefont {Lesniak},\ and\
  \citenamefont {Loiseau}}]{JPD_PRD89}%
  \BibitemOpen
  \bibfield  {author} {\bibinfo {author} {\bibfnamefont {J.~P.}\ \bibnamefont
  {Dedonder}}, \bibinfo {author} {\bibfnamefont {R.}~\bibnamefont {Kaminski}},
  \bibinfo {author} {\bibfnamefont {L.}~\bibnamefont {Lesniak}}, \ and\
  \bibinfo {author} {\bibfnamefont {B.}~\bibnamefont {Loiseau}},\ }\href
  {\doibase 10.1103/PhysRevD.89.094018} {\bibfield  {journal} {\bibinfo
  {journal} {Phys. Rev.}\ }\textbf {\bibinfo {volume} {D89}},\ \bibinfo {pages}
  {094018} (\bibinfo {year} {2014})},\ \Eprint {http://arxiv.org/abs/1403.2971}
  {arXiv:1403.2971 [hep-ph]} \BibitemShut {NoStop}%
%%CITATION = ARXIV:1403.2971;%%
\bibitem [{\citenamefont {El-Bennich}\ \emph {et~al.}(2009)\citenamefont
  {El-Bennich}, \citenamefont {Furman}, \citenamefont {Kaminski}, \citenamefont
  {Lesniak}, \citenamefont {Loiseau},\ and\ \citenamefont
  {Moussallam}}]{ElBennichetal09}%
  \BibitemOpen
  \bibfield  {author} {\bibinfo {author} {\bibfnamefont {B.}~\bibnamefont
  {El-Bennich}}, \bibinfo {author} {\bibfnamefont {A.}~\bibnamefont {Furman}},
  \bibinfo {author} {\bibfnamefont {R.}~\bibnamefont {Kaminski}}, \bibinfo
  {author} {\bibfnamefont {L.}~\bibnamefont {Lesniak}}, \bibinfo {author}
  {\bibfnamefont {B.}~\bibnamefont {Loiseau}}, \ and\ \bibinfo {author}
  {\bibfnamefont {B.}~\bibnamefont {Moussallam}},\ }\href {\doibase
  10.1103/PhysRevD.83.039903, 10.1103/PhysRevD.79.094005} {\bibfield  {journal}
  {\bibinfo  {journal} {Phys. Rev.}\ }\textbf {\bibinfo {volume} {D79}},\
  \bibinfo {pages} {094005} (\bibinfo {year} {2009})},\ \bibinfo {note}
  {[Erratum: Phys. Rev.D83,039903(2011)]},\ \Eprint
  {http://arxiv.org/abs/0902.3645} {arXiv:0902.3645 [hep-ph]} \BibitemShut
  {NoStop}%
%%CITATION = ARXIV:0902.3645;%%
\bibitem [{\citenamefont {Buras}(1995)}]{BurasNPB434_606}%
  \BibitemOpen
  \bibfield  {author} {\bibinfo {author} {\bibfnamefont {A.~J.}\ \bibnamefont
  {Buras}},\ }\href {\doibase 10.1016/0550-3213(94)00482-T} {\bibfield
  {journal} {\bibinfo  {journal} {Nucl. Phys.}\ }\textbf {\bibinfo {volume}
  {B434}},\ \bibinfo {pages} {606} (\bibinfo {year} {1995})},\ \Eprint
  {http://arxiv.org/abs/hep-ph/9409309} {arXiv:hep-ph/9409309 [hep-ph]}
  \BibitemShut {NoStop}%
%%CITATION = HEP-PH/9409309;%%
\bibitem [{\citenamefont {Gardner}\ and\ \citenamefont
  {Meissner}(2002{\natexlab{b}})}]{MG02}%
  \BibitemOpen
  \bibfield  {author} {\bibinfo {author} {\bibfnamefont {S.}~\bibnamefont
  {Gardner}}\ and\ \bibinfo {author} {\bibfnamefont {U.-G.}\ \bibnamefont
  {Meissner}},\ }\href {\doibase 10.1103/PhysRevD.65.094004} {\bibfield
  {journal} {\bibinfo  {journal} {Phys. Rev.}\ }\textbf {\bibinfo {volume}
  {D65}},\ \bibinfo {pages} {094004} (\bibinfo {year} {2002}{\natexlab{b}})},\
  \Eprint {http://arxiv.org/abs/hep-ph/0112281} {arXiv:hep-ph/0112281 [hep-ph]}
  \BibitemShut {NoStop}%
%%CITATION = HEP-PH/0112281;%%
\bibitem [{\citenamefont {Jamin}\ \emph {et~al.}(2006)\citenamefont {Jamin},
  \citenamefont {Oller},\ and\ \citenamefont {Pich}}]{FFs1}%
  \BibitemOpen
  \bibfield  {author} {\bibinfo {author} {\bibfnamefont {M.}~\bibnamefont
  {Jamin}}, \bibinfo {author} {\bibfnamefont {J.~A.}\ \bibnamefont {Oller}}, \
  and\ \bibinfo {author} {\bibfnamefont {A.}~\bibnamefont {Pich}},\ }\href
  {\doibase 10.1103/PhysRevD.74.074009} {\bibfield  {journal} {\bibinfo
  {journal} {Phys. Rev.}\ }\textbf {\bibinfo {volume} {D74}},\ \bibinfo {pages}
  {074009} (\bibinfo {year} {2006})},\ \Eprint
  {http://arxiv.org/abs/hep-ph/0605095} {arXiv:hep-ph/0605095 [hep-ph]}
  \BibitemShut {NoStop}%
%%CITATION = HEP-PH/0605095;%%
\bibitem [{\citenamefont {Moussallam}(2008)}]{FFs2}%
  \BibitemOpen
  \bibfield  {author} {\bibinfo {author} {\bibfnamefont {B.}~\bibnamefont
  {Moussallam}},\ }\href {\doibase 10.1140/epjc/s10052-007-0464-7} {\bibfield
  {journal} {\bibinfo  {journal} {Eur. Phys. J.}\ }\textbf {\bibinfo {volume}
  {C53}},\ \bibinfo {pages} {401} (\bibinfo {year} {2008})},\ \Eprint
  {http://arxiv.org/abs/0710.0548} {arXiv:0710.0548 [hep-ph]} \BibitemShut
  {NoStop}%
%%CITATION = ARXIV:0710.0548;%%
\bibitem [{\citenamefont {Boito}\ \emph {et~al.}(2010)\citenamefont {Boito},
  \citenamefont {Escribano},\ and\ \citenamefont {Jamin}}]{FFs3}%
  \BibitemOpen
  \bibfield  {author} {\bibinfo {author} {\bibfnamefont {D.~R.}\ \bibnamefont
  {Boito}}, \bibinfo {author} {\bibfnamefont {R.}~\bibnamefont {Escribano}}, \
  and\ \bibinfo {author} {\bibfnamefont {M.}~\bibnamefont {Jamin}},\ }\href
  {\doibase 10.1007/JHEP09(2010)031} {\bibfield  {journal} {\bibinfo  {journal}
  {JHEP}\ }\textbf {\bibinfo {volume} {09}},\ \bibinfo {pages} {031} (\bibinfo
  {year} {2010})},\ \Eprint {http://arxiv.org/abs/1007.1858} {arXiv:1007.1858
  [hep-ph]} \BibitemShut {NoStop}%
%%CITATION = ARXIV:1007.1858;%%
\bibitem [{\citenamefont {Melikhov}(2002)}]{Melikhov}%
  \BibitemOpen
  \bibfield  {author} {\bibinfo {author} {\bibfnamefont {D.}~\bibnamefont
  {Melikhov}},\ }\href@noop {} {\bibfield  {journal} {\bibinfo  {journal} {Eur.
  Phys. J.direct}\ }\textbf {\bibinfo {volume} {C4}},\ \bibinfo {pages} {2}
  (\bibinfo {year} {2002})},\ \Eprint {http://arxiv.org/abs/hep-ph/0110087}
  {arXiv:hep-ph/0110087 [hep-ph]} \BibitemShut {NoStop}%
%%CITATION = HEP-PH/0110087;%%
\bibitem [{\citenamefont {Bhattacharya}\ \emph
  {et~al.}(2014{\natexlab{a}})\citenamefont {Bhattacharya}, \citenamefont
  {Gronau}, \citenamefont {Imbeault}, \citenamefont {London},\ and\
  \citenamefont {Rosner}}]{BPPP_FS}%
  \BibitemOpen
  \bibfield  {author} {\bibinfo {author} {\bibfnamefont {B.}~\bibnamefont
  {Bhattacharya}}, \bibinfo {author} {\bibfnamefont {M.}~\bibnamefont
  {Gronau}}, \bibinfo {author} {\bibfnamefont {M.}~\bibnamefont {Imbeault}},
  \bibinfo {author} {\bibfnamefont {D.}~\bibnamefont {London}}, \ and\ \bibinfo
  {author} {\bibfnamefont {J.~L.}\ \bibnamefont {Rosner}},\ }\href {\doibase
  10.1103/PhysRevD.89.074043} {\bibfield  {journal} {\bibinfo  {journal} {Phys.
  Rev.}\ }\textbf {\bibinfo {volume} {D89}},\ \bibinfo {pages} {074043}
  (\bibinfo {year} {2014}{\natexlab{a}})},\ \Eprint
  {http://arxiv.org/abs/1402.2909} {arXiv:1402.2909 [hep-ph]} \BibitemShut
  {NoStop}%
%%CITATION = ARXIV:1402.2909;%%
\bibitem [{\citenamefont {Gronau}(2000)}]{MGUspin}%
  \BibitemOpen
  \bibfield  {author} {\bibinfo {author} {\bibfnamefont {M.}~\bibnamefont
  {Gronau}},\ }\href {\doibase 10.1016/S0370-2693(00)01119-9} {\bibfield
  {journal} {\bibinfo  {journal} {Phys. Lett.}\ }\textbf {\bibinfo {volume}
  {B492}},\ \bibinfo {pages} {297} (\bibinfo {year} {2000})},\ \Eprint
  {http://arxiv.org/abs/hep-ph/0008292} {arXiv:hep-ph/0008292 [hep-ph]}
  \BibitemShut {NoStop}%
%%CITATION = HEP-PH/0008292;%%
\bibitem [{\citenamefont {Rey-Le~Lorier}\ and\ \citenamefont
  {London}(2012)}]{BPPPgamma1}%
  \BibitemOpen
  \bibfield  {author} {\bibinfo {author} {\bibfnamefont {N.}~\bibnamefont
  {Rey-Le~Lorier}}\ and\ \bibinfo {author} {\bibfnamefont {D.}~\bibnamefont
  {London}},\ }\href {\doibase 10.1103/PhysRevD.85.016010} {\bibfield
  {journal} {\bibinfo  {journal} {Phys. Rev.}\ }\textbf {\bibinfo {volume}
  {D85}},\ \bibinfo {pages} {016010} (\bibinfo {year} {2012})},\ \Eprint
  {http://arxiv.org/abs/1109.0881} {arXiv:1109.0881 [hep-ph]} \BibitemShut
  {NoStop}%
%%CITATION = ARXIV:1109.0881;%%
\bibitem [{\citenamefont {Bhattacharya}\ \emph
  {et~al.}(2014{\natexlab{b}})\citenamefont {Bhattacharya}, \citenamefont
  {Imbeault},\ and\ \citenamefont {London}}]{BPPPgamma2}%
  \BibitemOpen
  \bibfield  {author} {\bibinfo {author} {\bibfnamefont {B.}~\bibnamefont
  {Bhattacharya}}, \bibinfo {author} {\bibfnamefont {M.}~\bibnamefont
  {Imbeault}}, \ and\ \bibinfo {author} {\bibfnamefont {D.}~\bibnamefont
  {London}},\ }\href {\doibase 10.1016/j.physletb.2013.11.038} {\bibfield
  {journal} {\bibinfo  {journal} {Phys. Lett.}\ }\textbf {\bibinfo {volume}
  {B728}},\ \bibinfo {pages} {206} (\bibinfo {year} {2014}{\natexlab{b}})},\
  \Eprint {http://arxiv.org/abs/1303.0846} {arXiv:1303.0846 [hep-ph]}
  \BibitemShut {NoStop}%
%%CITATION = ARXIV:1303.0846;%%
\bibitem [{\citenamefont {Lorier}\ \emph {et~al.}(2011)\citenamefont {Lorier},
  \citenamefont {Imbeault},\ and\ \citenamefont {London}}]{BPPPdiagrams}%
  \BibitemOpen
  \bibfield  {author} {\bibinfo {author} {\bibfnamefont {N.~R.-L.}\
  \bibnamefont {Lorier}}, \bibinfo {author} {\bibfnamefont {M.}~\bibnamefont
  {Imbeault}}, \ and\ \bibinfo {author} {\bibfnamefont {D.}~\bibnamefont
  {London}},\ }\href {\doibase 10.1103/PhysRevD.84.034040} {\bibfield
  {journal} {\bibinfo  {journal} {Phys. Rev.}\ }\textbf {\bibinfo {volume}
  {D84}},\ \bibinfo {pages} {034040} (\bibinfo {year} {2011})},\ \Eprint
  {http://arxiv.org/abs/1011.4972} {arXiv:1011.4972 [hep-ph]} \BibitemShut
  {NoStop}%
%%CITATION = ARXIV:1011.4972;%%
\bibitem [{\citenamefont {Oller}\ and\ \citenamefont {Oset}(1997)}]{npa}%
  \BibitemOpen
  \bibfield  {author} {\bibinfo {author} {\bibfnamefont {J.~A.}\ \bibnamefont
  {Oller}}\ and\ \bibinfo {author} {\bibfnamefont {E.}~\bibnamefont {Oset}},\
  }\href {\doibase 10.1016/S0375-9474(97)00160-7} {\bibfield  {journal}
  {\bibinfo  {journal} {Nucl. Phys.}\ }\textbf {\bibinfo {volume} {A620}},\
  \bibinfo {pages} {438} (\bibinfo {year} {1997})},\ \bibinfo {note} {[Erratum:
  Nucl. Phys.A652,407(1999)]},\ \Eprint {http://arxiv.org/abs/hep-ph/9702314}
  {arXiv:hep-ph/9702314 [hep-ph]} \BibitemShut {NoStop}%
%%CITATION = HEP-PH/9702314;%%
\bibitem [{\citenamefont {Oller}\ \emph {et~al.}(1999)\citenamefont {Oller},
  \citenamefont {Oset},\ and\ \citenamefont {Pelaez}}]{ramonet}%
  \BibitemOpen
  \bibfield  {author} {\bibinfo {author} {\bibfnamefont {J.~A.}\ \bibnamefont
  {Oller}}, \bibinfo {author} {\bibfnamefont {E.}~\bibnamefont {Oset}}, \ and\
  \bibinfo {author} {\bibfnamefont {J.~R.}\ \bibnamefont {Pelaez}},\ }\href
  {\doibase 10.1103/PhysRevD.59.074001, 10.1103/PhysRevD.60.099906,
  10.1103/PhysRevD.75.099903} {\bibfield  {journal} {\bibinfo  {journal} {Phys.
  Rev.}\ }\textbf {\bibinfo {volume} {D59}},\ \bibinfo {pages} {074001}
  (\bibinfo {year} {1999})},\ \bibinfo {note} {[Erratum: Phys.
  Rev.D75,099903(2007)]},\ \Eprint {http://arxiv.org/abs/hep-ph/9804209}
  {arXiv:hep-ph/9804209 [hep-ph]} \BibitemShut {NoStop}%
%%CITATION = HEP-PH/9804209;%%
\bibitem [{\citenamefont {Liang}\ and\ \citenamefont
  {Oset}(2014{\natexlab{a}})}]{liang}%
  \BibitemOpen
  \bibfield  {author} {\bibinfo {author} {\bibfnamefont {W.~H.}\ \bibnamefont
  {Liang}}\ and\ \bibinfo {author} {\bibfnamefont {E.}~\bibnamefont {Oset}},\
  }\href {\doibase 10.1016/j.physletb.2014.08.030} {\bibfield  {journal}
  {\bibinfo  {journal} {Phys. Lett.}\ }\textbf {\bibinfo {volume} {B737}},\
  \bibinfo {pages} {70} (\bibinfo {year} {2014}{\natexlab{a}})},\ \Eprint
  {http://arxiv.org/abs/1406.7228} {arXiv:1406.7228 [hep-ph]} \BibitemShut
  {NoStop}%
%%CITATION = ARXIV:1406.7228;%%
\bibitem [{\citenamefont {Xie}\ \emph {et~al.}(2015)\citenamefont {Xie},
  \citenamefont {Dai},\ and\ \citenamefont {Oset}}]{dai}%
  \BibitemOpen
  \bibfield  {author} {\bibinfo {author} {\bibfnamefont {J.-J.}\ \bibnamefont
  {Xie}}, \bibinfo {author} {\bibfnamefont {L.~R.}\ \bibnamefont {Dai}}, \ and\
  \bibinfo {author} {\bibfnamefont {E.}~\bibnamefont {Oset}},\ }\href {\doibase
  10.1016/j.physletb.2015.02.006} {\bibfield  {journal} {\bibinfo  {journal}
  {Phys. Lett.}\ }\textbf {\bibinfo {volume} {B742}},\ \bibinfo {pages} {363}
  (\bibinfo {year} {2015})},\ \Eprint {http://arxiv.org/abs/1409.0401}
  {arXiv:1409.0401 [hep-ph]} \BibitemShut {NoStop}%
%%CITATION = ARXIV:1409.0401;%%
\bibitem [{\citenamefont {Weinberg}(1979)}]{Weinberg:1978kz}%
  \BibitemOpen
  \bibfield  {author} {\bibinfo {author} {\bibfnamefont {S.}~\bibnamefont
  {Weinberg}},\ }\href {\doibase 10.1016/0378-4371(79)90223-1} {\bibfield
  {journal} {\bibinfo  {journal} {Physica}\ }\textbf {\bibinfo {volume}
  {A96}},\ \bibinfo {pages} {327} (\bibinfo {year} {1979})}\BibitemShut
  {NoStop}%
%%CITATION = PHYSA,A96,327;%%
\bibitem [{\citenamefont {Gasser}\ and\ \citenamefont
  {Leutwyler}(1984)}]{Gasser:1983yg}%
  \BibitemOpen
  \bibfield  {author} {\bibinfo {author} {\bibfnamefont {J.}~\bibnamefont
  {Gasser}}\ and\ \bibinfo {author} {\bibfnamefont {H.}~\bibnamefont
  {Leutwyler}},\ }\href {\doibase 10.1016/0003-4916(84)90242-2} {\bibfield
  {journal} {\bibinfo  {journal} {Annals Phys.}\ }\textbf {\bibinfo {volume}
  {158}},\ \bibinfo {pages} {142} (\bibinfo {year} {1984})}\BibitemShut
  {NoStop}%
%%CITATION = APNYA,158,142;%%
\bibitem [{\citenamefont {Bando}\ \emph {et~al.}(1988)\citenamefont {Bando},
  \citenamefont {Kugo},\ and\ \citenamefont {Yamawaki}}]{bando}%
  \BibitemOpen
  \bibfield  {author} {\bibinfo {author} {\bibfnamefont {M.}~\bibnamefont
  {Bando}}, \bibinfo {author} {\bibfnamefont {T.}~\bibnamefont {Kugo}}, \ and\
  \bibinfo {author} {\bibfnamefont {K.}~\bibnamefont {Yamawaki}},\ }\href
  {\doibase 10.1016/0370-1573(88)90019-1} {\bibfield  {journal} {\bibinfo
  {journal} {Phys. Rept.}\ }\textbf {\bibinfo {volume} {164}},\ \bibinfo
  {pages} {217} (\bibinfo {year} {1988})}\BibitemShut {NoStop}%
%%CITATION = PRPLC,164,217;%%
\bibitem [{\citenamefont {van Beveren}\ \emph
  {et~al.}(1986{\natexlab{a}})\citenamefont {van Beveren}, \citenamefont
  {Rijken}, \citenamefont {Metzger}, \citenamefont {Dullemond}, \citenamefont
  {Rupp},\ and\ \citenamefont {Ribeiro}}]{vanBeveren:1986ea}%
  \BibitemOpen
  \bibfield  {author} {\bibinfo {author} {\bibfnamefont {E.}~\bibnamefont {van
  Beveren}}, \bibinfo {author} {\bibfnamefont {T.~A.}\ \bibnamefont {Rijken}},
  \bibinfo {author} {\bibfnamefont {K.}~\bibnamefont {Metzger}}, \bibinfo
  {author} {\bibfnamefont {C.}~\bibnamefont {Dullemond}}, \bibinfo {author}
  {\bibfnamefont {G.}~\bibnamefont {Rupp}}, \ and\ \bibinfo {author}
  {\bibfnamefont {J.~E.}\ \bibnamefont {Ribeiro}},\ }\href {\doibase
  10.1007/BF01571811} {\bibfield  {journal} {\bibinfo  {journal} {Z. Phys.}\
  }\textbf {\bibinfo {volume} {C30}},\ \bibinfo {pages} {615} (\bibinfo {year}
  {1986}{\natexlab{a}})},\ \Eprint {http://arxiv.org/abs/0710.4067}
  {arXiv:0710.4067 [hep-ph]} \BibitemShut {NoStop}%
%%CITATION = ARXIV:0710.4067;%%
\bibitem [{\citenamefont {Tornqvist}\ and\ \citenamefont
  {Roos}(1996)}]{Tornqvist:1995ay}%
  \BibitemOpen
  \bibfield  {author} {\bibinfo {author} {\bibfnamefont {N.~A.}\ \bibnamefont
  {Tornqvist}}\ and\ \bibinfo {author} {\bibfnamefont {M.}~\bibnamefont
  {Roos}},\ }\href {\doibase 10.1103/PhysRevLett.76.1575} {\bibfield  {journal}
  {\bibinfo  {journal} {Phys. Rev. Lett.}\ }\textbf {\bibinfo {volume} {76}},\
  \bibinfo {pages} {1575} (\bibinfo {year} {1996})},\ \Eprint
  {http://arxiv.org/abs/hep-ph/9511210} {arXiv:hep-ph/9511210 [hep-ph]}
  \BibitemShut {NoStop}%
%%CITATION = HEP-PH/9511210;%%
\bibitem [{\citenamefont {Fariborz}\ \emph
  {et~al.}(2009{\natexlab{a}})\citenamefont {Fariborz}, \citenamefont {Jora},\
  and\ \citenamefont {Schechter}}]{Fariborz:2009cq}%
  \BibitemOpen
  \bibfield  {author} {\bibinfo {author} {\bibfnamefont {A.~H.}\ \bibnamefont
  {Fariborz}}, \bibinfo {author} {\bibfnamefont {R.}~\bibnamefont {Jora}}, \
  and\ \bibinfo {author} {\bibfnamefont {J.}~\bibnamefont {Schechter}},\ }\href
  {\doibase 10.1103/PhysRevD.79.074014} {\bibfield  {journal} {\bibinfo
  {journal} {Phys. Rev.}\ }\textbf {\bibinfo {volume} {D79}},\ \bibinfo {pages}
  {074014} (\bibinfo {year} {2009}{\natexlab{a}})},\ \Eprint
  {http://arxiv.org/abs/0902.2825} {arXiv:0902.2825 [hep-ph]} \BibitemShut
  {NoStop}%
%%CITATION = ARXIV:0902.2825;%%
\bibitem [{\citenamefont {Fariborz}\ \emph
  {et~al.}(2009{\natexlab{b}})\citenamefont {Fariborz}, \citenamefont {Park},
  \citenamefont {Schechter},\ and\ \citenamefont
  {Naeem~Shahid}}]{Fariborz:2009wf}%
  \BibitemOpen
  \bibfield  {author} {\bibinfo {author} {\bibfnamefont {A.~H.}\ \bibnamefont
  {Fariborz}}, \bibinfo {author} {\bibfnamefont {N.~W.}\ \bibnamefont {Park}},
  \bibinfo {author} {\bibfnamefont {J.}~\bibnamefont {Schechter}}, \ and\
  \bibinfo {author} {\bibfnamefont {M.}~\bibnamefont {Naeem~Shahid}},\ }\href
  {\doibase 10.1103/PhysRevD.80.113001} {\bibfield  {journal} {\bibinfo
  {journal} {Phys. Rev.}\ }\textbf {\bibinfo {volume} {D80}},\ \bibinfo {pages}
  {113001} (\bibinfo {year} {2009}{\natexlab{b}})},\ \Eprint
  {http://arxiv.org/abs/0907.0482} {arXiv:0907.0482 [hep-ph]} \BibitemShut
  {NoStop}%
%%CITATION = ARXIV:0907.0482;%%
\bibitem [{\citenamefont {Gamermann}\ \emph {et~al.}(2007)\citenamefont
  {Gamermann}, \citenamefont {Oset}, \citenamefont {Strottman},\ and\
  \citenamefont {Vicente~Vacas}}]{gamermann}%
  \BibitemOpen
  \bibfield  {author} {\bibinfo {author} {\bibfnamefont {D.}~\bibnamefont
  {Gamermann}}, \bibinfo {author} {\bibfnamefont {E.}~\bibnamefont {Oset}},
  \bibinfo {author} {\bibfnamefont {D.}~\bibnamefont {Strottman}}, \ and\
  \bibinfo {author} {\bibfnamefont {M.~J.}\ \bibnamefont {Vicente~Vacas}},\
  }\href {\doibase 10.1103/PhysRevD.76.074016} {\bibfield  {journal} {\bibinfo
  {journal} {Phys. Rev.}\ }\textbf {\bibinfo {volume} {D76}},\ \bibinfo {pages}
  {074016} (\bibinfo {year} {2007})},\ \Eprint
  {http://arxiv.org/abs/hep-ph/0612179} {arXiv:hep-ph/0612179 [hep-ph]}
  \BibitemShut {NoStop}%
%%CITATION = HEP-PH/0612179;%%
\bibitem [{\citenamefont {Gamermann}\ and\ \citenamefont
  {Oset}(2008)}]{danires}%
  \BibitemOpen
  \bibfield  {author} {\bibinfo {author} {\bibfnamefont {D.}~\bibnamefont
  {Gamermann}}\ and\ \bibinfo {author} {\bibfnamefont {E.}~\bibnamefont
  {Oset}},\ }\href {\doibase 10.1140/epja/i2007-10580-5} {\bibfield  {journal}
  {\bibinfo  {journal} {Eur. Phys. J.}\ }\textbf {\bibinfo {volume} {A36}},\
  \bibinfo {pages} {189} (\bibinfo {year} {2008})},\ \Eprint
  {http://arxiv.org/abs/0712.1758} {arXiv:0712.1758 [hep-ph]} \BibitemShut
  {NoStop}%
%%CITATION = ARXIV:0712.1758;%%
\bibitem [{\citenamefont {Albaladejo}\ \emph {et~al.}(2015)\citenamefont
  {Albaladejo}, \citenamefont {Nielsen},\ and\ \citenamefont {Oset}}]{miguel}%
  \BibitemOpen
  \bibfield  {author} {\bibinfo {author} {\bibfnamefont {M.}~\bibnamefont
  {Albaladejo}}, \bibinfo {author} {\bibfnamefont {M.}~\bibnamefont {Nielsen}},
  \ and\ \bibinfo {author} {\bibfnamefont {E.}~\bibnamefont {Oset}},\ }\href
  {\doibase 10.1016/j.physletb.2015.05.019} {\bibfield  {journal} {\bibinfo
  {journal} {Phys. Lett.}\ }\textbf {\bibinfo {volume} {B746}},\ \bibinfo
  {pages} {305} (\bibinfo {year} {2015})},\ \Eprint
  {http://arxiv.org/abs/1501.03455} {arXiv:1501.03455 [hep-ph]} \BibitemShut
  {NoStop}%
%%CITATION = ARXIV:1501.03455;%%
\bibitem [{\citenamefont {Navarra}\ \emph {et~al.}(2015)\citenamefont
  {Navarra}, \citenamefont {Nielsen}, \citenamefont {Oset},\ and\ \citenamefont
  {Sekihara}}]{sekihara}%
  \BibitemOpen
  \bibfield  {author} {\bibinfo {author} {\bibfnamefont {F.~S.}\ \bibnamefont
  {Navarra}}, \bibinfo {author} {\bibfnamefont {M.}~\bibnamefont {Nielsen}},
  \bibinfo {author} {\bibfnamefont {E.}~\bibnamefont {Oset}}, \ and\ \bibinfo
  {author} {\bibfnamefont {T.}~\bibnamefont {Sekihara}},\ }\href {\doibase
  10.1103/PhysRevD.92.014031} {\bibfield  {journal} {\bibinfo  {journal} {Phys.
  Rev.}\ }\textbf {\bibinfo {volume} {D92}},\ \bibinfo {pages} {014031}
  (\bibinfo {year} {2015})},\ \Eprint {http://arxiv.org/abs/1501.03422}
  {arXiv:1501.03422 [hep-ph]} \BibitemShut {NoStop}%
%%CITATION = ARXIV:1501.03422;%%
\bibitem [{\citenamefont {Liang}\ \emph {et~al.}(2015)\citenamefont {Liang},
  \citenamefont {Xie}, \citenamefont {Oset}, \citenamefont {Molina},\ and\
  \citenamefont {D{\"o}ring}}]{liangxyz}%
  \BibitemOpen
  \bibfield  {author} {\bibinfo {author} {\bibfnamefont {W.-H.}\ \bibnamefont
  {Liang}}, \bibinfo {author} {\bibfnamefont {J.-J.}\ \bibnamefont {Xie}},
  \bibinfo {author} {\bibfnamefont {E.}~\bibnamefont {Oset}}, \bibinfo {author}
  {\bibfnamefont {R.}~\bibnamefont {Molina}}, \ and\ \bibinfo {author}
  {\bibfnamefont {M.}~\bibnamefont {D{\"o}ring}},\ }\href {\doibase
  10.1140/epja/i2015-15058-3} {\bibfield  {journal} {\bibinfo  {journal} {Eur.
  Phys. J.}\ }\textbf {\bibinfo {volume} {A51}},\ \bibinfo {pages} {58}
  (\bibinfo {year} {2015})},\ \Eprint {http://arxiv.org/abs/1502.02932}
  {arXiv:1502.02932 [hep-ph]} \BibitemShut {NoStop}%
%%CITATION = ARXIV:1502.02932;%%
\bibitem [{\citenamefont {Aaij}\ \emph
  {et~al.}(2015{\natexlab{i}})\citenamefont {Aaij} \emph {et~al.}}]{penta}%
  \BibitemOpen
  \bibfield  {author} {\bibinfo {author} {\bibfnamefont {R.}~\bibnamefont
  {Aaij}} \emph {et~al.} (\bibinfo {collaboration} {LHCb collaboration}),\
  }\href {\doibase 10.1103/PhysRevLett.115.072001} {\bibfield  {journal}
  {\bibinfo  {journal} {Phys. Rev. Lett.}\ }\textbf {\bibinfo {volume} {115}},\
  \bibinfo {pages} {072001} (\bibinfo {year} {2015}{\natexlab{i}})},\ \Eprint
  {http://arxiv.org/abs/1507.03414} {arXiv:1507.03414 [hep-ex]} \BibitemShut
  {NoStop}%
%%CITATION = ARXIV:1507.03414;%%
\bibitem [{\citenamefont {Flatte}(1976)}]{PLB63p224}%
  \BibitemOpen
  \bibfield  {author} {\bibinfo {author} {\bibfnamefont {S.~M.}\ \bibnamefont
  {Flatte}},\ }\href {\doibase 10.1016/0370-2693(76)90654-7} {\bibfield
  {journal} {\bibinfo  {journal} {Phys. Lett.}\ }\textbf {\bibinfo {volume}
  {B63}},\ \bibinfo {pages} {224} (\bibinfo {year} {1976})}\BibitemShut
  {NoStop}%
%%CITATION = PHLTA,B63,224;%%
\bibitem [{\citenamefont {Aston}\ \emph
  {et~al.}(1988{\natexlab{b}})\citenamefont {Aston} \emph
  {et~al.}}]{NPB296p493}%
  \BibitemOpen
  \bibfield  {author} {\bibinfo {author} {\bibfnamefont {D.}~\bibnamefont
  {Aston}} \emph {et~al.},\ }\href {\doibase 10.1016/0550-3213(88)90028-4}
  {\bibfield  {journal} {\bibinfo  {journal} {Nucl. Phys.}\ }\textbf {\bibinfo
  {volume} {B296}},\ \bibinfo {pages} {493} (\bibinfo {year}
  {1988}{\natexlab{b}})}\BibitemShut {NoStop}%
%%CITATION = NUPHA,B296,493;%%
\bibitem [{\citenamefont {Appel}(2001)}]{Appel01}%
  \BibitemOpen
  \bibfield  {author} {\bibinfo {author} {\bibfnamefont {J.~A.}\ \bibnamefont
  {Appel}},\ }\href@noop {} {\enquote {\bibinfo {title} {private
  communications},}\ } (\bibinfo {year} {2001})\BibitemShut {NoStop}%
\bibitem [{\citenamefont {van Beveren}\ and\ \citenamefont
  {Rupp}(1999)}]{EPJC10p469}%
  \BibitemOpen
  \bibfield  {author} {\bibinfo {author} {\bibfnamefont {E.}~\bibnamefont {van
  Beveren}}\ and\ \bibinfo {author} {\bibfnamefont {G.}~\bibnamefont {Rupp}},\
  }\href {\doibase 10.1007/s100520050769, 10.1007/s100529900126} {\bibfield
  {journal} {\bibinfo  {journal} {Eur. Phys. J.}\ }\textbf {\bibinfo {volume}
  {C10}},\ \bibinfo {pages} {469} (\bibinfo {year} {1999})},\ \Eprint
  {http://arxiv.org/abs/hep-ph/9806246} {arXiv:hep-ph/9806246 [hep-ph]}
  \BibitemShut {NoStop}%
%%CITATION = HEP-PH/9806246;%%
\bibitem [{\citenamefont {Tornqvist}(1995)}]{ZPC68p647}%
  \BibitemOpen
  \bibfield  {author} {\bibinfo {author} {\bibfnamefont {N.~A.}\ \bibnamefont
  {Tornqvist}},\ }\href {\doibase 10.1007/BF01565264} {\bibfield  {journal}
  {\bibinfo  {journal} {Z. Phys.}\ }\textbf {\bibinfo {volume} {C68}},\
  \bibinfo {pages} {647} (\bibinfo {year} {1995})},\ \Eprint
  {http://arxiv.org/abs/hep-ph/9504372} {arXiv:hep-ph/9504372 [hep-ph]}
  \BibitemShut {NoStop}%
%%CITATION = HEP-PH/9504372;%%
\bibitem [{\citenamefont {van Beveren}\ \emph
  {et~al.}(1986{\natexlab{b}})\citenamefont {van Beveren}, \citenamefont
  {Rijken}, \citenamefont {Metzger}, \citenamefont {Dullemond}, \citenamefont
  {Rupp},\ and\ \citenamefont {Ribeiro}}]{ZPC30p615}%
  \BibitemOpen
  \bibfield  {author} {\bibinfo {author} {\bibfnamefont {E.}~\bibnamefont {van
  Beveren}}, \bibinfo {author} {\bibfnamefont {T.~A.}\ \bibnamefont {Rijken}},
  \bibinfo {author} {\bibfnamefont {K.}~\bibnamefont {Metzger}}, \bibinfo
  {author} {\bibfnamefont {C.}~\bibnamefont {Dullemond}}, \bibinfo {author}
  {\bibfnamefont {G.}~\bibnamefont {Rupp}}, \ and\ \bibinfo {author}
  {\bibfnamefont {J.~E.}\ \bibnamefont {Ribeiro}},\ }\href {\doibase
  10.1007/BF01571811} {\bibfield  {journal} {\bibinfo  {journal} {Z. Phys.}\
  }\textbf {\bibinfo {volume} {C30}},\ \bibinfo {pages} {615} (\bibinfo {year}
  {1986}{\natexlab{b}})},\ \Eprint {http://arxiv.org/abs/0710.4067}
  {arXiv:0710.4067 [hep-ph]} \BibitemShut {NoStop}%
%%CITATION = ARXIV:0710.4067;%%
\bibitem [{\citenamefont {van Beveren}\ and\ \citenamefont
  {Rupp}(2001)}]{EPJC22p493}%
  \BibitemOpen
  \bibfield  {author} {\bibinfo {author} {\bibfnamefont {E.}~\bibnamefont {van
  Beveren}}\ and\ \bibinfo {author} {\bibfnamefont {G.}~\bibnamefont {Rupp}},\
  }\href {\doibase 10.1007/s100520100823} {\bibfield  {journal} {\bibinfo
  {journal} {Eur. Phys. J.}\ }\textbf {\bibinfo {volume} {C22}},\ \bibinfo
  {pages} {493} (\bibinfo {year} {2001})},\ \Eprint
  {http://arxiv.org/abs/hep-ex/0106077} {arXiv:hep-ex/0106077 [hep-ex]}
  \BibitemShut {NoStop}%
%%CITATION = HEP-EX/0106077;%%
\bibitem [{\citenamefont {van Beveren}\ and\ \citenamefont
  {Rupp}(2006{\natexlab{a}})}]{IJTPGTNLO11p179}%
  \BibitemOpen
  \bibfield  {author} {\bibinfo {author} {\bibfnamefont {E.}~\bibnamefont {van
  Beveren}}\ and\ \bibinfo {author} {\bibfnamefont {G.}~\bibnamefont {Rupp}},\
  }\href@noop {} {\bibfield  {journal} {\bibinfo  {journal} {Int. J. Theor.
  Phys. Group Theor. Nonlin. Opt.}\ }\textbf {\bibinfo {volume} {11}},\
  \bibinfo {pages} {179} (\bibinfo {year} {2006}{\natexlab{a}})},\ \Eprint
  {http://arxiv.org/abs/hep-ph/0304105} {arXiv:hep-ph/0304105 [hep-ph]}
  \BibitemShut {NoStop}%
%%CITATION = HEP-PH/0304105;%%
\bibitem [{\citenamefont {van Beveren}\ and\ \citenamefont
  {Rupp}(2003)}]{PRL91p012003}%
  \BibitemOpen
  \bibfield  {author} {\bibinfo {author} {\bibfnamefont {E.}~\bibnamefont {van
  Beveren}}\ and\ \bibinfo {author} {\bibfnamefont {G.}~\bibnamefont {Rupp}},\
  }\href {\doibase 10.1103/PhysRevLett.91.012003} {\bibfield  {journal}
  {\bibinfo  {journal} {Phys. Rev. Lett.}\ }\textbf {\bibinfo {volume} {91}},\
  \bibinfo {pages} {012003} (\bibinfo {year} {2003})},\ \Eprint
  {http://arxiv.org/abs/hep-ph/0305035} {arXiv:hep-ph/0305035 [hep-ph]}
  \BibitemShut {NoStop}%
%%CITATION = HEP-PH/0305035;%%
\bibitem [{\citenamefont {van Beveren}\ \emph {et~al.}(2006)\citenamefont {van
  Beveren}, \citenamefont {Bugg}, \citenamefont {Kleefeld},\ and\ \citenamefont
  {Rupp}}]{PLB641p265}%
  \BibitemOpen
  \bibfield  {author} {\bibinfo {author} {\bibfnamefont {E.}~\bibnamefont {van
  Beveren}}, \bibinfo {author} {\bibfnamefont {D.~V.}\ \bibnamefont {Bugg}},
  \bibinfo {author} {\bibfnamefont {F.}~\bibnamefont {Kleefeld}}, \ and\
  \bibinfo {author} {\bibfnamefont {G.}~\bibnamefont {Rupp}},\ }\href {\doibase
  10.1016/j.physletb.2006.08.051} {\bibfield  {journal} {\bibinfo  {journal}
  {Phys. Lett.}\ }\textbf {\bibinfo {volume} {B641}},\ \bibinfo {pages} {265}
  (\bibinfo {year} {2006})},\ \Eprint {http://arxiv.org/abs/hep-ph/0606022}
  {arXiv:hep-ph/0606022 [hep-ph]} \BibitemShut {NoStop}%
%%CITATION = HEP-PH/0606022;%%
\bibitem [{\citenamefont {van Beveren}\ and\ \citenamefont
  {Rupp}(2006{\natexlab{b}})}]{PRL97p202001_a}%
  \BibitemOpen
  \bibfield  {author} {\bibinfo {author} {\bibfnamefont {E.}~\bibnamefont {van
  Beveren}}\ and\ \bibinfo {author} {\bibfnamefont {G.}~\bibnamefont {Rupp}},\
  }\href {\doibase 10.1103/PhysRevLett.97.202001} {\bibfield  {journal}
  {\bibinfo  {journal} {Phys. Rev. Lett.}\ }\textbf {\bibinfo {volume} {97}},\
  \bibinfo {pages} {202001} (\bibinfo {year} {2006}{\natexlab{b}})},\ \Eprint
  {http://arxiv.org/abs/hep-ph/0606110} {arXiv:hep-ph/0606110 [hep-ph]}
  \BibitemShut {NoStop}%
%%CITATION = HEP-PH/0606110;%%
\bibitem [{\citenamefont {van Beveren}\ and\ \citenamefont
  {Rupp}(2010)}]{PRL97p202001_b}%
  \BibitemOpen
  \bibfield  {author} {\bibinfo {author} {\bibfnamefont {E.}~\bibnamefont {van
  Beveren}}\ and\ \bibinfo {author} {\bibfnamefont {G.}~\bibnamefont {Rupp}},\
  }\href {\doibase 10.1103/PhysRevD.81.118101} {\bibfield  {journal} {\bibinfo
  {journal} {Phys. Rev.}\ }\textbf {\bibinfo {volume} {D81}},\ \bibinfo {pages}
  {118101} (\bibinfo {year} {2010})},\ \Eprint {http://arxiv.org/abs/0908.1142}
  {arXiv:0908.1142 [hep-ph]} \BibitemShut {NoStop}%
%%CITATION = ARXIV:0908.1142;%%
\bibitem [{\citenamefont {van Beveren}\ and\ \citenamefont
  {Rupp}(2009{\natexlab{a}})}]{AP324p1620}%
  \BibitemOpen
  \bibfield  {author} {\bibinfo {author} {\bibfnamefont {E.}~\bibnamefont {van
  Beveren}}\ and\ \bibinfo {author} {\bibfnamefont {G.}~\bibnamefont {Rupp}},\
  }\href {\doibase 10.1016/j.aop.2009.03.013} {\bibfield  {journal} {\bibinfo
  {journal} {Annals Phys.}\ }\textbf {\bibinfo {volume} {324}},\ \bibinfo
  {pages} {1620} (\bibinfo {year} {2009}{\natexlab{a}})},\ \Eprint
  {http://arxiv.org/abs/0809.1149} {arXiv:0809.1149 [hep-ph]} \BibitemShut
  {NoStop}%
%%CITATION = ARXIV:0809.1149;%%
\bibitem [{\citenamefont {Coito}\ \emph
  {et~al.}(2011{\natexlab{a}})\citenamefont {Coito}, \citenamefont {Rupp},\
  and\ \citenamefont {van Beveren}}]{PRD84p094020}%
  \BibitemOpen
  \bibfield  {author} {\bibinfo {author} {\bibfnamefont {S.}~\bibnamefont
  {Coito}}, \bibinfo {author} {\bibfnamefont {G.}~\bibnamefont {Rupp}}, \ and\
  \bibinfo {author} {\bibfnamefont {E.}~\bibnamefont {van Beveren}},\ }\href
  {\doibase 10.1103/PhysRevD.84.094020} {\bibfield  {journal} {\bibinfo
  {journal} {Phys. Rev.}\ }\textbf {\bibinfo {volume} {D84}},\ \bibinfo {pages}
  {094020} (\bibinfo {year} {2011}{\natexlab{a}})},\ \Eprint
  {http://arxiv.org/abs/1106.2760} {arXiv:1106.2760 [hep-ph]} \BibitemShut
  {NoStop}%
%%CITATION = ARXIV:1106.2760;%%
\bibitem [{\citenamefont {Coito}\ \emph
  {et~al.}(2011{\natexlab{b}})\citenamefont {Coito}, \citenamefont {Rupp},\
  and\ \citenamefont {van Beveren}}]{EPJC71p1762}%
  \BibitemOpen
  \bibfield  {author} {\bibinfo {author} {\bibfnamefont {S.}~\bibnamefont
  {Coito}}, \bibinfo {author} {\bibfnamefont {G.}~\bibnamefont {Rupp}}, \ and\
  \bibinfo {author} {\bibfnamefont {E.}~\bibnamefont {van Beveren}},\ }\href
  {\doibase 10.1140/epjc/s10052-011-1762-7} {\bibfield  {journal} {\bibinfo
  {journal} {Eur. Phys. J.}\ }\textbf {\bibinfo {volume} {C71}},\ \bibinfo
  {pages} {1762} (\bibinfo {year} {2011}{\natexlab{b}})},\ \Eprint
  {http://arxiv.org/abs/1008.5100} {arXiv:1008.5100 [hep-ph]} \BibitemShut
  {NoStop}%
%%CITATION = ARXIV:1008.5100;%%
\bibitem [{\citenamefont {van Beveren}(1984)}]{ZPC21p291}%
  \BibitemOpen
  \bibfield  {author} {\bibinfo {author} {\bibfnamefont {E.}~\bibnamefont {van
  Beveren}},\ }\href {\doibase 10.1007/BF01577044} {\bibfield  {journal}
  {\bibinfo  {journal} {Z. Phys.}\ }\textbf {\bibinfo {volume} {C21}},\
  \bibinfo {pages} {291} (\bibinfo {year} {1984})},\ \Eprint
  {http://arxiv.org/abs/hep-ph/0602246} {arXiv:hep-ph/0602246 [hep-ph]}
  \BibitemShut {NoStop}%
%%CITATION = HEP-PH/0602246;%%
\bibitem [{\citenamefont {Bali}\ \emph {et~al.}(2005)\citenamefont {Bali},
  \citenamefont {Neff}, \citenamefont {Duessel}, \citenamefont {Lippert},\ and\
  \citenamefont {Schilling}}]{PRD71p114513}%
  \BibitemOpen
  \bibfield  {author} {\bibinfo {author} {\bibfnamefont {G.~S.}\ \bibnamefont
  {Bali}}, \bibinfo {author} {\bibfnamefont {H.}~\bibnamefont {Neff}}, \bibinfo
  {author} {\bibfnamefont {T.}~\bibnamefont {Duessel}}, \bibinfo {author}
  {\bibfnamefont {T.}~\bibnamefont {Lippert}}, \ and\ \bibinfo {author}
  {\bibfnamefont {K.}~\bibnamefont {Schilling}} (\bibinfo {collaboration}
  {SESAM}),\ }\href {\doibase 10.1103/PhysRevD.71.114513} {\bibfield  {journal}
  {\bibinfo  {journal} {Phys. Rev.}\ }\textbf {\bibinfo {volume} {D71}},\
  \bibinfo {pages} {114513} (\bibinfo {year} {2005})},\ \Eprint
  {http://arxiv.org/abs/hep-lat/0505012} {arXiv:hep-lat/0505012 [hep-lat]}
  \BibitemShut {NoStop}%
%%CITATION = HEP-LAT/0505012;%%
\bibitem [{\citenamefont {Olive}\ \emph
  {et~al.}(2014{\natexlab{a}})\citenamefont {Olive} \emph {et~al.}}]{PDG2014}%
  \BibitemOpen
  \bibfield  {author} {\bibinfo {author} {\bibfnamefont {K.~A.}\ \bibnamefont
  {Olive}} \emph {et~al.} (\bibinfo {collaboration} {Particle Data Group}),\
  }\href {\doibase 10.1088/1674-1137/38/9/090001} {\bibfield  {journal}
  {\bibinfo  {journal} {Chin. Phys.}\ }\textbf {\bibinfo {volume} {C38}},\
  \bibinfo {pages} {090001} (\bibinfo {year} {2014}{\natexlab{a}})}\BibitemShut
  {NoStop}%
\bibitem [{\citenamefont {van Beveren}\ and\ \citenamefont
  {Rupp}(2008{\natexlab{a}})}]{AP323p1215}%
  \BibitemOpen
  \bibfield  {author} {\bibinfo {author} {\bibfnamefont {E.}~\bibnamefont {van
  Beveren}}\ and\ \bibinfo {author} {\bibfnamefont {G.}~\bibnamefont {Rupp}},\
  }\href {\doibase 10.1016/j.aop.2007.11.012} {\bibfield  {journal} {\bibinfo
  {journal} {Annals Phys.}\ }\textbf {\bibinfo {volume} {323}},\ \bibinfo
  {pages} {1215} (\bibinfo {year} {2008}{\natexlab{a}})},\ \Eprint
  {http://arxiv.org/abs/0706.4119} {arXiv:0706.4119 [hep-ph]} \BibitemShut
  {NoStop}%
%%CITATION = ARXIV:0706.4119;%%
\bibitem [{\citenamefont {van Beveren}\ and\ \citenamefont
  {Rupp}(2008{\natexlab{b}})}]{EPL81p61002}%
  \BibitemOpen
  \bibfield  {author} {\bibinfo {author} {\bibfnamefont {E.}~\bibnamefont {van
  Beveren}}\ and\ \bibinfo {author} {\bibfnamefont {G.}~\bibnamefont {Rupp}},\
  }\href {\doibase 10.1209/0295-5075/81/61002} {\bibfield  {journal} {\bibinfo
  {journal} {Europhys. Lett.}\ }\textbf {\bibinfo {volume} {81}},\ \bibinfo
  {pages} {61002} (\bibinfo {year} {2008}{\natexlab{b}})},\ \Eprint
  {http://arxiv.org/abs/0710.5823} {arXiv:0710.5823 [hep-ph]} \BibitemShut
  {NoStop}%
%%CITATION = ARXIV:0710.5823;%%
\bibitem [{\citenamefont {van Beveren}\ and\ \citenamefont
  {Rupp}(2008{\natexlab{c}})}]{EPL84p51002}%
  \BibitemOpen
  \bibfield  {author} {\bibinfo {author} {\bibfnamefont {E.}~\bibnamefont {van
  Beveren}}\ and\ \bibinfo {author} {\bibfnamefont {G.}~\bibnamefont {Rupp}},\
  }\href {\doibase 10.1209/0295-5075/84/51002} {\bibfield  {journal} {\bibinfo
  {journal} {Europhys. Lett.}\ }\textbf {\bibinfo {volume} {84}},\ \bibinfo
  {pages} {51002} (\bibinfo {year} {2008}{\natexlab{c}})}\BibitemShut {NoStop}%
%%CITATION = EULEE,84,51002;%%
\bibitem [{\citenamefont {Pennington}\ and\ \citenamefont
  {Wilson}(2008)}]{EPL84p51001}%
  \BibitemOpen
  \bibfield  {author} {\bibinfo {author} {\bibfnamefont {M.~R.}\ \bibnamefont
  {Pennington}}\ and\ \bibinfo {author} {\bibfnamefont {D.~J.}\ \bibnamefont
  {Wilson}},\ }\href {\doibase 10.1209/0295-5075/84/51001} {\bibfield
  {journal} {\bibinfo  {journal} {Europhys. Lett.}\ }\textbf {\bibinfo {volume}
  {84}},\ \bibinfo {pages} {51001} (\bibinfo {year} {2008})},\ \Eprint
  {http://arxiv.org/abs/0711.3521} {arXiv:0711.3521 [hep-ph]} \BibitemShut
  {NoStop}%
%%CITATION = ARXIV:0711.3521;%%
\bibitem [{\citenamefont {van Beveren}\ \emph {et~al.}(2009)\citenamefont {van
  Beveren}, \citenamefont {Liu}, \citenamefont {Coimbra},\ and\ \citenamefont
  {Rupp}}]{EPL85p61002}%
  \BibitemOpen
  \bibfield  {author} {\bibinfo {author} {\bibfnamefont {E.}~\bibnamefont {van
  Beveren}}, \bibinfo {author} {\bibfnamefont {X.}~\bibnamefont {Liu}},
  \bibinfo {author} {\bibfnamefont {R.}~\bibnamefont {Coimbra}}, \ and\
  \bibinfo {author} {\bibfnamefont {G.}~\bibnamefont {Rupp}},\ }\href {\doibase
  10.1209/0295-5075/85/61002} {\bibfield  {journal} {\bibinfo  {journal}
  {Europhys. Lett.}\ }\textbf {\bibinfo {volume} {85}},\ \bibinfo {pages}
  {61002} (\bibinfo {year} {2009})},\ \Eprint {http://arxiv.org/abs/0809.1151}
  {arXiv:0809.1151 [hep-ph]} \BibitemShut {NoStop}%
%%CITATION = ARXIV:0809.1151;%%
\bibitem [{\citenamefont {van Beveren}\ and\ \citenamefont
  {Rupp}(2009{\natexlab{b}})}]{PRD80p074001}%
  \BibitemOpen
  \bibfield  {author} {\bibinfo {author} {\bibfnamefont {E.}~\bibnamefont {van
  Beveren}}\ and\ \bibinfo {author} {\bibfnamefont {G.}~\bibnamefont {Rupp}},\
  }\href {\doibase 10.1103/PhysRevD.80.074001} {\bibfield  {journal} {\bibinfo
  {journal} {Phys. Rev.}\ }\textbf {\bibinfo {volume} {D80}},\ \bibinfo {pages}
  {074001} (\bibinfo {year} {2009}{\natexlab{b}})},\ \Eprint
  {http://arxiv.org/abs/0908.0242} {arXiv:0908.0242 [hep-ph]} \BibitemShut
  {NoStop}%
%%CITATION = ARXIV:0908.0242;%%
\bibitem [{\citenamefont {Adolph}\ \emph {et~al.}(2015)\citenamefont {Adolph}
  \emph {et~al.}}]{Adolph:2015tqa}%
  \BibitemOpen
  \bibfield  {author} {\bibinfo {author} {\bibfnamefont {C.}~\bibnamefont
  {Adolph}} \emph {et~al.},\ }\href@noop {} {\  (\bibinfo {year} {2015})},\
  \Eprint {http://arxiv.org/abs/1509.00992} {arXiv:1509.00992 [hep-ex]}
  \BibitemShut {NoStop}%
%%CITATION = ARXIV:1509.00992;%%
\bibitem [{\citenamefont {Palano}(2015)}]{AntimoPalanoHadron2015}%
  \BibitemOpen
  \bibfield  {author} {\bibinfo {author} {\bibfnamefont {A.}~\bibnamefont
  {Palano}} (\bibinfo {collaboration} {on behalf of the BaBar collaboration}),\
  }\href
  {https://www.jlab.org/conferences/hadron2015/talks/thursday/parallel/session3/3C1_AntimoPalano.pdf}
  {\enquote {\bibinfo {title} {Dalitz plot analysis of three-body charmonium
  decays at {B}a{B}ar},}\ } (\bibinfo {year} {2015}),\ \bibinfo {note} {{Talk
  at \href{https://www.jlab.org/conferences/hadron2015/}{XVI International
  Conference on Hadron Spectroscopy},
  \href{https://www.jlab.org/conferences/hadron2015/talks/thursday/parallel/session3/3C1_AntimoPalano.pdf}{https://www.jlab.org/conferences/hadron2015/talks/thursday/parallel/session3/3C1\_AntimoPalano.pdf}}}\BibitemShut
  {NoStop}%
\bibitem [{\citenamefont {Battaglieri}\ \emph {et~al.}(2015)\citenamefont
  {Battaglieri} \emph {et~al.}}]{Battaglieri:2014gca}%
  \BibitemOpen
  \bibfield  {author} {\bibinfo {author} {\bibfnamefont {M.}~\bibnamefont
  {Battaglieri}} \emph {et~al.},\ }\href {\doibase 10.5506/APhysPolB.46.257}
  {\bibfield  {journal} {\bibinfo  {journal} {Acta Phys. Polon.}\ }\textbf
  {\bibinfo {volume} {B46}},\ \bibinfo {pages} {257} (\bibinfo {year}
  {2015})},\ \Eprint {http://arxiv.org/abs/1412.6393} {arXiv:1412.6393
  [hep-ph]} \BibitemShut {NoStop}%
%%CITATION = ARXIV:1412.6393;%%
\bibitem [{\citenamefont {Marshak}\ \emph {et~al.}(1969)\citenamefont
  {Marshak}, \citenamefont {Riazuddin},\ and\ \citenamefont {Ryan}}]{Marshak}%
  \BibitemOpen
  \bibfield  {author} {\bibinfo {author} {\bibfnamefont {R.~E.}\ \bibnamefont
  {Marshak}}, \bibinfo {author} {\bibnamefont {Riazuddin}}, \ and\ \bibinfo
  {author} {\bibfnamefont {C.~P.}\ \bibnamefont {Ryan}},\ }\href@noop {} {\emph
  {\bibinfo {title} {{Theory of Weak Interactions in Particle Physics}}}}\
  (\bibinfo  {publisher} {{Wiley-Interscience}},\ \bibinfo {year}
  {1969})\BibitemShut {NoStop}%
\bibitem [{\citenamefont {Branco}\ \emph {et~al.}(1999)\citenamefont {Branco},
  \citenamefont {Lavoura},\ and\ \citenamefont {Silva}}]{Branco}%
  \BibitemOpen
  \bibfield  {author} {\bibinfo {author} {\bibfnamefont {G.~C.}\ \bibnamefont
  {Branco}}, \bibinfo {author} {\bibfnamefont {L.}~\bibnamefont {Lavoura}}, \
  and\ \bibinfo {author} {\bibfnamefont {J.~P.}\ \bibnamefont {Silva}},\
  }\href@noop {} {\emph {\bibinfo {title} {{CP Violation}}}},\ \bibinfo
  {series} {Int. Ser. Monogr. Phys.}, Vol.\ \bibinfo {volume} {103}\ (\bibinfo
  {publisher} {{Oxford University Press}},\ \bibinfo {year} {1999})\BibitemShut
  {NoStop}%
\bibitem [{\citenamefont {Bigi}\ and\ \citenamefont {Sanda}(2000)}]{bigibook}%
  \BibitemOpen
  \bibfield  {author} {\bibinfo {author} {\bibfnamefont {I.~I.~Y.}\
  \bibnamefont {Bigi}}\ and\ \bibinfo {author} {\bibfnamefont {A.~I.}\
  \bibnamefont {Sanda}},\ }\href@noop {} {\emph {\bibinfo {title} {{CP
  violation}}}},\ \bibinfo {edition} {2nd}\ ed.,\ \bibinfo {series} {Camb.
  Monogr. Part. Phys. Nucl. Phys. Cosmol.}, Vol.~\bibinfo {volume} {9}\
  (\bibinfo  {publisher} {{Cambridge University Press}},\ \bibinfo {year}
  {2000})\BibitemShut {NoStop}%
%%CITATION = CMPCE,9,1;%%
\bibitem [{\citenamefont {Bander}\ \emph {et~al.}(1979)\citenamefont {Bander},
  \citenamefont {Silverman},\ and\ \citenamefont {Soni}}]{BSS}%
  \BibitemOpen
  \bibfield  {author} {\bibinfo {author} {\bibfnamefont {M.}~\bibnamefont
  {Bander}}, \bibinfo {author} {\bibfnamefont {D.}~\bibnamefont {Silverman}}, \
  and\ \bibinfo {author} {\bibfnamefont {A.}~\bibnamefont {Soni}},\ }\href
  {\doibase 10.1103/PhysRevLett.43.242} {\bibfield  {journal} {\bibinfo
  {journal} {Phys. Rev. Lett.}\ }\textbf {\bibinfo {volume} {43}},\ \bibinfo
  {pages} {242} (\bibinfo {year} {1979})}\BibitemShut {NoStop}%
%%CITATION = PRLTA,43,242;%%
\bibitem [{\citenamefont {Atwood}\ and\ \citenamefont
  {Soni}(1998)}]{Soni2005_a}%
  \BibitemOpen
  \bibfield  {author} {\bibinfo {author} {\bibfnamefont {D.}~\bibnamefont
  {Atwood}}\ and\ \bibinfo {author} {\bibfnamefont {A.}~\bibnamefont {Soni}},\
  }\href {\doibase 10.1103/PhysRevD.58.036005} {\bibfield  {journal} {\bibinfo
  {journal} {Phys. Rev.}\ }\textbf {\bibinfo {volume} {D58}},\ \bibinfo {pages}
  {036005} (\bibinfo {year} {1998})},\ \Eprint
  {http://arxiv.org/abs/hep-ph/9712287} {arXiv:hep-ph/9712287 [hep-ph]}
  \BibitemShut {NoStop}%
%%CITATION = HEP-PH/9712287;%%
\bibitem [{\citenamefont {Cheng}\ \emph {et~al.}(2005)\citenamefont {Cheng},
  \citenamefont {Chua},\ and\ \citenamefont {Soni}}]{Soni2005_b}%
  \BibitemOpen
  \bibfield  {author} {\bibinfo {author} {\bibfnamefont {H.-Y.}\ \bibnamefont
  {Cheng}}, \bibinfo {author} {\bibfnamefont {C.-K.}\ \bibnamefont {Chua}}, \
  and\ \bibinfo {author} {\bibfnamefont {A.}~\bibnamefont {Soni}},\ }\href
  {\doibase 10.1103/PhysRevD.71.014030} {\bibfield  {journal} {\bibinfo
  {journal} {Phys. Rev.}\ }\textbf {\bibinfo {volume} {D71}},\ \bibinfo {pages}
  {014030} (\bibinfo {year} {2005})},\ \Eprint
  {http://arxiv.org/abs/hep-ph/0409317} {arXiv:hep-ph/0409317 [hep-ph]}
  \BibitemShut {NoStop}%
%%CITATION = HEP-PH/0409317;%%
\bibitem [{\citenamefont {Aaij}\ \emph
  {et~al.}(2013{\natexlab{d}})\citenamefont {Aaij} \emph {et~al.}}]{LHCbPRL13}%
  \BibitemOpen
  \bibfield  {author} {\bibinfo {author} {\bibfnamefont {R.}~\bibnamefont
  {Aaij}} \emph {et~al.} (\bibinfo {collaboration} {LHCb collaboration}),\
  }\href {\doibase 10.1103/PhysRevLett.111.101801} {\bibfield  {journal}
  {\bibinfo  {journal} {Phys. Rev. Lett.}\ }\textbf {\bibinfo {volume} {111}},\
  \bibinfo {pages} {101801} (\bibinfo {year} {2013}{\natexlab{d}})},\ \Eprint
  {http://arxiv.org/abs/1306.1246} {arXiv:1306.1246 [hep-ex]} \BibitemShut
  {NoStop}%
%%CITATION = ARXIV:1306.1246;%%
\bibitem [{\citenamefont {Cohen}\ \emph {et~al.}(1980)\citenamefont {Cohen},
  \citenamefont {Ayres}, \citenamefont {Diebold}, \citenamefont {Kramer},
  \citenamefont {Pawlicki},\ and\ \citenamefont {Wicklund}}]{Cohen1980}%
  \BibitemOpen
  \bibfield  {author} {\bibinfo {author} {\bibfnamefont {D.~H.}\ \bibnamefont
  {Cohen}}, \bibinfo {author} {\bibfnamefont {D.~S.}\ \bibnamefont {Ayres}},
  \bibinfo {author} {\bibfnamefont {R.}~\bibnamefont {Diebold}}, \bibinfo
  {author} {\bibfnamefont {S.~L.}\ \bibnamefont {Kramer}}, \bibinfo {author}
  {\bibfnamefont {A.~J.}\ \bibnamefont {Pawlicki}}, \ and\ \bibinfo {author}
  {\bibfnamefont {A.~B.}\ \bibnamefont {Wicklund}},\ }\href {\doibase
  10.1103/PhysRevD.22.2595} {\bibfield  {journal} {\bibinfo  {journal} {Phys.
  Rev.}\ }\textbf {\bibinfo {volume} {D22}},\ \bibinfo {pages} {2595} (\bibinfo
  {year} {1980})}\BibitemShut {NoStop}%
%%CITATION = PHRVA,D22,2595;%%
\bibitem [{\citenamefont {Pel\'aez}\ and\ \citenamefont
  {Yndur\'ain}(2005)}]{pelaprd05_a}%
  \BibitemOpen
  \bibfield  {author} {\bibinfo {author} {\bibfnamefont {J.~R.}\ \bibnamefont
  {Pel\'aez}}\ and\ \bibinfo {author} {\bibfnamefont {F.~J.}\ \bibnamefont
  {Yndur\'ain}},\ }\href {\doibase 10.1103/PhysRevD.71.074016} {\bibfield
  {journal} {\bibinfo  {journal} {Phys. Rev.}\ }\textbf {\bibinfo {volume}
  {D71}},\ \bibinfo {pages} {074016} (\bibinfo {year} {2005})},\ \Eprint
  {http://arxiv.org/abs/hep-ph/0411334} {arXiv:hep-ph/0411334 [hep-ph]}
  \BibitemShut {NoStop}%
%%CITATION = HEP-PH/0411334;%%
\bibitem [{\citenamefont {Garc\'ia-Mart\'in}\ \emph {et~al.}(2011)\citenamefont
  {Garc\'ia-Mart\'in}, \citenamefont {Kam\'inski}, \citenamefont {Pel\'aez},
  \citenamefont {Ruiz~de Elvira},\ and\ \citenamefont
  {Yndur\'ain}}]{pelaprd05_b}%
  \BibitemOpen
  \bibfield  {author} {\bibinfo {author} {\bibfnamefont {R.}~\bibnamefont
  {Garc\'ia-Mart\'in}}, \bibinfo {author} {\bibfnamefont {R.}~\bibnamefont
  {Kam\'inski}}, \bibinfo {author} {\bibfnamefont {J.~R.}\ \bibnamefont
  {Pel\'aez}}, \bibinfo {author} {\bibfnamefont {J.}~\bibnamefont {Ruiz~de
  Elvira}}, \ and\ \bibinfo {author} {\bibfnamefont {F.~J.}\ \bibnamefont
  {Yndur\'ain}},\ }\href {\doibase 10.1103/PhysRevD.83.074004} {\bibfield
  {journal} {\bibinfo  {journal} {Phys. Rev.}\ }\textbf {\bibinfo {volume}
  {D83}},\ \bibinfo {pages} {074004} (\bibinfo {year} {2011})},\ \Eprint
  {http://arxiv.org/abs/1102.2183} {arXiv:1102.2183 [hep-ph]} \BibitemShut
  {NoStop}%
%%CITATION = ARXIV:1102.2183;%%
\bibitem [{\citenamefont {Brambilla}\ \emph {et~al.}(2014)\citenamefont
  {Brambilla} \emph {et~al.}}]{ref8}%
  \BibitemOpen
  \bibfield  {author} {\bibinfo {author} {\bibfnamefont {N.}~\bibnamefont
  {Brambilla}} \emph {et~al.},\ }\href {\doibase
  10.1140/epjc/s10052-014-2981-5} {\bibfield  {journal} {\bibinfo  {journal}
  {Eur. Phys. J.}\ }\textbf {\bibinfo {volume} {C74}},\ \bibinfo {pages} {2981}
  (\bibinfo {year} {2014})},\ \Eprint {http://arxiv.org/abs/1404.3723}
  {arXiv:1404.3723 [hep-ph]} \BibitemShut {NoStop}%
%%CITATION = ARXIV:1404.3723;%%
\bibitem [{\citenamefont {Martin}\ and\ \citenamefont
  {Spearman}(1970)}]{MartinSpearman}%
  \BibitemOpen
  \bibfield  {author} {\bibinfo {author} {\bibfnamefont {A.}~\bibnamefont
  {Martin}}\ and\ \bibinfo {author} {\bibfnamefont {T.}~\bibnamefont
  {Spearman}},\ }\href@noop {} {\emph {\bibinfo {title} {{Elementary Particle
  Theory}}}}\ (\bibinfo  {publisher} {North-Holland, Amsterdam, and Elsevier,
  New York},\ \bibinfo {year} {1970})\BibitemShut {NoStop}%
\bibitem [{\citenamefont {Danilkin}\ \emph {et~al.}(2015)\citenamefont
  {Danilkin}, \citenamefont {Fernández-Ramírez}, \citenamefont {Guo},
  \citenamefont {Mathieu}, \citenamefont {Schott}, \citenamefont {Shi},\ and\
  \citenamefont {Szczepaniak}}]{Danilkin:2014cra}%
  \BibitemOpen
  \bibfield  {author} {\bibinfo {author} {\bibfnamefont {I.~V.}\ \bibnamefont
  {Danilkin}}, \bibinfo {author} {\bibfnamefont {C.}~\bibnamefont
  {Fernández-Ramírez}}, \bibinfo {author} {\bibfnamefont {P.}~\bibnamefont
  {Guo}}, \bibinfo {author} {\bibfnamefont {V.}~\bibnamefont {Mathieu}},
  \bibinfo {author} {\bibfnamefont {D.}~\bibnamefont {Schott}}, \bibinfo
  {author} {\bibfnamefont {M.}~\bibnamefont {Shi}}, \ and\ \bibinfo {author}
  {\bibfnamefont {A.~P.}\ \bibnamefont {Szczepaniak}},\ }\href {\doibase
  10.1103/PhysRevD.91.094029} {\bibfield  {journal} {\bibinfo  {journal} {Phys.
  Rev.}\ }\textbf {\bibinfo {volume} {D91}},\ \bibinfo {pages} {094029}
  (\bibinfo {year} {2015})},\ \Eprint {http://arxiv.org/abs/1409.7708}
  {arXiv:1409.7708 [hep-ph]} \BibitemShut {NoStop}%
%%CITATION = ARXIV:1409.7708;%%
\bibitem [{\citenamefont {Pham}\ and\ \citenamefont
  {Truong}(1977)}]{Pham:1976yi}%
  \BibitemOpen
  \bibfield  {author} {\bibinfo {author} {\bibfnamefont {T.~N.}\ \bibnamefont
  {Pham}}\ and\ \bibinfo {author} {\bibfnamefont {T.~N.}\ \bibnamefont
  {Truong}},\ }\href {\doibase 10.1103/PhysRevD.16.896} {\bibfield  {journal}
  {\bibinfo  {journal} {Phys. Rev.}\ }\textbf {\bibinfo {volume} {D16}},\
  \bibinfo {pages} {896} (\bibinfo {year} {1977})}\BibitemShut {NoStop}%
%%CITATION = PHRVA,D16,896;%%
\bibitem [{\citenamefont {Gribov}(2009)}]{gribov}%
  \BibitemOpen
  \bibfield  {author} {\bibinfo {author} {\bibfnamefont {V.~N.}\ \bibnamefont
  {Gribov}},\ }\href@noop {} {\emph {\bibinfo {title} {{Strong interactions of
  hadrons at high emnergies: Gribov lectures on theoretical physics}}}},\
  edited by\ \bibinfo {editor} {\bibfnamefont {Y.~L.}\ \bibnamefont
  {Dokshitzer}}\ and\ \bibinfo {editor} {\bibfnamefont {J.}~\bibnamefont
  {Nyiri}},\ Cambridge Monographs on Particle Physics, Nuclear Physics and
  Cosmology\ (\bibinfo  {publisher} {Cambridge University Press},\ \bibinfo
  {year} {2009})\BibitemShut {NoStop}%
%%CITATION = INSPIRE-833953;%%
\bibitem [{\citenamefont {Schmid}(1967)}]{schmid}%
  \BibitemOpen
  \bibfield  {author} {\bibinfo {author} {\bibfnamefont {C.}~\bibnamefont
  {Schmid}},\ }\href {\doibase 10.1103/PhysRev.154.1363} {\bibfield  {journal}
  {\bibinfo  {journal} {Phys. Rev.}\ }\textbf {\bibinfo {volume} {154}},\
  \bibinfo {pages} {1363} (\bibinfo {year} {1967})}\BibitemShut {NoStop}%
\bibitem [{\citenamefont
  {Szczepaniak}(2015{\natexlab{b}})}]{Szczepaniak:2015eza}%
  \BibitemOpen
  \bibfield  {author} {\bibinfo {author} {\bibfnamefont {A.~P.}\ \bibnamefont
  {Szczepaniak}},\ }\href {\doibase 10.1016/j.physletb.2015.06.029} {\bibfield
  {journal} {\bibinfo  {journal} {Phys. Lett.}\ }\textbf {\bibinfo {volume}
  {B747}},\ \bibinfo {pages} {410} (\bibinfo {year} {2015}{\natexlab{b}})},\
  \Eprint {http://arxiv.org/abs/1501.01691} {arXiv:1501.01691 [hep-ph]}
  \BibitemShut {NoStop}%
%%CITATION = ARXIV:1501.01691;%%
\bibitem [{\citenamefont {Pelaez}(2015)}]{Pelaez:2015qba}%
  \BibitemOpen
  \bibfield  {author} {\bibinfo {author} {\bibfnamefont {J.~R.}\ \bibnamefont
  {Pelaez}},\ }\href@noop {} {\  (\bibinfo {year} {2015})},\ \Eprint
  {http://arxiv.org/abs/1510.00653} {arXiv:1510.00653 [hep-ph]} \BibitemShut
  {NoStop}%
%%CITATION = ARXIV:1510.00653;%%
\bibitem [{\citenamefont {Aaij}\ \emph
  {et~al.}(2013{\natexlab{e}})\citenamefont {Aaij} \emph
  {et~al.}}]{Aaij:2013zpt}%
  \BibitemOpen
  \bibfield  {author} {\bibinfo {author} {\bibfnamefont {R.}~\bibnamefont
  {Aaij}} \emph {et~al.} (\bibinfo {collaboration} {LHCb collaboration}),\
  }\href {\doibase 10.1103/PhysRevD.87.052001} {\bibfield  {journal} {\bibinfo
  {journal} {Phys. Rev.}\ }\textbf {\bibinfo {volume} {D87}},\ \bibinfo {pages}
  {052001} (\bibinfo {year} {2013}{\natexlab{e}})},\ \Eprint
  {http://arxiv.org/abs/1301.5347} {arXiv:1301.5347 [hep-ex]} \BibitemShut
  {NoStop}%
%%CITATION = ARXIV:1301.5347;%%
\bibitem [{\citenamefont {Stone}\ and\ \citenamefont {Zhang}(2013)}]{Stone}%
  \BibitemOpen
  \bibfield  {author} {\bibinfo {author} {\bibfnamefont {S.}~\bibnamefont
  {Stone}}\ and\ \bibinfo {author} {\bibfnamefont {L.}~\bibnamefont {Zhang}},\
  }\href {\doibase 10.1103/PhysRevLett.111.062001} {\bibfield  {journal}
  {\bibinfo  {journal} {Phys. Rev. Lett.}\ }\textbf {\bibinfo {volume} {111}},\
  \bibinfo {pages} {062001} (\bibinfo {year} {2013})},\ \Eprint
  {http://arxiv.org/abs/1305.6554} {arXiv:1305.6554 [hep-ex]} \BibitemShut
  {NoStop}%
%%CITATION = ARXIV:1305.6554;%%
\bibitem [{\citenamefont {Fleischer}\ \emph {et~al.}(2011)\citenamefont
  {Fleischer}, \citenamefont {Knegjens},\ and\ \citenamefont
  {Ricciardi}}]{Fleischer:2011au}%
  \BibitemOpen
  \bibfield  {author} {\bibinfo {author} {\bibfnamefont {R.}~\bibnamefont
  {Fleischer}}, \bibinfo {author} {\bibfnamefont {R.}~\bibnamefont {Knegjens}},
  \ and\ \bibinfo {author} {\bibfnamefont {G.}~\bibnamefont {Ricciardi}},\
  }\href {\doibase 10.1140/epjc/s10052-011-1832-x} {\bibfield  {journal}
  {\bibinfo  {journal} {Eur. Phys. J.}\ }\textbf {\bibinfo {volume} {C71}},\
  \bibinfo {pages} {1832} (\bibinfo {year} {2011})},\ \Eprint
  {http://arxiv.org/abs/1109.1112} {arXiv:1109.1112 [hep-ph]} \BibitemShut
  {NoStop}%
%%CITATION = ARXIV:1109.1112;%%
\bibitem [{\citenamefont {Olive}\ \emph
  {et~al.}(2014{\natexlab{b}})\citenamefont {Olive} \emph {et~al.}}]{PDG}%
  \BibitemOpen
  \bibfield  {author} {\bibinfo {author} {\bibfnamefont {K.~A.}\ \bibnamefont
  {Olive}} \emph {et~al.} (\bibinfo {collaboration} {Particle Data Group}),\
  }\href {\doibase 10.1088/1674-1137/38/9/090001} {\bibfield  {journal}
  {\bibinfo  {journal} {Chin. Phys.}\ }\textbf {\bibinfo {volume} {C38}},\
  \bibinfo {pages} {090001} (\bibinfo {year} {2014}{\natexlab{b}})}\BibitemShut
  {NoStop}%
%%CITATION = CHPHD,C38,090001;%%
\bibitem [{\citenamefont {Cohen}\ \emph {et~al.}(2014)\citenamefont {Cohen},
  \citenamefont {Llanes-Estrada}, \citenamefont {Pelaez},\ and\ \citenamefont
  {Ruiz~de Elvira}}]{Cohen:2014vta}%
  \BibitemOpen
  \bibfield  {author} {\bibinfo {author} {\bibfnamefont {T.}~\bibnamefont
  {Cohen}}, \bibinfo {author} {\bibfnamefont {F.~J.}\ \bibnamefont
  {Llanes-Estrada}}, \bibinfo {author} {\bibfnamefont {J.~R.}\ \bibnamefont
  {Pelaez}}, \ and\ \bibinfo {author} {\bibfnamefont {J.}~\bibnamefont {Ruiz~de
  Elvira}},\ }\href {\doibase 10.1103/PhysRevD.90.036003} {\bibfield  {journal}
  {\bibinfo  {journal} {Phys. Rev.}\ }\textbf {\bibinfo {volume} {D90}},\
  \bibinfo {pages} {036003} (\bibinfo {year} {2014})},\ \Eprint
  {http://arxiv.org/abs/1405.4831} {arXiv:1405.4831 [hep-ph]} \BibitemShut
  {NoStop}%
%%CITATION = ARXIV:1405.4831;%%
\bibitem [{\citenamefont {Liang}\ and\ \citenamefont
  {Oset}(2014{\natexlab{b}})}]{Liang:2014tia}%
  \BibitemOpen
  \bibfield  {author} {\bibinfo {author} {\bibfnamefont {W.~H.}\ \bibnamefont
  {Liang}}\ and\ \bibinfo {author} {\bibfnamefont {E.}~\bibnamefont {Oset}},\
  }\href {\doibase 10.1016/j.physletb.2014.08.030} {\bibfield  {journal}
  {\bibinfo  {journal} {Phys. Lett.}\ }\textbf {\bibinfo {volume} {B737}},\
  \bibinfo {pages} {70} (\bibinfo {year} {2014}{\natexlab{b}})},\ \Eprint
  {http://arxiv.org/abs/1406.7228} {arXiv:1406.7228 [hep-ph]} \BibitemShut
  {NoStop}%
%%CITATION = ARXIV:1406.7228;%%
\bibitem [{\citenamefont {Weinberg}(1963{\natexlab{a}})}]{weinbergdeuterium_a}%
  \BibitemOpen
  \bibfield  {author} {\bibinfo {author} {\bibfnamefont {S.}~\bibnamefont
  {Weinberg}},\ }\href {\doibase 10.1103/PhysRev.130.776} {\bibfield  {journal}
  {\bibinfo  {journal} {Phys. Rev.}\ }\textbf {\bibinfo {volume} {130}},\
  \bibinfo {pages} {776} (\bibinfo {year} {1963}{\natexlab{a}})}\BibitemShut
  {NoStop}%
%%CITATION = PHRVA,130,776;%%
\bibitem [{\citenamefont {Weinberg}(1963{\natexlab{b}})}]{weinbergdeuterium_b}%
  \BibitemOpen
  \bibfield  {author} {\bibinfo {author} {\bibfnamefont {S.}~\bibnamefont
  {Weinberg}},\ }\href {\doibase 10.1103/PhysRev.131.440} {\bibfield  {journal}
  {\bibinfo  {journal} {Phys. Rev.}\ }\textbf {\bibinfo {volume} {131}},\
  \bibinfo {pages} {440} (\bibinfo {year} {1963}{\natexlab{b}})}\BibitemShut
  {NoStop}%
%%CITATION = PHRVA,131,440;%%
\bibitem [{\citenamefont {Weinberg}(1965)}]{weinbergdeuterium_c}%
  \BibitemOpen
  \bibfield  {author} {\bibinfo {author} {\bibfnamefont {S.}~\bibnamefont
  {Weinberg}},\ }\href {\doibase 10.1103/PhysRev.137.B672} {\bibfield
  {journal} {\bibinfo  {journal} {Phys. Rev.}\ }\textbf {\bibinfo {volume}
  {137}},\ \bibinfo {pages} {B672} (\bibinfo {year} {1965})}\BibitemShut
  {NoStop}%
%%CITATION = PHRVA,137,B672;%%
\bibitem [{\citenamefont {Baru}\ \emph {et~al.}(2004)\citenamefont {Baru},
  \citenamefont {Haidenbauer}, \citenamefont {Hanhart}, \citenamefont
  {Kalashnikova},\ and\ \citenamefont {Kudryavtsev}}]{Baru:2003qq}%
  \BibitemOpen
  \bibfield  {author} {\bibinfo {author} {\bibfnamefont {V.}~\bibnamefont
  {Baru}}, \bibinfo {author} {\bibfnamefont {J.}~\bibnamefont {Haidenbauer}},
  \bibinfo {author} {\bibfnamefont {C.}~\bibnamefont {Hanhart}}, \bibinfo
  {author} {\bibfnamefont {{\relax Yu}.}~\bibnamefont {Kalashnikova}}, \ and\
  \bibinfo {author} {\bibfnamefont {A.~E.}\ \bibnamefont {Kudryavtsev}},\
  }\href {\doibase 10.1016/j.physletb.2004.01.088} {\bibfield  {journal}
  {\bibinfo  {journal} {Phys. Lett.}\ }\textbf {\bibinfo {volume} {B586}},\
  \bibinfo {pages} {53} (\bibinfo {year} {2004})},\ \Eprint
  {http://arxiv.org/abs/hep-ph/0308129} {arXiv:hep-ph/0308129 [hep-ph]}
  \BibitemShut {NoStop}%
%%CITATION = HEP-PH/0308129;%%
\bibitem [{\citenamefont {Daub}\ \emph {et~al.}(2016)\citenamefont {Daub},
  \citenamefont {Hanhart},\ and\ \citenamefont {Kubis}}]{Daub:2015xja}%
  \BibitemOpen
  \bibfield  {author} {\bibinfo {author} {\bibfnamefont {J.~T.}\ \bibnamefont
  {Daub}}, \bibinfo {author} {\bibfnamefont {C.}~\bibnamefont {Hanhart}}, \
  and\ \bibinfo {author} {\bibfnamefont {B.}~\bibnamefont {Kubis}},\ }\href
  {\doibase 10.1007/JHEP02(2016)009} {\bibfield  {journal} {\bibinfo  {journal}
  {JHEP}\ }\textbf {\bibinfo {volume} {02}},\ \bibinfo {pages} {009} (\bibinfo
  {year} {2016})},\ \Eprint {http://arxiv.org/abs/1508.06841} {arXiv:1508.06841
  [hep-ph]} \BibitemShut {NoStop}%
%%CITATION = ARXIV:1508.06841;%%
\bibitem [{\citenamefont {Close}\ and\ \citenamefont {Kirk}(2015)}]{Close}%
  \BibitemOpen
  \bibfield  {author} {\bibinfo {author} {\bibfnamefont {F.~E.}\ \bibnamefont
  {Close}}\ and\ \bibinfo {author} {\bibfnamefont {A.}~\bibnamefont {Kirk}},\
  }\href {\doibase 10.1103/PhysRevD.91.114015} {\bibfield  {journal} {\bibinfo
  {journal} {Phys. Rev.}\ }\textbf {\bibinfo {volume} {D91}},\ \bibinfo {pages}
  {114015} (\bibinfo {year} {2015})},\ \Eprint
  {http://arxiv.org/abs/1503.06942} {arXiv:1503.06942 [hep-ex]} \BibitemShut
  {NoStop}%
%%CITATION = ARXIV:1503.06942;%%
\bibitem [{\citenamefont {Close}\ and\ \citenamefont
  {Tornqvist}(2002)}]{Close:2002zu}%
  \BibitemOpen
  \bibfield  {author} {\bibinfo {author} {\bibfnamefont {F.~E.}\ \bibnamefont
  {Close}}\ and\ \bibinfo {author} {\bibfnamefont {N.~A.}\ \bibnamefont
  {Tornqvist}},\ }\href {\doibase 10.1088/0954-3899/28/10/201} {\bibfield
  {journal} {\bibinfo  {journal} {J. Phys.}\ }\textbf {\bibinfo {volume}
  {G28}},\ \bibinfo {pages} {R249} (\bibinfo {year} {2002})},\ \Eprint
  {http://arxiv.org/abs/hep-ph/0204205} {arXiv:hep-ph/0204205 [hep-ph]}
  \BibitemShut {NoStop}%
%%CITATION = HEP-PH/0204205;%%
\bibitem [{\citenamefont {Howarth}\ and\ \citenamefont
  {Giedt}(2015)}]{Howarth:2015caa}%
  \BibitemOpen
  \bibfield  {author} {\bibinfo {author} {\bibfnamefont {D.}~\bibnamefont
  {Howarth}}\ and\ \bibinfo {author} {\bibfnamefont {J.}~\bibnamefont
  {Giedt}},\ }\href@noop {} {\  (\bibinfo {year} {2015})},\ \Eprint
  {http://arxiv.org/abs/1508.05658} {arXiv:1508.05658 [hep-lat]} \BibitemShut
  {NoStop}%
%%CITATION = ARXIV:1508.05658;%%
\bibitem [{\citenamefont {Dedonder}\ \emph {et~al.}(2011)\citenamefont
  {Dedonder}, \citenamefont {Furman}, \citenamefont {Kaminski}, \citenamefont
  {Lesniak},\ and\ \citenamefont {Loiseau}}]{BenoitBppp}%
  \BibitemOpen
  \bibfield  {author} {\bibinfo {author} {\bibfnamefont {J.~P.}\ \bibnamefont
  {Dedonder}}, \bibinfo {author} {\bibfnamefont {A.}~\bibnamefont {Furman}},
  \bibinfo {author} {\bibfnamefont {R.}~\bibnamefont {Kaminski}}, \bibinfo
  {author} {\bibfnamefont {L.}~\bibnamefont {Lesniak}}, \ and\ \bibinfo
  {author} {\bibfnamefont {B.}~\bibnamefont {Loiseau}},\ }\href {\doibase
  10.5506/APhysPolB.42.2013} {\bibfield  {journal} {\bibinfo  {journal} {Acta
  Phys. Polon.}\ }\textbf {\bibinfo {volume} {B42}},\ \bibinfo {pages} {2013}
  (\bibinfo {year} {2011})},\ \Eprint {http://arxiv.org/abs/1011.0960}
  {arXiv:1011.0960 [hep-ph]} \BibitemShut {NoStop}%
%%CITATION = ARXIV:1011.0960;%%
\bibitem [{\citenamefont {Burdman}\ and\ \citenamefont {Donoghue}(1992)}]{BD}%
  \BibitemOpen
  \bibfield  {author} {\bibinfo {author} {\bibfnamefont {G.}~\bibnamefont
  {Burdman}}\ and\ \bibinfo {author} {\bibfnamefont {J.~F.}\ \bibnamefont
  {Donoghue}},\ }\href {\doibase 10.1016/0370-2693(92)90068-F} {\bibfield
  {journal} {\bibinfo  {journal} {Phys. Lett.}\ }\textbf {\bibinfo {volume}
  {B280}},\ \bibinfo {pages} {287} (\bibinfo {year} {1992})}\BibitemShut
  {NoStop}%
%%CITATION = PHLTA,B280,287;%%
\bibitem [{\citenamefont {Wise}(1992)}]{wise}%
  \BibitemOpen
  \bibfield  {author} {\bibinfo {author} {\bibfnamefont {M.~B.}\ \bibnamefont
  {Wise}},\ }\href {\doibase 10.1103/PhysRevD.45.R2188} {\bibfield  {journal}
  {\bibinfo  {journal} {Phys. Rev.}\ }\textbf {\bibinfo {volume} {D45}},\
  \bibinfo {pages} {2188} (\bibinfo {year} {1992})}\BibitemShut {NoStop}%
%%CITATION = PHRVA,D45,2188;%%
\bibitem [{\citenamefont {Gasser}\ and\ \citenamefont
  {Leutwyler}(1985)}]{GL_a}%
  \BibitemOpen
  \bibfield  {author} {\bibinfo {author} {\bibfnamefont {J.}~\bibnamefont
  {Gasser}}\ and\ \bibinfo {author} {\bibfnamefont {H.}~\bibnamefont
  {Leutwyler}},\ }\href {\doibase 10.1016/0550-3213(85)90492-4} {\bibfield
  {journal} {\bibinfo  {journal} {Nucl. Phys.}\ }\textbf {\bibinfo {volume}
  {B250}},\ \bibinfo {pages} {465} (\bibinfo {year} {1985})}\BibitemShut
  {NoStop}%
%%CITATION = NUPHA,B250,465;%%
\bibitem [{\citenamefont {Ecker}\ \emph {et~al.}(1989)\citenamefont {Ecker},
  \citenamefont {Gasser}, \citenamefont {Pich},\ and\ \citenamefont
  {de~Rafael}}]{EGPR}%
  \BibitemOpen
  \bibfield  {author} {\bibinfo {author} {\bibfnamefont {G.}~\bibnamefont
  {Ecker}}, \bibinfo {author} {\bibfnamefont {J.}~\bibnamefont {Gasser}},
  \bibinfo {author} {\bibfnamefont {A.}~\bibnamefont {Pich}}, \ and\ \bibinfo
  {author} {\bibfnamefont {E.}~\bibnamefont {de~Rafael}},\ }\href {\doibase
  10.1016/0550-3213(89)90346-5} {\bibfield  {journal} {\bibinfo  {journal}
  {Nucl. Phys.}\ }\textbf {\bibinfo {volume} {B321}},\ \bibinfo {pages} {311}
  (\bibinfo {year} {1989})}\BibitemShut {NoStop}%
%%CITATION = NUPHA,B321,311;%%
\bibitem [{\citenamefont {Bernard}\ \emph {et~al.}(1991)\citenamefont
  {Bernard}, \citenamefont {Kaiser},\ and\ \citenamefont {Meissner}}]{Bernard}%
  \BibitemOpen
  \bibfield  {author} {\bibinfo {author} {\bibfnamefont {V.}~\bibnamefont
  {Bernard}}, \bibinfo {author} {\bibfnamefont {N.}~\bibnamefont {Kaiser}}, \
  and\ \bibinfo {author} {\bibfnamefont {U.~G.}\ \bibnamefont {Meissner}},\
  }\href {\doibase 10.1016/0550-3213(91)90461-6} {\bibfield  {journal}
  {\bibinfo  {journal} {Nucl. Phys.}\ }\textbf {\bibinfo {volume} {B357}},\
  \bibinfo {pages} {129} (\bibinfo {year} {1991})}\BibitemShut {NoStop}%
%%CITATION = NUPHA,B357,129;%%
\bibitem [{\citenamefont {Colangelo}\ \emph {et~al.}(2001)\citenamefont
  {Colangelo}, \citenamefont {Gasser},\ and\ \citenamefont {Leutwyler}}]{CGL}%
  \BibitemOpen
  \bibfield  {author} {\bibinfo {author} {\bibfnamefont {G.}~\bibnamefont
  {Colangelo}}, \bibinfo {author} {\bibfnamefont {J.}~\bibnamefont {Gasser}}, \
  and\ \bibinfo {author} {\bibfnamefont {H.}~\bibnamefont {Leutwyler}},\ }\href
  {\doibase 10.1016/S0550-3213(01)00147-X} {\bibfield  {journal} {\bibinfo
  {journal} {Nucl. Phys.}\ }\textbf {\bibinfo {volume} {B603}},\ \bibinfo
  {pages} {125} (\bibinfo {year} {2001})},\ \Eprint
  {http://arxiv.org/abs/hep-ph/0103088} {arXiv:hep-ph/0103088 [hep-ph]}
  \BibitemShut {NoStop}%
%%CITATION = HEP-PH/0103088;%%
\bibitem [{\citenamefont {Kaminski}\ \emph {et~al.}(2006)\citenamefont
  {Kaminski}, \citenamefont {Pelaez},\ and\ \citenamefont {Yndurain}}]{Pelaez}%
  \BibitemOpen
  \bibfield  {author} {\bibinfo {author} {\bibfnamefont {R.}~\bibnamefont
  {Kaminski}}, \bibinfo {author} {\bibfnamefont {J.~R.}\ \bibnamefont
  {Pelaez}}, \ and\ \bibinfo {author} {\bibfnamefont {F.~J.}\ \bibnamefont
  {Yndurain}},\ }\href {\doibase 10.1103/PhysRevD.74.014001,
  10.1103/PhysRevD.74.079903} {\bibfield  {journal} {\bibinfo  {journal} {Phys.
  Rev.}\ }\textbf {\bibinfo {volume} {D74}},\ \bibinfo {pages} {014001}
  (\bibinfo {year} {2006})},\ \bibinfo {note} {[Erratum: Phys.
  Rev.D74,079903(2006)]},\ \Eprint {http://arxiv.org/abs/hep-ph/0603170}
  {arXiv:hep-ph/0603170 [hep-ph]} \BibitemShut {NoStop}%
%%CITATION = HEP-PH/0603170;%%
\bibitem [{\citenamefont {Buettiker}\ \emph {et~al.}(2004)\citenamefont
  {Buettiker}, \citenamefont {Descotes-Genon},\ and\ \citenamefont
  {Moussallam}}]{Moussalam}%
  \BibitemOpen
  \bibfield  {author} {\bibinfo {author} {\bibfnamefont {P.}~\bibnamefont
  {Buettiker}}, \bibinfo {author} {\bibfnamefont {S.}~\bibnamefont
  {Descotes-Genon}}, \ and\ \bibinfo {author} {\bibfnamefont {B.}~\bibnamefont
  {Moussallam}},\ }\href {\doibase 10.1140/epjc/s2004-01591-1} {\bibfield
  {journal} {\bibinfo  {journal} {Eur. Phys. J.}\ }\textbf {\bibinfo {volume}
  {C33}},\ \bibinfo {pages} {409} (\bibinfo {year} {2004})},\ \Eprint
  {http://arxiv.org/abs/hep-ph/0310283} {arXiv:hep-ph/0310283 [hep-ph]}
  \BibitemShut {NoStop}%
%%CITATION = HEP-PH/0310283;%%
\bibitem [{\citenamefont {Guimar\~aes}\ \emph {et~al.}(2010)\citenamefont
  {Guimar\~aes}, \citenamefont {Bediaga}, \citenamefont {Delfino},
  \citenamefont {Frederico}, \citenamefont {dos Reis},\ and\ \citenamefont
  {Tomio}}]{lc09}%
  \BibitemOpen
  \bibfield  {author} {\bibinfo {author} {\bibfnamefont {K.~S. F.~F.}\
  \bibnamefont {Guimar\~aes}}, \bibinfo {author} {\bibfnamefont
  {I.}~\bibnamefont {Bediaga}}, \bibinfo {author} {\bibfnamefont
  {A.}~\bibnamefont {Delfino}}, \bibinfo {author} {\bibfnamefont
  {T.}~\bibnamefont {Frederico}}, \bibinfo {author} {\bibfnamefont {A.~C.}\
  \bibnamefont {dos Reis}}, \ and\ \bibinfo {author} {\bibfnamefont
  {L.}~\bibnamefont {Tomio}},\ }\bibfield  {booktitle} {\emph {\bibinfo
  {booktitle} {{Proceedings, International Workshop on Relativistic hadronic
  and particle physics (Light Cone 2009)}}},\ }\href {\doibase
  10.1016/j.nuclphysbps.2010.02.056} {\bibfield  {journal} {\bibinfo  {journal}
  {Nucl. Phys. Proc. Suppl.}\ }\textbf {\bibinfo {volume} {199}},\ \bibinfo
  {pages} {341} (\bibinfo {year} {2010})}\BibitemShut {NoStop}%
%%CITATION = NUPHZ,199,341;%%
\bibitem [{\citenamefont {Boito}\ and\ \citenamefont
  {Robilotta}(2007)}]{Diogo}%
  \BibitemOpen
  \bibfield  {author} {\bibinfo {author} {\bibfnamefont {D.~R.}\ \bibnamefont
  {Boito}}\ and\ \bibinfo {author} {\bibfnamefont {M.~R.}\ \bibnamefont
  {Robilotta}},\ }\href {\doibase 10.1103/PhysRevD.76.094011} {\bibfield
  {journal} {\bibinfo  {journal} {Phys. Rev.}\ }\textbf {\bibinfo {volume}
  {D76}},\ \bibinfo {pages} {094011} (\bibinfo {year} {2007})},\ \Eprint
  {http://arxiv.org/abs/0705.3260} {arXiv:0705.3260 [hep-ph]} \BibitemShut
  {NoStop}%
%%CITATION = ARXIV:0705.3260;%%
\bibitem [{\citenamefont {Magalh\~aes}(2014)}]{PatThesis}%
  \BibitemOpen
  \bibfield  {author} {\bibinfo {author} {\bibfnamefont {P.~C.}\ \bibnamefont
  {Magalh\~aes}},\ }\href@noop {} {Ph.D. thesis},\ \bibinfo  {school}
  {University of S\~ao Paulo} (\bibinfo {year} {2014})\BibitemShut {NoStop}%
\bibitem [{\citenamefont {Breit}\ and\ \citenamefont {Wigner}(1936)}]{BW}%
  \BibitemOpen
  \bibfield  {author} {\bibinfo {author} {\bibfnamefont {G.}~\bibnamefont
  {Breit}}\ and\ \bibinfo {author} {\bibfnamefont {E.}~\bibnamefont {Wigner}},\
  }\href {\doibase 10.1103/PhysRev.49.519} {\bibfield  {journal} {\bibinfo
  {journal} {Phys. Rev.}\ }\textbf {\bibinfo {volume} {49}},\ \bibinfo {pages}
  {519} (\bibinfo {year} {1936})}\BibitemShut {NoStop}%
%%CITATION = PHRVA,49,519;%%
\bibitem [{\citenamefont {Bjorken}\ and\ \citenamefont
  {Drell}(1964)}]{BjorkenDrell}%
  \BibitemOpen
  \bibfield  {author} {\bibinfo {author} {\bibfnamefont {J.~D.}\ \bibnamefont
  {Bjorken}}\ and\ \bibinfo {author} {\bibfnamefont {S.~D.}\ \bibnamefont
  {Drell}},\ }\href@noop {} {\emph {\bibinfo {title} {{Relativistic Quantum
  Mechanics}}}}\ (\bibinfo  {publisher} {McGraw-Hill},\ \bibinfo {address} {New
  York},\ \bibinfo {year} {1964})\BibitemShut {NoStop}%
\bibitem [{\citenamefont {Scherer}(2003)}]{Scherer}%
  \BibitemOpen
  \bibfield  {author} {\bibinfo {author} {\bibfnamefont {S.}~\bibnamefont
  {Scherer}},\ }\href@noop {} {\bibfield  {journal} {\bibinfo  {journal} {Adv.
  Nucl. Phys.}\ }\textbf {\bibinfo {volume} {27}},\ \bibinfo {pages} {277}
  (\bibinfo {year} {2003})},\ \Eprint {http://arxiv.org/abs/hep-ph/0210398}
  {arXiv:hep-ph/0210398 [hep-ph]} \BibitemShut {NoStop}%
%%CITATION = HEP-PH/0210398;%%
\bibitem [{\citenamefont {Brown}(1994)}]{LSB}%
  \BibitemOpen
  \bibfield  {author} {\bibinfo {author} {\bibfnamefont {L.~S.}\ \bibnamefont
  {Brown}},\ }\href@noop {} {\emph {\bibinfo {title} {{Quantum field
  theory}}}}\ (\bibinfo  {publisher} {Cambridge University Press},\ \bibinfo
  {year} {1994})\BibitemShut {NoStop}%
%%CITATION = INSPIRE-340704;%%
\bibitem [{\citenamefont {Gounaris}\ and\ \citenamefont {Sakurai}(1968)}]{GS}%
  \BibitemOpen
  \bibfield  {author} {\bibinfo {author} {\bibfnamefont {G.~J.}\ \bibnamefont
  {Gounaris}}\ and\ \bibinfo {author} {\bibfnamefont {J.~J.}\ \bibnamefont
  {Sakurai}},\ }\href {\doibase 10.1103/PhysRevLett.21.244} {\bibfield
  {journal} {\bibinfo  {journal} {Phys. Rev. Lett.}\ }\textbf {\bibinfo
  {volume} {21}},\ \bibinfo {pages} {244} (\bibinfo {year} {1968})}\BibitemShut
  {NoStop}%
%%CITATION = PRLTA,21,244;%%
\bibitem [{\citenamefont {Oller}\ and\ \citenamefont {Oset}(1999)}]{OO}%
  \BibitemOpen
  \bibfield  {author} {\bibinfo {author} {\bibfnamefont {J.~A.}\ \bibnamefont
  {Oller}}\ and\ \bibinfo {author} {\bibfnamefont {E.}~\bibnamefont {Oset}},\
  }\bibfield  {booktitle} {\emph {\bibinfo {booktitle} {{Hadron spectroscopy.
  Proceedings, Workshop, Frascati, Italy, March 8-12, 1999}}},\ }\href
  {\doibase 10.1103/PhysRevD.60.074023} {\bibfield  {journal} {\bibinfo
  {journal} {Phys. Rev.}\ }\textbf {\bibinfo {volume} {D60}},\ \bibinfo {pages}
  {074023} (\bibinfo {year} {1999})},\ \bibinfo {note} {(see
  also~\cite{npa})},\ \Eprint {http://arxiv.org/abs/hep-ph/9809337}
  {arXiv:hep-ph/9809337 [hep-ph]} \BibitemShut {NoStop}%
%%CITATION = HEP-PH/9809337;%%
\end{thebibliography}%

\end{document}